# Fatigue Behavior of High-Entropy Alloys

Shiyi Chen[1], Xuesong Fan[1], Hugh Shortt[1], Baldur Steingrimsson[2,3,4*], Weidong Li[1], and Peter K Liaw[1]

[1] Department of Materials Science and Engineering, The University of Tennessee, Knoxville, TN 37996, USA
[2] Imagars LLC, Hillsboro, OR 97124, USA
[3] School of Mechanical, Industrial, and Manufacturing Engineering, Oregon State University, 204 Rogers Hall, Corvallis OR 97331, USA
[4] Department of Manufacturing, Mechanical Engineering and Technology, Oregon Institute of Technology, Wilsonville, OR 97070, USA

## Abstract

High-entropy alloys (HEAs) refer to alloys composed of five or more elements in equal or near-equal amounts or in an atomic concentration range of 5 to 35 atomic percent (at%). Different elemental ratios will affect the microstructures of HEAs and provide them with unique properties. Based on past research, HEAs have exhibited superior performance, relative to most conventional alloys, with respect to many properties, such as strength, toughness, corrosion resistance, magnetic behavior, etc. Among them, fatigue behavior has been a topic of focus, due to its importance in industrial applications. In this article, we summarized the research progress in the HEA-fatigue behavior in the past ten years, including experimental results and theoretical studies in subdivisions, such as high-cycle fatigue, low-cycle fatigue, fatigue-crack growth, fatigue mechanisms, etc. The influence of the processing and test methods on HEAs is described. The accuracy of several commonly used prediction models is also outlined. Finally, unresolved issues and suggestions on the direction of future research efforts are presented.

[*] Corresponding author: baldur.steingrimsson@oregonstate.edu, baldur.steingrimsson@oit.edu, baldur@imagars.com.

## ABBREVIATIONS

| | |
|---|---|
| 3PB | Three-point bending |
| 4PB | Four-point bending |
| AC | Alternating current |
| AM | Arc-melting |
| ANN | Artificial neural network |
| BCC | Body-centered-cubic |
| BM | Ball milling |
| BSE | Back scattered electron |
| CDR | Cyclic-deformation response |
| CG | Coarse grained |
| CMOD | Crack mouth opening displacement |
| CPFEM | Crystal-plasticity finite-element method |
| CR | Cold rolling |
| CRSS | Critical resolved shear stress |
| CSR | Cyclic-stress response |
| CT | Compact tension |
| CRT | Cryogenic temperature |
| DC | Direct current |
| DC(T) | Disc-shaped compact-tension |
| DFT | Density functional theory |
| DSA | Dynamic strain ageing |
| DT | Deformation twinning |
| EA | Electromagnetic resonance axial |
| EBSD | Electron back-scattering diffraction |
| ECAP | Equal channel angular processing |
| ECC | Electron channeling contrast |
| EHEA | Eutectic high-entropy alloy |
| EVPSC | Elastic-visco-plastic self-consistent |
| FCC | Face-centered-cubic |
| FCGR | Fatigue-crack-growth rate |
| FCP | Fatigue-crack propagation |
| FEM | Finite-element modeling |
| FG | Fine grained |
| FIB | Focused ion beam |
| FL | Fatigue limit |
| GSFE | Generalized stacking fault energy |
| HAADF-STEM | High-angle annular dark-field scanning-transmission electron microscopy |
| HCF | High-cycle fatigue |
| HCP | Hexagonal-close-packed |
| HE | Hot extruded |
| HEA | High-entropy alloy |
| HEM | Homogeneous effective medium |
| HIP | Hot iso-static pressing |
| HT | High-temperature |
| IPF | Inverse pole figure |
| KAM | Kernel average misorientation |
| LAGB | Low-angle grain boundary |
| LBW | Laser-beam welding |
| LCF | Low-cycle fatigue |
| LEFM | Linear-elastic fracture mechanics |
| ML | Machine learning |
| PDF | Probability-density function |
| PSB | Persistent slip band |
| RT | Room temperature |
| SB | Slip-band |
| SCH | Secondary cyclic-hardening |
| SEB | Single-edge notch bending |
| SENB | Single-edge notch bending |
| SENT | Single-edge notch tension |
| SEM | Scanning-electron microscopy |
| SF | Stacking fault |
| SFE | Stacking-fault energy |
| SLM | Selective laser melting |
| SIF | Stress intensity factor |
| SMT | Surface mechanical treatment |
| SPD | Severe plastic deformation |
| SPS | Spark-plasma sintering |
| SSD | Striation spacing decrease |
| TB | Twin-boundary |
| TEM | Transmission-electron microscopy |
| TMF | Thermo-mechanical fatigue |
| TRIP | Transformation-induced plasticity |
| TWIP | Twinning-induced plasticity |
| UA | Ultrasonic axial |
| UFG | Ultrafine grained |
| UTS | Ultimate tensile strength |
| VAM | Vacuum-arc melting |
| VASP | Vienna ab initio simulation package |
| VHCF | Very high-cycle fatigue |
| VIM | Vacuum-induction melting |
| WR | Warm-rolled |
| XRD | X-ray diffraction |
| YS | Yield strength |

# 1. Introduction

## 1.1. What are high-entropy alloys (HEAs)?

The field of high-entropy alloys (HEAs) is relatively "young". The HEA concept was first introduced in the 1990s, and the first related paper was published in 2004 by Yeh, et al. [1]. Moreover, multi-principal element alloys are also studied by Cantor, et al. [2]. The earliest definition of HEAs is based on the elemental composition, which regards any alloy composed of five or more main elements as a HEA [1]. Initially, the composition of HEA was strictly limited to equimolar ratios. But in subsequent studies, the limitation of the "equimolar ratio" has been diluted [3]. Now the full composition-based definition is "the alloys composed of five or more elements with the concentrations between 5 - 35 atomic percent (at%)" [1]. This definition is easier to identify and measure.

The other common definition of HEA is the entropy-based definition [1, 3-6]. The phrase 'entropy' here refers to the total configurational molar entropy. The total configurational molar entropy of an ideal solution can be estimated, using the Boltzmann Eq. (1.1.1):

$$S^{SS,ideal} = -R_0 \sum_i x_i(lnx_i) \tag{1.1.1}$$

Here, $x$ represents the fraction of the composition, and $R_0$ the gas constant. Using this definition, an alloy with $S^{SS,ideal} < 0.5\ R_0$ is defined as a low-entropy alloy, an alloy with $0.5\ R_0 < S^{SS,ideal} < 1.5\ R_0$ is considered as a medium-entropy alloy, and an alloy with $S^{SS,ideal} > 1.5\ R_0$ is defined as a high-entropy alloy. However, the metal solid solution is usually not ideal. Because the entropy of the alloy changes with temperature, the total configurational molar entropy of many alloys is difficult to calculate accurately. Therefore, in most studies, the composition-based definition has higher applicability than the entropy-based definition [3].

Compared with conventional alloys, HEAs are characterized by their unique structures. In most cases, the atoms of different elements occupy the lattice position randomly, which promotes the formation of disordered solid-solution structures with lattice distortion in HEAs, as shown in Figure 1.1.1. Common solid-solution HEAs are generally face-centered-cubic (FCC) and/or body-centered-cubic (BCC) HEAs. A small amount of rare-earth element-based HEAs exhibit a single-phase hexagonal-close-packing (HCP) structure. Structural differences endow these HEAs with unique mechanical and functional properties, including fatigue properties [3].

### *1.1.1. FCC HEAs*

Typically, there are two structures present in HEAs with FCC microstructures. When the interaction between atoms is strong enough, certain positions tend to be occupied by the same element, so that the FCC HEA forms a superlattice FCC structure, such as the $L1_2$ structure.

### *1.1.2. BCC HEAs*

Figure 1.1.1 shows the structural difference between conventional BCC alloys and BCC HEAs. Random occupation of atoms and lattice distortion are prominent features in HEAs. Similar to the FCC HEA, BCC solid solutions can also be divided into ordered and disordered structures according to the arrangement of atoms [7].

### *1.1.3. Multiphase HEAs*

HEAs with multiphase structures, including two or more disordered solid-solution phases, ordered phases, or intermetallic compounds phases, can be defined as multiphase HEAs. When the interaction between elements is very complex, multiphase HEAs are formed [7]. In fatigue-related studies, the more common multiphase HEAs are dual-phase HEAs with both FCC and BCC phases, and some HEAs with three phases.

### *1.1.4. Metastable HEAs*

A "metastable" system is defined as a system that remains stable under the interference of an energy less than a certain threshold value, but transitions to another state under the interference of energy beyond a critical value. In this paper, the definition of metastable HEAs is that the phase transition occurs when disturbed by the external energy (mostly the FCC → HCP phase transition).

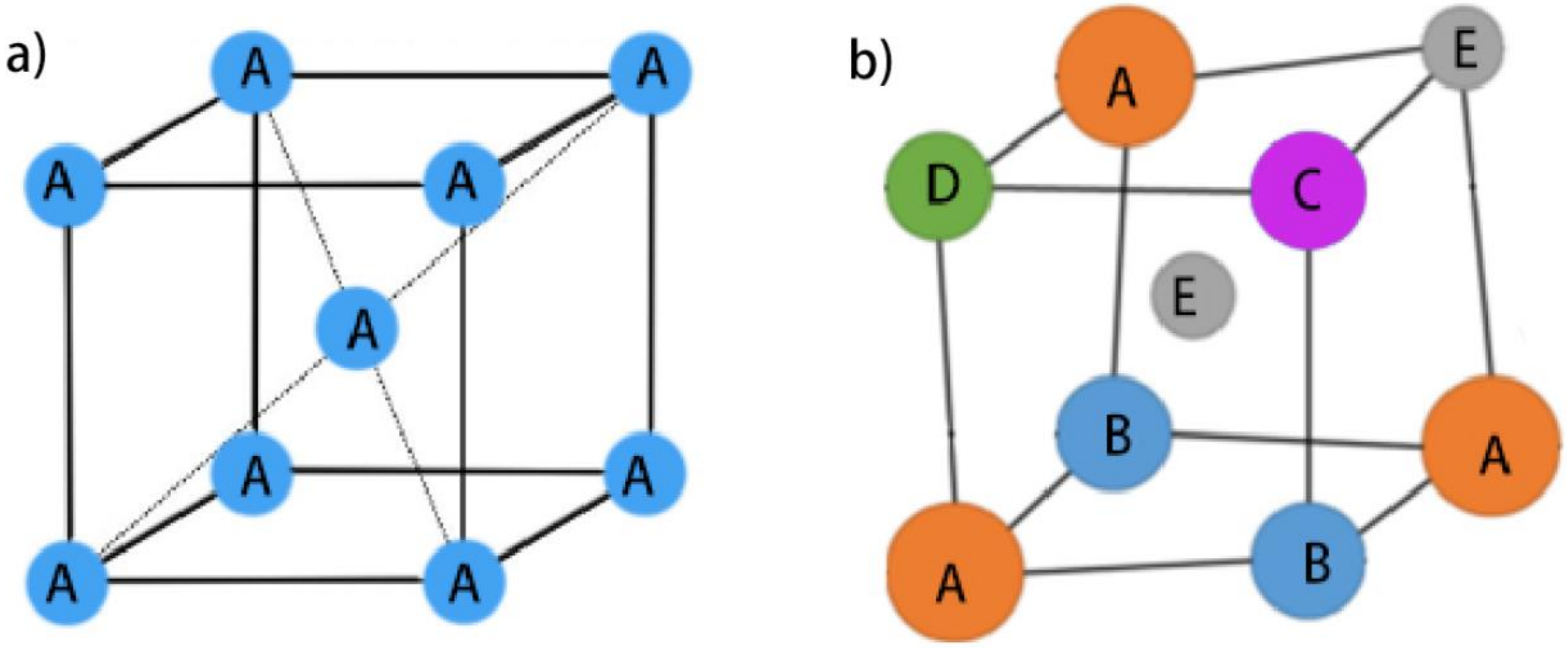

**Figure 1.1.1:** Schematic diagram of (a) typical single-component BCC crystal lattice and (b) HEA with a BCC microstructure [3].

## 1.2.What is fatigue?

HEAs possess many desirable mechanical properties and are therefore highly valued in materials science and engineering. Among their many outstanding mechanical properties, fatigue behavior entails a subject worthy of an in-depth study [8]. As early as the industrial revolution in the 19th century, material fatigue was an important cause of many large-scale industrial accidents (e.g., railway, bridge, and axle fracture). Therefore, research on fatigue behavior began to attract people's attention [9]. To this day, fatigue behavior remains an important criterion in the design of engineering structures. Fatigue refers to the entire process of an object breaking and failing under cyclic loading [10]. The fatigue process certainly includes microstructural changes, e.g., formation of dislocation cells, slip-band formation, etc. As shown in Figure 1.2.1, fatigue can be subdivided into thermal, corrosion, fretting, and mechanical fatigue, according to different factors influencing the behavior observed [11]. When mentioned, fatigue usually refers to mechanical fatigue, that is, to the process of material fracture and failure under cyclic loading. Mechanical failure of a piece of material may be the result of a complex interaction of factors, such as the applied load, the duration of the load being applied, and the environment. It is very important to study the failure mechanisms as well as the underlying principles.

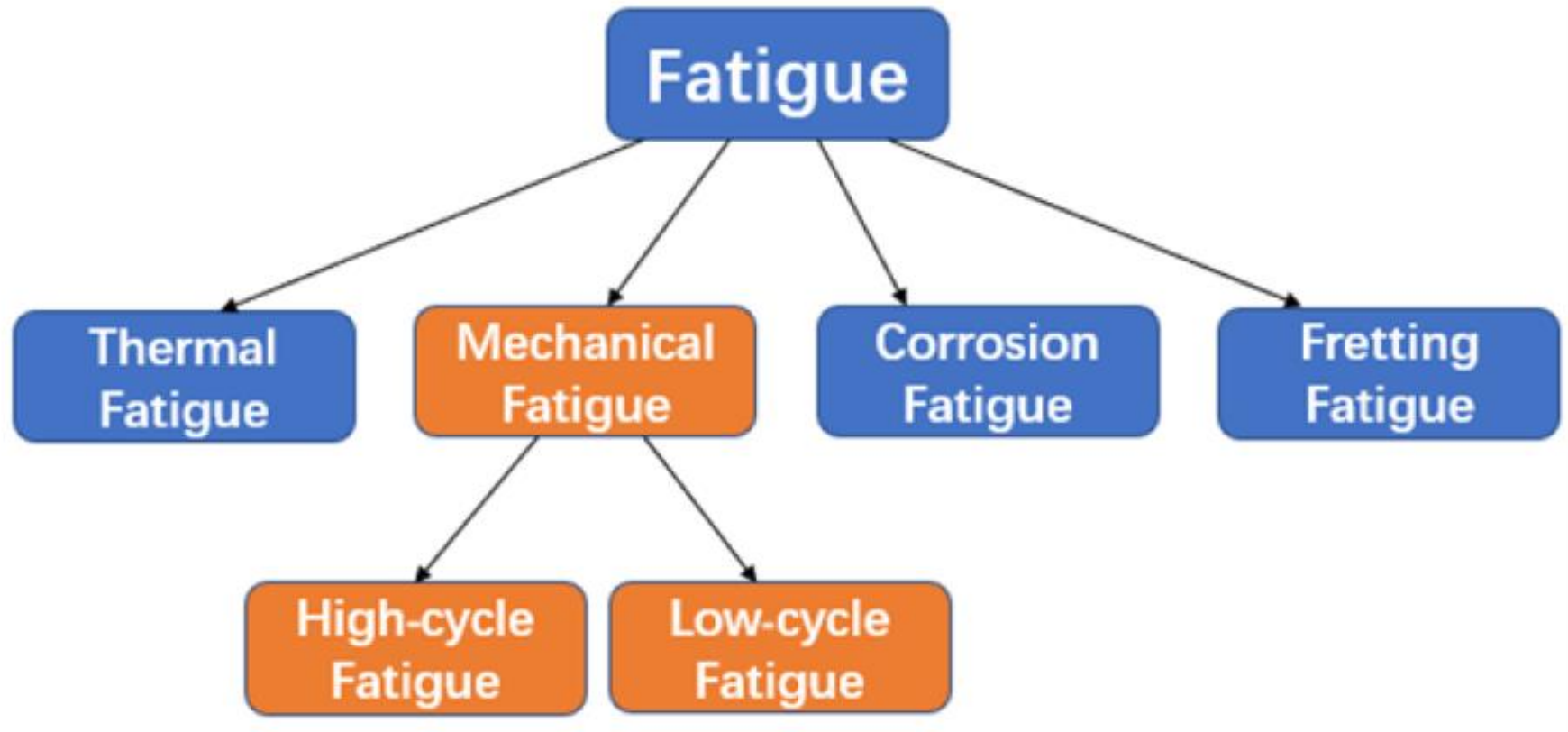

**Figure 1.2.1:** Classification of different fatigue behavior.

The earliest and most commonly used the fatigue-behavior test method was proposed by Auguste in the 1860s [9]. Here, a cyclic stress is applied to the material under study with a certain constant stress ratio ($R = \frac{\sigma_{min}}{\sigma_{max}}$, where $\sigma_{min}$ and $\sigma_{max}$ refer to the minimum and maximum stresses applied under cyclic loading, respectively) or strain ratio ($R = \frac{\varepsilon_{min}}{\varepsilon_{max}}$, where $\varepsilon_{min}$ and $\varepsilon_{max}$ refer to the minimum and maximum strains applied under cyclic loading, respectively), and the relationship is recorded between the stress/strain amplitude ($\sigma_a, \varepsilon_a$) and the number of cycles until failure ($N_f$). According to this criterion, mechanical fatigue can be divided into high-cycle fatigue (HCF) and low-cycle fatigue (LCF) regimes, as presented in Figure 1.2.2. Here, the HCF regime includes the very high-cycle fatigue (VHCF) regime, i.e., fatigue failures occurring after more than $10^7$ load cycles [12]. In HCF, the amplitude of cyclic loading is stress, and the stress amplitude is often lower than the yield strength of the material studied. Under HCF, the material can usually withstand stress cycles within the range of 10,000 to 100,000. The LCF fatigue refers to the fatigue failure generally caused by large cyclic plastic strains; the fatigue life is usually less than 10,000 cycles. Consistent with the different experimental purposes, there are three commonly used theoretical modeling methods employed, namely stress-life modeling of high-cycle fatigue [13], strain-life modeling of low-cycle fatigue (LCF) [13], and modeling of fatigue-crack-growth [14].

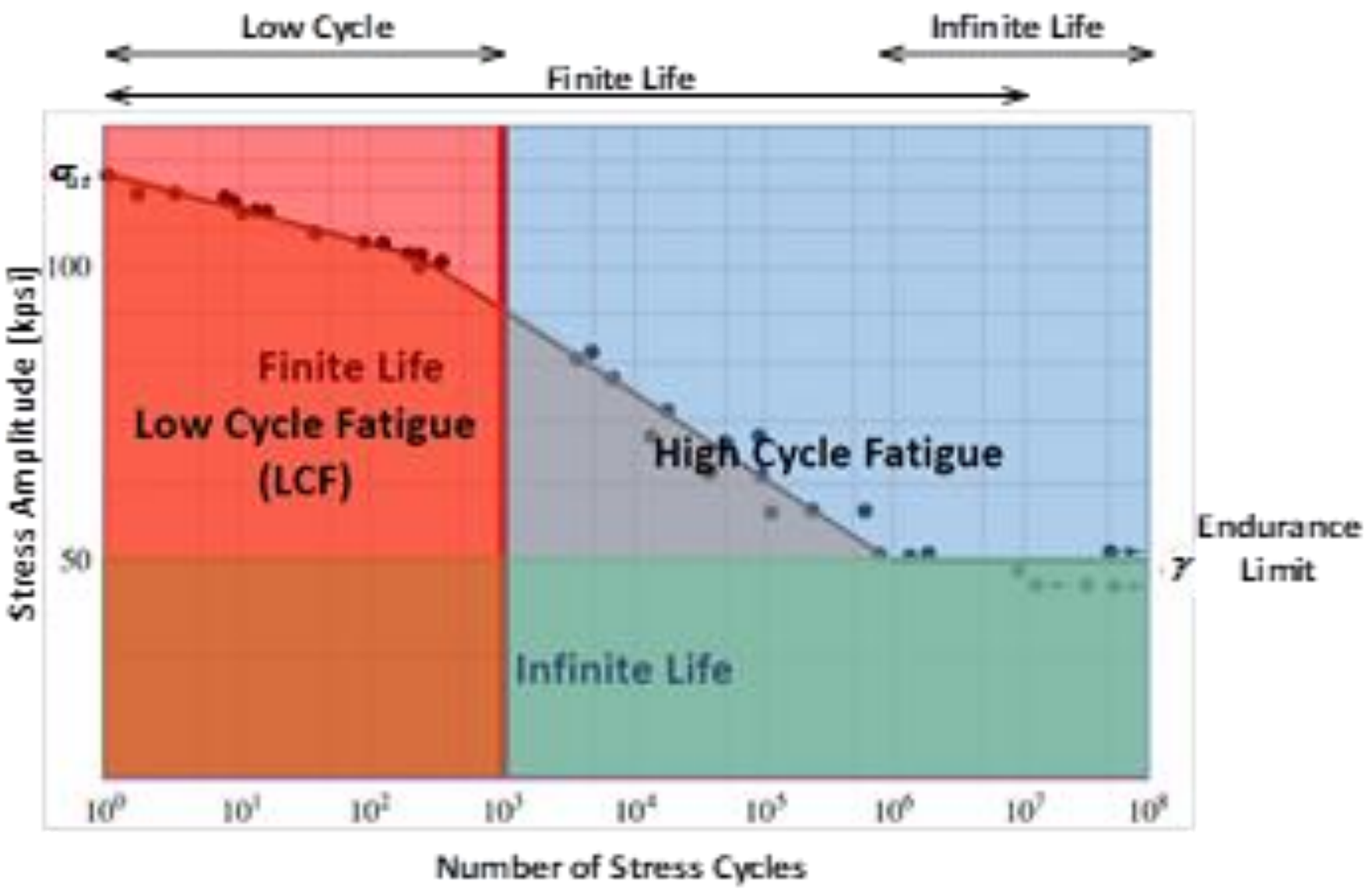

**Figure 1.2.2:** Definition of key-operation regimes for their stress/life analysis (adapted from [13]). Here, the HCF regime includes the very high-cycle fatigue (VHCF) regime [12].

## 1.3.Towards a roadmap: Summary of fatigue properties for HEAs

In an effort to offer a clear, overall picture, and to facilitate overall storytelling, Table 1.3.1 presents a high-level overview (summary) of fatigue properties and mechanisms in HEAs observed. Individual aspects of Table 1.3.1 are covered in Sections 2 - 4 below. Suffice to say that, per Supplementary Figure S8, fatigue degradation involves a multiscale process, with complex interplays at the microstructural level between grains, between different types of dislocations, and different types of dislocation cell structures. Section 2 also covers some of the essential background on stress-controlled HCF tests, strain-controlled LCF and on testing of fatigue crack growth.

**Table 1.3.1:** High-level overview (summary) of fatigue properties and mechanisms in HEAs observed. The intent of the Table is to offer a roadmap through the paper, provide a clear high-level picture of fatigue mechanisms in HEAs, and to facilitate understanding. Majority of high-cycle fatigue experiments was carried out in air and at ambient temperature.

| **Category** | **HEA Type** | | | | |
|---|---|---|---|---|---|
| | **FCC HEAs** | **BCC HEAs** | **Multiphase HEAs** | **Metastable HEAs** | **HEAs with Hierarchical Microstructure** |
| High-Cycle Fatigue | * Fatigue strength significantly enhanced by reducing the grain size [15]<br>* Dislocation-blocking feature by hard B2 precipitates in FCC phase [16] | * Fatigue strength of ~ 430 MPa and a fatigue ratio of 0.43 were exhibited in HfNbTaTiZr, presumably due to limited cyclic strain hardening behavior caused by limited multi-plication between impeded dislocations [17].<br>* Observed intergranular fatigue crack initiation mechanism, observed at the specimen surface, also contributed to superior fatigue strength [18]. | * Grain refinement in wrought $AlCoCrFeNi_{2.1}$ caused by a cold-rolling process greatly increases strength of this HEA [16].<br>* Heat-treatment process results in the B2 precipitation and Cr precipitation in the microstructure and these precipitation structures become dislocation barriers to a certain extent [16]. | * TRIP effect can retard fatigue-crack propagation at high stress amplitudes, because it can alter mean free path for crack propagation due to localized variation in work hardening. [19]<br>* Due to synergetic contributions from grain refinement and TRIP, $Fe_{38.5}Mn_{20}Co_{20}Cr_{15}Cu_{1.5}Si_{5}$ metastable and UFG HEA shows exceptional fatigue resistance with a fatigue strength of 700 MPa and a fatigue ratio of 0.62 [20]. | * Extraordinary HCF resistance of the AM CoCrFeMnNi HEA is primarily attributed to unique heterogeneous microstructure in the AM-built samples, including heterogeneous grain structures, dislocation networks, in-situ formed oxides, and fatigue induced deform-ation twinning [21].<br>* Microstructural hierarchy can be engineered to improve endurance limit of eutectic HEAs, e.g., due to different crack-initiation behavior between different microstructural morphologies [16].<br>* Can be combined with grain refinement [22] |
| Low-Cycle Fatigue | * Refining grains is more efficient for improving LCF resistance at low strain amplitudes due to the enhanced fatigue crack-initiation resistance by grain refinement but does not work well at high strain amplitudes because of accelerated fatigue-crack propagation along high-density grain boundaries [19]. | * There is very limited LCF data available for BCC HEAs, presumably due to high rate of crack growth.<br>* Compared to FCC structure, a BCC lattice is not a non-close-packed structure.<br>* Hence, the dislocations of the BCC phase do not strictly slip along the close-packed plane, which makes the dislocation | * A multi-component B2 precipitates-strengthened $Al_{0.5}CoCrFeNi$ HEA exhibits outstanding LCF resistance at low plastic strain amplitudes ($< 10^{-3}$) [25].<br>* Since strains between the FCC and B2 phases are not compatible, dislocations form first in the FCC matrix and accumulate at the boundary of the BCC phase during cycling. | * History-independent cyclic response and enhanced LCF life can be attained in TRIP HEAs by incorporating slip reversibility and reversible martensitic transformation [19].<br>* Localized cyclic-plastic strain and stress near deformation twins in CoCrNi MEA at RT contribute to the activation of TWIP and TRIP plasticity | * Similar to the HCF case, microstructures containing hierarchical features can be engineered to bring out favorable LCF resistance [19].<br>* Due to ductile-transformable nature of multi-component B2 precipitates, fatigue-crack-initiation resistance, and hence LCF life at low strain amplitudes, of $Al_{0.5}CoCrFeNi$ HEA with hierarchically distributed multi-component B2 precip- |

| | * Reduction in grain size of CoCrFeMnNi (from 65 to 5 μm) can slightly increase fatigue life at RT (from 0.3% strain amplitude to 0.4%) through promotion of a transition from slip-band cracking to twin-boundary cracking because of the reduced slip band spacing that increases impingement sites on twin boundaries [23].<br>* Superior fatigue life found in ultra-fine grained (<1 μm) CoCrFeMnNi sample prepared by ECAP at the low strain amplitude of 0.2%, while a better fatigue life was observed in a coarse-grained (~12 μm) CoCrFeMnNi sample at a higher strain amplitude of 0.6% [24]. | recovery of the BCC HEA easier during the stretching process, and it is not easy to form dislocation accumulation like in FCC HEA.<br>* On the other hand, the BCC structure has more slip systems (48, compared to 12 for FCC and 3 for HCP).<br>* This makes fatigue cracks in the BCC HEA almost always start in the form of intergranular cracking and expand in the form of trans-granular cracking [16].<br>* Therefore, grain boundaries are far less obstructive to fatigue in BCC HEA than in FCC HEA. | * Deformation only starts to occur in the BCC phase, as the cycle period and strain amplitude increase.<br>In the dual-phase * HEA consisting of FCC + BCC phases, cracks tend to nucleate from the interface of two phases.<br>* At the plastic strain amplitude of ~0.03%,the $Al_{0.5}$CoCrFeNi HEA shows at least four times longer life than conventional alloys [25].<br>* This shows that the HEA-design concept provides a feasible way to design fatigue-resistant structural materials [25]. | even at low stress amplitudes during LCF [26].<br>* Although distinct martensitic transformation occurred during push-pull loading in dual-phase TRIP $Fe_{50}Mn_{30}Co_{10}Cr_{10}$ HEA, cyclic hardening was hardly observed at given strain amplitudes, which is substantially different from the observed remarkable strain hardening, during monotonic tension [27].<br>* In the case of carbon-doped TRIP $Fe_{50}Mn_{30}Co_{10}Cr_{10}C_{0.4}$ HEA with different grain sizes, mediocre hardening occurred during LCF, unlike prolonged hardening observed during monotonic deformation, due to slip planarity and partial reversibility of deformation [28]. | itates incorporated into FCC matrix has been greatly improved [25].<br>* Ductile-transformable B2 precipitates can act as efficient crack arresters to inhibit crack coalescence and propagation during cyclic loading, whereas in conventional precipitation-strengthened alloys fatigue cracks easily initiate near boundaries of precipitates at low strain amplitudes [19, 25].<br>* Owing to enhanced strain hardening, LCF resistance was improved for CoCrFeMnNi HEA into which harmonic structures with a shell fraction between 20% and 40% had been introduced [29].<br>* Here, the local plastic deformation dominantly occurred in the coarse-grained core areas, whereas ultra-fine-grained shell areas exhibited good cyclic stability [29]. |
|---|---|---|---|---|---|
| Crack Growth | * FCC HEAs exhibit favorably comparable fatigue-crack-growth resistance to austenitic stainless steels [19].<br>* Fatigue-crack growth resistance and threshold for CoCrFeMnNi compares favorably with conventional structural alloys [30]<br>* Extremely high fatigue threshold values above 20 MPa $m^{1/2}$ were found in as-cast Al-containing CrFeNi HEA [31]. | * The BCC-type HfNbTaTiZr HEA shows a lower fatigue threshold level than other materials, which is mainly due to absence of dislocation induced crack-tip shielding effect [17].<br>* Unlike intergranular crack propagation behavior in FCC materials, only transgranular crack propagation was found in HfNbTaTiZr HEA [17]. | * As-cast multi-phase Al-CrFeNi-based HEAs (Al $CrFeNi_2Cu$ and $Al_{0.2}CrFeNiTi_{0.2}$) outperform crack-growth resistance of other materials [31].<br>* Very high fatigue thresholds with $\Delta K_{th}$ values above 20 MPa $m^{1/2}$ rouse doubts, and could be caused by very coarse as-cast structures, small size of the specimens tested, and excessive plasticity throughout their cross-section [30, 32]. | * Fatigue crack-growth results for TRIP HEAs are very limited to date,<br>* But limited results suggest comparable fatigue-crack-growth resistance as for TWIP HEAs.<br>* Intense HCP-martensite formed didn't show a negative effect on crack-growth resistance, in case of metastable dual-phase $Fe_{50}Mn_{30}Co_{10}Cr_{10}$ [33].<br>* Here, TRIP effect can be triggered due to low SFE [19]. | * Hierarchical microstructures show great potential in enhancing crack initiation resistance of HEAs [19].<br>* This insight is consistent with observations from infusing bone-like crack resistance into hierarchical metastable nanolaminate steels [34]. |

## 1.4.Current status

In this section, we use the Web of Science search engine as a tool and use high-entropy alloy, multi-principal element alloy, concentrated solid solution alloy, multi-component alloy, and complex concentrated alloy as key words to count the research trends related to HEAs from 2004 to October 2024, that is, the number of publications. After obtaining the results of the initial search, we performed a secondary screening of all publications using the definitions in Section 1.1 to ensure statistical accuracy, namely "alloys composed of five or more nearly equimolar elements" or “alloys with configuration entropy greater than 1.5 $R_0$” [1, 3-6].

Figure 1.4.1 (a) shows the statistical results of HEA-related studies in the past 20 years. Here, the two indexes that we used are the number of publications and the total number of citations of all publications up to the current year. As can be seen from Figure 1.4.1(a), the related research on HEAs shows an exponential growth trend. The number of publications in the last year has exceeded 3,000, and the cumulative citations of all publications have exceeded 600,000. These results show how hot the HEA field has become.

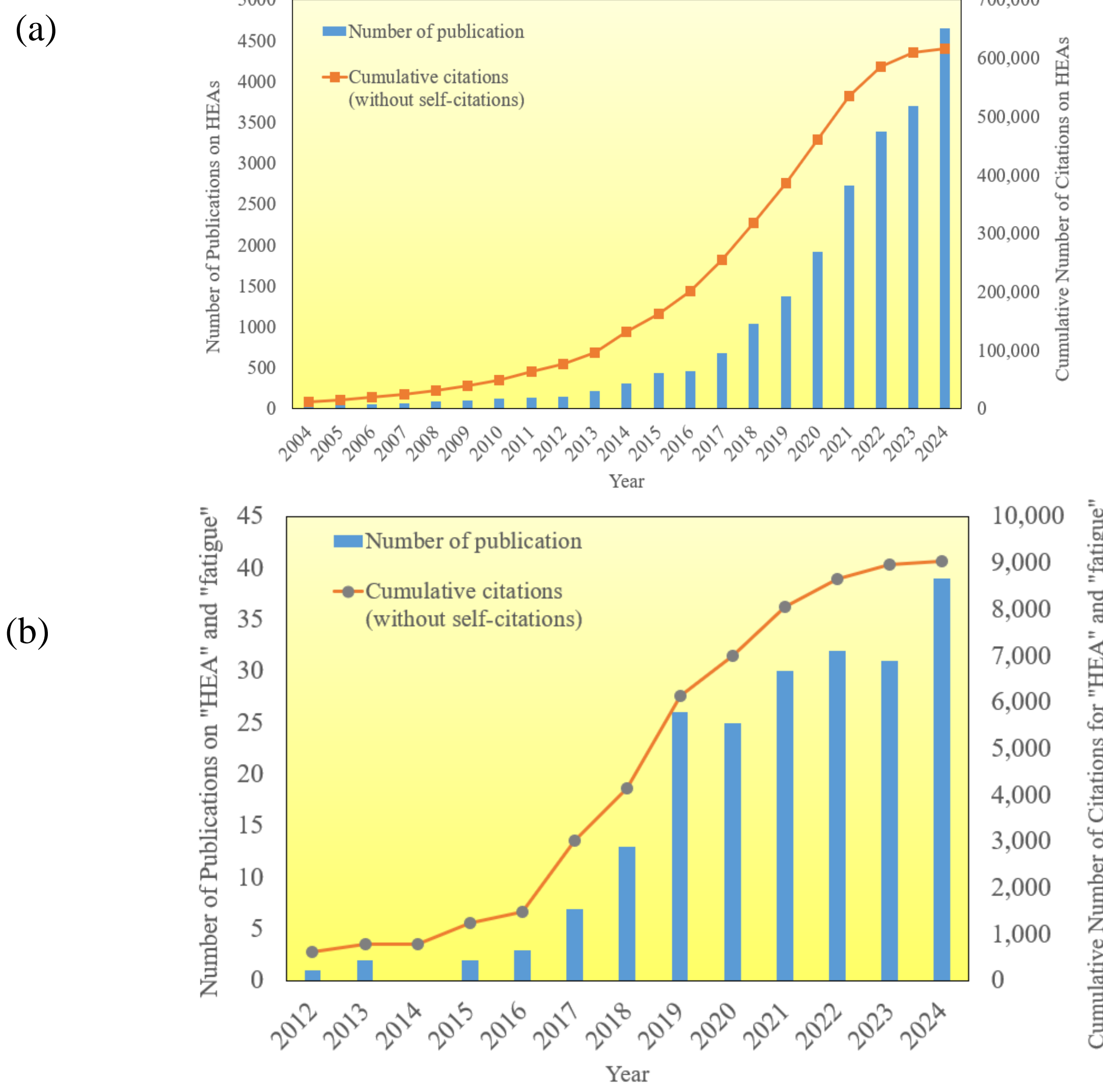

**Figure 1.4.1:** Number of HEA publications and total citations of (a) high-entropy alloys and (b) fatigue behavior of high-entropy alloys by year.

After adding fatigue as a keyword, on the basis of the above five keywords, we counted all the fatigue studies on high-entropy alloys.

Figure 1.4.1(b) presents studies related to the fatigue behavior of HEAs, based on the number of publications and cumulative citations prepared in the same way as Figure 1.4.1(a). Here, all preliminary search results have also been screened twice. At present, there are 78 papers specifically addressing the fatigue behavior of HEAs, and such papers have maintained a growing trend in recent years. It can be seen that fatigue behavior of HEAs has gradually developed into a field, which has begun to take shape, since the first paper came out in 2012 [35]. In addition, we divided the 78 papers specifically addressing the fatigue behavior of HEAs into the following, four categories: HCF, LCF, FCGR, and others, based on the modeling methods used by the authors. The last category, others, includes the mechanism research and theoretical review. The proportion of publications, for these four categories, is exhibited in Figure 1.4.2. At present, most research still focuses on HCF.

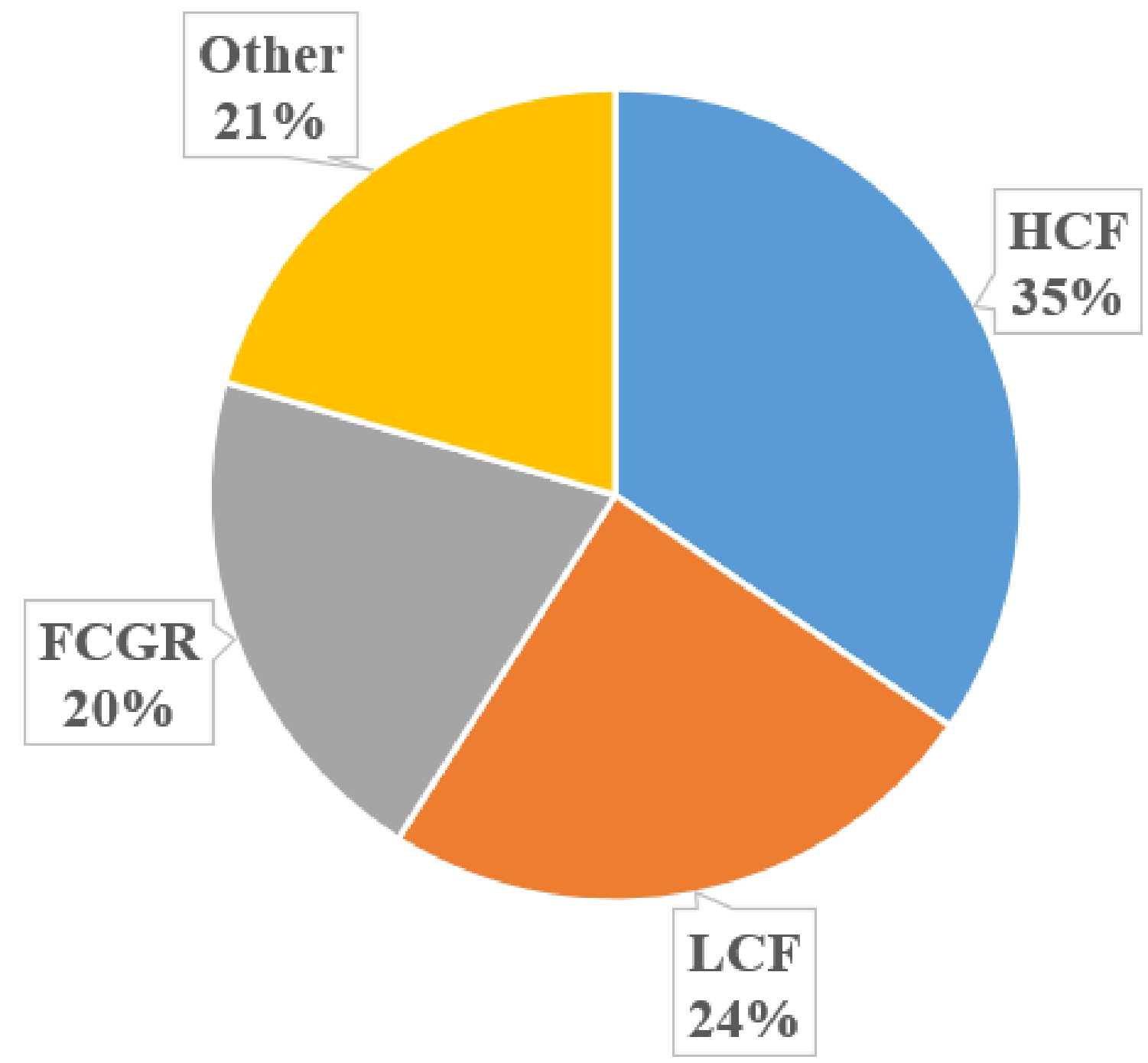

**Figure 1.4.2:** The percentage of publications for the four types of HEA fatigue studies considered.

# 2. Current Research Results

## 2.1.High-cycle fatigue of HEAs

### *2.1.1. Introduction*

The stress-life approach assumes the application of cyclic stress loads to smooth test specimens, such as shown in Figure 2.1.1. The stress is usually applied under an axial stress, with a constant stress amplitude ($\sigma_a$) and constant stress (load) ratio $R = \frac{\sigma_{min}}{\sigma_{max}}$ (often set at 0.1 or -1). Fatigue with the life ($N_f$) in the interval of $10^4$ - $10^7$ was defined as high-cycle fatigue, the stress amplitude corresponding to $10^7$ cycles is generally regarded as the fatigue limit of the material. In the HCF, where stress is low and predominantly elastic, fatigue data of materials are normally presented by the stress-life method, namely, the data are plotted as stress amplitude ($S$), $\sigma_a$, as a function of number of cycles or reversals to failure ($N$), $N_\mathrm{f}$ or $2N_\mathrm{f}$. The stress-life method is often referred to as the $S-N$ method and the plot as the $S-N$ plot. The $S-N$ curve of high-cycle fatigue is normally empirically modeled with the Basquin equation [36]:

$$\frac{\Delta\sigma}{2} = \sigma_\mathrm{a} = \sigma_\mathrm{f}'(2N_\mathrm{f})^b, \qquad (2.1.1)$$

where $\Delta\sigma = \sigma_\mathrm{max} - \sigma_\mathrm{min}$ is the stress range, with $\sigma_\mathrm{max}$ and $\sigma_\mathrm{min}$ being the maximum and minimum stresses, respectively, $\sigma_\mathrm{a}$ represents the stress amplitude equal to the half of $\Delta\sigma$, $2N_\mathrm{f}$ denotes the number of load reversals to failure, $\sigma_\mathrm{f}'$ is the fatigue-strength coefficient, and $b$ is the fatigue-strength exponent. In Eq. (2.1.1), $\sigma_\mathrm{f}'$ and $b$ are the unknown quantities that need to be determined in order to make the model useful. $\sigma_\mathrm{f}'$ and $b$ can be quantified relatively easily by first transforming Eq. (2.1.1) into a linear equation, by applying logarithm to both sides, and by then performing a linear fit on the measured $\frac{\Delta\sigma}{2}$ versus $2N_\mathrm{f}$ values. Such process is demonstrated in Figure 2.1.2 for the CoCrFeMnNi HEA with a grain size of $d = 0.65$ μm tested in air at room temperature (RT) [15].

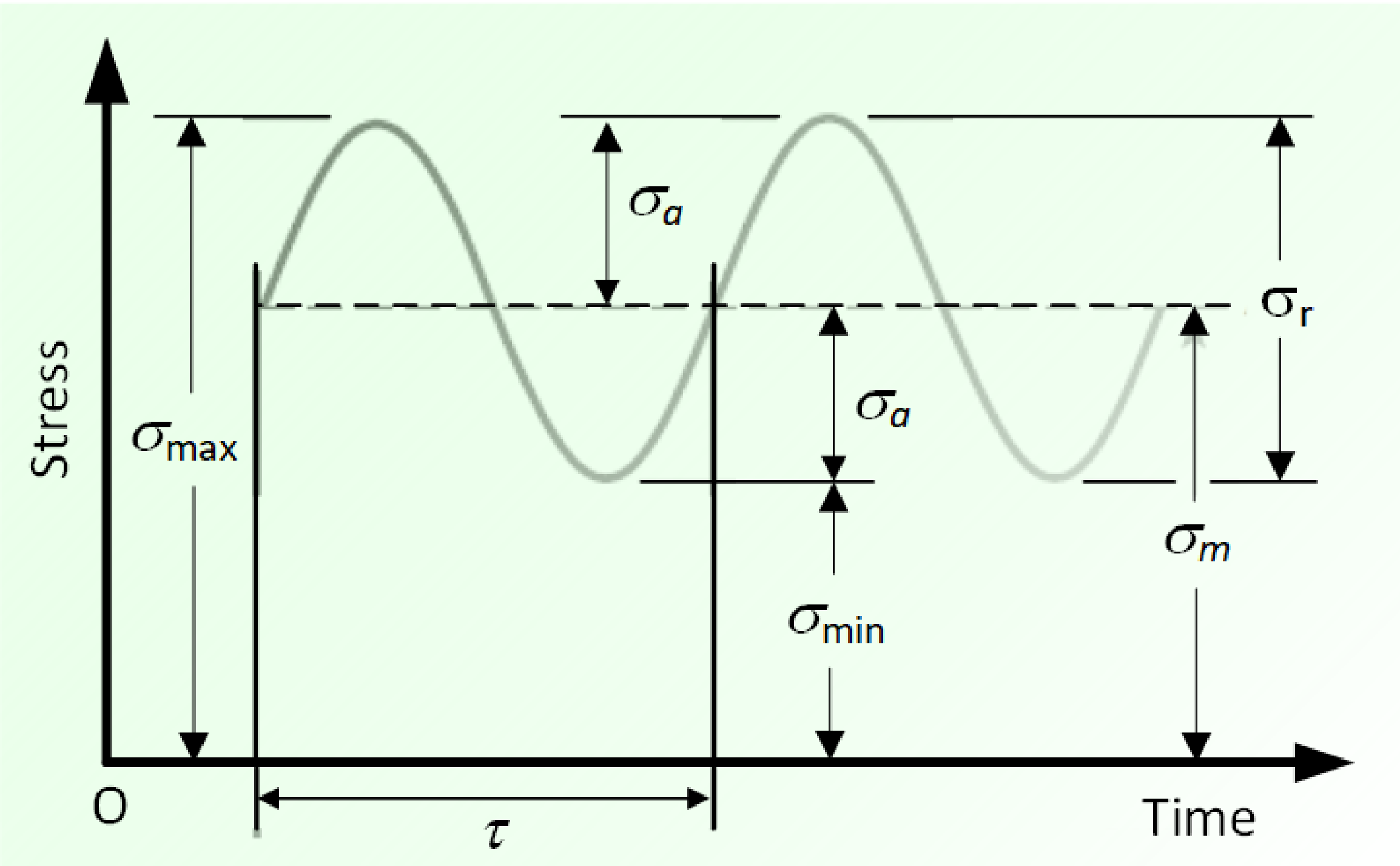

**Figure 2.1.1:** Key definitions related to the stress applied, in the case of constant amplitude stress control (adapted from [9]).

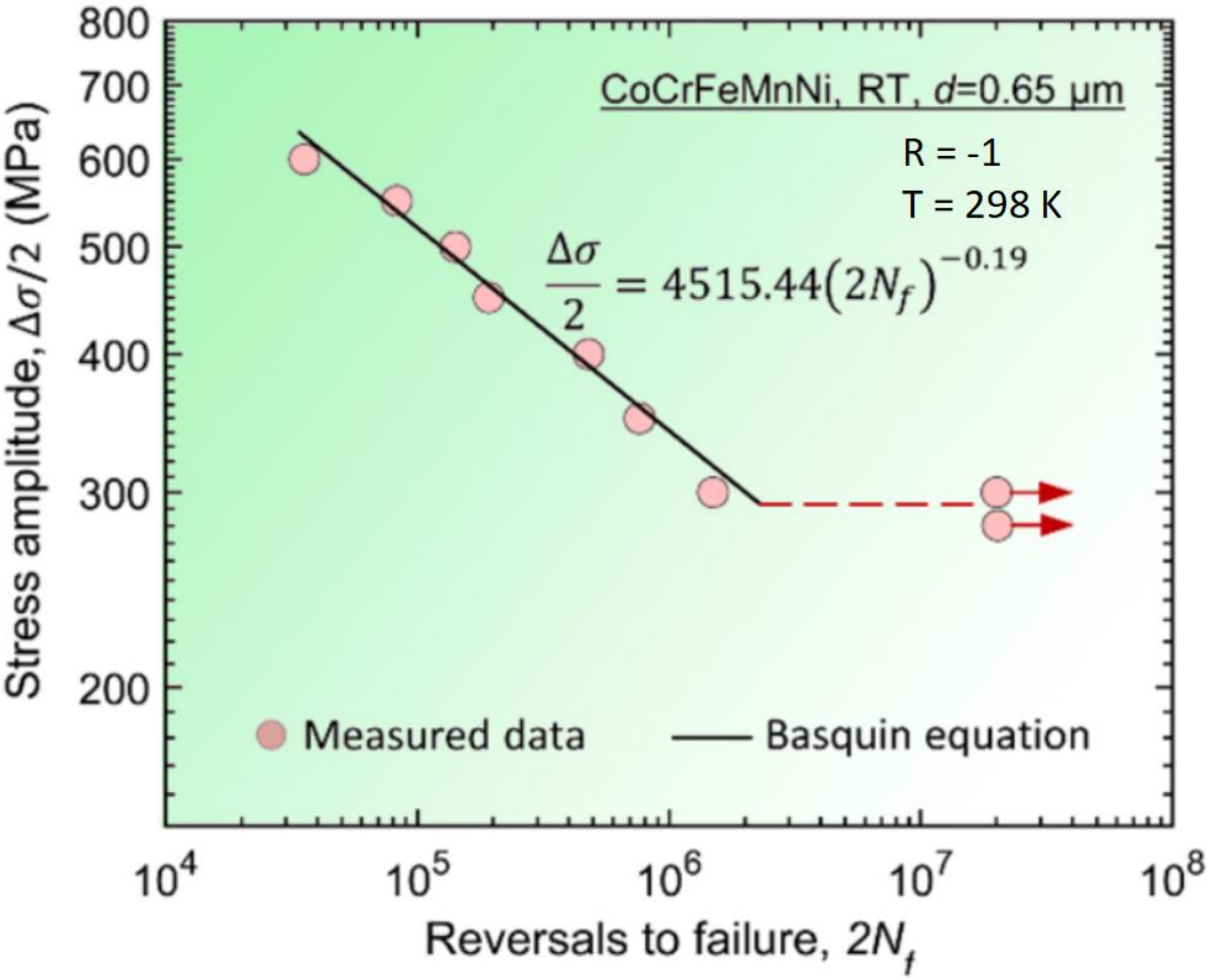

**Figure 2.1.2:** The stress amplitude versus the number of reversals to failure along with the Basquin fit (the S-N curve) of the CoCrFeMnNi HEA with a grain size of $d$ = 0.65 μm tested in air at room temperature and at $R$ = -1 [15].

Only the data under the same stress ratio can be compared. Therefore, for data obtained at different stress ratios, the Smith-Watson-Topper equation can be used to convert the data to the same stress ratio [37]:

$$\sigma_{a1} = \sigma_{a2}\sqrt{\frac{1 - R_1}{1 - R_2}} \tag{2.1.2}$$

Here, $\sigma_{a1}$ and $\sigma_{a2}$ represent the stress amplitudes under the stress ratios, $R_1$ and $R_2$, respectively.

### *2.1.2. Comparison between HEAs*

In HEA's HCF behavior, 18 studies were counted herein, including eight (8) on a FCC HEA (CoCrFeMnNi) [15, 21, 38-43], one (1) on a BCC HEA (TiHfZrNbTa) [18, 44, 52], five (5) on multiphase HEAs ( $Al_xCoCrFeNi_x$ ) [16, 22, 45], and two (2) on metastable HEAs ($Fe_{38.5}Mn_{20}Co_{20}Cr_{15}Si_5Cu_{1.5}$ and $Fe_{42}Mn_{28}Co_{10}Cr_{15}Si_5$) [20, 46]. All studies were carried out at room temperature and standard air pressure. More detailed information including the phase, average grain size, and ultimate tensile strength are reported in Table 2.1.1. All data were converted to the form of $R$ = - 1 using Eq. (2.1.2) and plotted in Figure 2.1.3 in the form of S-N and fatigue ratio ($\frac{\sigma_a}{\sigma_{UTS}}$) vs. fatigue life. Different types of HEAs are represented by different symbols.

Under the same stress amplitude, as shown in Figure 2.1.3(a), the fatigue life of the BCC HEA is the longest among all 4 HEA types; The fatigue resistance of the metastable HEA is also very good, only slightly less than that of the BCC one; the multiphase HEA of the $Al_xCoCrFeNi_x$ family is concentrated in the middle of the figure; the fatigue resistance of most FCC HEAs is relatively low, only a small number of specimens processed by a pre-strain treatment, spark-plasma sintering (SPS) process, or selective laser melting (SLM) process have extremely high fatigue performance, which are defined as specially processed FCC HEA in Figure 2.1.3. When using the fatigue ratio as the y-axis in Figure 2.1.3(b), four types of HEAs still show the same order of fatigue performance, which is BCC > Metastable > Multiphase > FCC. The three FCC HEAs with good performance are from the research of Lee et al. [38], Kim et al. [21], and Chlup et al. [42], respectively. Since the research of Chlup et al. did not mention the ultimate tensile strength of the HEA, the data from this group are not included in Figure 2.1.3b.

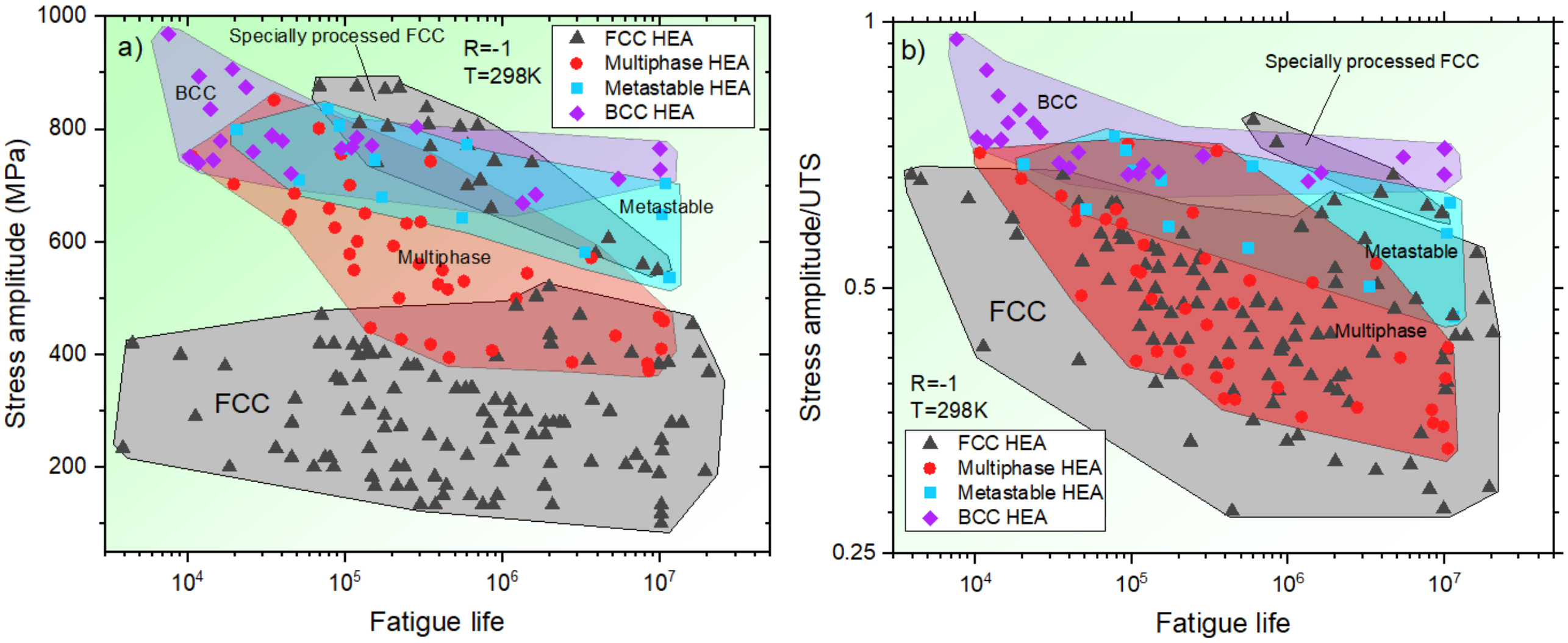

**Figure 2.1.3:** An overview of all (a) S-N data and (b) fatigue ratio vs. fatigue life data for FCC (black triangle), BCC (purple diamond), metastable (blue square), and multiphase (red dots) HEAs [21, 38, 42].

**Table 2.1.1:** Mechanical behavior of high-entropy alloys at room temperature, including $\sigma_{UTS}$, and fatigue limits at two stress ratios ($R$ = -1 and $R$ = 0.1).

*1FL=Fatigue limit, *2UTS=Ultimate tensile strength, *3VIM= Vacuum-induction melting, *4AM=Arc-melting, *5SLM=Selective laser melting, *6BM=Ball milling, *7SPS=Spark plasma sintering, *8LBW=Laser-beam welding, *9VAM=Vacuum arc-melting

| Micro-structure | Material | Manufacturing Method | Loading Method | Phase | Average Grain Size (μm) | UTS | Frequency (Hz) | Endurance Limit | | Fatigue Ratio (FL*1/UTS*2) | | Ref |
|---|---|---|---|---|---|---|---|---|---|---|---|---|
| | | | | | | | | $R$=-1 | $R$=0.1 | $R$=-1 | $R$=0.1 | |
| FCC | CoCrFeMnNi | VIM*3 + Pre strained | Rotary bending test | FCC | 14.3 ± 8.6 | 903.9 | 50 | 550 | 369 | 0.608 | 0.408 | [38] |
| | CoCrFeMnNi | AM*4 | Ultrasonic uniaxial test | FCC | 45 | 650 | 20,000 | 230 | 154 | 0.353 | 0.237 | [39] |
| | CoCrFeMnNi | VIM | Uniaxial test | FCC + $CrMn_2O_4$ particles | 245 | 626 | 20 | 210 | 141 | 0.335 | 0.225 | [40] |
| | CoCrFeMnNi | SLM*5 | Uniaxial test | FCC +$Mn_2O_3$ particles | 10 | 923 | 20 | 383 | 257 | 0.415 | 0.278 | [21] |
| | CoCrFeMnNi | VIM | Uniaxial test | FCC | 30 | 676 | 30 | 190 | 127 | 0.281 | 0.188 | [15] |
| | CoCrFeMnNi | VIM | Rotary bending test | FCC | 77 | 585 | 30 | 249 | 167 | 0.391 | 0.286 | [41] |
| | CoCrFeMnNi | BM*6 + SPS | 3PB test | FCC | 0.41 | N/A | N/A | 771 | 517 | N/A | N/A | [42] |
| | CoCrFeMnNi | BM + SPS*7 | 3PB test | FCC | 0.63 | N/A | N/A | 740 | 496 | N/A | N/A | [42] |
| | CoCrFeMnNi | LBW*8 | Uniaxial test | FCC | 100 | 205 | 130 | 134 | 90 | 0.653 | 0.44 | [43] |
| | CoCrFeNi | L-PBF | Uniaxial test | FCC | 30.57 | 601 | 20 | 101 | 67.68 | 0.167 | 0.112 | [47] |
| | $Al_{0.3}CoCrFeNi$ | VLM | 4PB test | FCC | 29.6 | 830 | 40 | 369 | 248 | 0.444 | 0.298 | [48] |
| BCC | HfNbTaTiZr | AM | 4PB test | BCC | 45 | 1,013 | 10 | 680 | 456 | 0.720 | 0.43 | [18] |
| | HfNbTaTiZr | AM | 4PB test | BCC | 45 | 1,139 | 10 | 766 | 512 | 0.673 | 0.451 | [44, 52] |
| Meta- | $Fe_{42}Mn_{28}Cr_{15}Co_{10}Si_5$ | VIM | Fully reversible | metastable | 2 | 1,158 | 20 | 537 | 360 | 0.464 | 0.311 | [46] |
| stable | | | bending test | FCC → HCP | | | | | | | | |
| | $Fe_{38.5}Mn_{20}Co_{20}Cr_{15}Si_5Cu_{1.5}$ | VIM | Fully reversible bending test | metastable FCC → HCP | 0.77 ± 0.35 | 1,126 | 20 | 649 | 435 | 0.352 | 0.386 | [20] |
| Multi-phase | $AlCoCrFeNi_{2.1}$ (As-cast) | Casting | Fully reversible bending test | L12 +B2 | | 1,057 | 20 | 372 | 249 | 0.576 | 0.236 | [16] |
| | $AlCoCrFeNi_{2.1}$ (Wrought) | Casting | Fully reversible bending test | L12 +B2 | | 1,340 | 20 | 469 | 315 | 0.35 | 0.235 | [16] |
| | $Al_{0.3}CoCrFeNi$ | VAM[*9] | Fully reversible bending test | FCC+B2 +σ | 0.71 ± 0.35 | 1,074 | 20 | 460 | 308 | 0.428 | 0.287 | [22] |
| | $Al_{0.5}CoCrCuFeNi$ (commercial-purity raw materials; 1st ingot) | VAM | 4PB test | FCC + FCC | | 1,344 | 10 | 460 | 309 | 0.342 | 0.230 | [35, 49] |
| | $Al_{0.5}CoCrCuFeNi$ (commercial-purity raw materials; 2nd ingot) | VAM | 4PB test | FCC + FCC | | 1,344 | 10 | 543 | 365 | 0.404 | 0.271 | [49] |
| | $Al_{0.5}CoCrCuFeNi$ (high-purity raw materials) | VAM | 4PB test | FCC + FCC | | 1,344 | 10 | 530 | 356 | 0.395 | 0.265 | [49] |
| | $Al_{0.7}CoCrFeNi$ | VAM | Fully reversible bending test | FCC+B2 | | 1,040 | 20 | 411 | 275 | 0.395 | 0.265 | [45] |
| | $Al_{0.7}CoCrFeNi$ | VAM | Fully reversible bending test | FCC+B2 +L12 | | 1,400 | 20 | 460 | 309 | 0.329 | 0.221 | [45] |

### *2.1.3. FCC HEAs*

Figure 2.1.4 presents a comprehensive comparison of all FCC HEAs with different grain sizes and second-phase particles labeled on the figure. Since all the materials studied are CoCrFeMnNi with an equal atomic ratio, this comparison is quite convincing. In these eight (8) studies, the factors affecting fatigue behavior can be classified into three types, namely 1) grain size, 2) inclusions, and 3) other microstructural aspects. In most cases, these three factors affect each other greatly. Hence, FCC HEAs with higher fatigue limit may have two or more characteristics at the same time.

<u>Grain size</u>

In previous alloy studies [50], the grain-size reduction was thought to improve the mechanical properties of the alloy. When focusing on the FCC HEAs, we can find that the fatigue behavior of HEAs also increases with the decrease of grain size. Figure 2.1.5(a) shows the relationship between the HCF limit and grain size of CoCrFeMnNi with an FCC structure at RT and with standard air pressure. All materials basically follow the law that the finer the grain size, the higher the fatigue limit. For HEAs with coarser grains (> 30 μm), the fatigue limit is in the range of about 100 - 250 MPa; the fatigue limit of the sample with a grain size close to 10 μm is also slightly higher than that of the coarse-grained sample. For specimens with very fine grains (> 1 μm), the fatigue limit is even as high as 700 - 800 MPa [42]. In Figure 2.1.5(b), the grain-size dependence of the fatigue ratios of different CoCrFeMnNi was not as significant as that of fatigue strength shown in Figure 2.1.5(a), a small number of CoCrFeMnNi with finer grains also has lower fatigue ratios. In general, under the same *R*, as the grain size increases, the fatigue stress and fatigue ratio that FCC HEA can withstand are reduced, but under the fatigue ratio standard, the data is more discrete.

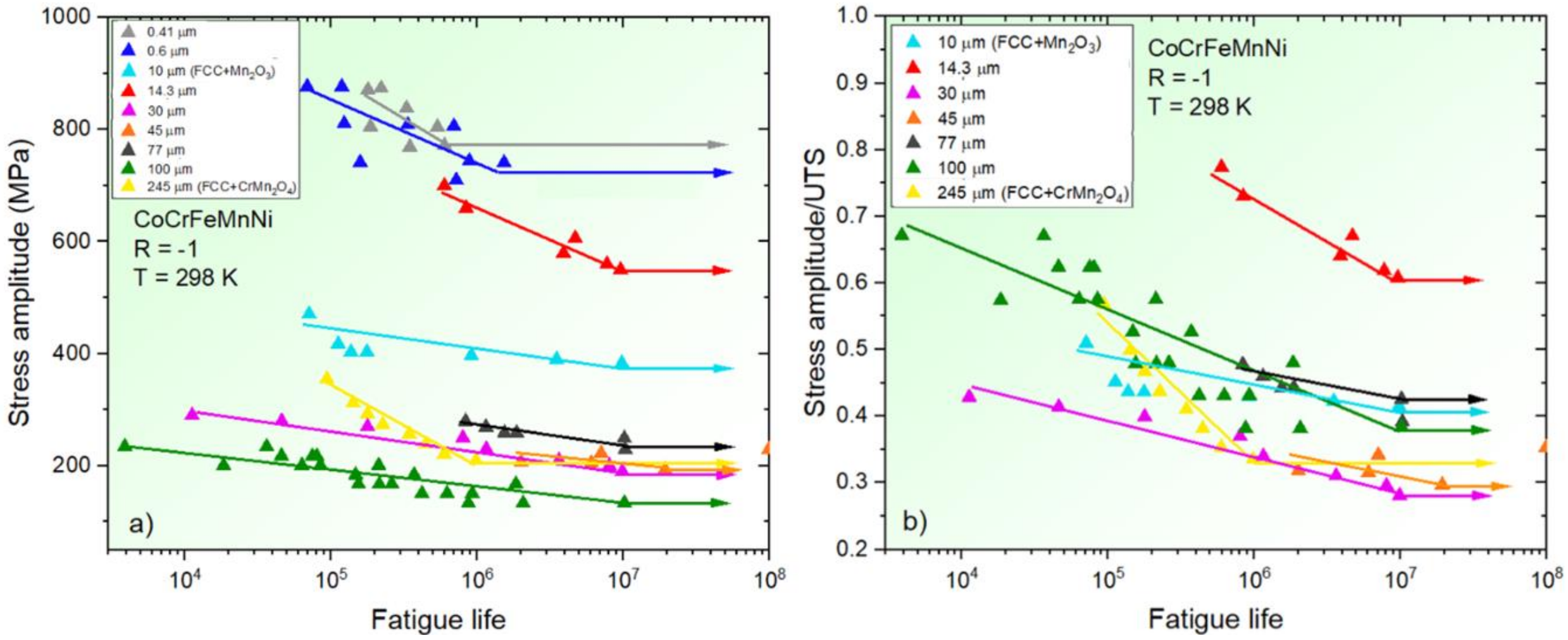

**Figure 2.1.4:** HCF performance of FCC HEAs under (a) S-N mode and (b) fatigue ratio vs. fatigue life [15, 21, 38-43].

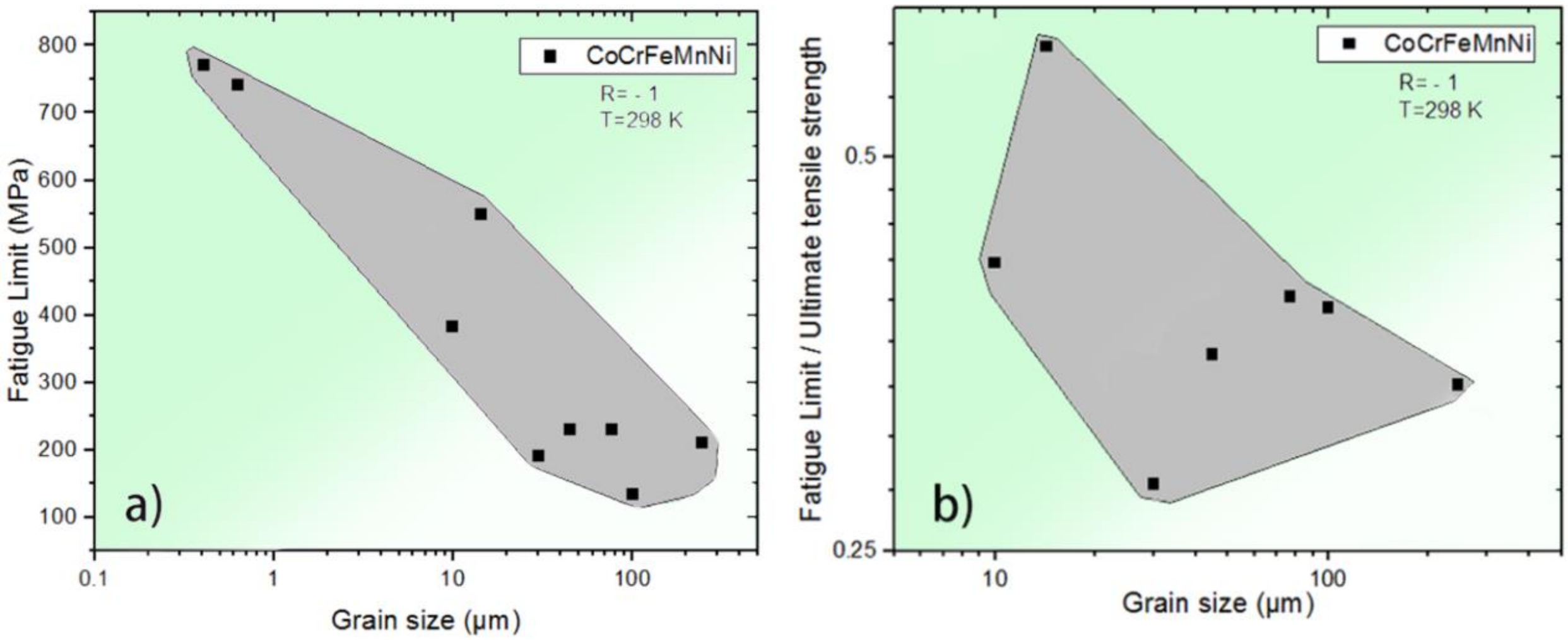

**Figure 2.1.5:** The relationship between the grain size and (a) the fatigue limit and (b) the fatigue ratio of CoCrFeMnNi.

Inclusions

In the study of CoCrFeMnNi, Kim et al. [21] and Tian et al. [15] demonstrated the positive impact of nanoscale inclusions on the FCC HEA. As shown in Figure 2.1.6, nanoscale inclusions in the FCC HEA are generally oxide particles, such as $CrMn_2O_4$ and $Mn_2O_3$. The uniform nano-scale inclusions can inhibit grain growth during the heat treatment, and make the material have higher fatigue resistance by refining the grains. This is the reason that nano-inclusions are often accompanied by extremely fine grains. However, the larger size inclusion particles have a negative impact on the material's fatigue behavior [51]. They affect the uniformity of the HEA microstructure, lead to stress concentration, cause the accumulation of dislocations, and accelerate the formation of cracks. In this case, microscale inclusions may be more important than the nanoscale inclusions, because of larger stress concentrations. In a high-temperature environment, the precipitates are more likely to grow into harmful impurity particles.

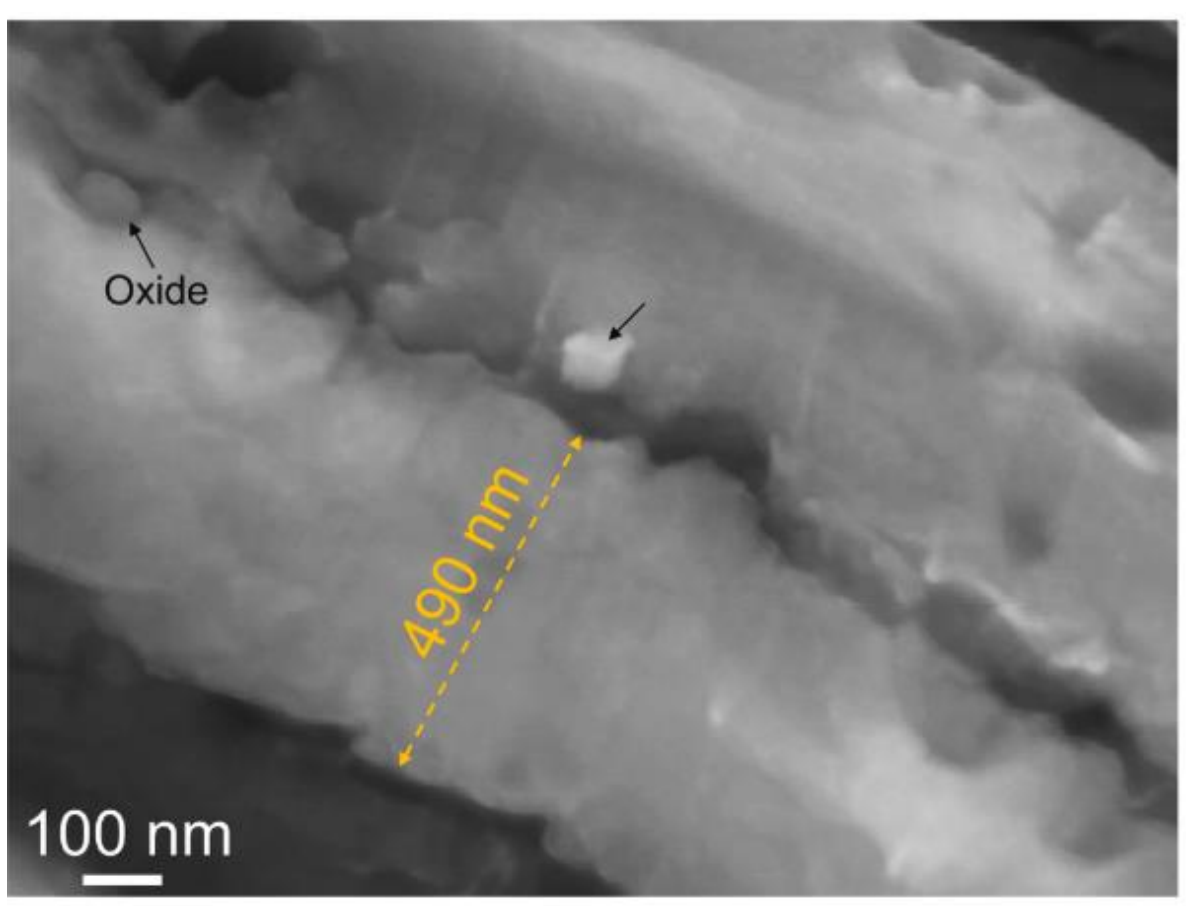

**Figure 2.1.6:** Nanoscale $Mn_2O_3$ particles observed in the SLM-built CoCrFeMnNi HEA [21].

Other microstructural aspects

In the FCC HEA, there are two other microstructural aspects that have been observed to improve the high-cycle fatigue endurance limit. One is the low-angle grain boundaries (LAGB) formed by the pre-straining process, which was observed in the study by Lee et al. [38]. At LAGBs, the difference in orientation between adjacent grains is small, which makes the movement of dislocations between different grains freer. The LAGBs structure alleviates the accumulation of dislocations and delays the rate of crack nucleation, which increases the fatigue limit of the specimen by ~ 300 MPa [38]. The other microstructural aspect involves the dislocation networks, mentioned in the research by Kim et al. [21]. The dislocation network is a sub-structure, only found in CoCrFeMnNi specimens fabricated by the SLM process, and formed during rapid solidification. As shown in Figure 2.1.7, the dislocation network has two forms: cellular and columnar. The role of the cellular and columnar dislocation networks is to act as a stable soft barrier, to inhibit the movement of dislocations, and thereby to change the mechanical properties of the material.

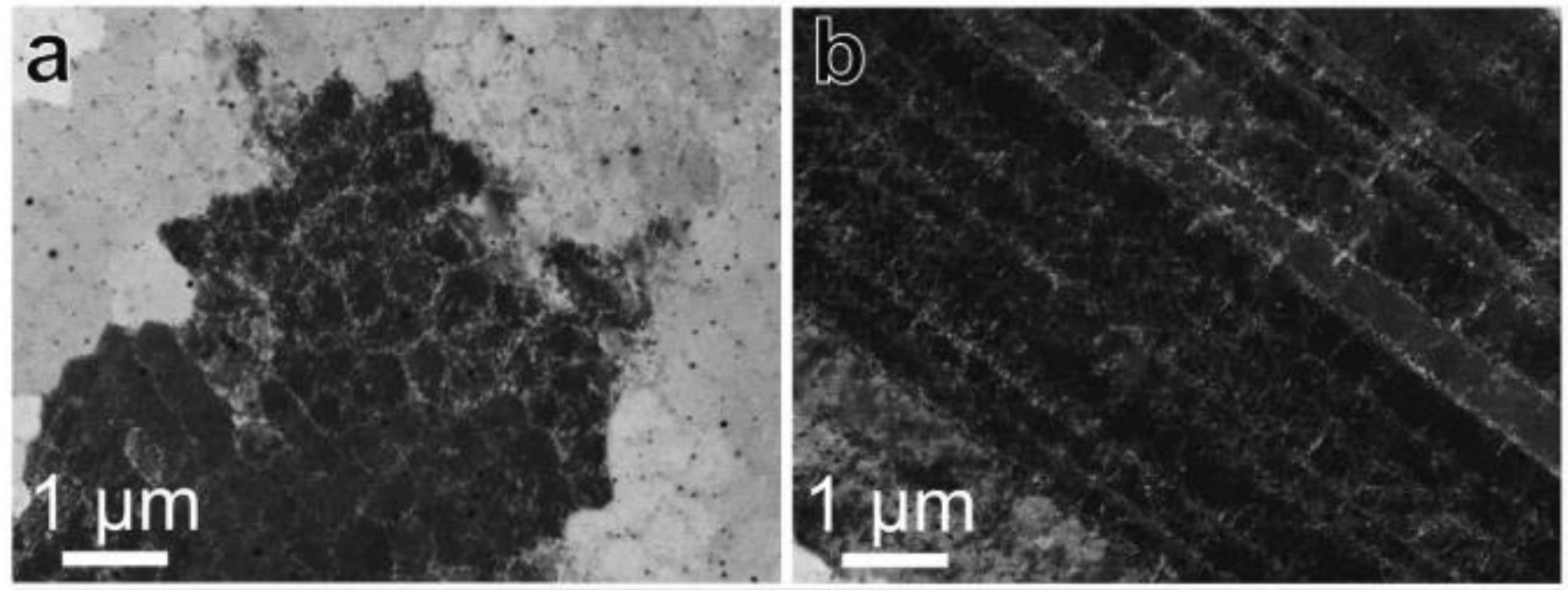

**Figure 2.1.7:** Structure of cellular and columnar dislocation networks in the CoCrFeMnNi HEA. These structures offer benefits, in terms of improving HCF performance under SLM processing [21].

### *2.1.4. BCC HEAs*

Two (2) HCF studies related to BCC HEAs have been performed through four-point flexural fatigue testing. Therefore, in order to be able to compare with other studies using conventional uniaxial fatigue testing, the maximum stress can be converted as follows, using classical theory for elastic beam bending [35, 52]:

$$\sigma_{\text{max,el}} = \frac{3P(S_o - S_i)}{2BW^2} \tag{2.1.3}$$

Here, $\sigma_{\text{max,el}}$ denotes the maximum stress on the tensile surface of a cyclically bent specimen, $P$ represents the applied load, $S_{\text{o}}$ is the outer span distance, $S_{\text{i}}$ is the inner span distance, $W$ is the specimen width, and $B$ is the specimen thickness, as shown in Figure 2.1.9(a). In Figure 2.1.9(b), the specimen thickness is denoted as $h$. The quantity $\sigma_{\text{max}}$ in Figure 2.1.9(b) represents the maximum load applied (consistent with Figure 2.1.1), $\sigma_0$ denotes the material's yield stress, whereas $\rho_P$ stands for the elastic zone width. Figure 2.1.9(b), and the associated elasto-plastic model of [18], shows how to calculate the elastic zone width $\rho_P$. It is mandatory to take into consideration the relaxation of the stress in zones where the material is plastically deformed [18]. Figure 2.1.9(b) shows how to account for the actual stress applied, during the four-point bending fatigue test, when the actual stress exceeds the yielding point.

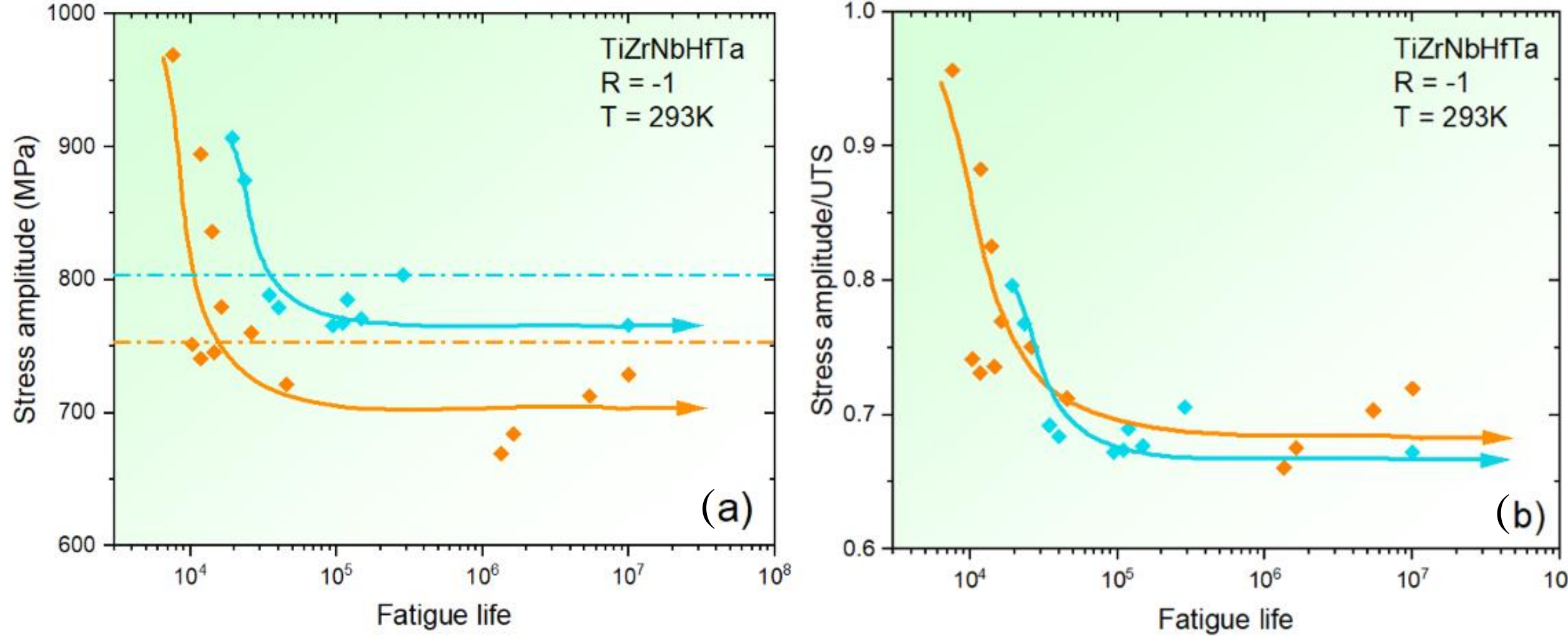

**Figure 2.1.8:** HCF performance of TiZrNbHfTa BCC HEAs under (a) S-N mode and (b) fatigue ratio vs. fatigue life, normalized to the stress ratio of $R$=-1 [18, 44, 52]. The cyan curve refers to as-annealed TiZrNbHfTa BCC HEA with average grain size of 45 ± 7 µm tested under four-point bending (4PB) (tested at the stress ratio of $R$ = 0.1 and frequency of 10 Hz) [25], The orange curve refers to a TiZrNbHfTa BCC HEA after heat treatment, with grain size ranging from ≈100 µm to ≈ 30 µm, also tested under four-point bending at a stress ratio of $R$ ≈ 0.1 [26].

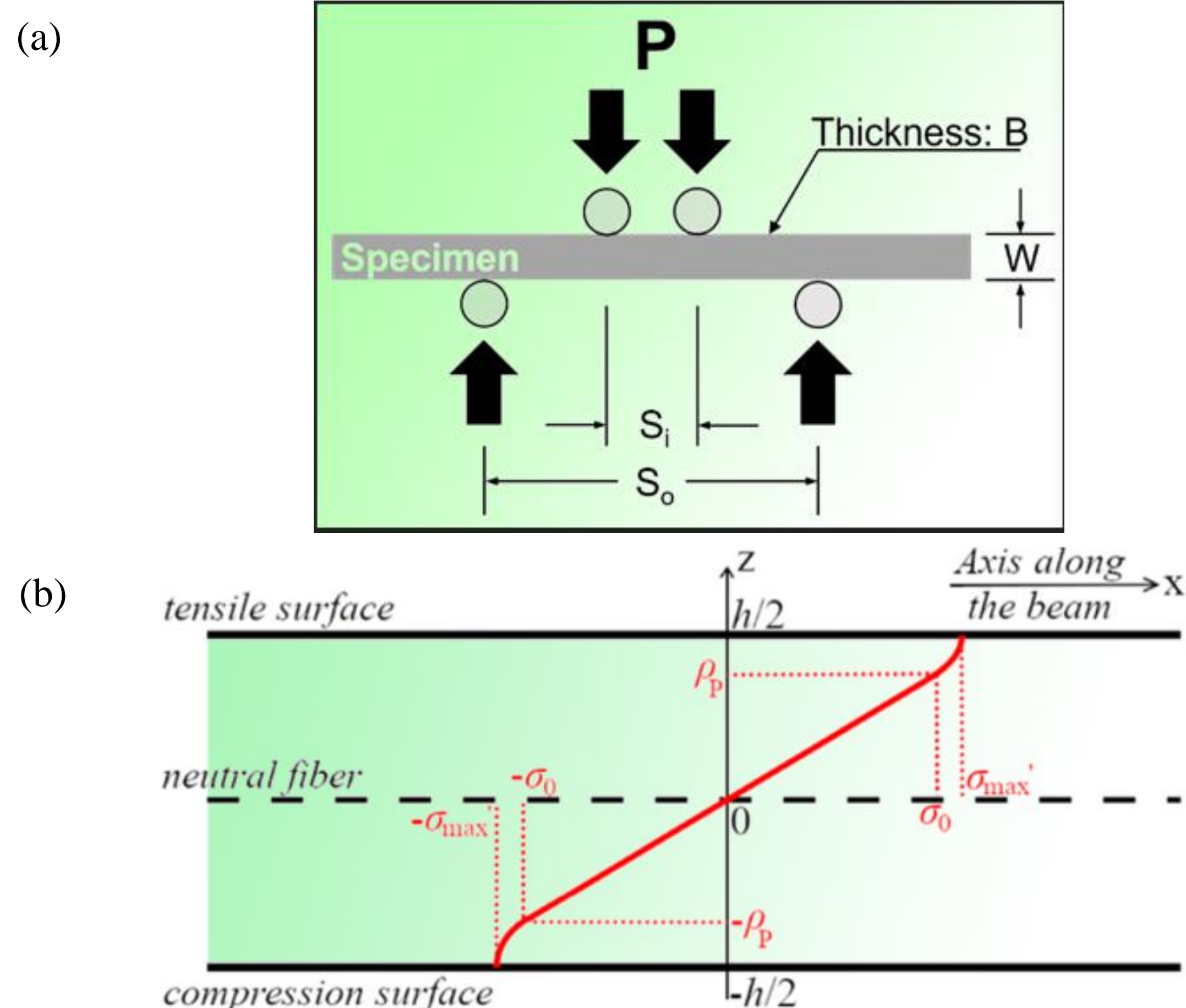

**Figure 2.1.9:** (a) A schematic of four-point bending fatigue tests and (b) a schematic of the stress distribution through the specimen's thickness [18].

Both the BCC HEAs studied are equimolar TiZrNbHfTa with similar grain sizes. Constituting brittle material, the BCC HEA specimens from neither study did exhibit significant plastic deformation. Figure 2.1.8(a) presents S-N data, normalized for $R$ = -1 in accordance with Eq. (2.1.2), whereas Figure 2.1.8(b) plots the fatigue ratio vs. the fatigue life. According to the results, the fatigue limit of the BCC HEA is the highest among the four types of HEAs. The fatigue limit of both studies is above 700 MPa. And the fatigue ratio under the same fatigue life is also high, which seems to be common for brittle materials. For the same lifetime, the specimen with the higher UTS (1,139 MPa) exhibited higher fatigue resistance than the specimen with the lower UTS (1,013 MPa). But in terms of the fatigue ratio, the difference was not significant, probably due to the grain size being similar (45 μm) and the testing method being the same (the four-point bending).

Another observation from Figure 2.1.8 pertains to the fact that the S-N relationship for the BCC HEA can be divided into two regions. When the true maximum stress ($\sigma_{max}$) applied on the specimen is lower than the true yield stress ($\sigma_y'$) for TiZrNbHfTa, the relationship between the fatigue life and stress amplitude tends to be flat. However, when the maximum stress exceeds this value, the fatigue life depicts a rather steep regression curve. The dotted line in Figure 2.1.8(a) roughly shows the stress amplitude of the fatigue stage transition. Variations in the fatigue-life in the high-stress region are mainly influenced by the plastic zone width ratio ($Pl_w$) [18]. For the specimen used in the four-point bending test, the plastic-zone width ratio can be calculated as [53]:

$$Pl_w = 1 - \frac{\rho_P}{h/2} \tag{2.1.4}$$

where $\rho_P$ represents the distance between the start of the plastic zone and the displacement indentation of neutral fibers due to plastic bending. But $h$ denotes the thickness of the specimen, as shown in Figure 2.1.9(b). As the stress increases, the plastic-zone width ratio becomes larger. The relationship between the plastic-zone width ratio and fatigue life is presented in Figure 2.1.10.

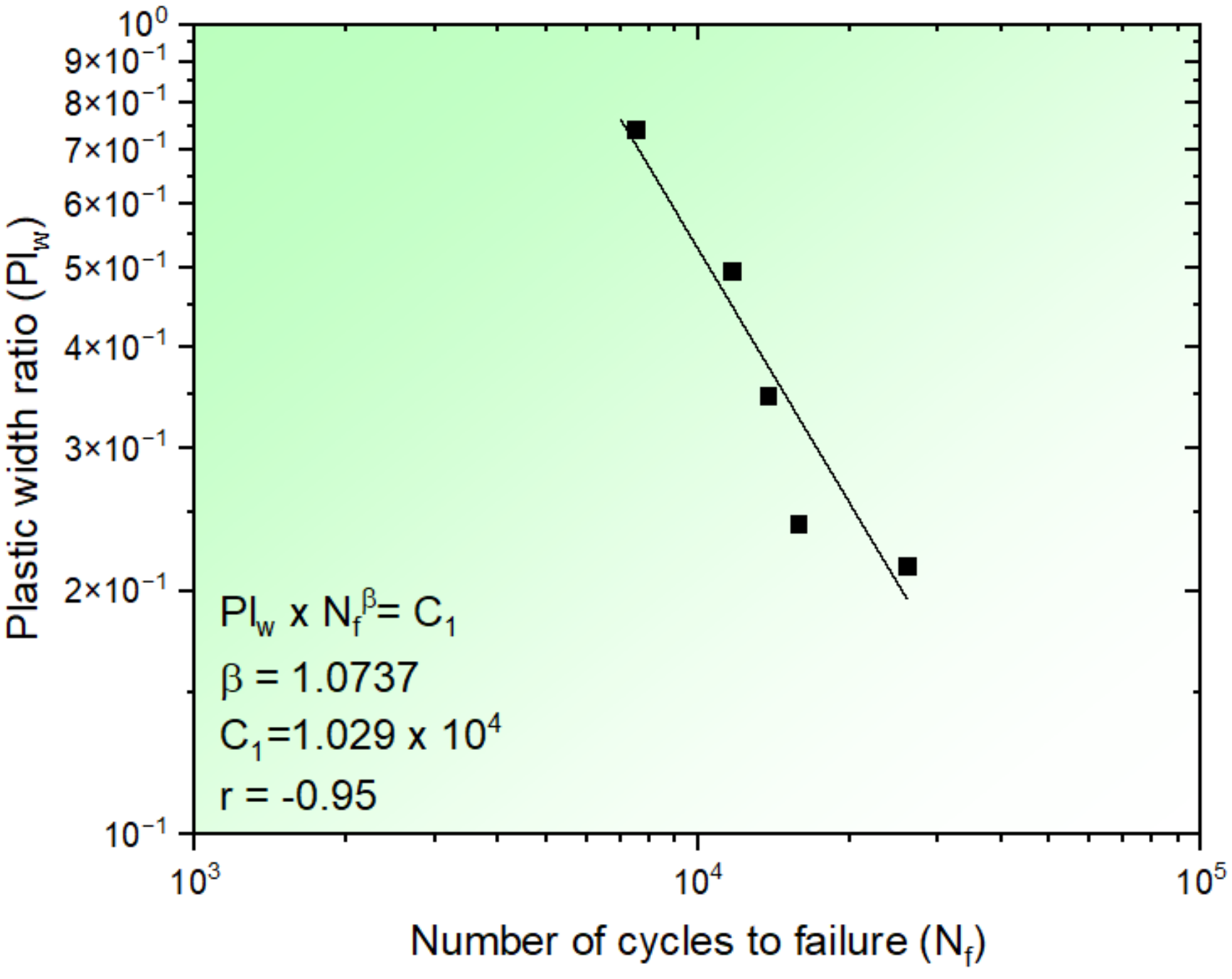

**Figure 2.1.10:** Relationship between the plastic-width ratio and fatigue life of the TiZrNbHfTa BCC HEA [18].

When the stress applied in the four-point bending fatigue test exceeds the yield strength of the specimen, i.e., when $\sigma_{max} > \sigma_0$ according to Figure 2.1.9(b), there are two methods to estimate the actual stress on the sample specimen (to correct for plastic deformation induced). The first method to correct for the plastic deformation induced, and to obtain close-to-realistic maximum stresses on the samples, entails the Neuber method [52]. Here, the strain energy density of a simplified linear elastic material is equated with that of an actual elastoplastic material, yielding [52]

$$\frac{\sigma_{LE}\cdot\varepsilon_{LE}}{2}=\frac{\sigma_{EP}\cdot\varepsilon_{EP}}{2} \quad (2.1.5)$$

with $\sigma_{LE}$ and $\varepsilon_{LE}$ denoting the elastic stress and strain, but $\sigma_{EP}$ and $\varepsilon_{EP}$ the plastic counterparts. The second method, to account for relaxation of the stress in zones, where the material is plastically deformed, is based on the work of Stok and Halilovic [53], and can be summarized by the equation:

$$\int_{\rho_P}^{h/2}\sigma_P(\varepsilon(z))zdz=\frac{Pl}{4b}-\frac{\sigma_0{\rho_p}^2}{3}. \quad (2.1.6)$$

Here, $\sigma_P$ represents the stress applied in the plastic zones, $\sigma_0$ denotes the yield stress of the material, $z$ is the axis through the beam′ s thickness, with a zero value at the neutral fiber in fully elastic loading, and $\varepsilon$ is the strain [26].

### ***2.1.5. Multiphase HEAs***

The three (3) studies on the multiphase HEA can be divided into two groups of $AlCoCrFeNi_{2.1}$ and $Al_xCoCrFeNi$. Shukla et al. report that the heat-treated wrought $AlCoCrFeNi_{2.1}$ HEA exhibits better HCF fatigue performance than the as-casted $AlCoCrFeNi_{2.1}$ HEA [16], according to Figure 2.1.11(a), due to the cold-rolling and heat-treatment processes involved. In the wrought $AlCoCrFeNi_{2.1}$ HEA, the grain refinement caused by the cold-rolling process greatly increases the strength of the HEA, and the heat-treatment process results in the B2 precipitation and Cr precipitation in the microstructure of the HEA. These B2 and Cr precipitation structures become dislocation barriers to a certain extent. In addition, heat treatment promotes the formation of a BCC phase, so the wrought $AlCoCrFeNi_{2.1}$ HEA contains larger fraction of the BCC phase (32%) than the as-cast $AlCoCrFeNi_{2.1}$ HEA (29%). The BCC is considered to be another microstructure favorable for HCF fatigue resistance. However, since some HCF studies of multiphase HEAs have not provided an accurate phase fraction for the BCC phase, additional studies are needed to corroborate this view.

In their studies of the $Al_xCoCrFeNi$ family, Kaimiao et al. demonstrated the effect of third-phase precipitates on fatigue behavior, captured in Figure 2.1.11(b). Compared to the dual-phase $Al_{0.7}CoCrFeNi$ without annealing (the green dots in Figure 2.1.11), the low-temperature annealed $Al_{0.7}CoCrFeNi$ (the purple dots) and $Al_{0.3}CoCrFeNi$ (the blue dots) produced $L1_2$-phase and σ-phase precipitates in their microstructures, as shown in Figure 2.1.12. These nano-sized precipitates are embedded in the FCC phase. The precipitates increase the material strength, while reducing ductility, and contribute to lower fatigue resistance in the high-stress amplitude region but higher fatigue resistance in the low-stress amplitude region. Therefore, in the region with greater plastic deformation, these three-phase HEAs perform worse than the dual-phase HEAs.

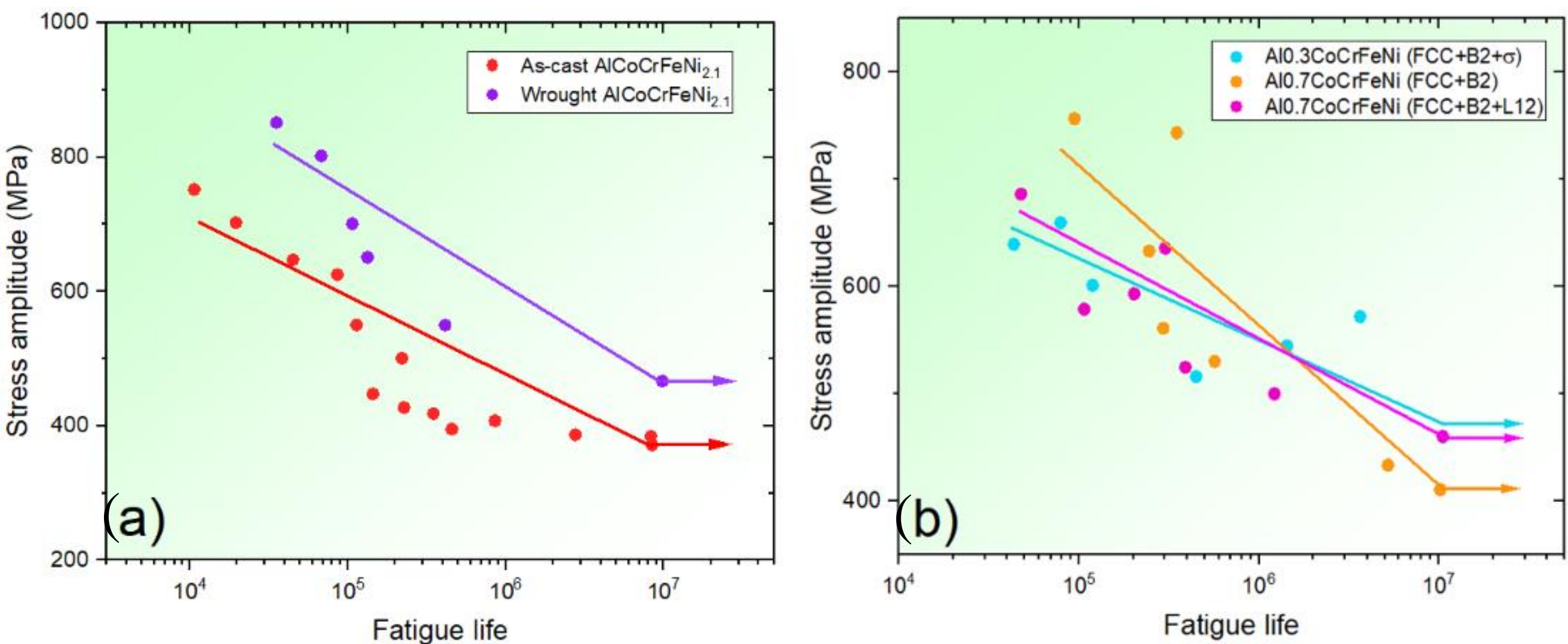

**Figure 2.1.11:** HCF performance of multiphase HEAs under a S-N mode [27-29].

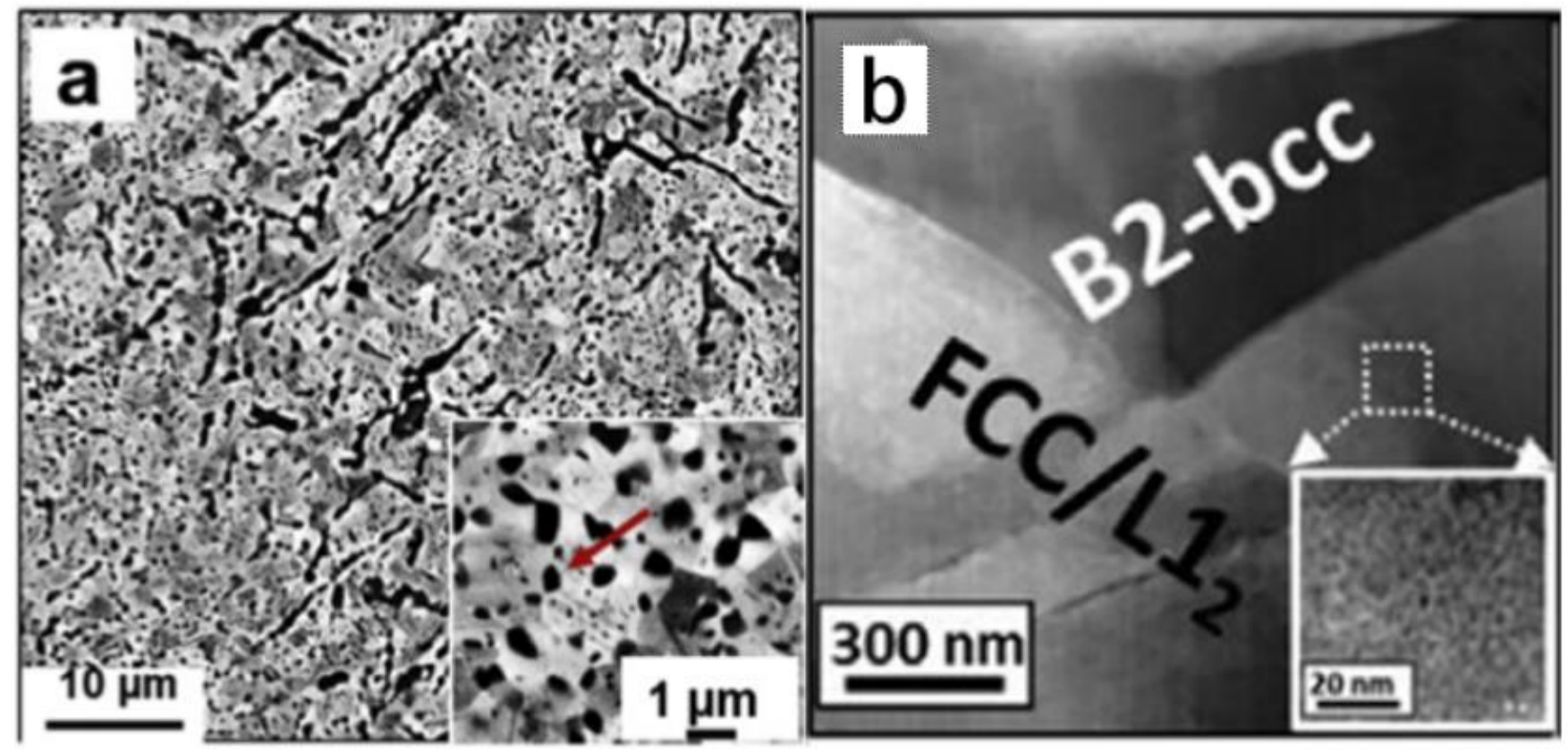

**Figure 2.1.12:** The scanning-electron microscopy (SEM) of precipitates embedded in (a) $Al_{0.3}CoCrFeNi$ [22] and (b) $Al_{0.7}CoCrFeNi$ [45].

### *2.1.6. Metastable HEAs*

Liu et al. [20] conducted fatigue studies on two types of Fe-based metastable HEAs processed by the friction stirring process (FSP) technique. Figure 2.1.14 shows the detailed microstructure of $Fe_{38.5}Mn_{20}Co_{20}Cr_{15}Si_5Cu_{1.5}$ by electron back-scattering diffraction (EBSD) and inverse pole figures (IPFs). The stirred Nugget region has a finer grain structure than the unstirred as-cast region. The improvement in work hardenability by grain refinement contributes to the metastable HEA possessing HCF behavior superior to most high-entropy alloys. In addition, FSP also changed the phase fraction of the $Fe_{38.5}Mn_{20}Co_{20}Cr_{15}Si_5Cu_{1.5}$ HEA. As shown in Figure 2.1.14($a_3$), the fraction of the HCP phase in the as-cast region is as high as 15%. In the Nugget region, the proportion of the HCP phase is reduced to 2%, which indicates that the Cu element in the Nugget region is more evenly distributed in the FCC phase, which could suggest solution strengthening [20]. It can be seen from Figure 2.1.13 that the HCF performance of $Fe_{38.5}Mn_{20}Co_{20}Cr_{15}Si_5Cu_{1.5}$, which is also a metastable HEA processed by the FSP technology, is superior to that of $Fe_{42}Mn_{28}Cr_{15}Co_{10}Si_5$. This difference indicates that the solid-solution strengthening presumably caused by the uniform distribution of copper elements effectively improves the fatigue resistance of the metastable HEA.

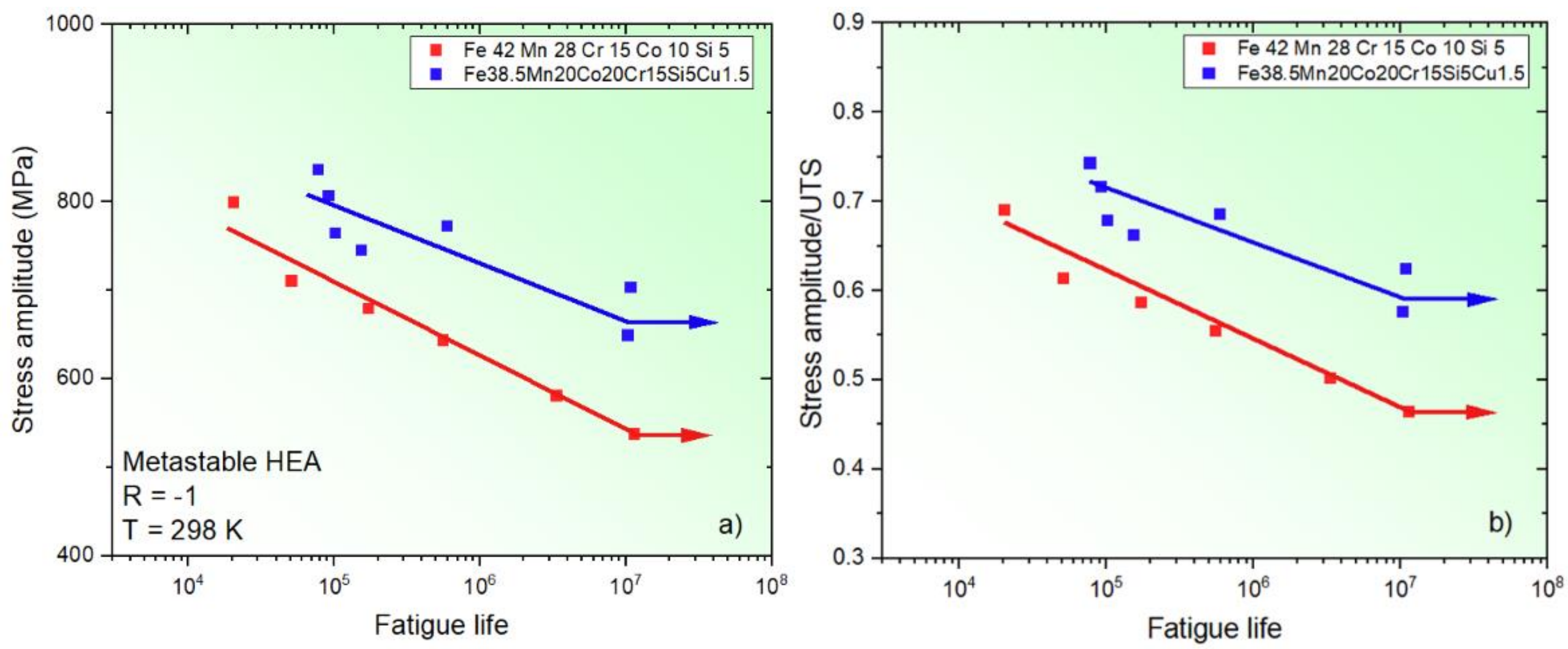

**Figure 2.1.13:** HCF performance of two Fe-based metastable HEAs [20].

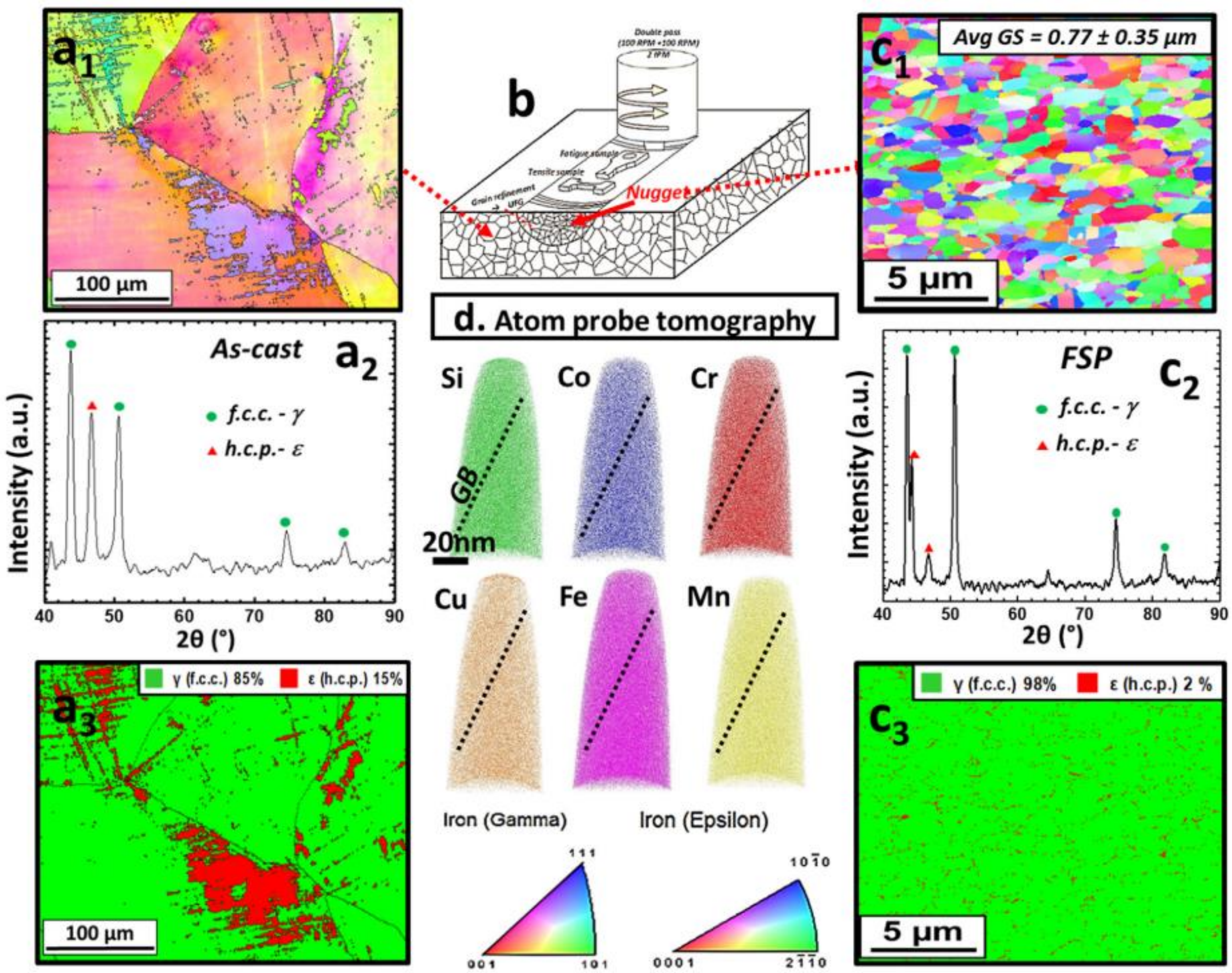

**Figure 2.1.14:** EBSD, x-ray diffraction (XRD), and IPF microstructural evolution in the $Fe_{38.5}Mn_{20}Co_{20}Cr_{15}Si_5Cu_{1.5}$ HEA [20].

### *2.1.7. Comparison between HEAs and conventional alloys*

In order to study the applicability of HEAs in the field of the industrial design, it is necessary to compare them with conventional alloys, especially with those metallic materials that have been widely used in industry. To this end, we retrieved extensive fatigue data on conventional alloys and plotted for comparison [54-59]. The reference data used in Figure 2.1.15 are obtained at room temperature and standard air pressure and have been converted to $R$ = -1.

Since the conventional alloys studied comprise of different compositions, microstructures, test conditions, etc., even within the same family of alloys, there is a large degree of dispersion in the fatigue data. This type of dispersion is acceptable. Due to the large number of conventional alloys, we use color shading in Figure 2.1.15 to illustrate the HCF performance of each alloy family, including magnesium alloys, copper alloys, titanium alloys, aluminum alloys, steels, bulk metallic glass, etc.

Specifically, Figure 2.1.15(a) shows a plot of fatigue strength vs. tensile strength for HEAs, BMGs, and conventional alloys. From Figure 2.1.15(a) the approximate HCF performance of HEAs can be observed and compared to that of BMGs and conventional alloys. The FCC HEAs exhibit fatigue endurance limit higher than the aluminum alloys, the magnesium alloys, and most copper alloys, but lower than most of the titanium alloys and the steels. For a given fatigue limit, the ultimate tensile strengths for the other three HEA categories (the BCC HEAs, the metastable HEAs and the multiphase HEAs) are slightly higher than for the FCC HEAs. In Figure 2.1.15(a), the multiphase HEA category is surrounded by titanium alloys, steels, bulk metallic glasses and their composites. The fatigue strengths of metastable HEAs are superior to most of the conventional alloys, including many of the titanium alloys and the steels. The fatigue endurance limit of the BCC HEA is the best, surpassing almost all the alloys accounted for in this paper. The BCC HEAs also exhibit the highest fatigue ratio of all materials.

Figure 2.1.15(b) shows the performance of the same research data under fatigue ratio vs. tensile strength, which corresponds to the slope in Figure 2.1.15a. From Figure 2.1.15(b), it is evident that the fatigue ratios of FCC, multiphase, and metastable HEAs are on the same level. Again, the fatigue ratio of the BCC HEAs far exceeds that of other HEA categories and of almost all the conventional alloys.

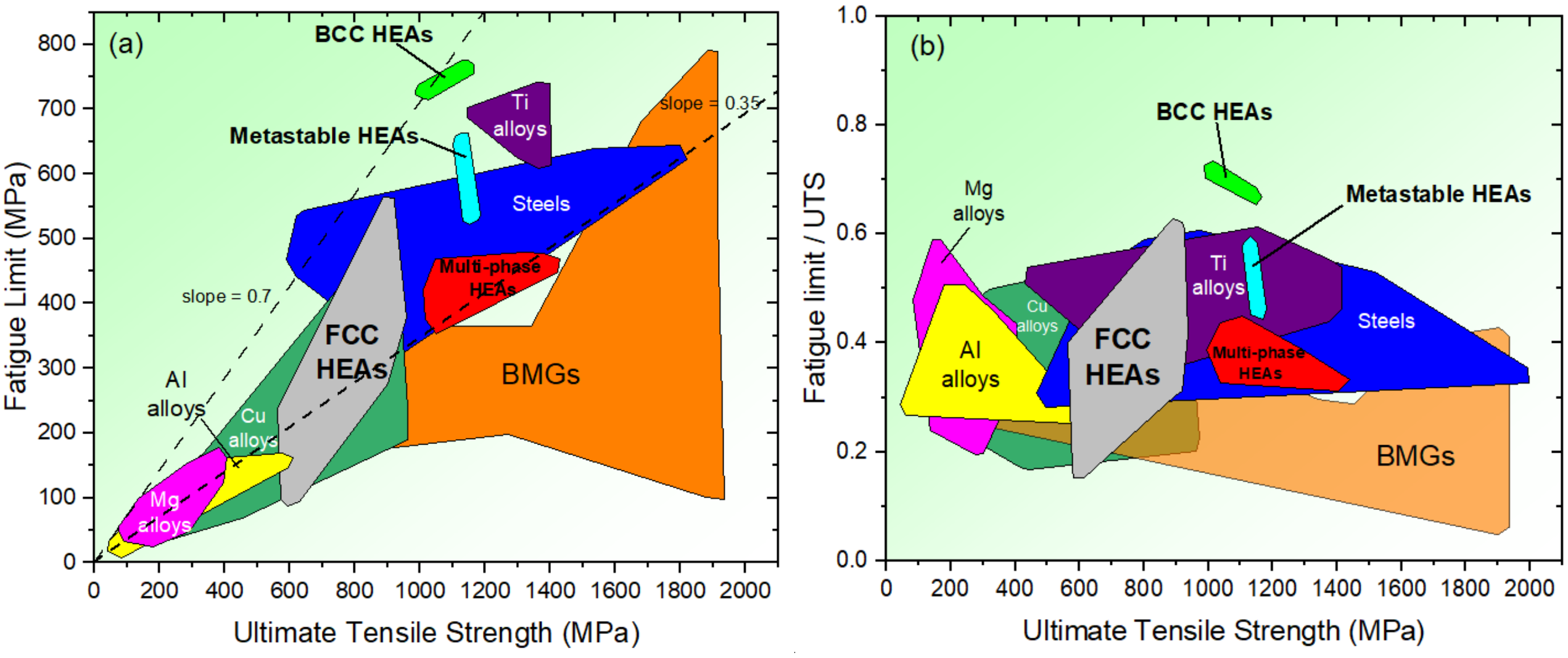

**Figure 2.1.15:** Comparison among HEAs, BMGs, and conventional alloys under a) Fatigue limit vs. Ultimate tensile strength and b) Fatigue ratio vs. Ultimate tensile strength [15, 21, 38, 40, 54-60].

## 2.2. Low-cycle fatigue of HEAs

### *2.2.1. Introduction*

Cumulative Degradation – Strain / Life (ε/*N*)

In the case of LCF, the Basquin model for the stress life may not prove accurate, due to its inability to account for plastic deformations. The strain-life (ε-*N*) model was first formulated independently and simultaneously in the early 1950s by Coffin and by Manson to describe the relationship between the LCF and cyclic plastic strain range [61, 62]. Coffin considered the constrained thermal fatigue of power-plant components [61], but Manson considered the isothermal fatigue of ground vehicles [62]. The strain-life model enables the detailed analysis of plastic deformation in localized regions. Fatigue failures almost always start at a local discontinuity. But the Coffin-Manson model compounds several idealizations, which leads to some uncertainty in the results.

The ε-*N* curves resemble the S-N curves but assume the application of cyclic strain to the material, as shown in Figure 1.1.1, as opposed to cyclic stress (as in case of the S-N curves). When the stress at a discontinuity exceeds the elastic limit, plastic strain occurs. The total strain amplitude can be resolved into elastic and plastic strain components from the steady-state hysteresis loops. The ε-*N* curve, presented in Figure 2.2.2, consists of a plastic-deformation zone, with a lower slope, together with an elastic-deformation zone, with a higher slope. Both the elastic and the plastic curves can be approximated as straight lines. At large strains or short lives, the plastic-strain component is predominant, but at small strains or longer lives, the elastic-strain component dominates.

In the case of the plastic-deformation zone (in the low-cycle regime) in Figure 2.2.2, the plastic-strain-based Coffin-Manson law can be used for the prediction of the fatigue life:

$$\frac{\Delta\varepsilon_{pl}}{2} = \varepsilon'_f\,(2\,N_f)^c. \qquad (2.2.1)$$

Here, $N_f$ represents the number of strain cycles to failure (the fatigue life), $\Delta\varepsilon_{pl}$ the plastic component of the strain range, $\varepsilon'_F$ the fatigue-ductility coefficient (which is approximately equal to the true fracture ductility, $\varepsilon_F$), and $c$, a fatigue ductility exponent (the slope of the plastic-strain line, see Figure 2.2.2).

But in the case of the elastic-deformation zone (the high-cycle regime) in Figure 2.2.2, the stress-based Basquin law can be used to determine the fatigue life:

$$\frac{\Delta\varepsilon_{el}}{2} = \frac{\sigma_f'}{E}(2N_f)^b. \tag{2.2.2}$$

Here, $\Delta\varepsilon_{el}$ represents the elastic component of the strain range, $\sigma'_f$ the fatigue-strength coefficient, which is approximately equal to the true fracture strength at fracture, $\sigma_f$, $E$ the modulus of elasticity (Young's modulus), and $b$ a fatigue strength exponent (the slope of the elastic strain line in Figure 2.2.2).

To account for both the plastic and elastic deformation zones, and to accurately predict the overall fatigue life, the Basquin and Coffin-Manson laws can be combined into a generalized Coffin-Manson model:

$$\varepsilon_a = \frac{\Delta\varepsilon}{2} = \frac{\Delta\varepsilon_{el}}{2} + \frac{\Delta\varepsilon_{pl}}{2} = \frac{\sigma_f'}{E}(2N_f)^b + \varepsilon_f'(2N_f)^c. \tag{2.2.3}$$

Here, $\Delta\varepsilon$ represents the total strain range, shown in Figure 2.2.1, and $\varepsilon_a$ the corresponding strain amplitude.

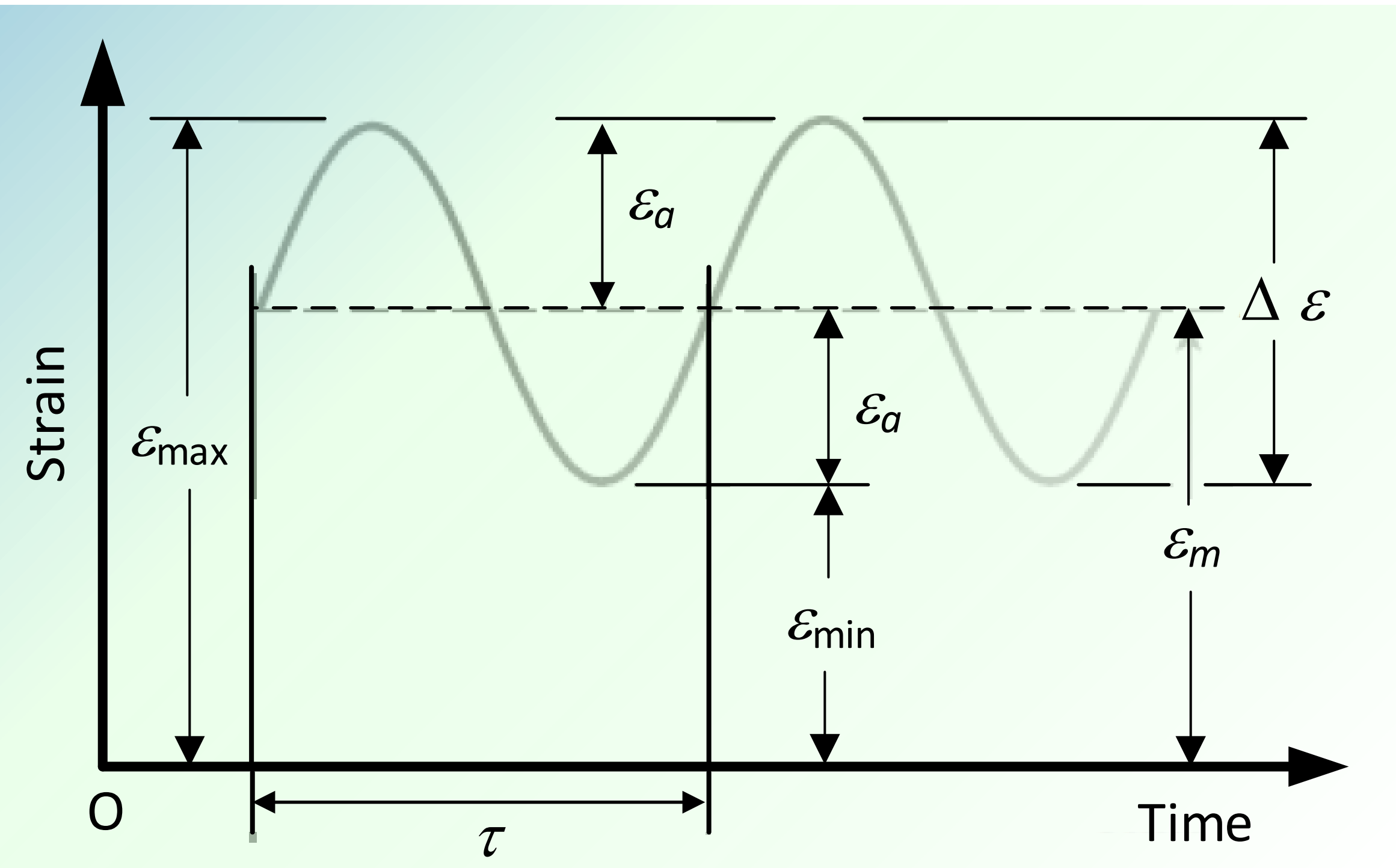

**Figure 2.2.1:** Key definitions related to the strain applied in the case of the constant amplitude-strain control (adapted from [63]).

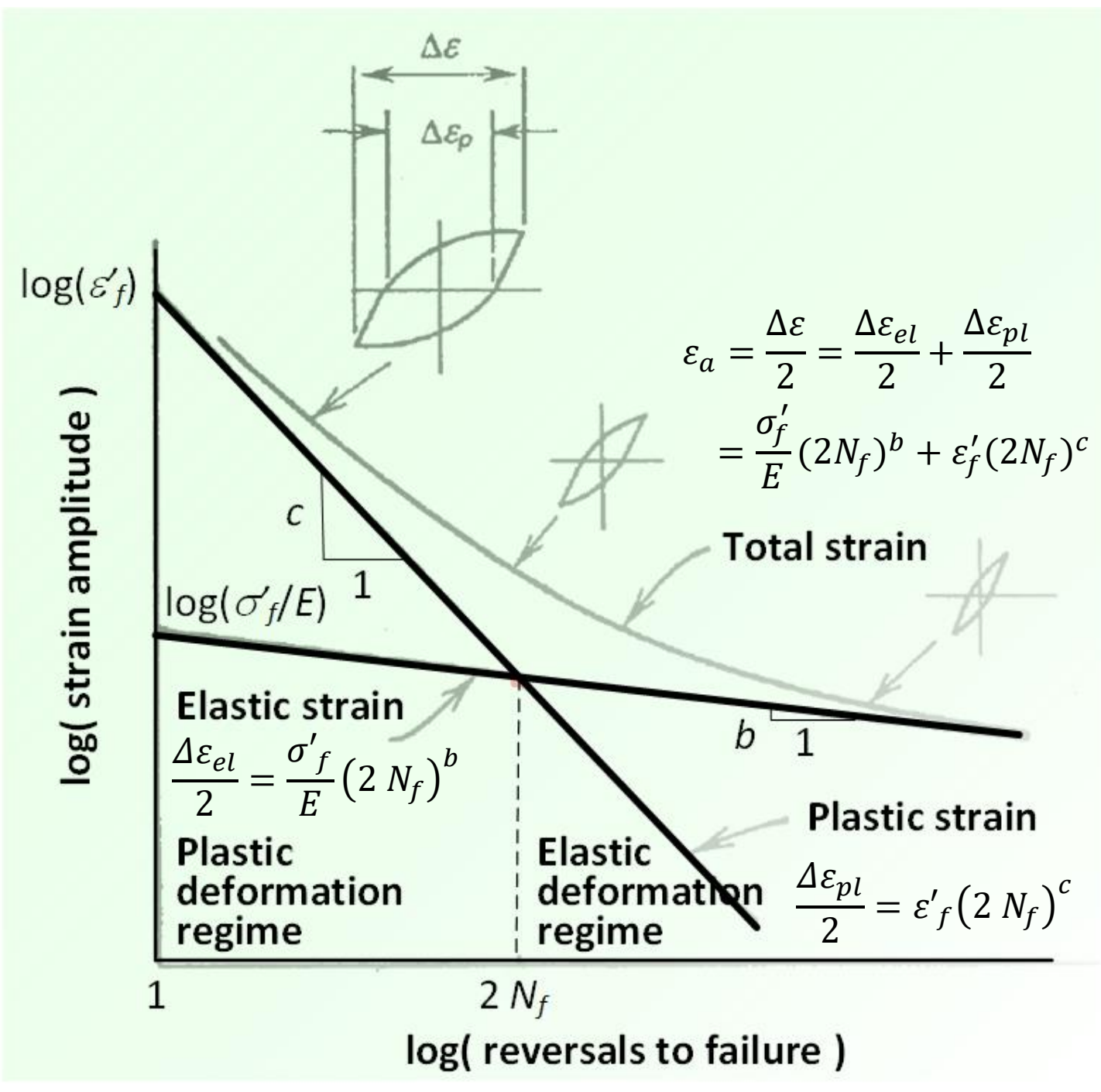

**Figure 2.2.2:** Strain / life curve featuring a plastic and an elastic deformation zone (adapted from [63]).

<u>Cyclic Stress-Strain Relation</u>

Cyclic stress-strain traces over cycles are normally presented in the form of hysteresis loops, as exemplified at the half life in Figure 2.2.3. The half width of the hysteresis loop at the zero stress gives the plastic-strain amplitude, $\frac{\Delta\varepsilon_{pl}}{2}$, while the half of the stress range defines the stress amplitude, $\frac{\Delta\sigma}{2}$.

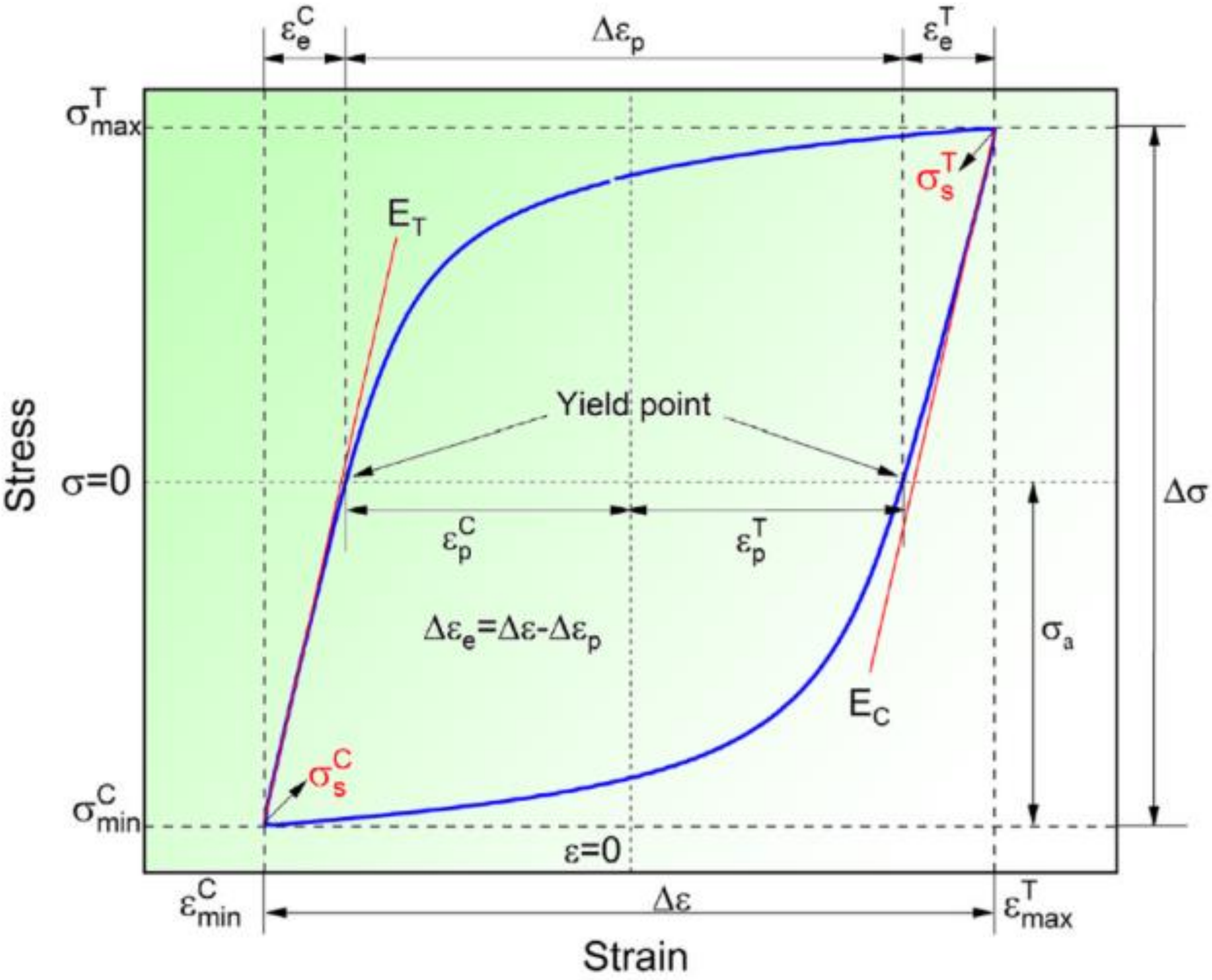

**Figure 2.2.3:** Schematic of a hysteresis loop under low-cycle fatigue [64].

The plastic data of LCF may also be presented in a stress-strain form like stress-strain curves for uniaxial tensile or compression data, resulting in the so-called cyclic stress-strain data. Specifically, $\frac{\Delta\sigma_t}{2}$ and $\frac{\Delta\varepsilon_{pl}}{2}$ from a LCF test are extracted and plotted against each other. The $\frac{\Delta\sigma_t}{2}$ versus $\frac{\Delta\varepsilon_{pl}}{2}$ data for a CoCrFeMnNi HEA with a grain size of $d = 60$ μm is presented on a bi-logarithmic scale in Figure 2.2.4. This cyclic stress-strain relation can be modeled by a power-law [65]:

$$\frac{\Delta\sigma_t}{2} = K'\left(\frac{\Delta\varepsilon_{pl}}{2}\right)^{n'} \tag{2.2.4}$$

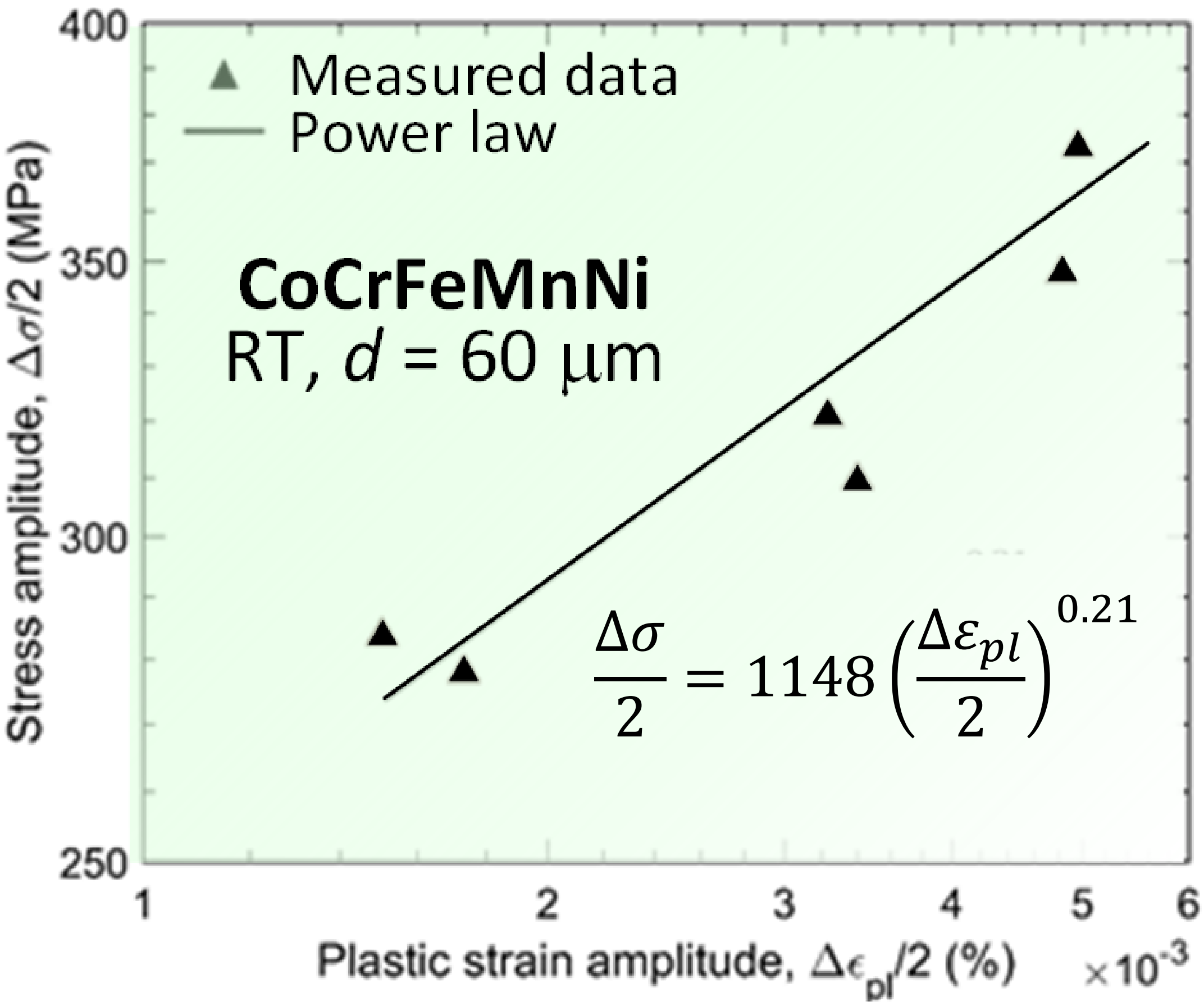

**Figure 2.2.4**: Cyclic stress-strain data along with the power-law fit of the CoCrFeMnNi with a grain size of $d$ = 60 μm, tested in air at room temperature [65].

Here, $K'$ represents the cyclic-strength coefficient, but $n'$ is the cyclic work-hardening exponent. The unknowns, $K'$ and $n'$, can be determined by fitting the logarithmic form of Eq. (2.2.4) to the measured $\frac{\Delta\sigma_t}{2}$ versus $\frac{\Delta\varepsilon_{pl}}{2}$ data, as noted by the straight (fitted) line in Figure 2.2.4.

### *2.2.2. Comparison between HEAs*

At present, there are relatively few studies on the LCF performance of HEAs, only 9 in total, including five on FCC HEAs [CoCrFeMnNi, CoCuFeMnNi, and $Al_5(CoCrFeMnNi)_{95}$] [24, 66-70], three on metastable HEAs ($Fe_{50}Mn_{30}Co_{10}Cr_{10}$, $Fe_{30}Cr_{30}Ni_{25}Mn_{10}Co_5$, $Fe_{48}Mn_{30}Co_{10}Cr_{10}Si_2$, $Fe_{46}Mn_{30}Co_{10}Cr_{10}Si_4$, and $Fe_{44}Mn_{30}Co_{10}Cr_{10}Si_6$) [27, 71, 72], and two on multiphase HEAs [$Al_{0.5}CoCrFeNi$ and $Al_{10}(CoCrFeMnNi)_{90}$] [25, 70]. Out of these, the investigation on CoCuFeMnNi [69] is not included in this paper, due to doubts about the accuracy of the data. Detailed information, including the grain size, ultimate tensile strength, yield strength, and testing frequency of all the studied HEAs, can be found in Table 2.2.1. All experiments were performed under conditions of full reversal and RT. Hence, the data obtained can be directly compared. Considering that the strain-life mode has the elastic-dominant and plastic-dominant stage, Figure 2.2.5 plots the data collected both in forms of $\Delta\varepsilon_{total}/2$ vs. $2N_f$ and $\Delta\varepsilon_{plastic}/2$ vs. $2N_f$, respectively. The plastic-strain amplitudes in Figure 2.2.5(b) are all calculated from hysteresis loops provided in the literature. Under the strain-life mode, except for some FCC HEAs with lower strain amplitudes, there is no significant difference among the data for different types of HEAs. Except for some FCC HEAs with lower fatigue life, the experimental data of the three types of HEAs show a large area of overlap, which is more noticeable under the condition of plastic-strain amplitude. This trend seems to suggest that the HEA-type effect on the LCF behavior is not as significant as in case of the HCF. Considering that the current research on both multiphase and metastable HEAs is quite limited, this conclusion still needs more research for support.

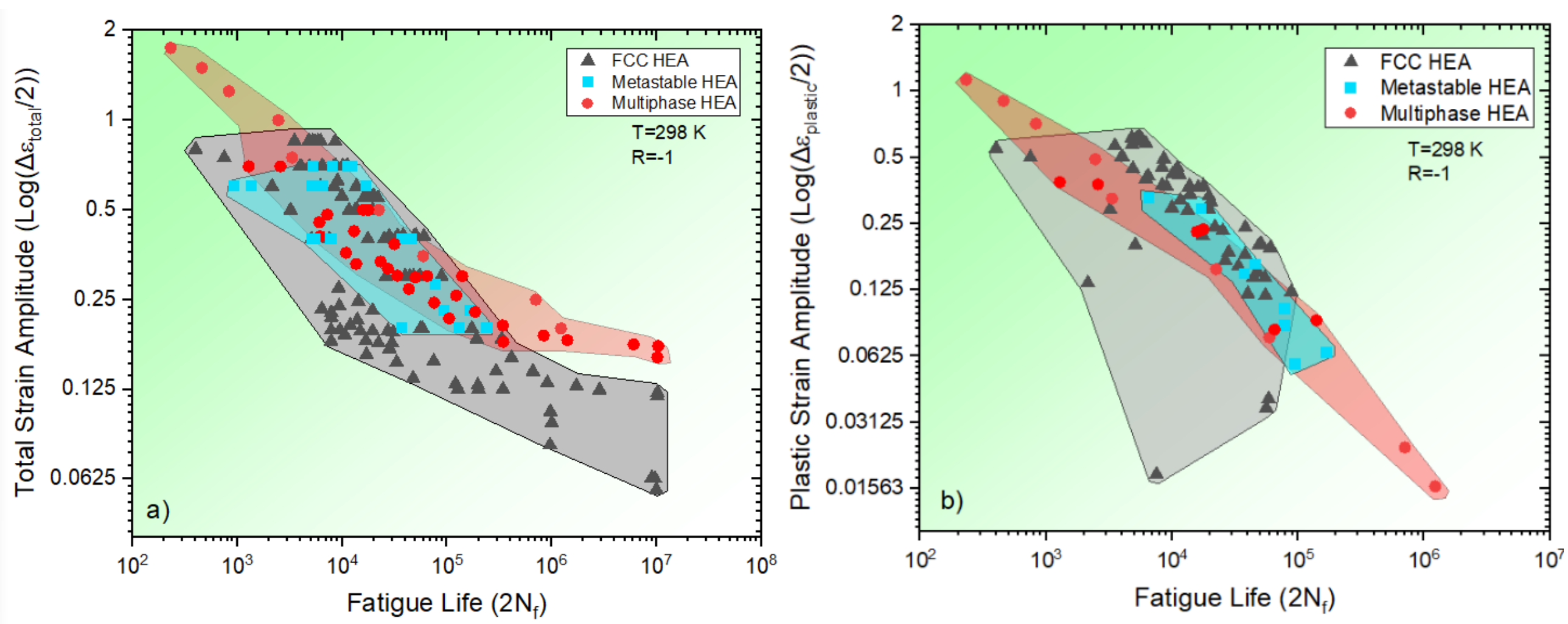

**Figure 2.2.5:** Strain-life data for three types of HEAs under a) total strain amplitude and b) plastic strain amplitude [46-55].

**Table 2.2.1:** Mechanical behavior of LCF-tested high-entropy alloys at room temperature, including $\sigma_{UTS}$ and $\sigma_{YS}$.

| **Composition** | **Testing Method** | **Phase** | **Grain Size [μm]** | **Tensile YS [MPa]** | **UTS [MPa]** | **Temperature [K]** | ***R*** | **Frequency** | **Ref** |
|---|---|---|---|---|---|---|---|---|---|
| CoCrFeMnNi | Uniaxial test | FCC | 6 ± 3 | N/A | N/A | 298 | -1 | N/A | [66] |
| CoCrFeMnNi | Uniaxial test | FCC | 60 | N/A | N/A | 298 | -1 | N/A | [66] |
| CoCrFeMnNi | Uniaxial test | FCC | 65 | 225 | 540 | 298 | -1 | 1Hz | [68] |
| CoCrFeMnNi | Uniaxial test | FCC | 10 ± 3 | 410 | 783 | 298 | -1 | 1Hz | [67] |
| CoCrFeMnNi | Uniaxial test | FCC | 15 ± 4 | 409 | 775 | 298 | -1 | 1Hz | [67] |
| CoCrFeMnNi | Uniaxial test | FCC | 66 ± 24 | 300 | 683 | 298 | -1 | 1Hz | [67] |
| CoCrFeMnNi | Uniaxial test | FCC+ $Cr_{23}C_6/Cr_7C_3$ | 4 ± 1 | 474 | 791 | 298 | -1 | 1Hz | [67] |

| CoCrFeMnNi | Uniaxial test | FCC | 12 | 255 | 800 | 298 | -1 | 0.25-0.75Hz | [24] |
|---|---|---|---|---|---|---|---|---|---|
| CoCrFeMnNi | Uniaxial test | FCC | 1 | 925 | 1,025 | 298 | -1 | 0.25-0.75Hz | [24] |
| CoCrFeMnNi | Bending test | FCC | 190 | 200 | 302 | 298 | -1 | 50 | [70] |
| $Al_5(CoCrFeMnNi)_{95}$ | Bending test | FCC | 229 | 211 | 323 | 298 | -1 | 50 | [70] |
| CoCuFeMnNi | Uniaxial test | FCC | 47 | 324 | 710 | 298 | -1 | N/A | [69] |
| $Fe_{50}Mn_{30}Co_{10}Cr_{10}$ | Uniaxial test | FCC + HCP | 5 | N/A | N/A | 298 | -1 | N/A | [27] |
| $Fe_{50}Mn_{30}Co_{10}Cr_{10}$ | Uniaxial test | FCC + HCP | 10 | N/A | N/A | 298 | -1 | N/A | [27] |
| $Fe_{30}Cr_{30}Ni_{25}Mn_{10}Co_5$ | Uniaxial test | FCC | 300 | 172 | 581 | 298 | -1 | N/A | [71] |
| $Fe_{30}Cr_{30}Ni_{25}Mn_{10}Co_5$ | Uniaxial test | FCC + BCC | 100 | 971 | 1,098 | 298 | -1 | N/A | [71] |
| $Fe_{30}Cr_{30}Ni_{25}Mn_{10}Co_5$ | Uniaxial test | FCC + BCC + σ | 1.5 | 617 | 906 | 298 | -1 | N/A | [71] |
| $Fe_{50}Mn_{30}Co_{10}Cr_{10}$ | Uniaxial test | γ + ε | 40±20 | 247 | 677 | 298 | -1 | 1 | [72] |
| $Fe_{48}Mn_{30}Co_{10}Cr_{10}Si_2$ | Uniaxial test | γ + ε | 51±28 | 247 | 670 | 298 | -1 | 1 | [72] |
| $Fe_{46}Mn_{30}Co_{10}Cr_{10}Si_4$ | Uniaxial test | γ + ε | 47±24 | 256 | 677 | 298 | -1 | 1 | [72] |
| $Fe_{44}Mn_{30}Co_{10}Cr_{10}Si_6$ | Uniaxial test | γ + ε | 67±35 | 258 | 683 | 298 | -1 | 1 | [72] |
| $Al_{10}(CoCrFeMnNi)_{90}$ | Bending test | FCC + BCC | 642 | 600 | 897 | 298 | -1 | 50 | [70] |
| $Al_{0.5}CoCrFeNi$ | Uniaxial test | FCC + B2 | 9 | 493 | 973 | 298 | -1 | N/A | [25] |

### *2.2.3. FCC HEAs*

In the strain-life mode, the grain-size effect in the FCC HEA is not as clear as in the stress-life mode, but still exists. In Figure 2.2.6, we plotted the strain-life data for all CoCrFeMnNi HEAs reported in the literature [46-51]. Based on the grain size, the data is divided into three levels from the finest to the coarsest: very fine grains ($d < 20$ μm), regular grains (20 μm $< d <$ 100 μm), and very coarse grains ($d > 100$ μm). In case of CoCrFeMnNi HEAs with very fine or regular grain sizes, the fatigue performance does not exhibit a large difference. But once the grain size exceeds a certain level (190 μm), the fatigue performance of the material is significantly reduced.

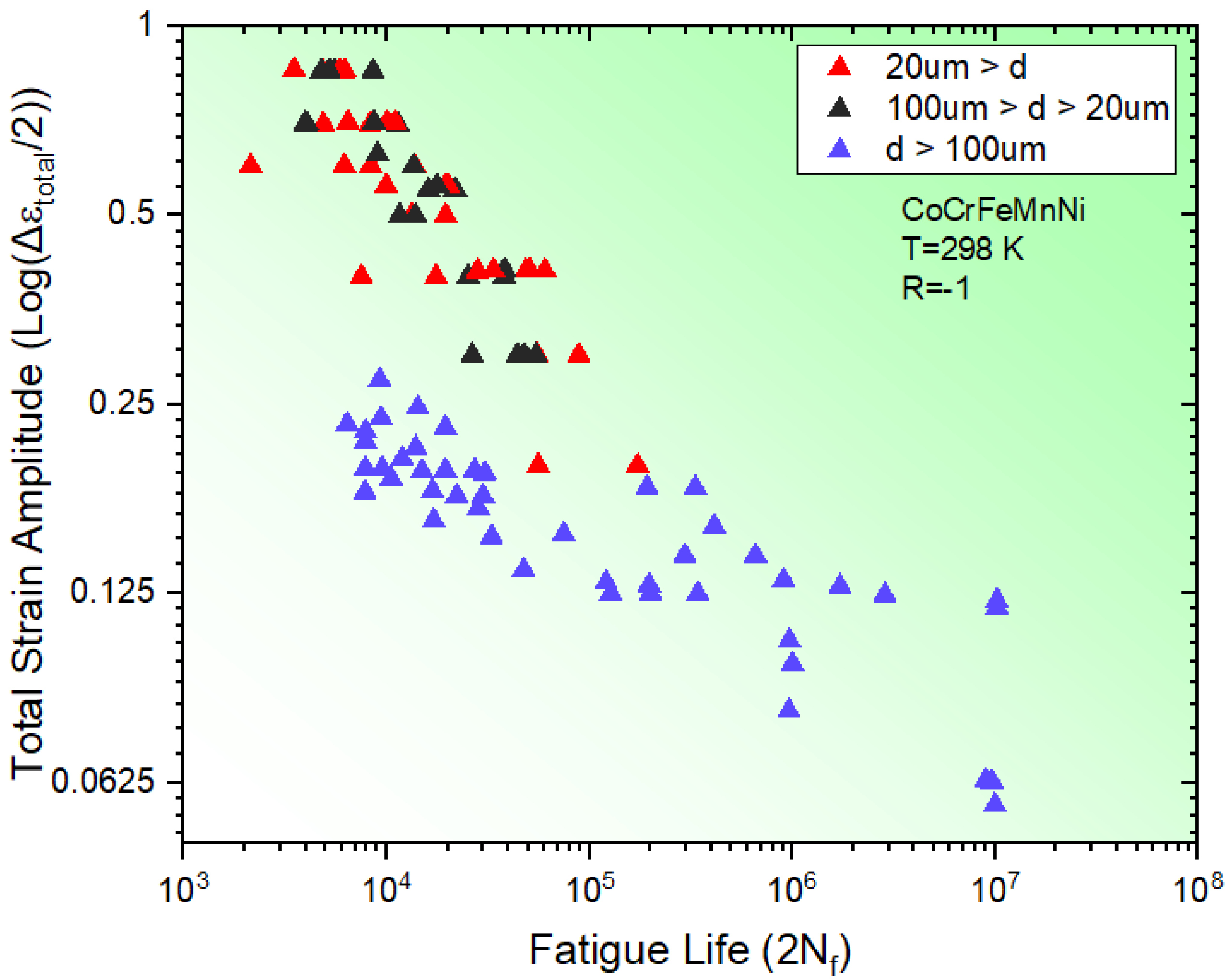

**Figure 2.2.6:** Fatigue performance of CoCrFeMnNi HEAs in the $\varepsilon_{total}$-$N$ mode [46-51].

The cyclic stress response involves another way to characterize the LCF resistance of materials. It is mainly used to record the force changes of materials under a given strain range. It can be seen from Figure 2.2.7 that the grain size exerts significant impact on cyclic stress response, under the same $\varepsilon_a$, in accordance with the following:

1. The smaller the grain size, the higher stress the material can withstand;
2. The smaller the grain size, the shorter cyclic hardening stage is observed;
3. The smaller the grain size, the higher the chance there is of observing a secondary cyclic-hardening (SCH) phenomenon.

Figure 2.2.7(a) shows data from different studies for the maximum strength vs. number of cycles for CoCrFeMnNi HEAs under the same strain amplitude [24, 51, 66-68]. By comparing the cyclic-stress response of the CoCrFeMnNi HEAs at the strain amplitude of 0.4%, the correspondence between the decrease in the grain size and the increase in the cyclic stress response is clear, especially for the specimens treated through equal channel angular pressing (ECAP).

Figure 2.2.7(b) and Figure 2.2.7(c) record the cyclic-stress responses of coarse-grained and fine-grained CoCrFeMnNi HEAs at strain amplitudes of 0.4%, 0.55%, 0.7%, and 0.85% [67], respectively. In general, the entire fatigue process is divided into two stages, a cyclic hardening stage and a cyclic softening stage. In the initial cycles, due to the rapid diffusion of dislocations and accumulation or interaction at grain boundaries, the material first experienced cyclic-hardening [68]. When the accumulation of dislocations reaches a critical point, annihilation of dislocations begins to occur, and the material enters the cyclic-softening stage. In the fine-grained CoCrFeMnNi HEAs, due to higher growth rate of dislocation density and lower twinning probability, the hardening peak appears earlier than in the coarse-grained HEAs, and the CoCrFeMnNi HEAs also enter the cyclic-softening stage sooner [68].

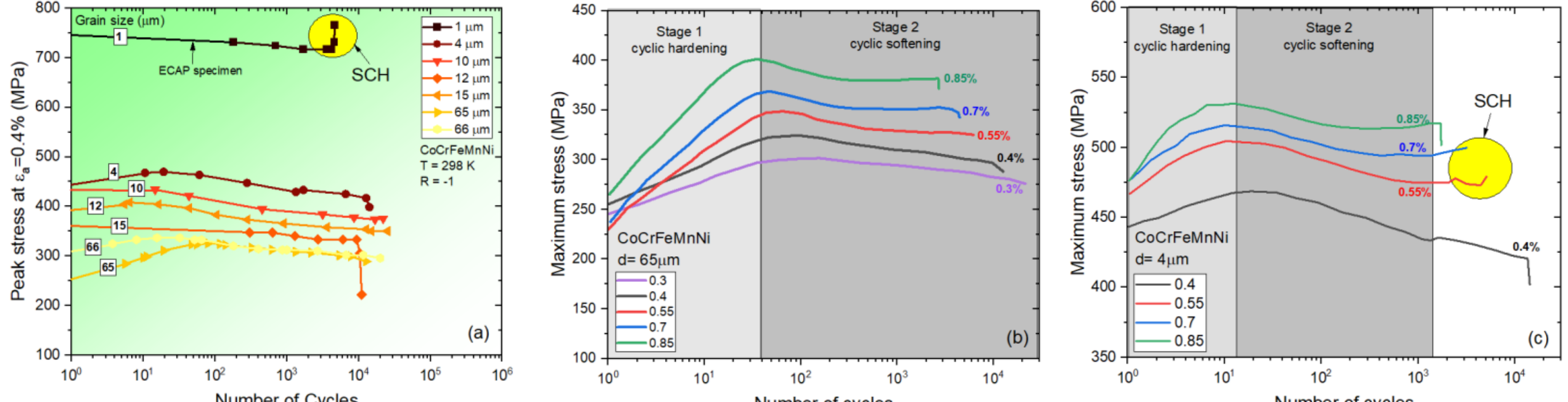

**Figure 2.2.7:** a) Cyclic-stress response of all CoCrFeMnNi HEAs [24, 51, 66-68]. b) Cyclic-stress response of fine-grain HEAs and c) coarse-grain CoCrFeMnNi HEAs under strain amplitudes of 0.3%, 0.4%, 0.55%, 0.7% and 0.85% [67].

The effect of strain amplitude on cyclic-stress responses of the CoCrFeMnNi HEAs is also very clear from Figure 2.2.7. The higher strain amplitude results in faster grain refinement and faster accumulation of dislocations, which leads to a more pronounced strength-hardening phenomenon and an earlier softening stage. Also, in the final stage near the occurrence of fracture, the fine-grained CoCrFeMnNi HEAs generate a large amount of dislocation slips under high strain amplitudes (under amplitudes of 0.55 % and 0.7%). The tendency of these slips to move back and forth along the grain boundary activates dislocation cross-slips and rearrangement into well-organized sub-crystalline structures, resulting in a unique secondary cyclic-hardening phenomenon [73]. The increase in the strain amplitude also impacts the structure of dislocation arrays. During low strain-amplitude cycling, the dislocation arrays are dominated by wall-like structures. As the amplitude increases, cell structures start to appear in the microstructure and become the dominant structures, as shown in Figure 2.2.8. Cell structures formed at high strain amplitudes are shown in Figure 2.2.9. Such structures greatly enhance the storage ability of dislocations and constitute one of the reasons for the secondary hardening phenomenon. But the hardening caused by this phenomenon is not as obvious or stable as the hardening caused by the grain refinement.

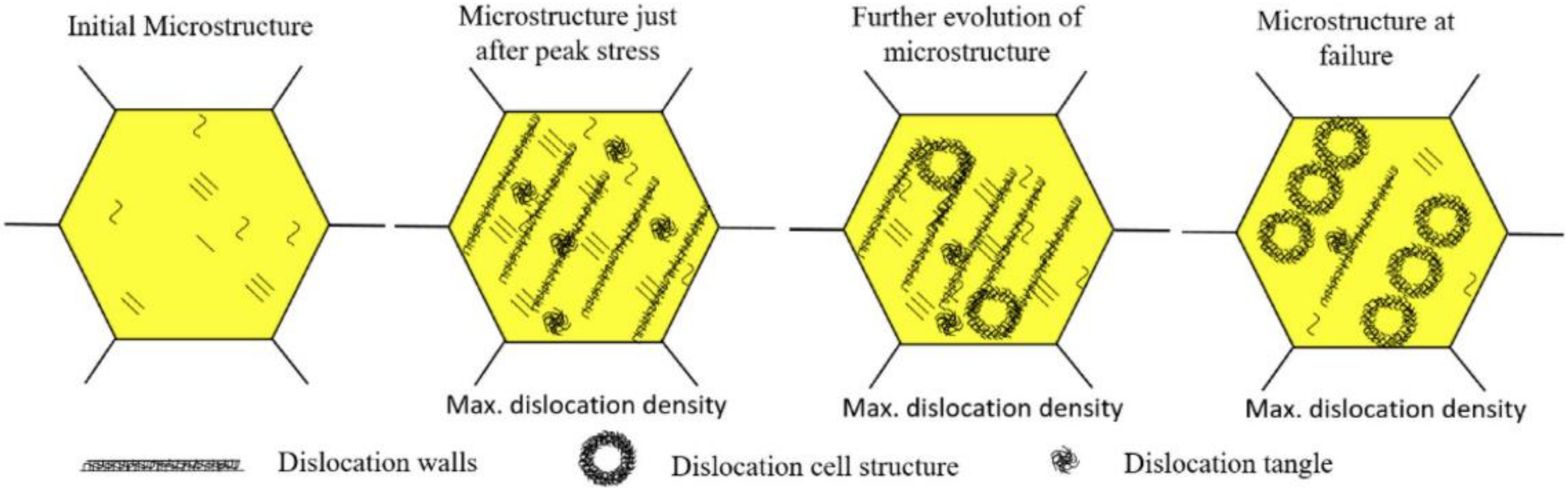

**Figure 2.2.8:** Schematic of the cell-structure formation in CoCrFeMnNi under high-strain amplitude [24].

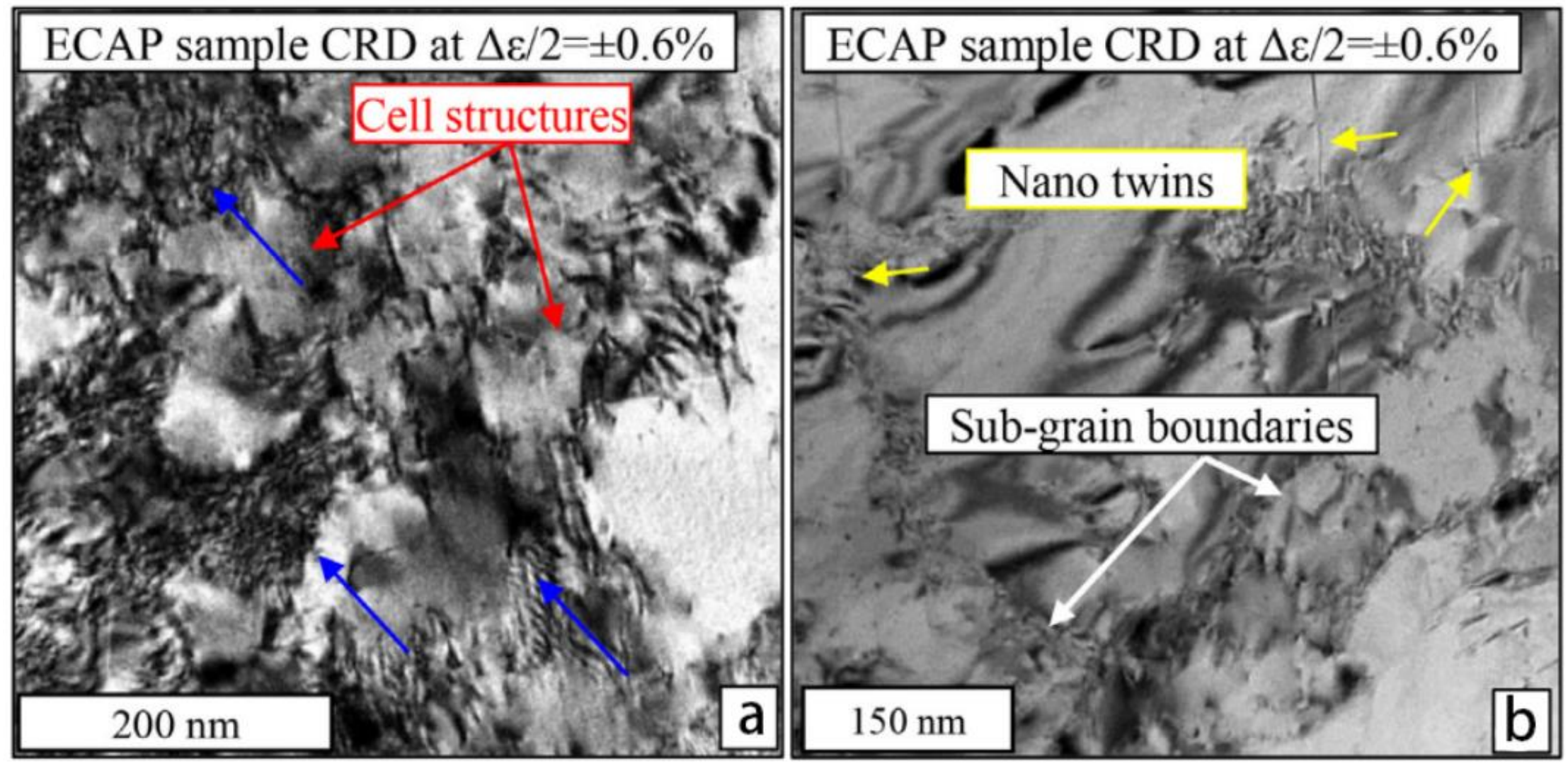

**Figure 2.2.9:** a) Cell structures and b) nano-twins found in an ECAP-processed CoCrFeMnNi at strain amplitude of 0.6% [24]. Nano-twin structures, lattice defects and cell structures are represented by yellow, blue and red arrows, respectively.

### *2.2.4. Multiphase HEAs*

Figure 2.2.10 shows LCF performance of the $Al_{0.5}CoCrFeNi$ and $Al_{10}(CrFeMnNi)_{90}$ dual-phase HEAs **[25, 70]**. Since both HEAs belong to the $Al_xCoCrFeNi$ family, they are both composed of an FCC matrix with a small amount of BCC (B2) phase, but at different volume fractions. Generally speaking, the dual-phase HEA with larger volume fraction of the BCC phase is considered to possess greater fatigue resistance. The $Al_{10}(CrCoFeMnNi)_{90}$ HEA without grain refinement has extremely coarse grains of 642 μm, whereas the $Al_{0.5}CoCrFeNi$ HEA has much finer grains (grain size of 9 μm). This excessive gap in grain size greatly reduces the difference in the LCF result between the two dual-phase HEAs.

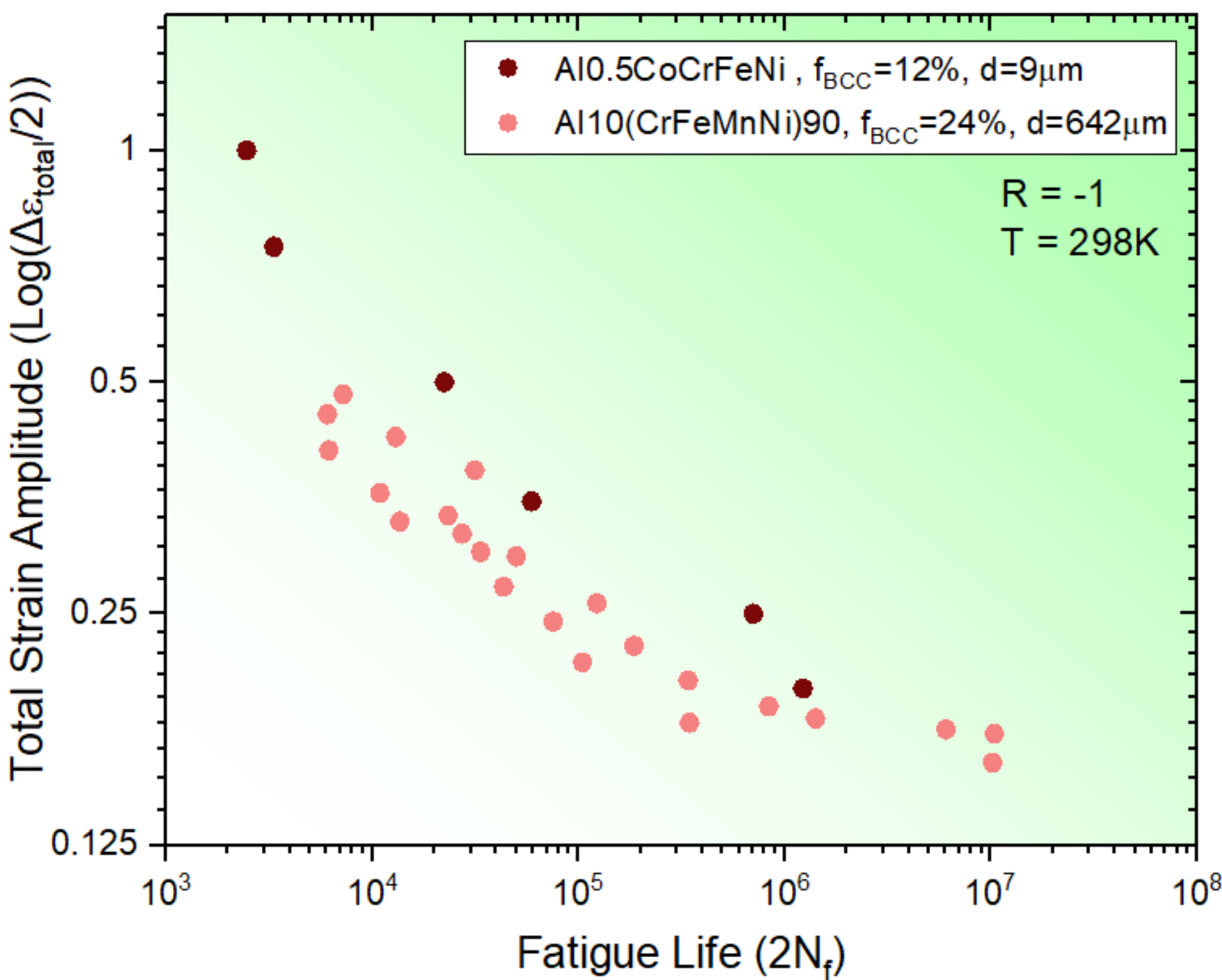

**Figure 2.2.10:** Strain-life fatigue data for the $Al_{0.5}CoCrFeNi$ and $Al_{10}(CrFeMnNi)_{90}$ dual-phase HEAs [25, 70].

Under low strain amplitudes, the dislocation structure in the FCC matrix of the $Al_xCoCrFeNi$ system is dominated by cross-slips and planar arrays. But as the strain amplitude increases, the dislocation structures gradually turn into cell structures. At the high strain amplitude of 1.75%, these cell structures get pulled into long strips. Twin formation is also observed at this strain amplitude, due to large concentration of dislocations resulting in local shear stresses exceeding the critical stress required for twin formation. Since strains between the FCC and BCC (B2) phases are not compatible, dislocations form first in the FCC matrix and accumulate at the boundary with the BCC phase during cycling. Deformation only starts to occur in the BCC phase, as the cycle period and strain amplitude increase. In the dual-phase HEAs comprising of FCC and BCC phases, cracks tend to nucleate at the interface between the two phases.

### *2.2.5. Metastable HEAs*

Based on the $\varepsilon$-$N$ data in Figure 2.2.11(a), it appears there is no significant difference in the LCF performance of $Fe_{50}Mn_{30}Co_{10}Cr$ metastable HEAs with slightly different grain sizes. In the Fe-based metastable HEAs studied [24, 27, 51, 66-68, 71, 72], transformation-induced plasticity (TRIP) is a widely observed phenomenon. The TRIP effect is manifested as a hardening phenomenon in the metastable HEAs under an external force. TRIP comprises a deformation-driven transformation from the FCC phase to HCP martensite phase [20]. This phenomenon is considered to be an important mechanism for improving HCF behavior in metastable HEAs. However, in Niendorf's study [27], presented in Figure 2.2.11(b), the cyclic-stress response of the $Fe_{50}Mn_{30}Co_{10}Cr_{10}$ metastable HEA under a uniform strain amplitude is at a level close to that of the CoCrFeMnNi FCC HEA. Also, no clear strain-hardening effect is observed in either coarse-grained (10 μm) or fine-grained (5 μm) specimens. One hypothesis for the TRIP effect is that it results from the planar nature of slips and from the partial reversibility of deformations weakening the interactions between the HCP martensite phase and dislocations. However, this hypothesis still needs more research in order to validate.

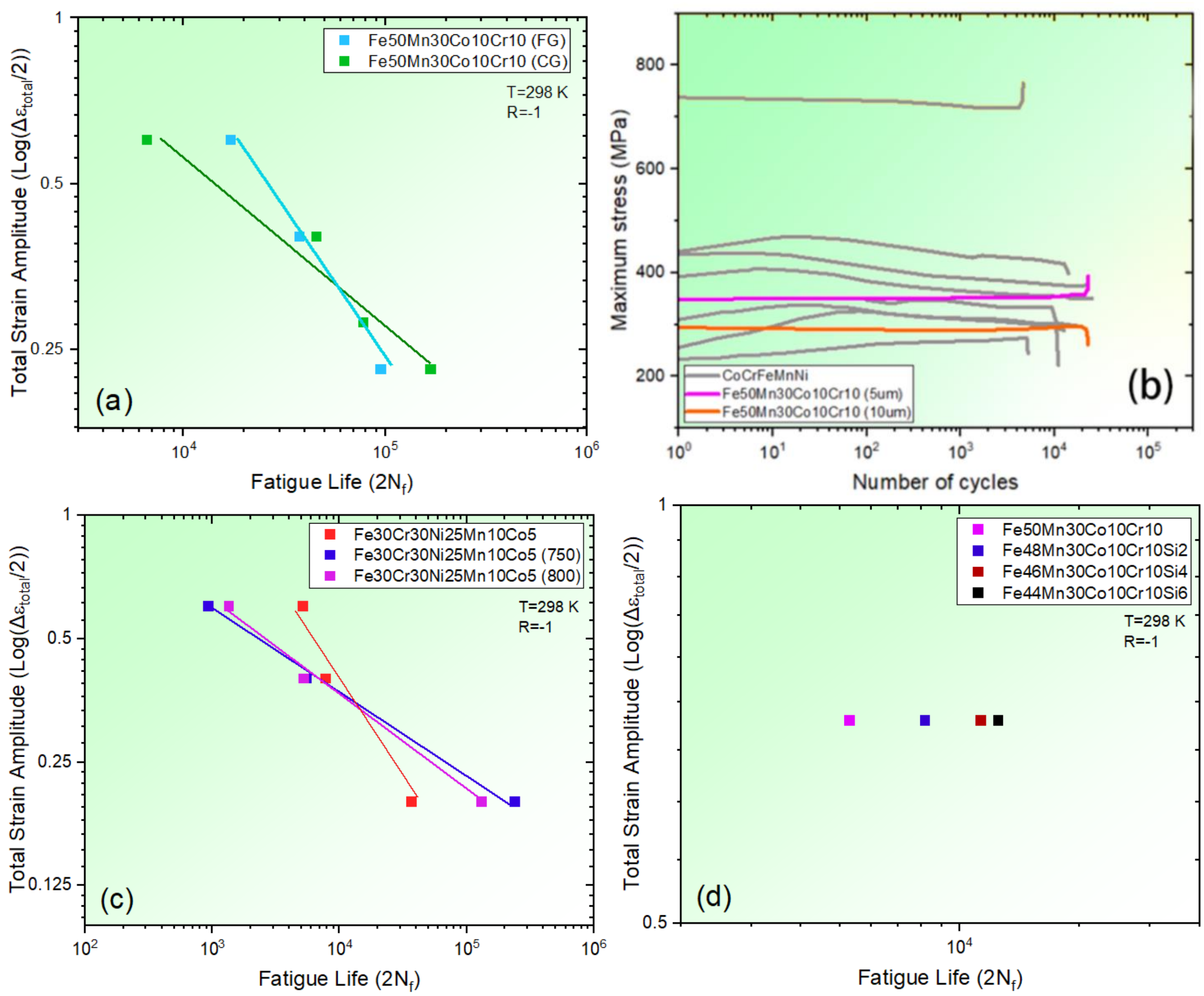

**Figure 2.2.11:** a) LCF performance of the $Fe_{50}Mn_{30}Co_{10}Cr$ and b) comparison of cyclic-stress response between the $Fe_{50}Mn_{30}Co_{10}Cr_{10}$ metastable HEA and CoCrFeMnNi used in Figure 2.2.7(a) at $\varepsilon_a = 0.4\%$ [24, 27, 51, 66-68], (c) [71], and (d) [72].

### *2.2.6. Comparison between HEAs and conventional alloys*

In this paper, the strain-life, i.e., the relationship between a given strain amplitude and fatigue life, is used for comparison of LCF performance between alloys. Because the strain-life data for the FCC, multiphase and metastable HEAs studied have much overlap, the LCF performance for these three categories of HEAs is collectively presented in gray in Figure 2.2.12. The strain-life data for the conventional alloys are obtained from references [20, 74-85] and presented for comparison. According to Figure 2.2.12, the LCF performance of the HEAs is overall superior to that of copper alloys, most steels, and aluminum alloys, and basically matches that of magnesium alloys and titanium alloys. The LCF performance of the HEAs is only inferior to that of the tin alloys.

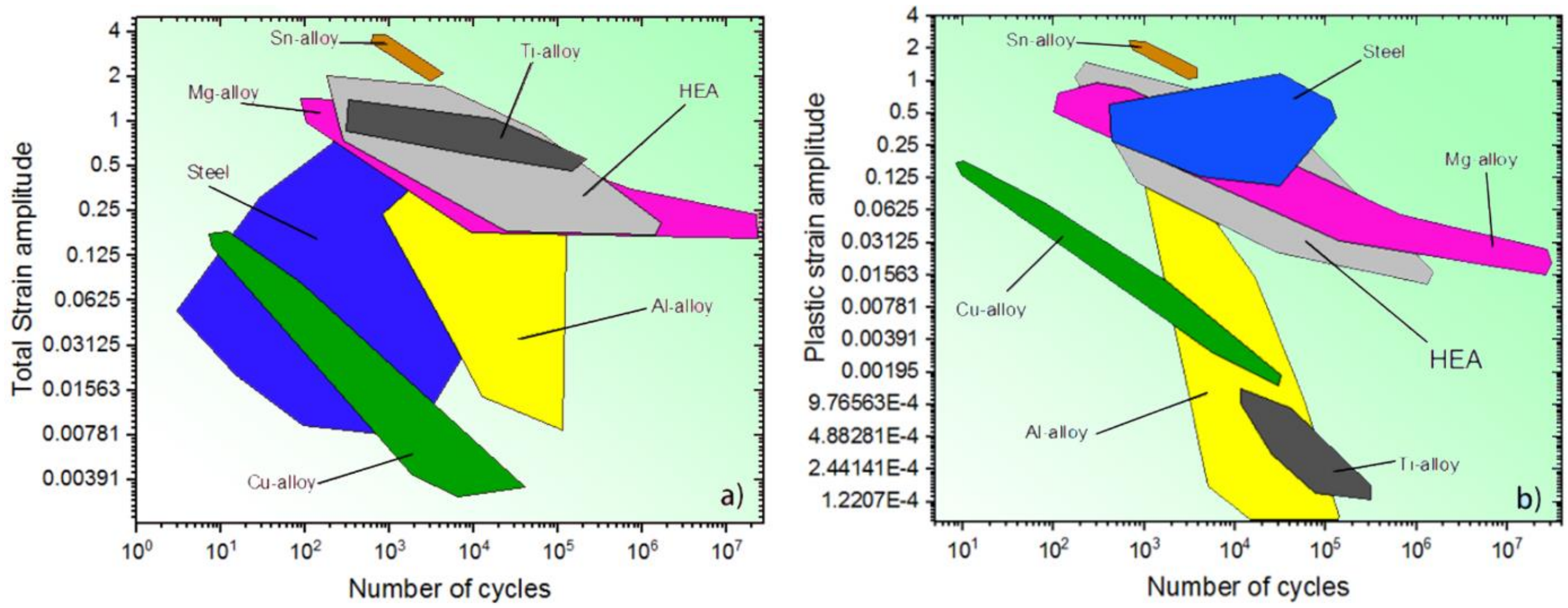

**Figure 2.2.12:** Comparison of LCF performance between HEAs and conventional alloys under a) total strain amplitude and b) plastic strain amplitude [20, 74-85].

## 2.3. Fatigue-crack-growth in HEAs

### *2.3.1. Introduction*

Cumulative Degradation – Crack Formation and Growth - Key Concepts, Assumptions, and Historical Perspective

Griffith is the “early father” of linear elastic fracture mechanics (LEFM) [14, 86]. In the 1920s, he found that the strength of a glass depended on the size of microscopic cracks. He, further, found that

$$\sigma \sqrt{a} = \text{Constant} \tag{2.3.1}$$

where $\sigma$ represents the glass strength, but *a* the crack size. In 1948, Irwin extended the work of Griffith by extending theories to ductile materials through inclusion of the energy dissipated by the local plastic flow [86]. In 1956, Irwin used analysis by Westergaard to introduce the concept of the *stress intensity factor* (SIF), *K*, as the amplitude of the crack-tip-stress field [86, 87]. The stress intensity factor, *K*, embodies loading and geometry conditions. In 1957, Irwin derived a relationship between the crack extension force, introduced by Griffith, and *K*, establishing the *K*-based fracture mechanics on firm footing, and ushering the era of modern fracture mechanics [86, 88]. Irwin discovered that a crack subjected to arbitrary loading could be resolved into three types of linearly independent cracking modes [8]. These modes are termed Modes I, II, and III and shown in Figure 2.3.1. Mode I comprises an opening (tensile) mode, where the crack surfaces move directly apart. Mode II involves a sliding (in-plane shear) mode, where the crack surfaces slide over one another in a direction perpendicular to the leading edge of the crack. Mode III consists of a tearing (anti-plane shear) mode, where the crack surfaces move relative to one another and parallel to the leading edge of the crack. The stress-intensity factors for these three modes can be defined as [89]

$$K_{\text{Mode I}} \equiv \lim_{r \to 0} \sqrt{2\pi r}\,\sigma_{yy}(r, 0). \tag{2.3.2}$$

$$K_{\text{Mode II}} \equiv \lim_{r \to 0} \sqrt{2\pi r}\,\sigma_{yx}(r, 0). \tag{2.3.3}$$

$$K_{\text{Mode III}} \equiv \lim_{r \to 0} \sqrt{2\pi r}\,\sigma_{yz}(r, 0). \tag{2.3.4}$$

For definitions of the polar radius, *r*, of the polar angle, *θ*, as well as of the stress fields near the crack tip, refer to Figure 2.3.2.

In 1961, Paris discovered that the fatigue-crack-growth rate, *da* / *dN*, is related to the *stress intensity factor range*, $\Delta K$:

$$\frac{da}{dN} = C\ (\Delta K)^{m}. \tag{2.3.5}$$

Here, we assume that *a* represents the crack length at the load cycle, *N* (cyclic stress loading), that an initial crack, $a_0$, already exists, and that $m$ is a constant. The stress intensity factor range is defined as

$$\Delta K = K_{max} - K_{min}, \tag{2.3.6}$$

where $K_{max}$ represents the maximum stress intensity factor in a load cycle but $K_{min}$ the minimum factor. Paris further proposed the concept of the threshold value of $\Delta K$ (threshold stress intensity factor range, $\Delta K_{\mathrm{th}}$), investigated the effects of the stress ratio, $R$, and made seminal contributions to the development of characterizing fatigue crack propagation and elastic-plastic fracture mechanics. Irwin is by many regarded as "Father", and Paris as the "God" of modern fracture mechanics.

The study of crack formation and growth is an important link in the study of the microstructural-degradation process that leads to fatigue fracture and failure. Research on the crack formation and growth can be divided into Crack Growth Analysis and Fracture Surface Analysis.

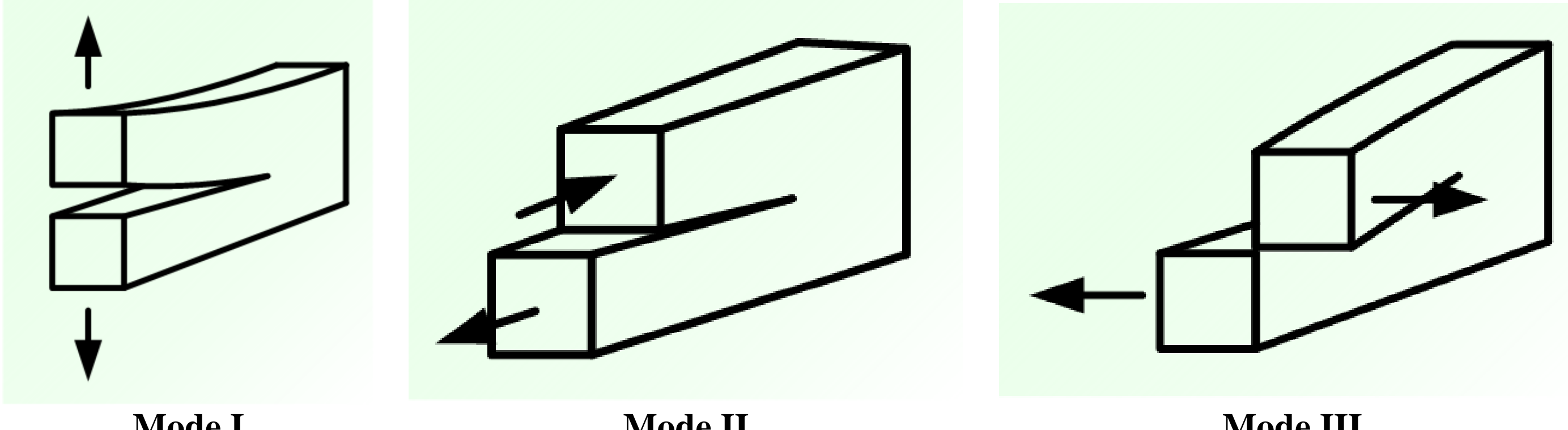

**Mode I** **Mode II** **Mode III**

**Figure 2.3.1:** Definition of stress-intensity modes for crack growth. Left: Mode I (Opening). Center: Mode II (In-plane shear). Right: Mode III (Out-of-plane shear) [8].

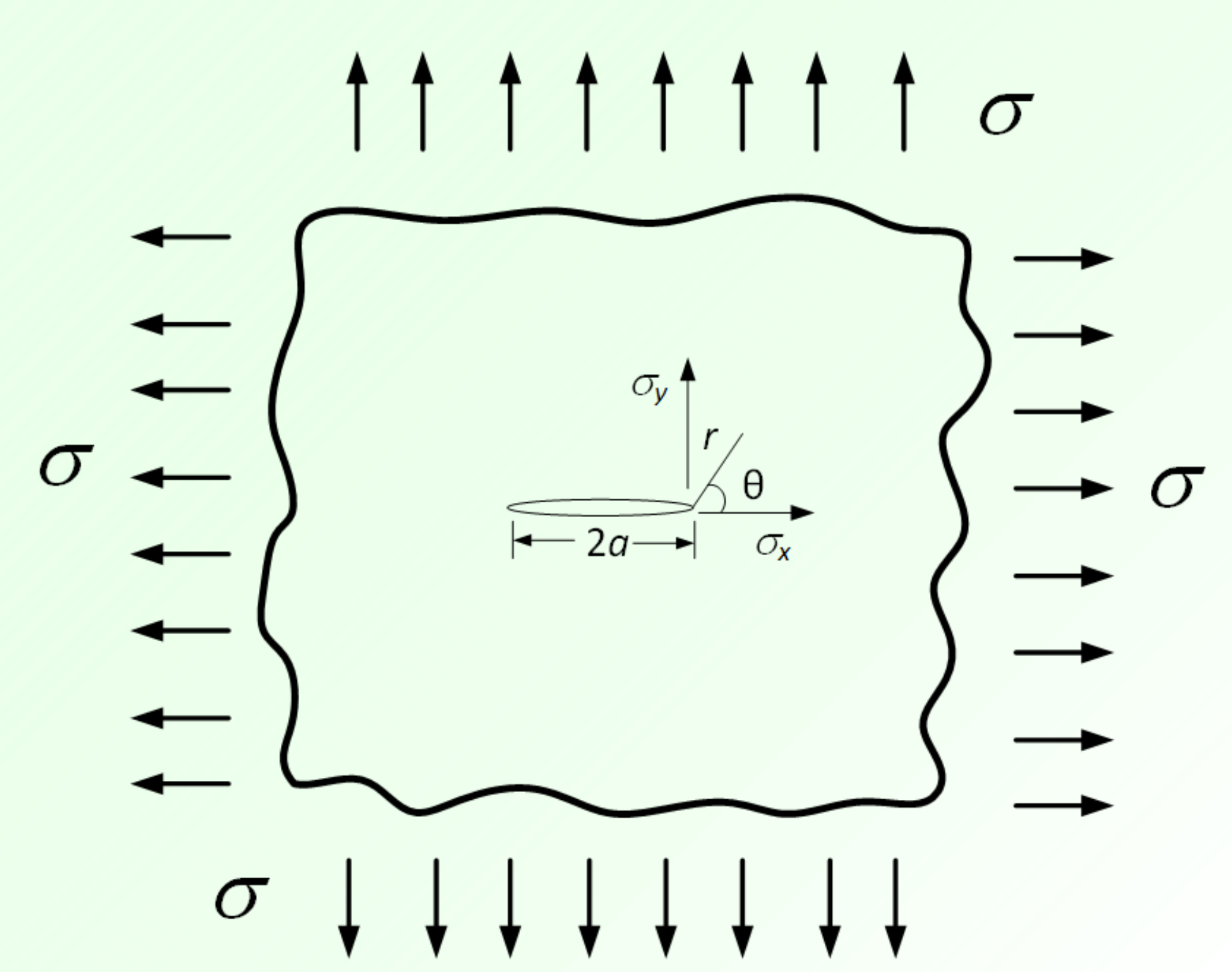

**Figure 2.3.2:** Definition of polar coordinates and stress components at the crack tip for a crack of length, $2a$, in an infinite plate [90].

Analysis of Fatigue Crack Growth

Fatigue crack-growth rate (FCGR) analysis under cyclic-stress loading is inherently a meso-structural (multi-scale) process [91]. The microstructural-degradation process usually goes through three stages that lead up to a fracture:

1. Stage I (crack nucleation): Initiation of a micro-crack, due to cyclic plastic deformation [91].
2. Stage II (crack growth): Progresses to a macro-crack that repeatedly opens and closes, creating the striation and beach marks [91].
3. Stage III (final fracture): The crack has propagated far enough that the remaining material is insufficient to carry the load, and fails by simple ultimate failure [91].

Figure 2.3.3 visualizes the crack-growth process for Stages I – III. In qualitative terms, dislocations can move back and forth during high-cycle loading. Then the dislocations may combine and form slip planes. Crack growth involving a single slip plane is considered Stage-I growth, but crack growth involving two slip planes is termed Stage-II growth. The dislocations and slip planes tend to combine, move to the surface of the sample, and form intrusions or extrusions there. Intrusions and extrusions comprise surface defects from which a crack can initiate. In the case of HCF, the cracks usually start from a surface defect. The HCF behavior is sensitive to surface finish (surface scratches). Here, one only tends to observe elastic deformations. Plasticity may only be observed in localized regions. In the case of LCF, one is looking at a similar process. But there the crack initiation will be faster, because it is assisted by great plasticity. In the case of LCF, the plasticity kicks in right away.

Further along these slip lines of dislocations, the crack occurs, in Stage I. When the stress intensity factor range is lower than a certain limit ($\Delta K_{th}$),

$$\Delta K < \Delta K_{th}, \tag{2.3.7}$$

the crack will not propagate. In Stage II, the crack grows at a linear rate in log ($da/dN$) - log ($\Delta K$) domain, in accordance with the Paris law [Eq. (2.3.5)], and as shown in Figure 2.3.4. In Stage III, the material has completely fractured. The factor, $K_c$, in Figure 2.3.4 represents the critical stress intensity factor (and the cyclic fracture toughness) of the material. The critical crack length, $a_c$, is the length at which a crack becomes unstable at a certain applied stress. In case of an unstable crack (unstable Stage-III growth), crack propagation continues spontaneously, once started, without an increase in the magnitude of the applied stress. The critical crack length, $a_c$, indicates the transition from a stable crack-growth regime to unstable crack growth that leads to catastrophic fracture or failure.
In regard to the S-N curve, Figure 2.3.5 offers a representation of the crack-initiation, crack-growth, and final failure process. The overall crack-growth fatigue life can be obtained as

$$\text{Total fatigue stress life} = N = \int dN = \int_{a_{min}}^{a_{max}} \frac{dN}{da}\, da = \int_{a_{min}}^{a_{max}} \frac{1}{\left(\frac{da}{dN}\right)}\, da, \qquad (2.3.8)$$

The S-N approach described in Section 2.1.1 measures the total fatigue life (the time to final failure) and cannot distinguish between the individual stages of fatigue crack initiation and growth.

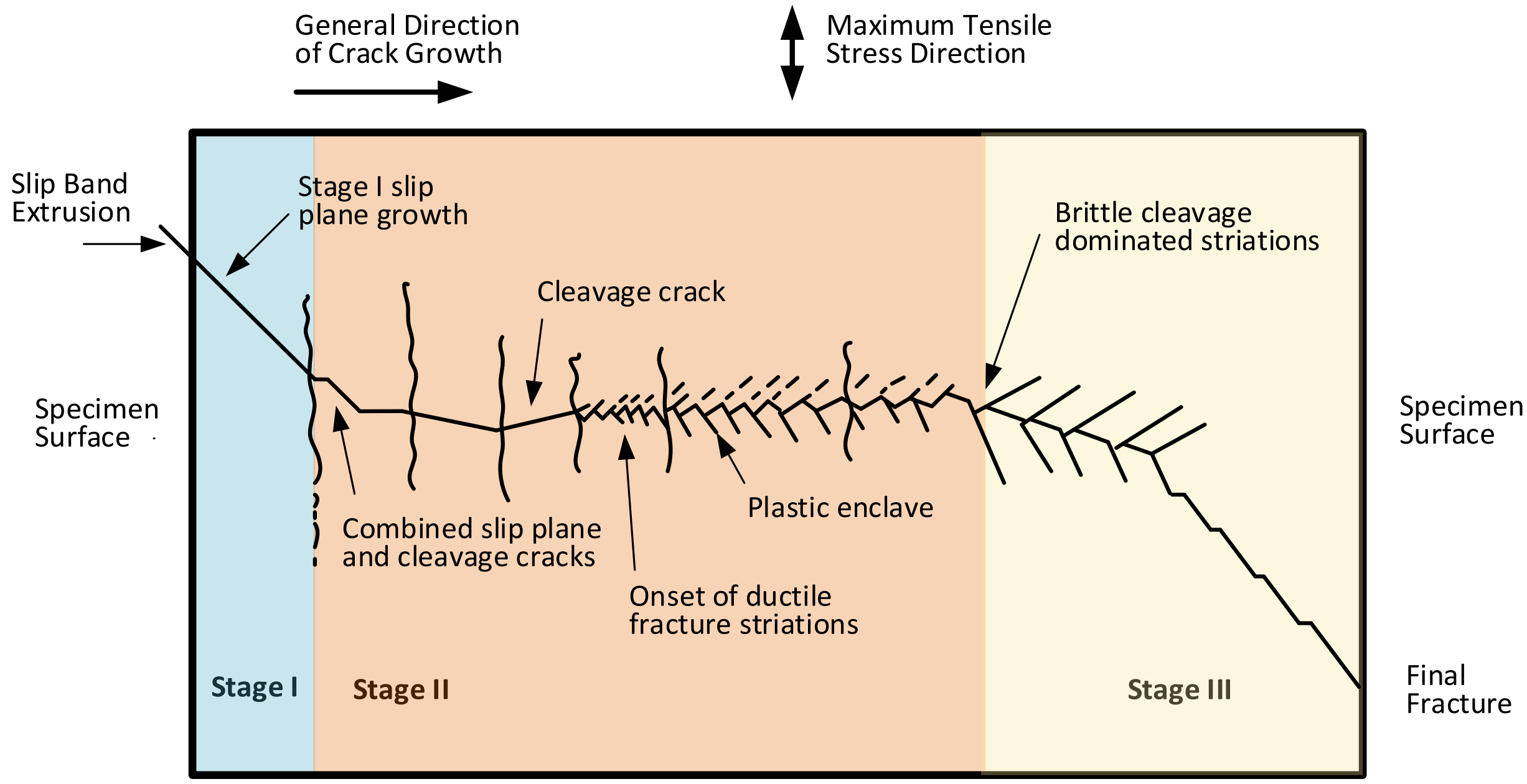

**Figure 2.3.3:** The crack-growth process for Stages I - III. The figure has been adapted from [92, 93].

<u>Measurements of the Stress Intensity Factor</u>

The stress intensity factor, $K$, is a theoretical construct characterizing the amplitude of the crack-tip-stress field [86, 87]. The SIF is a dimensionless measurement of the intensity of stresses and displacements near the trip of a fracture, one that varies with geometry and load. For a given geometry for a test specimen, the SIF can be measured using a variety of methods, including:

1. An analytical method

Being dependent on the specimen geometry, the size and location of the crack, and the magnitude and the distribution of loads in the specimen, the SIF can be written as [94, 95]:

$$K = \sigma \sqrt{\pi a}\; f(a/W), \qquad (2.3.9)$$

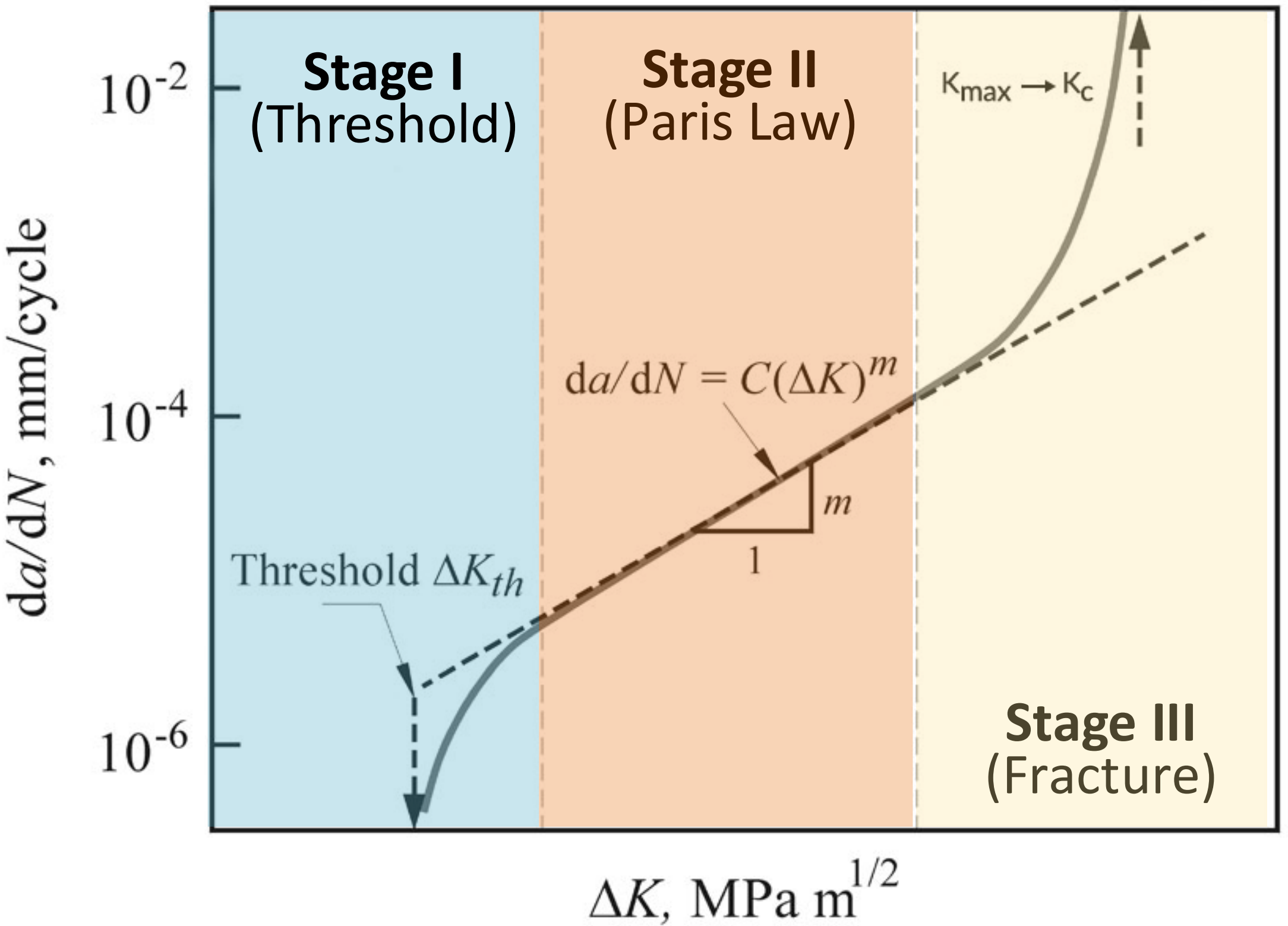

**Figure 2.3.4:** Typical behavior of the crack-growth rate (d*a*/d*N*) in relation to the stress intensity factor range ($\Delta K$). Adapted from [93].

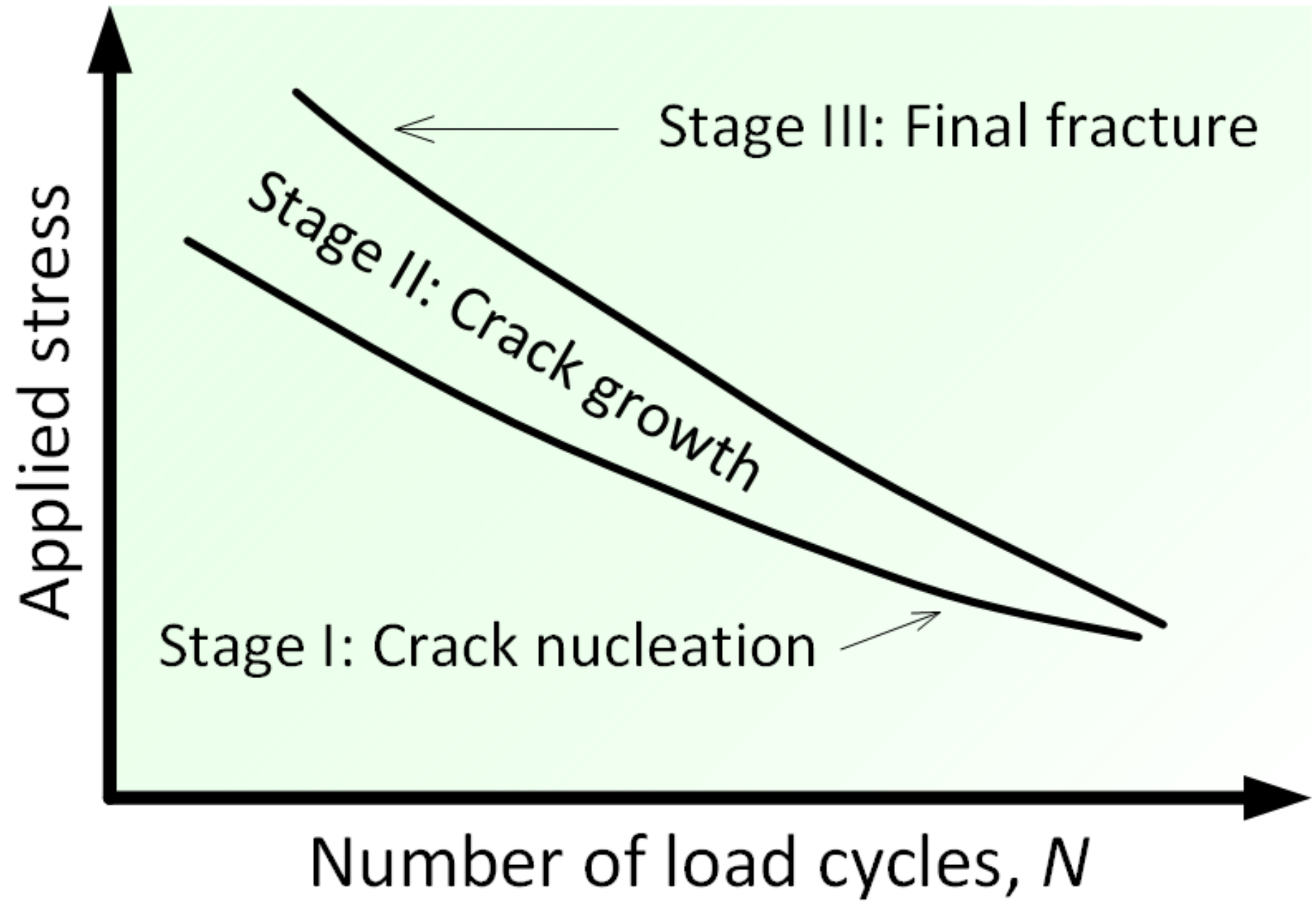

**Figure 2.3.5:** The three stages of crack growth represented in a S-N diagram. Adapted from [13].

Here, $f(a/W)$ is a function that depends on the geometry of the test specimen through the ratio $a/W$, where $a$ represents the crack length and $W$ the specimen width, and $\sigma$ denotes the stress applied.

2. A numerical method

In vicinity of the crack, the stress and displacement field can be calculated numerically using methods like the finite-element or the boundary element method.

3. A weight function method

The weight function method involves calculations of the stress and shear stress in an uncracked component, then by multiplying a resulting weight function by an internal stress function, and then integrating the product along the crack length.

<u>Measurements of Crack Growth</u>

Crack-growth analysis is usually carried out by cyclically loading pre-cracked samples and by observing the cracks using an optical microscope (OM), a scanning electron microscope (SEM), a transmission electron Microscope (TEM), unloading compliance, or electrical potential difference methods, as described below. During this process, a fatigue-crack-growth test equipment compliant with the American Society for Testing and Materials (ASTM) E647 standard for measurements of fatigue crack growth rate [96] will record the crack length, the number of cycles, the stress range, and calculate the crack-growth rate, $da/dN$, the stress intensity factor range, $\Delta K$, and will draw them as a curve similar to Figure 2.3.4.

The two most commonly used methods for measuring crack sizes (growth) involve the use of

1. Unloading compliance [97], [98]
2. Electrical potential difference [97], [99]

to determine the crack size.

The unloading compliance is defined as the reciprocal of the slope on a load-displacement curve normalized for the elastic modulus and specimen thickness, as shown in Figure 2.3.6 [97]. By partially unloading a test specimen at a certain deformation level, it is possible to calculate the slope of the load versus the crack mouth opening displacement (CMOD). When the crack grows, the sample stiffness will reduce, resulting in an increase of compliance. The relationship between the compliance and crack size has been analytically derived for a number of standard test specimens [100]. Such relationships are usually expressed in terms of the dimensionless quantities of compliance, $E\ v\ B\ /\ P$, and the normalized crack size, $a\ /\ W$, where $E$ represents the elastic modulus, $v$ denotes the displacement between measurement points, $B$ is the specimen thickness, $P$ represents a force (the load applied), $a$ is the crack size, and $W$ is the width of the test specimen [97].

The electric potential difference method for determining the crack size relies on the principle that the electrical field in a cracked specimen, with a current flowing through it, is a function of the specimen geometry, and in particular, the crack size [97]. The fundamental approach, outlined in Figure 2.3.7, utilizes the principle.

$$\begin{aligned}&(\text{Voltage across crack length})\\&\qquad = (\text{Current through sample}) * (\text{Resistance across crack length})\end{aligned} \tag{2.3.10}$$

In the case of a direct current (DC) approach, the current through the sample is kept constant. So as the crack extends, the resistance across the crack length increases, and hence, the voltage across the crack length. The relationship between the crack length and the voltage observed across the crack length can be expressed as [97]

$$a = f(\ V\ /\ V_r, a_r). \tag{2.3.11}$$

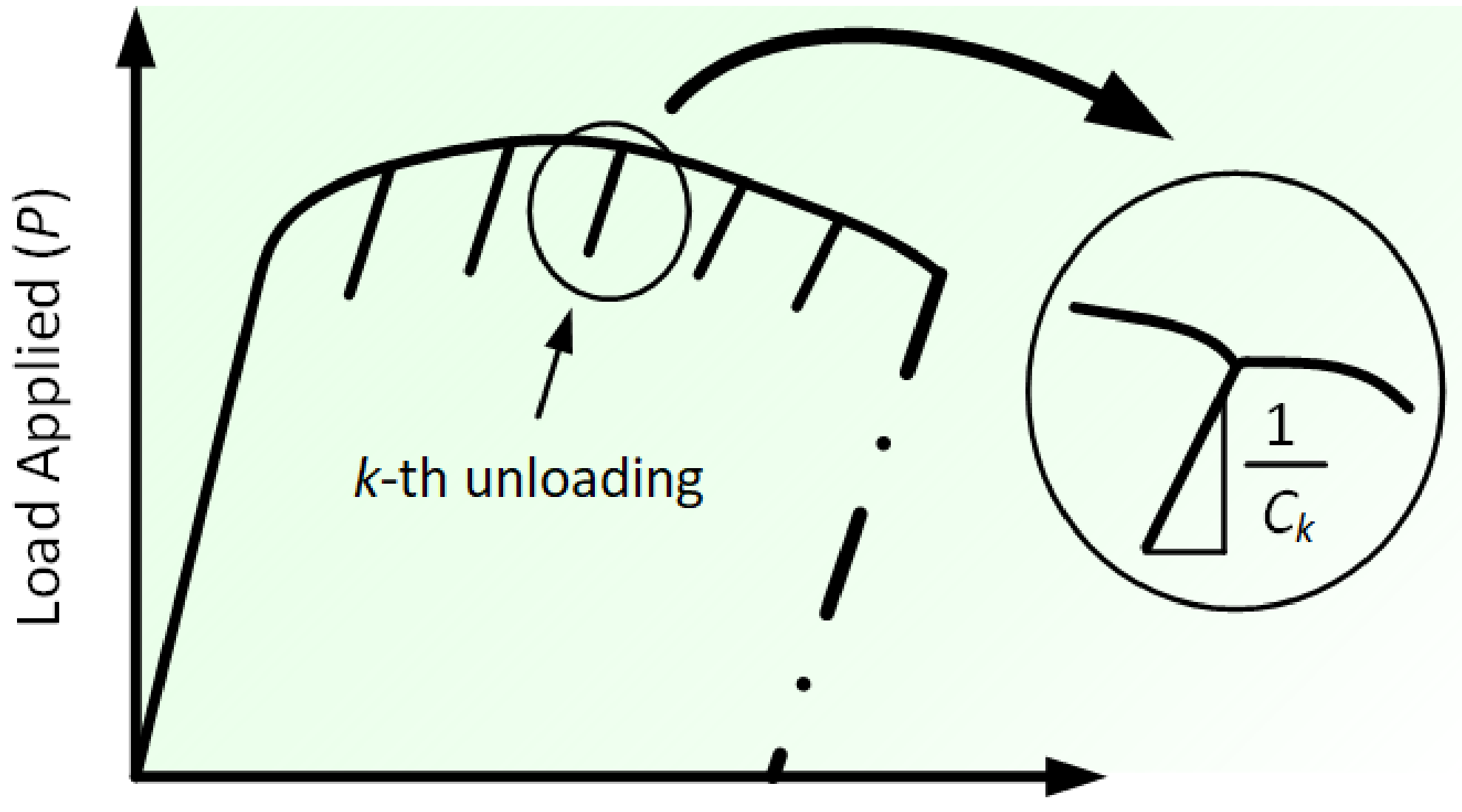

**Figure 2.3.6:** Compliance at the *k*-th unloading cycle [98, 101].

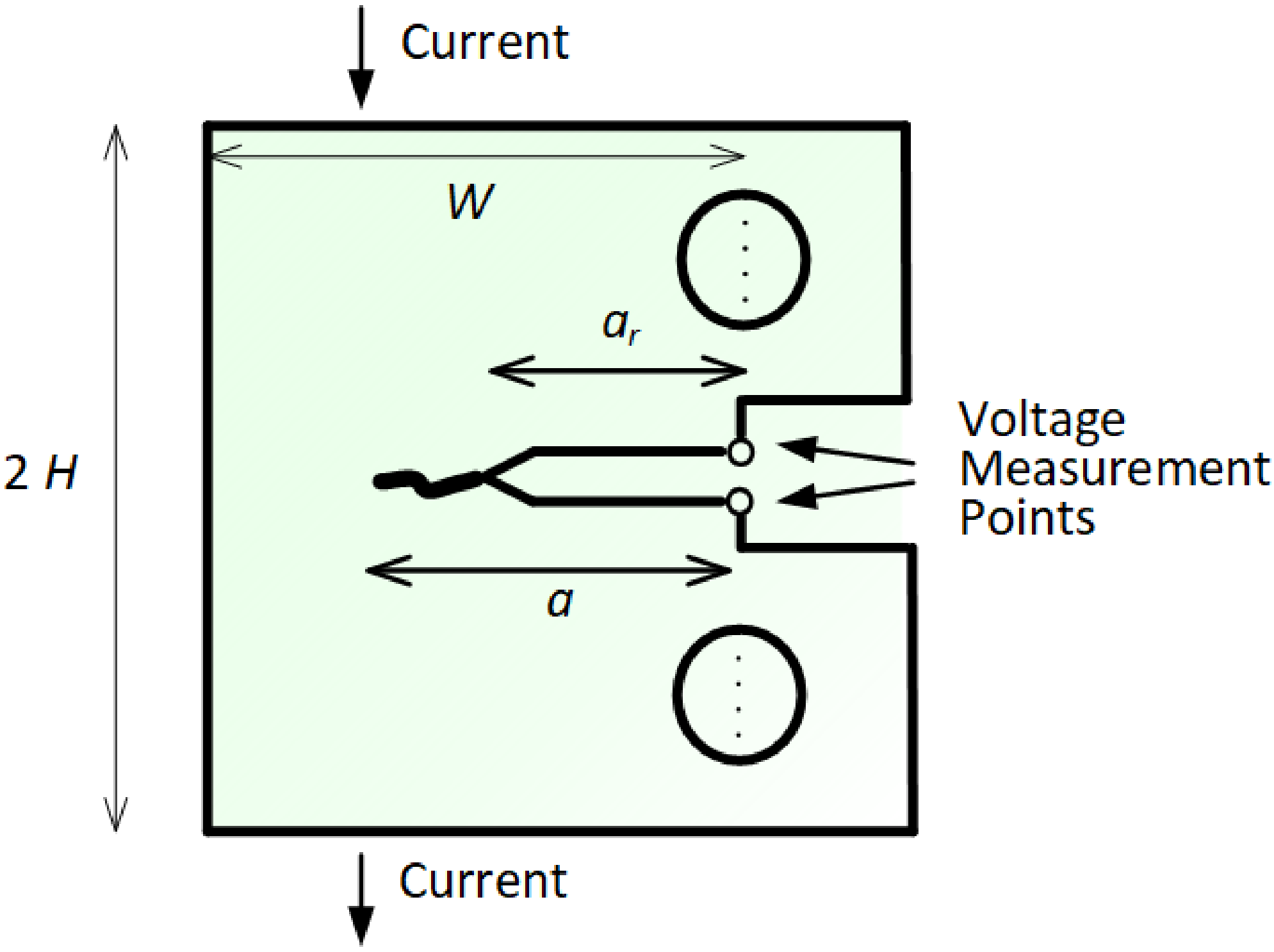

**Figure 2.3.7:** Sample application of the electrical-potential difference method for the measurement of the crack size (adapted from [99]).

Here $a$ represents the present crack length, $V$ the voltage presently observed across the crack length, $a_r$ a reference crack length, and $V_r$ a voltage observed across the crack length for the reference condition. The specimen width, $W$, and height, $H$, are defined in Figure 2.3.7.

The electric-potential difference procedures for the determination of the crack size are applicable to virtually any electrically conducting material and for a wide range of testing environments [97]. Both DC and alternating current (AC) techniques can be applied to measure the crack size, but the DC approach is more common.

Material scientists have done great work, in recent years, in terms of studying and measuring crack-growth rates. Recent progress includes in-situ fatigue measurements of subcritical crack growth with SEM or TEM [93].

Crack and Fracture-Surface Analysis

Crack and fracture surface analysis entails a study of the causes of material failure. Progression marks (also referred to as beach marks) are visible marks that show how a crack has grown (how a crack face has progressed across a surface), which allows the crack origin to be determined [102]. Chevrons (also referred to as herringbone patterns) resemble the beach marks, indicate the direction of crack growth, and allow the determination of the crack origin [103]. The fatigue striations are marks produced on a fracture surface that show the incremental growth of a fatigue crack. A fatigue striation indicates each stress cycle experienced by a component, is usually visible only under high magnification, and is typically only observed in Stage-II growth.

Fractures are sometimes referred to as being brittle or ductile. In the case of a ductile fracture, one typically observes dimples on the fracture surface. Figure 2.3.9(a) shows high density of dimple-like structures distributed on the fracture surface of the $CoCrFeNiTiAl_{0.5}$ HEA [104]. However, as shown in Figure 2.3.9(b), typical cleavage fracture with river-like patterns and cleavage steps can be observed from the fracture surface of the $CoCrFeNiTiAl_{0.5}$ alloy [104].

**Stage I: Crack Initiation and Slip Plane Growth**

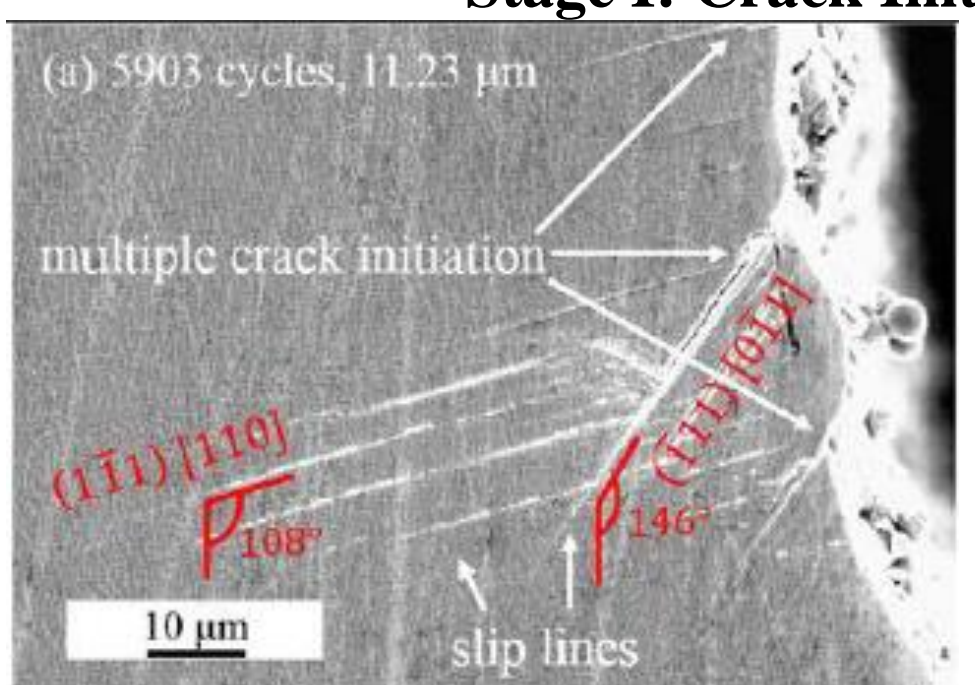

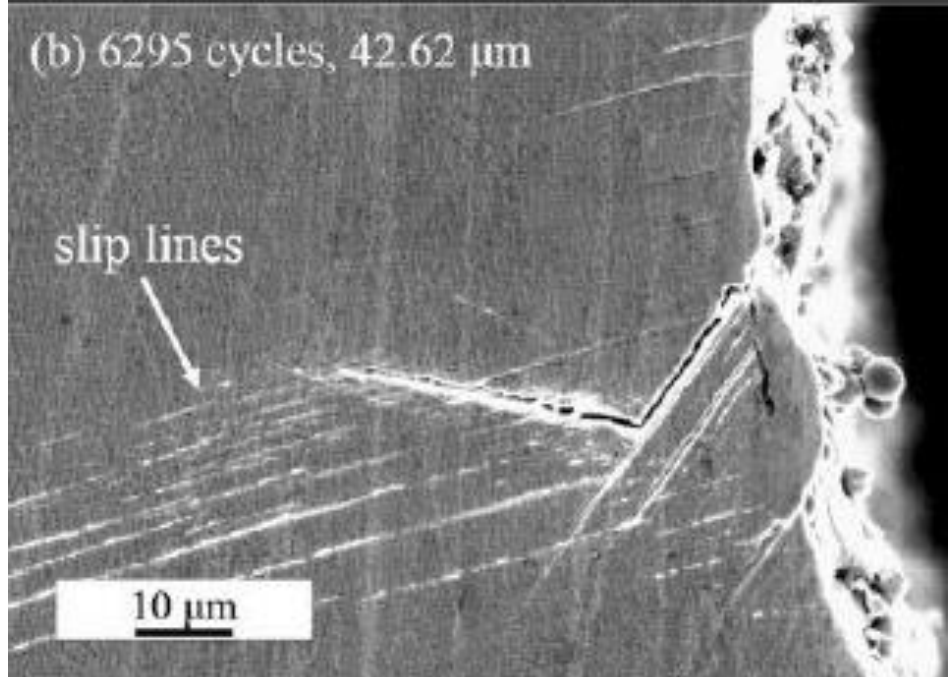

**Stage II: Secondary Crack Initiation**

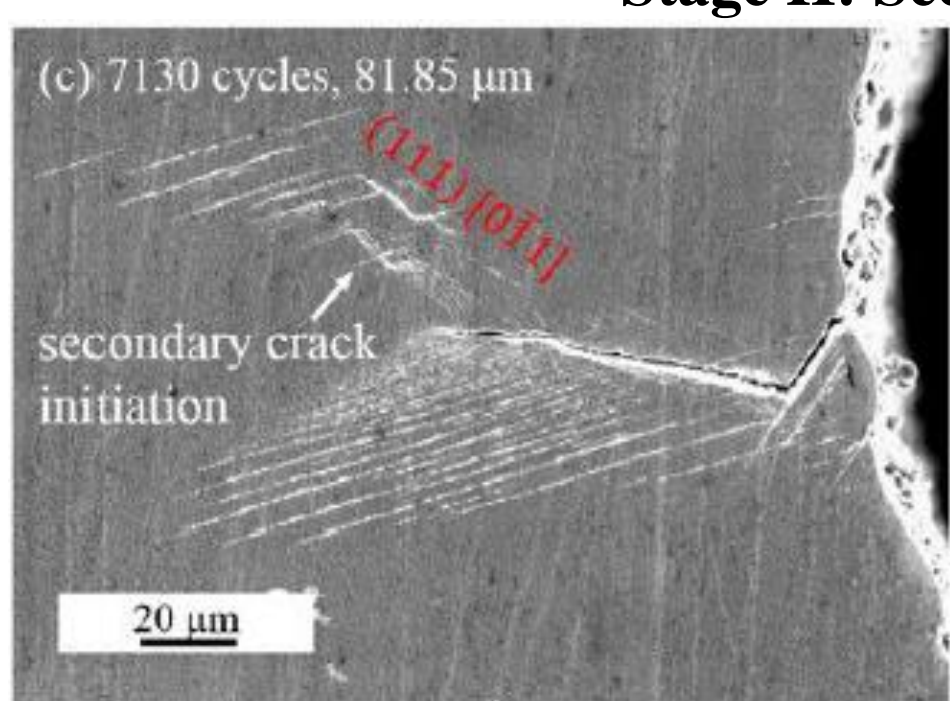

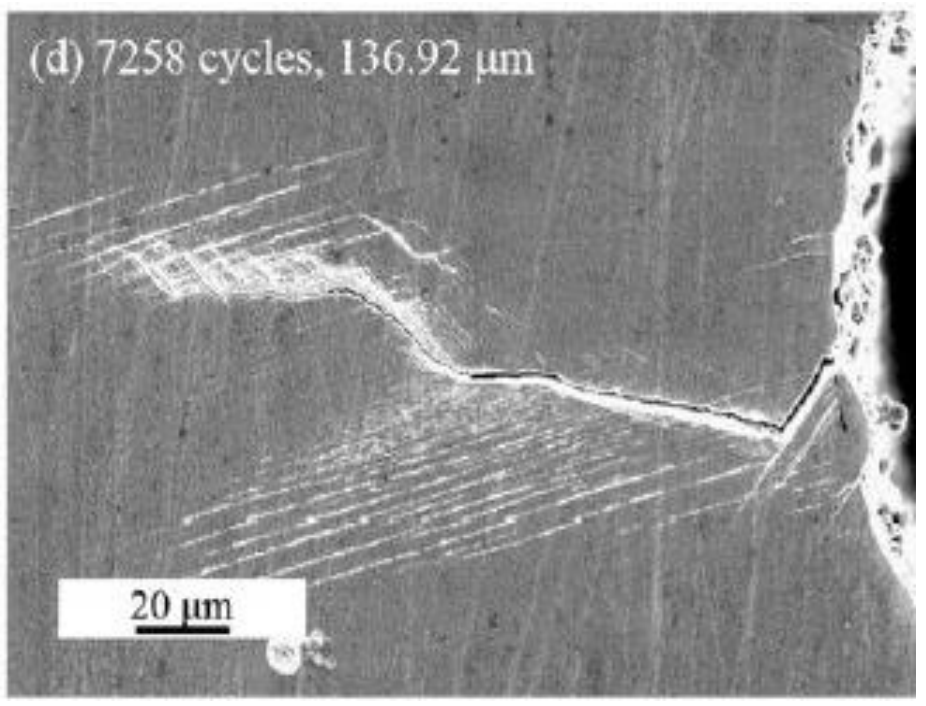

**Stage III: Large Zig-Zag Path and Crack-Tip Opening Displacement**

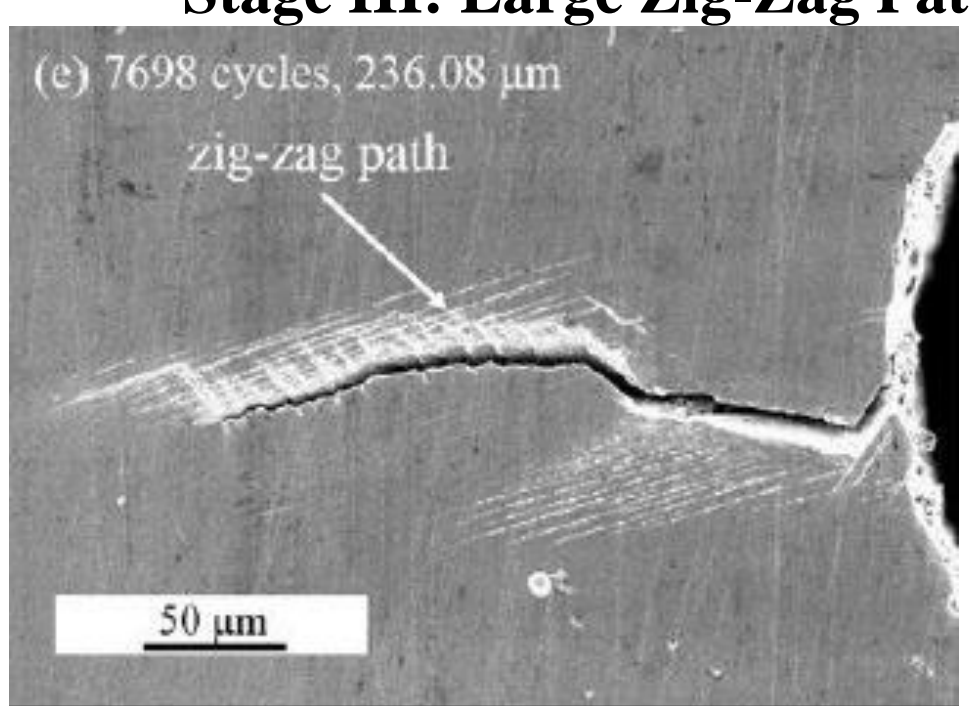

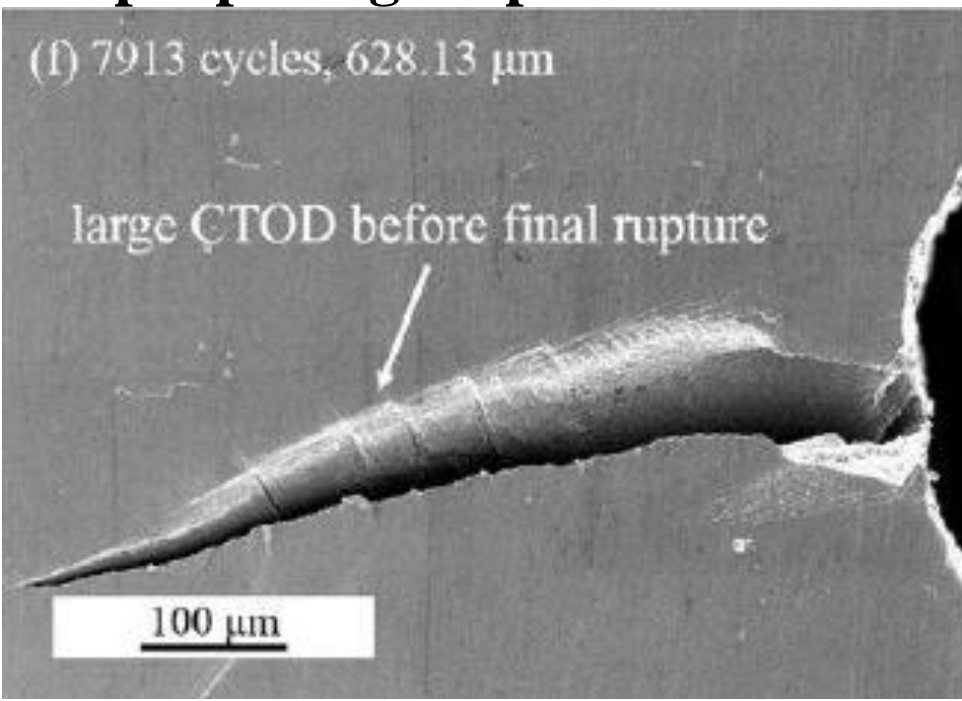

**Figure 2.3.8:** In-situ SEM observation of fatigue-crack initiation and slip-controlled crack growth for a [001]-oriented single-crystal Ni superalloy specimen loaded at RT and with a vertical loading direction [93, 105].

**Fracture surface with dimples (ductile fracture)** **Fracture surface with cleavages (brittle fracture)**

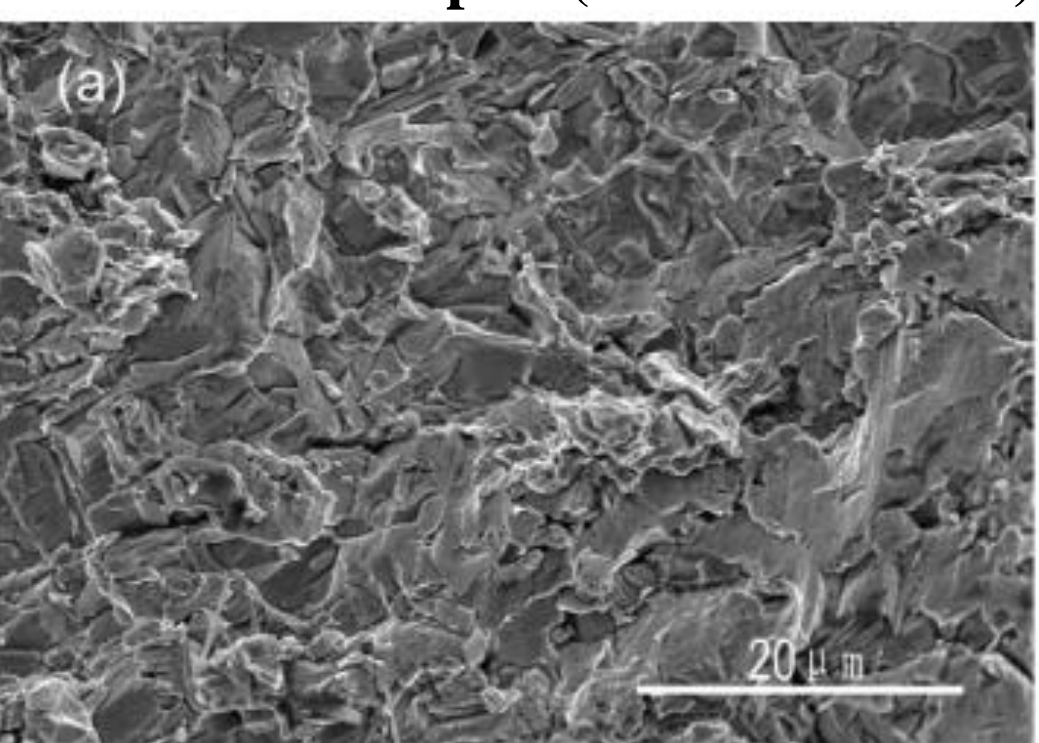

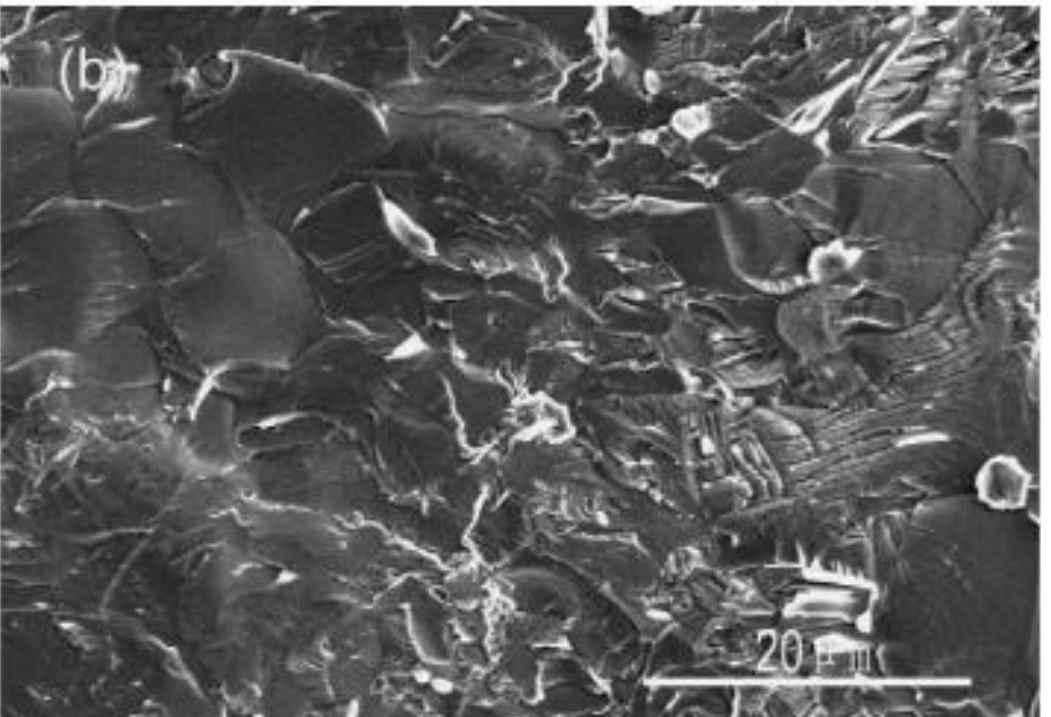

**Figure 2.3.9:** Representative SEM images of ductile and brittle HEA fracture surfaces [104].

### *2.3.2. FCGR of HEAs*

In terms of fatigue-crack growth, we collected data from 10 studies, including three studies on FCC HEAs (CoCrFeMnNi, CoCrFeNi, and $CoCrFeNiMo_{0.2}$) [30, 106, 107], two studies on multiphase HEAs ($AlCrFeNi_2Cu$ and$Al_{0.2}CrFeNiTi_{0.2}$) [31], one study on metastable HEAs ($Fe_{30}Mn_{10}Cr_{10}Co$) [33], and one study on BCC HEAs (HfNbTaTiZr) [17]. All the experiments were done at RT and standard air pressure, and the stress ratio and frequency were within the same standard range ($R$ = 0.1 and $f$ = 1 – 25 Hz). Detailed information, including the grain size, the ultimate tensile strength as well as experimentally fitted $\Delta K_{th}$ and $m$ values, is recorded in Table 2.3.1.

**Table 2.3.1:** FCGR behavior of HEAs, including threshold stress intensity factor range and Paris index, obtained under different test conditions (methods, frequencies, etc.) at room temperature.

[*1]SENT = Single-edge notch tension [*2]CT = Compact tension, [*3]DC(T) =Disc-shaped compact-tension, [*4]SENB = Single-edge notch bend.

| Material | Phase | Grain Size [μm] | *E* | YS [MPa] | UTS [MPa] | Temperature [K] | *R* | Frequency [Hz] | Testing samples and methods | $\Delta K_{th}$ | *m* | Ref. |
|---|---|---|---|---|---|---|---|---|---|---|---|---|
| CoCrFeNi | FCC | 15 | 221 | 250 | 650 | 298 | 0.1 | 6Hz | SENT[*1] | N/A | 6.2 | [106] |
| $CoCrFeNiMo_{0.2}$ | FCC | 12 | 228 | 375 | 780 | 298 | 0.1 | 6Hz | SENT | N/A | 7.4 | [106] |
| CrMnFeCoNi | FCC | 7±3 | 202 | 410 | 760 | 293 | 0.1 | 25Hz | DC(T) [*3] | 4.8 | 3.5 | [30] |
| CrMnFeCoNi | FCC | 7±3 | 209 | 520 | 925 | 198 | 0.1 | 25Hz | DC(T) | 6.3 | 4.5 | [30] |
| CoCrFeMnNi | FCC | N/A | N/A | N/A | N/A | 298 | 0.1 | 10Hz | CT | N/A | N/A | [108] |
| CoCrFeMnNi<111> | FCC | N/A | N/A | 170 | 650 | 298 | 0.05 | 1Hz | N/A | N/A | 3.21 | [109] |
| CoCrFeMnNi<001> | FCC | N/A | N/A | 150 | 440 | 298 | 0.05 | 1Hz | N/A | N/A | 2.51 | [109] |
| $AlCrFeNi_2Cu$ | FCC + BCC + undefined phase | N/A | N/A | N/A | N/A | 293 | 0.1 | 20Hz | SENB[*4] | 17 | 3.4 | [31] |
| $Al_{0.2}CrFeNiTi_{0.2}$ | FCC + BCC + undefined phase | N/A | N/A | N/A | N/A | 293 | 0.1 | 20Hz | SENB | 16 | 4.9 | [31] |
| $Fe_{30}Mn_{10}Cr_{10}Co$ | FCC → HCP | N/A | N/A | 172 | 733 | 298 | 0.1 | 1Hz | CT[*2] | N/A | 3.5 | [33] |
| HfNbTaTiZr | BCC | N/A | 92 | 953 | 1,095 | 298 | 0.1 | 10Hz | SENB | 2.5 | 2.2 | [17] |

The crack-growth profiles for the four categories of HEAs are presented in Figure 2.3.10. These four categories of HEAs are labeled as follows: Triangle (FCC HEAs), square (metastable HEAs), circle (multiphase HEAs), and diamond (BCC HEAs). Since some of the crack-growth profiles only intercept the data of the Paris-region, the point of the starting data cannot simply be regarded as the $\Delta K_{th}$, but should be based on the data in Table 2.3.1. In terms of the fatigue crack growth, the following two criteria are used for assessing the fatigue resistance: (1) A higher $\Delta K_{th}$ value and (2) a lower crack growth rate under the same $\Delta K$. Both criteria indicate that the material has better fatigue-crack-growth resistance. According to these two criteria, the fatigue strengths of the four categories of HEAs are ranked as follows: The metastable HEAs ≈ multiphase HEAs > FCC HEAs > BCC HEAs. To this effect, the $\Delta K_{th}$ values for the metastable HEAs and the multiphase HEAs are the largest, the $\Delta K_{th}$ value of the FCC HEA are the second largest, and the $\Delta K_{th}$ values for the BCC HEAs are the smallest. In the Paris region (Stage II), the multiphase HEA exhibit the lowest crack-growth-rate, followed by the FCC HEAs and then by the metastable HEAs. The BCC HEAs exhibit the highest crack-growth rate.

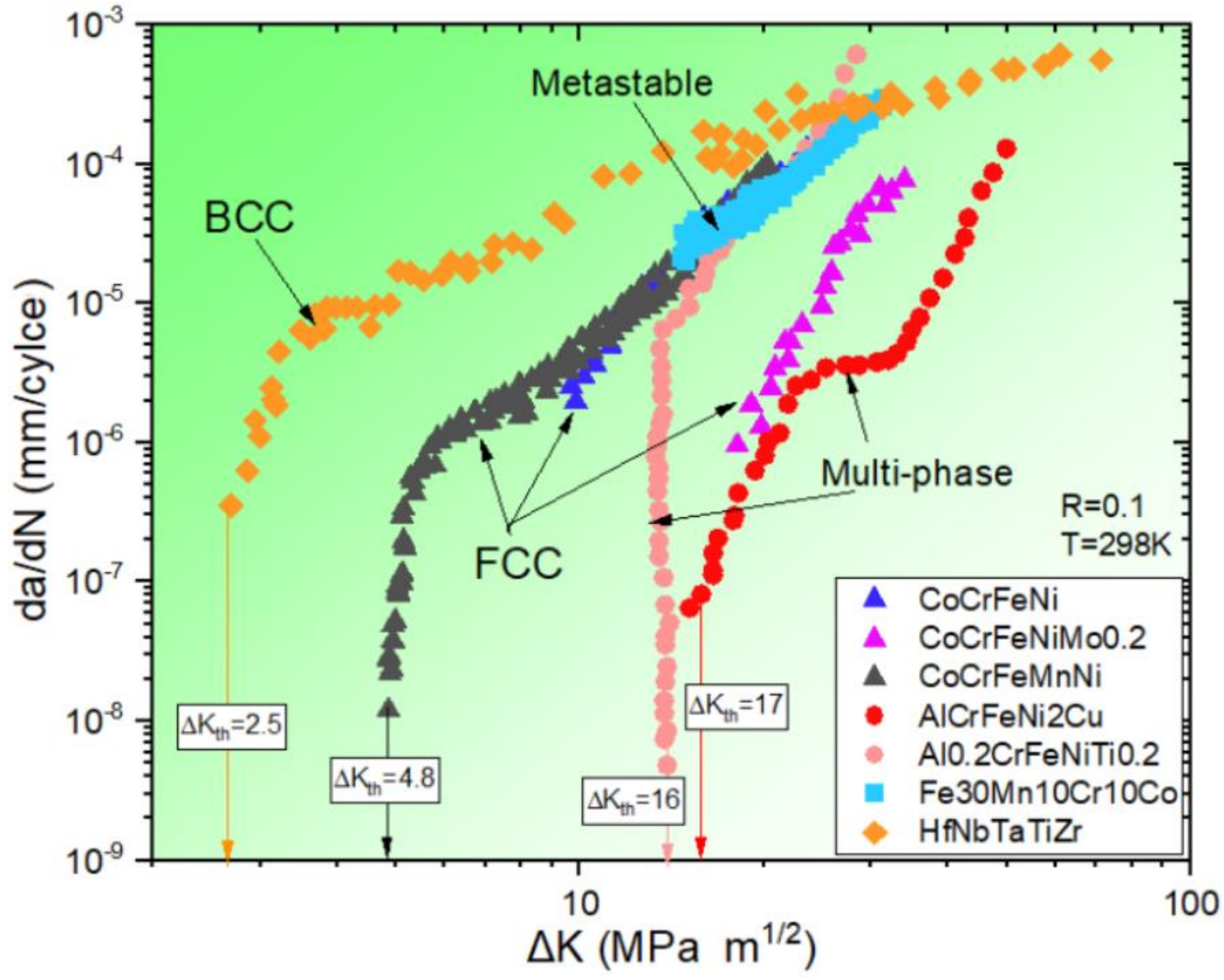

**Figure 2.3.10:** Rate of fatigue crack growth for four categories of HEAs [17, 30, 31, 33, 106, 109-111].

These results nicely illustrate the difference between the BCC HEAs, the FCC HEAs, the multiphase HEAs, and the metastable HEAs, in terms of the fatigue crack growth. The source of the extremely high fatigue strength for the multiphase and metastable HEAs pertains to original or TRIP-induced second phase compositions in these HEAs [17, 31, 33]. Such structures can passivate and deflect cracks, while improving the strengths of HEAs, hence contributing to favorable FCGR performance of the multiphase and metastable HEAs. In contrast, the source of the strength for the BCC HEAs relates to their unstable stacking fault energy and multi-slip systems, rather than the interaction between constituent phases. In the case of the BCC HEAs, more energy is needed to induce planar slips, compared to the other HEA categories. But once the planar slips have been initiated, the rate of slip is higher, which results in the BCC HEAs possessing higher strength but weaker FCGR resistance, compared to the other HEA categories [112]. The differences in the fatigue mechanisms between the BCC and other HEAs will be described in detail in Section 3.1.

### *2.3.3. FCC HEAs*

Among the three FCC HEAs studied, CoCrFeMnNi, CoCrFeNi and $CoCrFeNiMo_{0.2}$, the FCGR performance of the CoCrFeMnNi HEA is closer to that of the CoCrFeNi HEA than to that of the $CoCrFeNiMo_{0.2}$ HEA. Figure 2.3.11(a) offers comparison of the FCGR performance of the CoCrFeNi and the $CoCrFeNiMo_{0.2}$ HEAs. The inhibitory effect of Mo on the cracking of CoCrFeNi HEA can be seen from the lower crack propagation rate of the $CoCrFeNiMo_{0.2}$ HEA. This is because the larger atomic radius of Mo distorts the lattice, which inhibits the movement of dislocations. In addition, the activation energy for nucleation of dislocations at the crack tip of $CoCrFeNiMo_{0.2}$ affects the plastic strain of the material and thus the crack propagation rate. Li et al. [106] measured the average irreversible strain increment of $CoCrFeNiMo_{0.2}$ and CoCrFeNi, under a given effective $\Delta K$ value, by the focused ion beam (FIB)-TEM technique. The results show that the $CoCrFeNiMo_{0.2}$ HEA exhibits lower averaged increment of the irreversible strain than the CoCrFeNi HEA. This suggests that the $CoCrFeNiMo_{0.2}$ HEA has higher deformation reversibility, as presented in Figure 2.3.12, which delays the failure of unruptured ligaments by slowing the accumulation of damage before the crack tip, thereby slowing the rate of crack growth.

The microstructures of the $CoCrFeNiMo_{0.2}$ and the CoCrFeNi HEAs are shown in Figure 2.3.13. The $CoCrFeNiMo_{0.2}$ HEA produces a large amount of deformation twins (DTs) and stacking faults (SFs) during the fatigue process. These structures are caused by low stacking fault energy (SFE), and their interaction with dislocations contributes to the fatigue resistance of the HEAs. In the CoCrFeNi HEA, only dislocation bands with higher dislocation densities were observed, and no DT or SF structures were found.

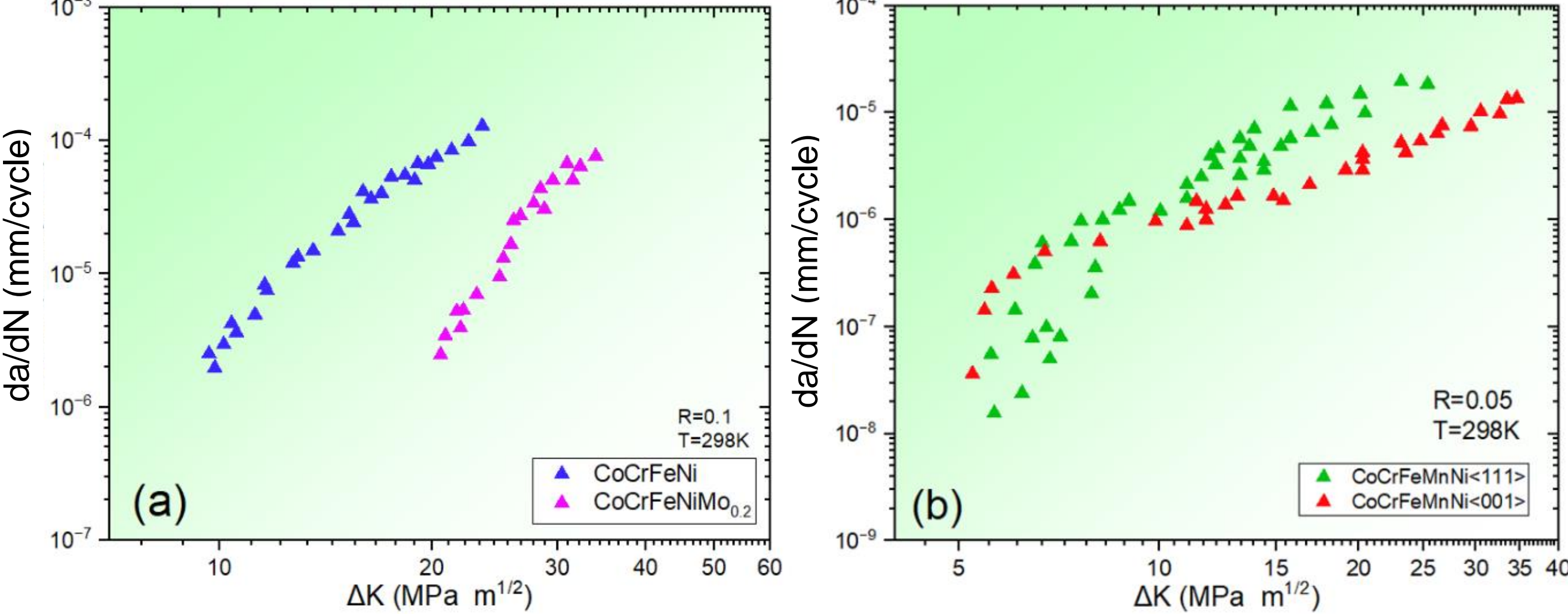

**Figure 2.3.11**: Comparison of the rate of fatigue crack growth for a) CoCrFeNi and $CoCrFeNiMo_{0.2}$ HEAs [106] and b) CoCrFeMnNi HEA in different crystal directions [109].

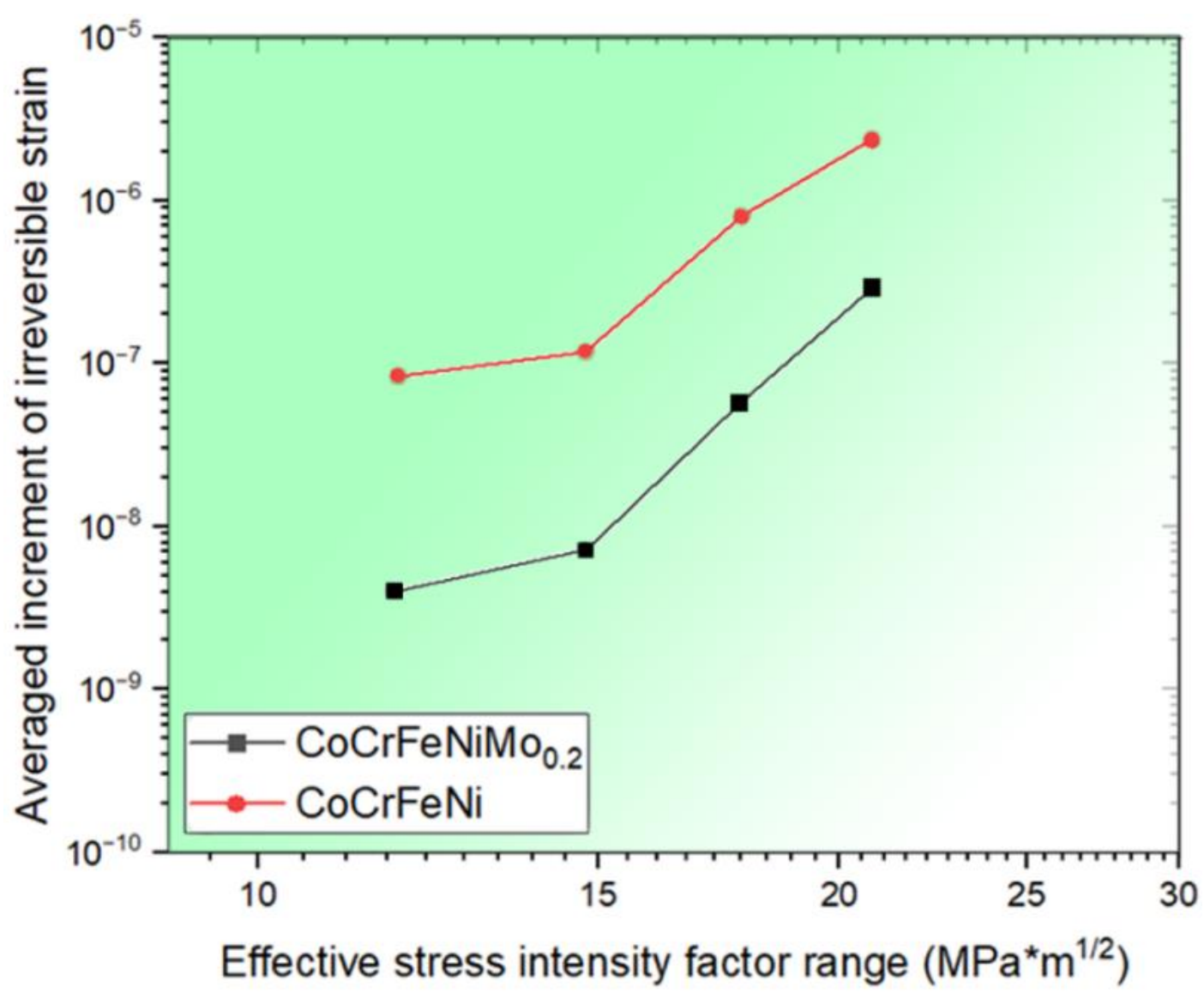

**Figure 2.3.12:** Relationship between the averaged increment of the irreversible strain, during a complete loading cycle, and effective stress intensity factor range ($\Delta K_{eff}$) of the CoCrFeNi and $CoCrFeNiMo_{0.2}$ HEAs [106].

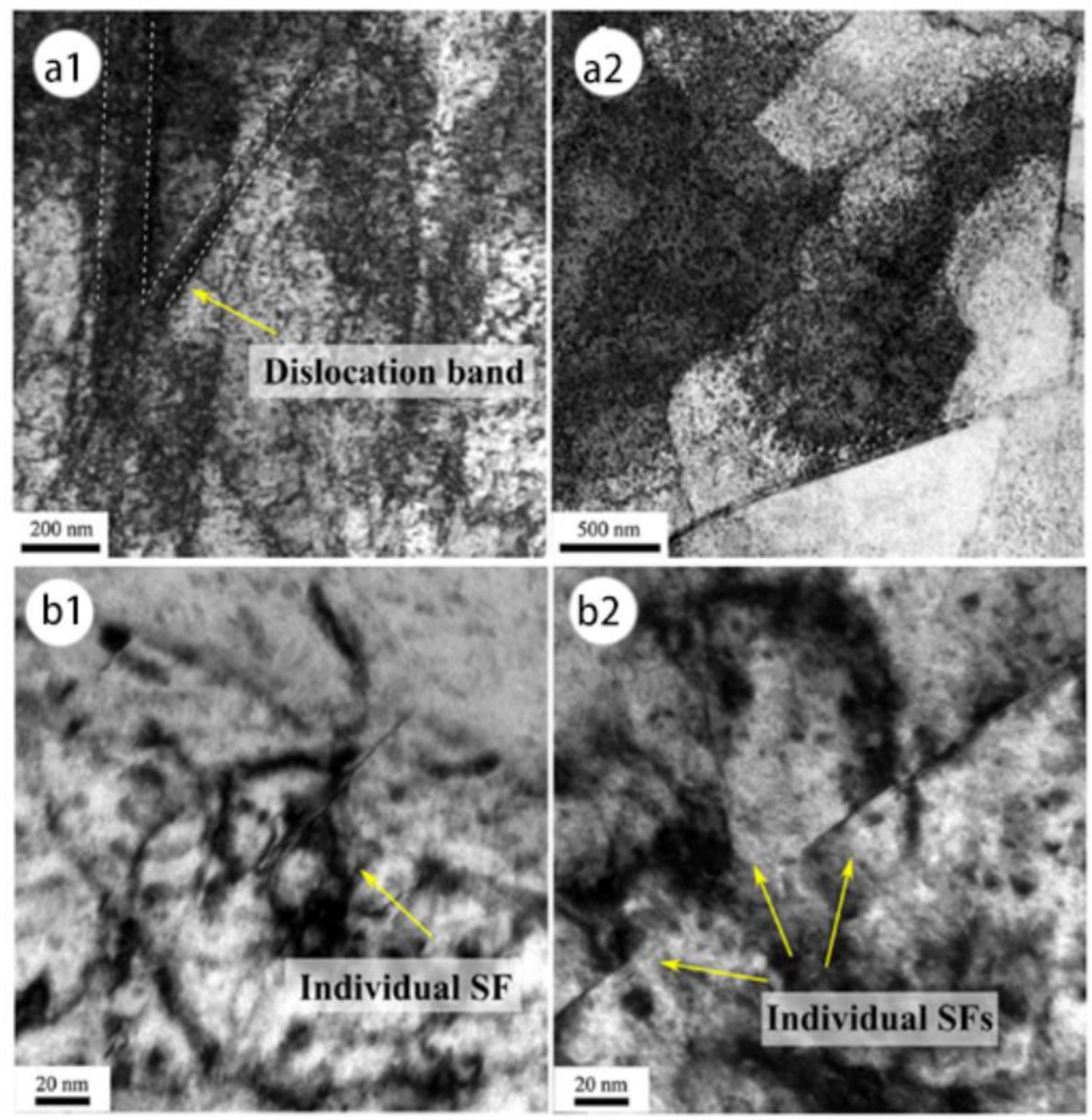

**Figure 2.3.13:** Dislocation structures in the a) CoCrFeNi and b) $CoCrFeNiMo_{0.2}$ HEAs [106].

Figure 2.3.11(b) compares the rate of fatigue crack growth for the CoCrFeMnNi HEA with the tensile axes parallel to the <001> crystal system to that of CoCrFeMnNi HEA with the tensile axes parallel to the <111> crystal system. Such FCGR data is not included in Figure 2.3.10 due to the different stress ratios involved. In the crack initiation stage, the $\Delta K_{\mathrm{th}}$ values were not affected. However, in the subsequent stretching, the crack tip stress of the <001> (350 nm) specimen reaches the critical shear stress of twinning formation earlier than the crack tip stress of the <111> specimen, which results in earlier twinning and higher thickness. The obstruction of dislocation movement by the twin formation causes a slow increase in the rate of crack extension in the <001> specimens and delays the start of Stage III crack growth. In addition, the cracking of <001> specimens is basically all Mode I cracking, per Figure 2.3.1, whereas the <111> specimens feature both Mode I and Mode II cracking.

### ***2.3.4. BCC HEAs***

With the caveat that there is only a single study on the FCGR of BCC HEAs, the FCGR of this one BCC HEA (HfNbTaTiZr) is higher than that of the other three categories of HEAs. Figure 2.3.14 shows the fracture surface of the HfNbTaTiZr single-edge notch bending (SENB) specimen after a four-point bending test. In the region with a low stress intensity factor range (3.6 MPa $m^{1/2}$), the fracture surface exhibits many neat flat facets. These facets represent different crystal planes. The transition trace between crystal planes is relatively obvious in Figure 2.3.14(a). However, in the region with a high stress intensity factor range (10.4 MPa $m^{1/2}$), the fracture expansion began to shift to the crystal plane with a relatively low Schmid factor, which resulted in a large number of tear patterns on the fracture surface, as shown in Figure 2.3.14(b).

The EBSD results of the crack path, captured in Figure 2.3.15, illustrate that the mechanism of crack propagation in the HfNbTaTiZr BCC HEA is trans-granular. Keep in mind that the crystal plane with a smaller deflection angle is preferred during crack propagation. Furthermore, the difference between the close-packed plane and the sub-close-packed plane is smaller for the BCC crystal structure than for the FCC crystal structure. Therefore, in Figure 2.3.15, no inter-granular cracking was observed.

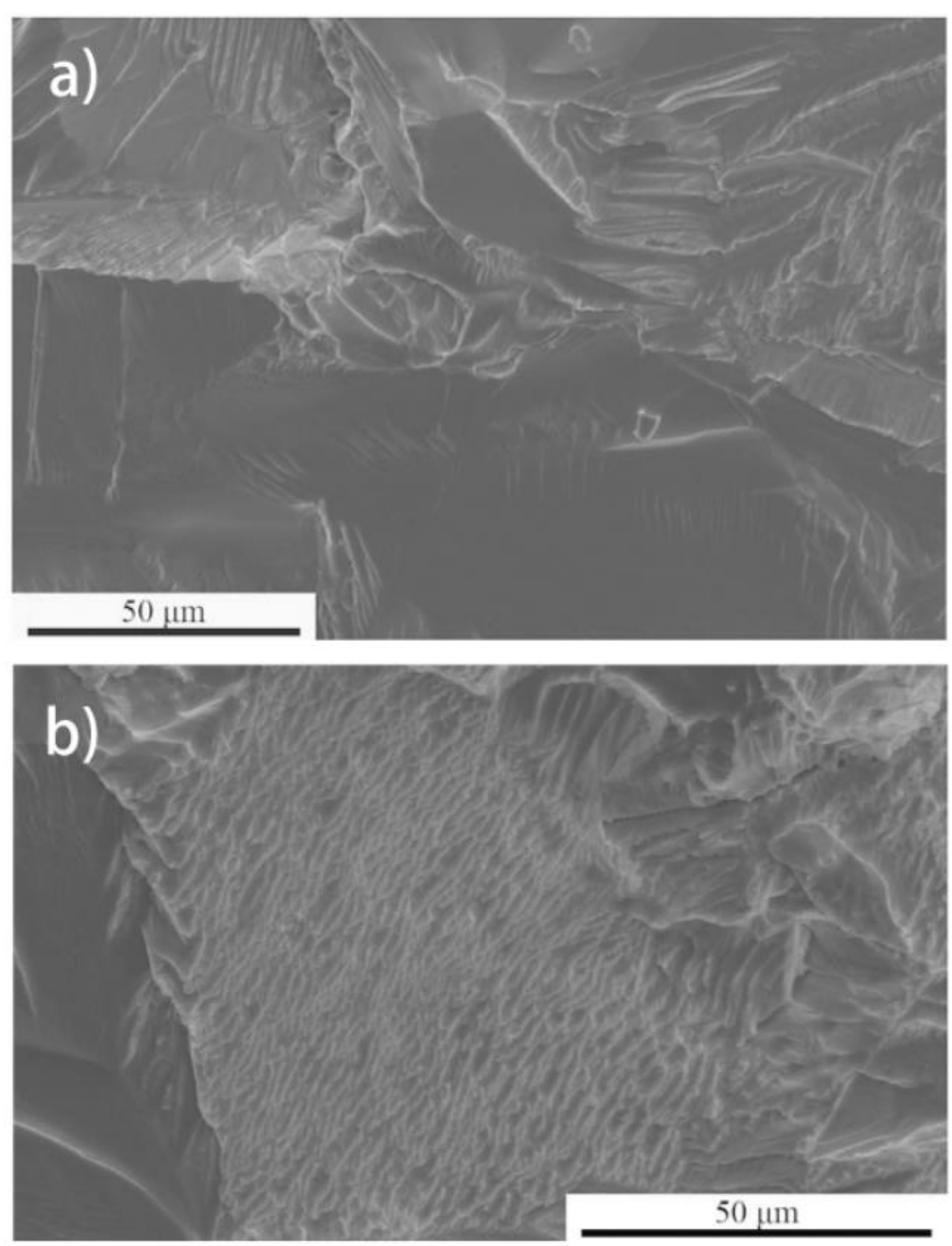

**Figure 2.3.14:** The fracture surface of the HfNbTaTiZr HEA in a) the low Δ*K* (3.6 MPa $m^{1/2}$) region and b) the high Δ*K* region (10.4 MPa $m^{1/2}$) [17].

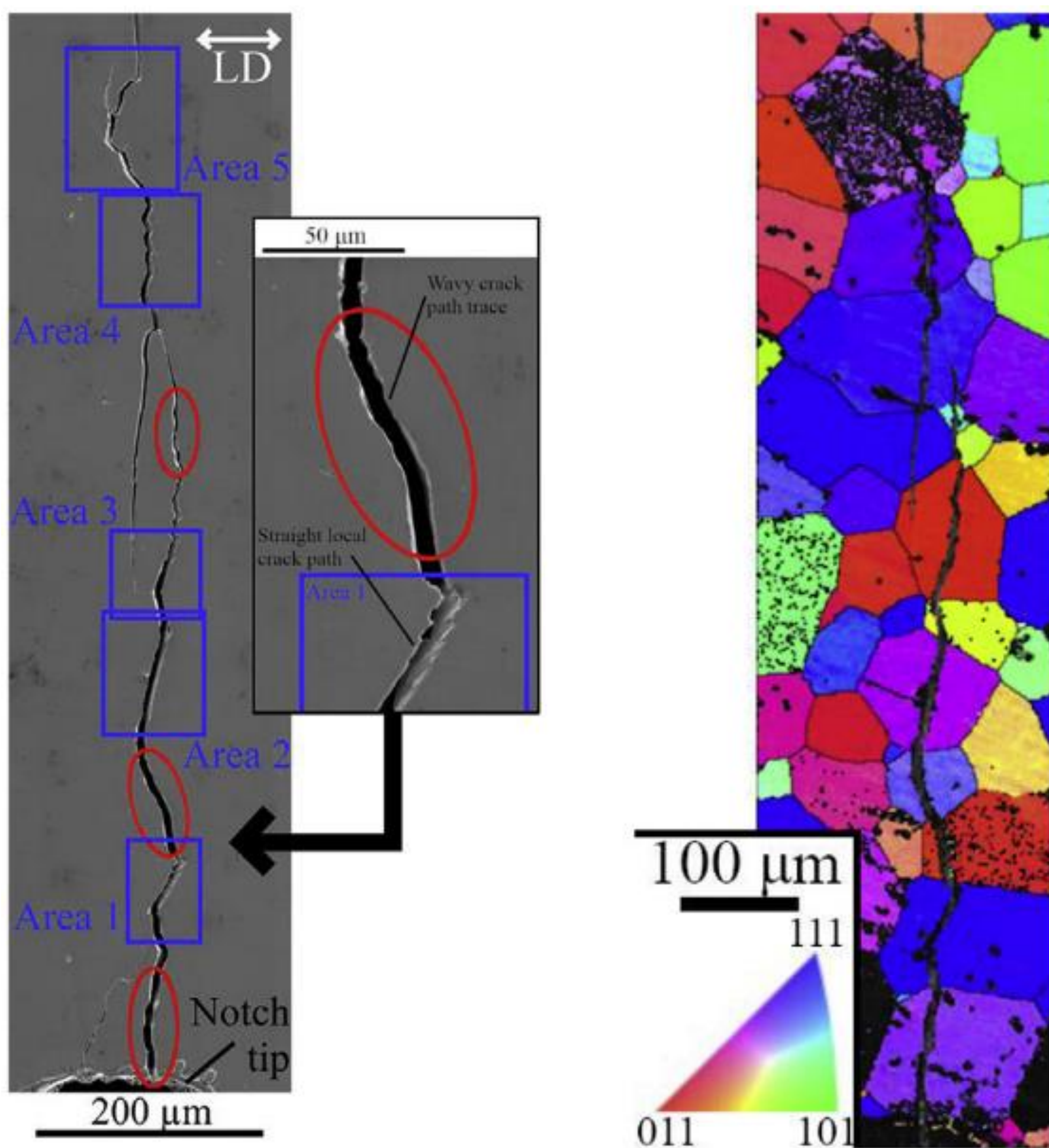

**Figure 2.3.15:** Microstructure of the crack on the HfNbTaTiZr SENB specimen in the form of a) SEM micrograph and b) EBSD micrograph [17].

### *2.3.5. Multiphase HEAs*

The two multiphase HEA systems studied, $Al_{0.2}CrFeNiTi_{0.2}$ and $AlCrFeNi_2Cu$, comprise both of FCC and BCC phases. Figure 2.3.16 shows the microstructures of the $Al_{0.2}CrFeNiTi_{0.2}$ and $AlCrFeNi_2Cu$ HEAs [31]. The matrix part is rich in Cr and Fe (regions labeled as white in Figure 2.3.16), and represents the FCC phase, whereas the precipitates in the inter-dendritic region are rich in Al, Ni or Cu [31]. From the micrographs in Figure 2.3.16, we can tell that the extent of the white region for the $AlCrFeNi_2Cu$ HEA is larger than that of the $Al_{0.2}CrFeNiTi_{0.2}$ HEA. This observation seems to suggest that in case of the $AlCrFeNi_2Cu$ HEA, the FCC phase dominates, while the $Al_{0.2}CrFeNiTi_{0.2}$ HEA features more BCC components. This observation is also corroborated by hardness and strength data. This observation partly explains why the rate of fatigue crack growth is larger for the $AlCrFeNi_2Cu$ HEA than for the $Al_{0.2}CrFeNiTi_{0.2}$ HEA. In a multiphase HEA comprised of FCC and BCC phases, crack propagation mostly occurs in the more densely packed FCC matrix. When extending to the interface between the phases, most cracks are deflected along the interface. Cracks only penetrate the BCC phase, when the stress at the crack tip is large enough. Therefore, in the two multiphase HEAs, the rate of crack growth in the HEA with the smaller volume fraction of BCC phase is relatively slow.

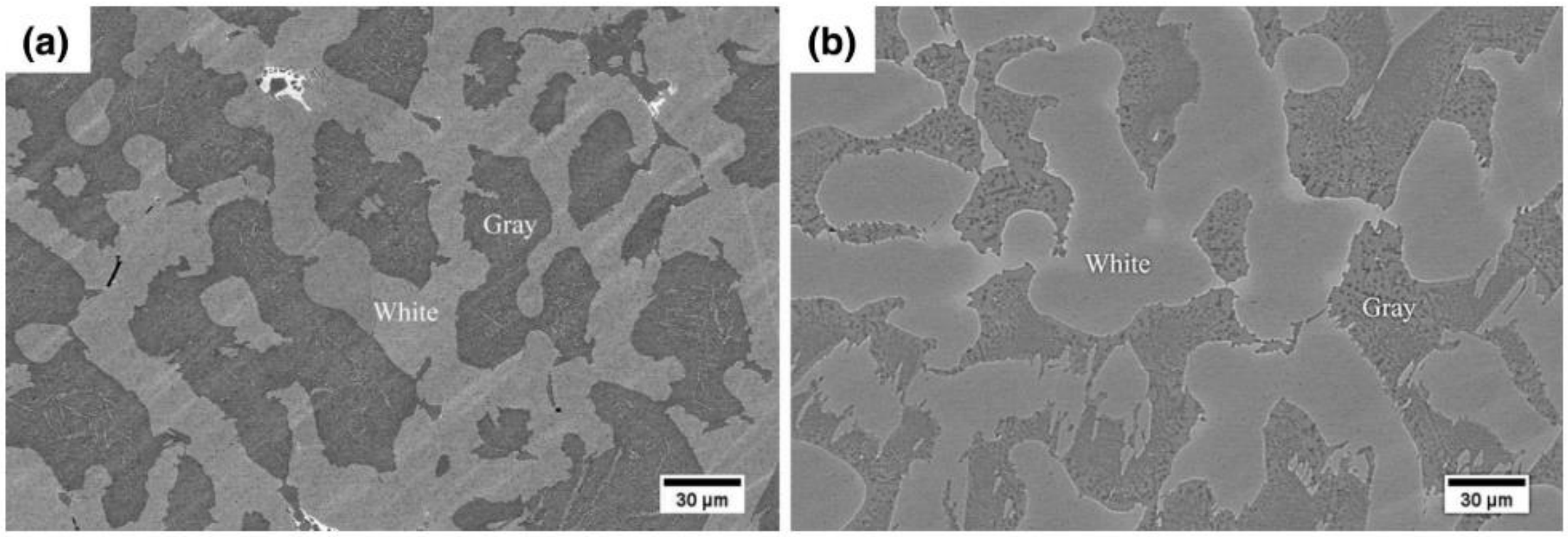

**Figure 2.3.16:** Microstructures of (a) $Al_{0.2}CrFeNiTi_{0.2}$ and (b) $AlCrFeNi_2Cu$ [31].

### *2.3.6. Metastable HEAs*

Eguchi et al. completed the one study available on metastable HEAs, specifically on the iron-based $Fe_{30}Mn_{10}Cr_{10}Co$ HEA [33]. As shown in Figure 2.3.17, the fracture surface of the $Fe_{30}Mn_{10}Cr_{10}Co$ specimen consists of a main crack and a small number of secondary cracks. Compared to conventional metastable alloys, the number of secondary cracks in the $Fe_{30}Mn_{10}Cr_{10}Co$ is much smaller. Furthermore, the shape of cracks resembles the shape one can expect from a conventional FCC alloy (e.g., 316L steel). An EBSD micrograph of the crack beneath the fracture surface is shown in Figure 2.3.18. The microstructure of this $Fe_{30}Mn_{10}Cr_{10}Co$ HEA shows features of a typical metastable alloy: The crack is surrounded by an HCP martensite phase with a distinct orientation gradient. In addition, no twins were observed in this specimen, which differs from the tensile test results for the $Fe_{30}Mn_{10}Cr_{10}Co$ HEA by Li et al [113]. This observation seems to indicate that tensile and fatigue fractures of the $Fe_{30}Mn_{10}Cr_{10}Co$ metastable HEA may be subjected to different mechanisms.

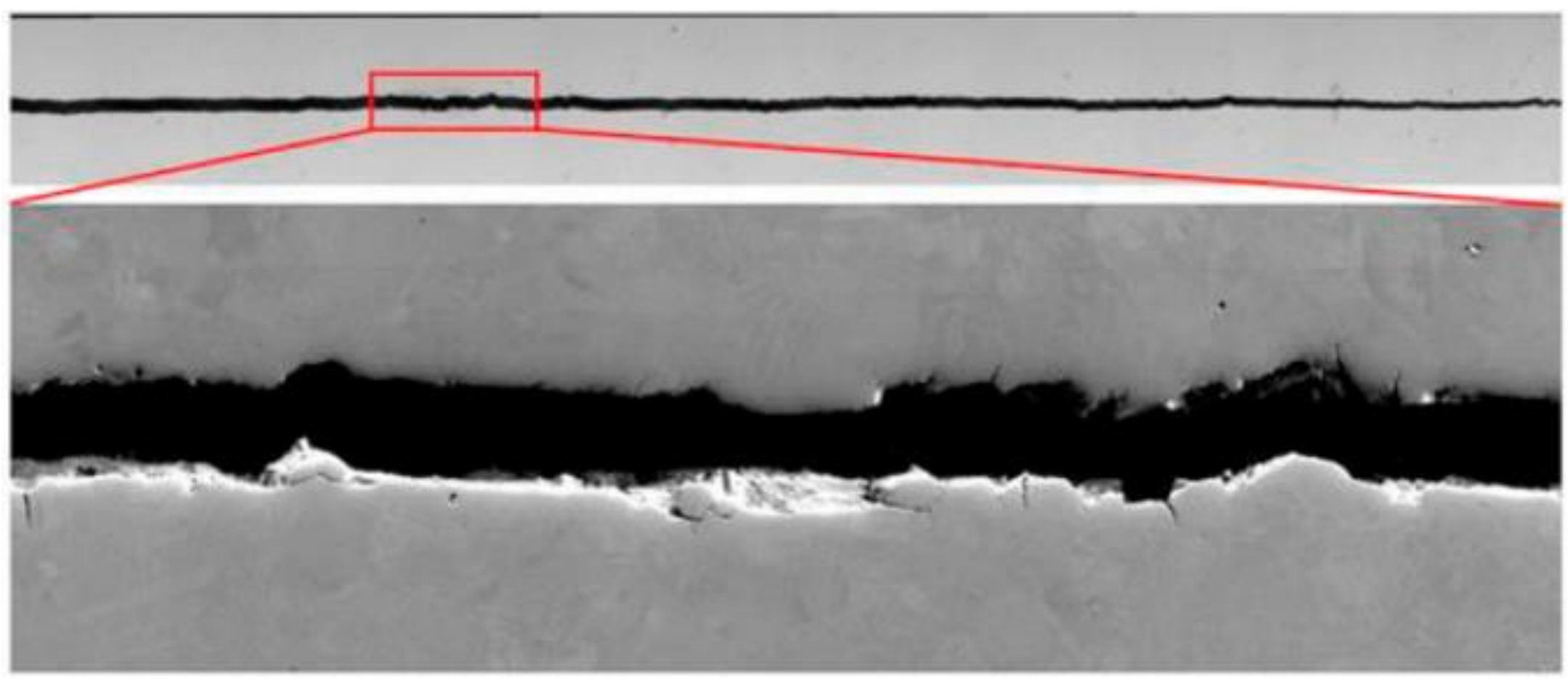
**Figure 2.3.17:** Fatigue-crack-propagation paths in the $Fe_{30}Mn_{10}Cr_{10}Co$ HEA [33].

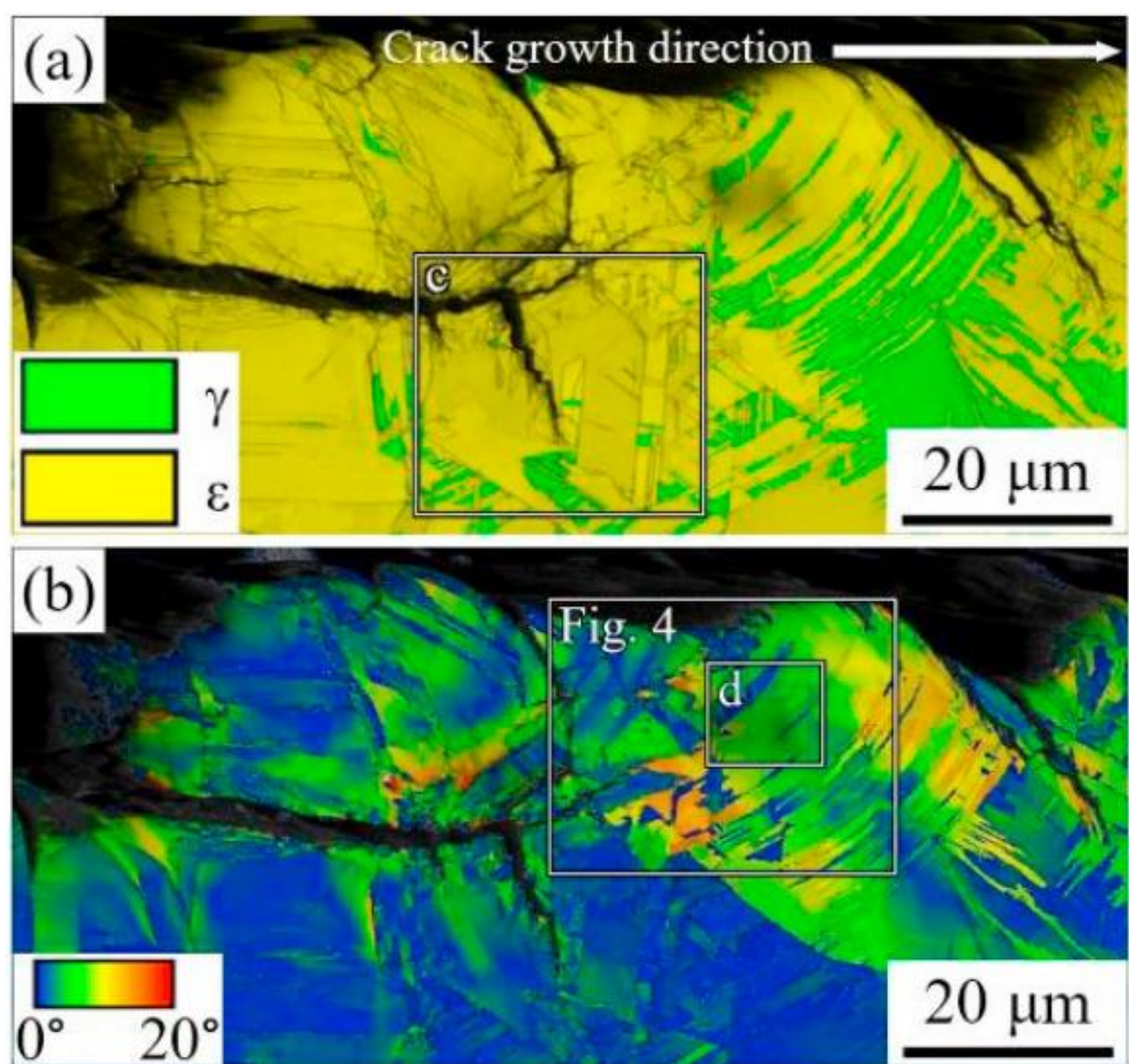

**Figure 2.3.18:** a) EBSD and b) Kernel average misorientation (KAM) micrographs beneath the fracture surface of the $Fe_{30}Mn_{10}Cr_{10}Co$ HEA at $\Delta K$ of 18 MPa·m$^{1/2}$ [33].

### *2.3.7. Comparison between HEAs and conventional alloys*

The comparison of FCGR is carried out from two perspectives, namely (1) the fatigue crack-growth rate in the Paris region and (2) the threshold stress intensity factor $\Delta K_{th}$ under the same $R$ ratio. Figure 2.3.19 summarizes the FCGR comparison. The conventional alloys, which include steels, aluminum alloys, magnesium alloys, titanium alloys and copper alloys. are presented for reference [114-119]. The selection of representative alloys for comparison was based on a wide range of industrial applications and test conditions consistent with the HEAs ($R \approx 0.1$, $T \approx 298$K, and $f \approx 25$ Hz). It is not difficult to see that in the Paris region, the performance of the BCC HEA is slightly stronger than that of aluminum alloys and magnesium alloys, but lower than that of other conventional alloys. The performance of the FCC HEAs, in this respect, is close to that of traditional steels and copper alloys. Both the one metastable HEA and the two multiphase HEAs exhibit better FCGR performance than most of the conventional alloys, with the one exception of titanium.

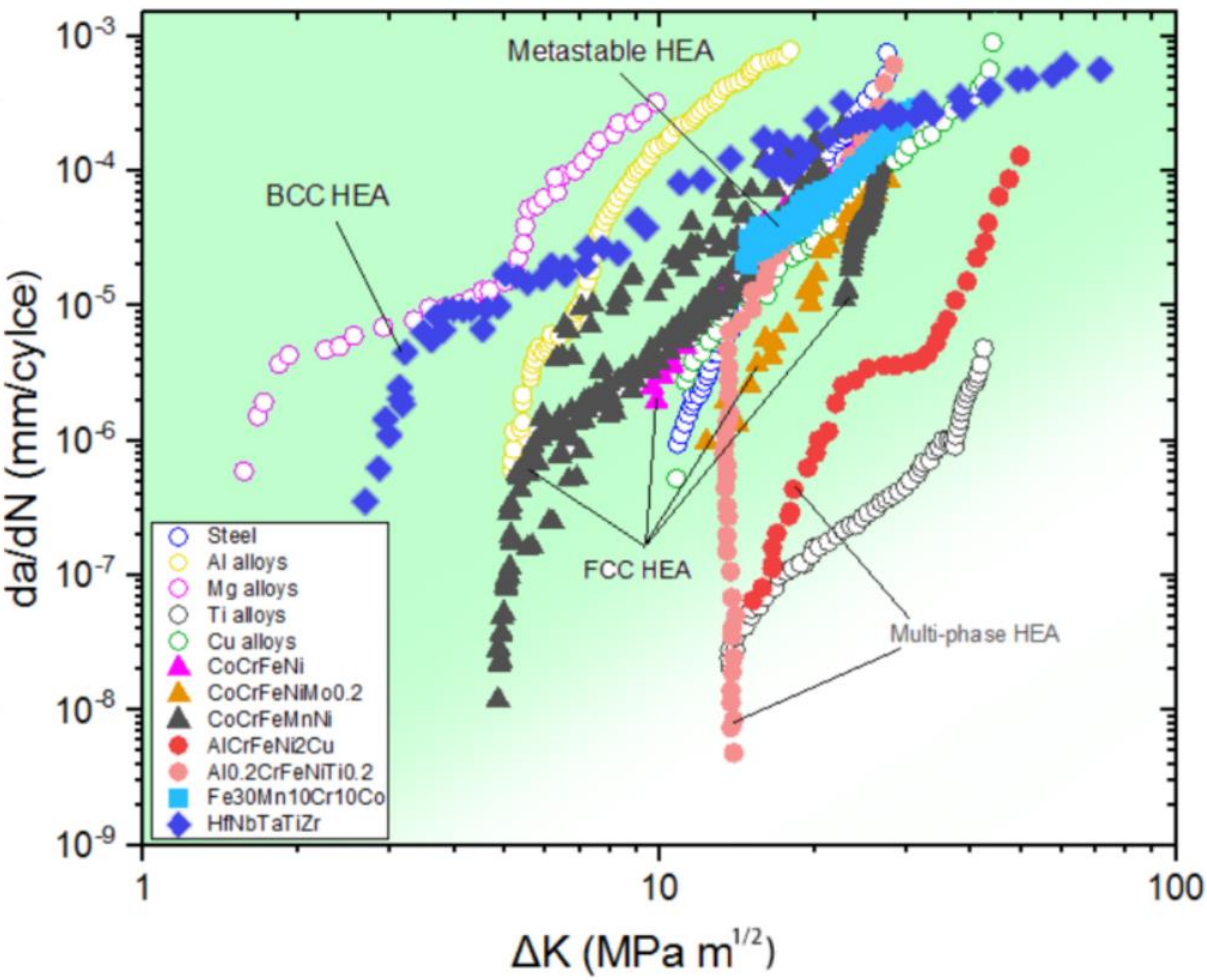

**Figure 2.3.19:** Comparison of FCGR behavior between HEAs and conventional alloys.

In the comparison of the threshold stress intensity factor $\Delta K_{\mathrm{th}}$, we additionally use the stress ratio, *R*, as a variable, as shown in Figure 2.3.20. Figure 2.3.20 facilitates convenient comparison of the effect of the stress ratio, mentioned in Section 3.4, on the HEAs and on the conventional alloys. This effect of the stress ratio is more pronounced for materials with high $\Delta K_{\mathrm{th}}$ values. In this regard, the performance of the BCC HEAs is similar to that of the lowest magnesium alloys and of the aluminum alloys. According to Figure 2.3.20, the $\Delta K_{\mathrm{th}}$ values of the FCC HEAs are in a similar range as those of the titanium alloys. The one metastable HEA exhibits a higher $\Delta K_{\mathrm{th}}$ level than all the conventional alloys. The $\Delta K_{\mathrm{th}}$ values for the multiphase alloys also appear quite large, according to Figure 2.3.20. In the low-*R* region, $\Delta K_{\mathrm{th}}$ values of the multiphase HEAs appear much higher than for the conventional alloys. In the high-*R* region, the $\Delta K_{\mathrm{th}}$ values of the multiphase HEAs also appear the highest, up to $R \approx 0.7$.

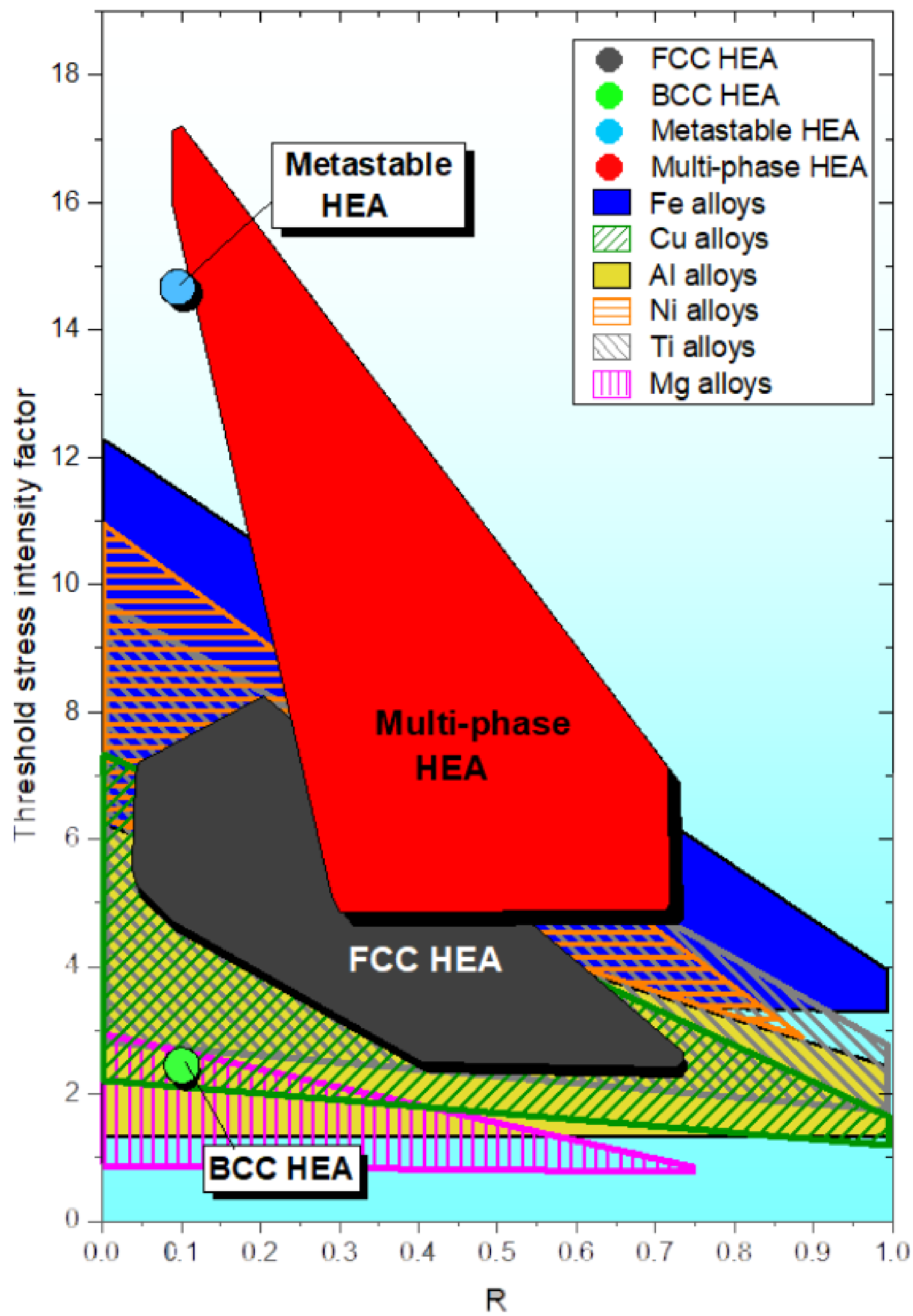

**Figure 2.3.20:** Comparison of $\Delta K_{th}$ values under different stress ratios between HEAs and conventional alloys [114-119].

To sum up, the FCC HEAs may not be a good choice when designing new fatigue-resistant HEAs, due to their mediocre performance in terms of accumulate damage and crack growth. Based on the fatigue data currently available, metastable HEAs and multiphase HEAs look the most promising, because both their fatigue resistance and ultimate tensile strength look medium-to-high compared to the conventional alloys. Moreover, in terms of the fatigue crack growth, the metastable and multiphase HEAs outperform most of the conventional alloys. Excellent fatigue resistance of the metastable and multiphase HEAs originates from their unique fatigue mechanisms, which are described in Section 3.1 below. From this point of view, through further control of the composition and the microstructure of the metastable and multiphase HEAs, one may be able to further improve their fatigue resistance in a targeted manner. In addition, the BCC HEA is also HEA with some potential for future research and development, because it exhibits high fatigue limit compared to the other HEAs and conventional alloys. Based on the high fatigue limit and poor crack growth resistance of the one BCC HEA studied, targeted applications for BCC HEAs may possibly be conceivable.

# 3. Mechanistic Understanding

## 3.1. Fatigue mechanisms

The fatigue-failure process of metallic materials comprise of three stages: crack nucleation, crack propagation, and final failure. Overall, the process of fatigue degradation in HEAs also subjected to these three stages. Stress concentration and accumulation of dislocations causes cracks to nucleate (initiate) at inclusions or large-angle grain boundaries, expand along the crystal slip planes, and then rapidly grow in a direction perpendicular to the applied cyclic load, leading to final failure [120]. But for all four categories of HEAs (FCC, BCC, multiphase, and metastable HEAs), differences in microstructures result in various fatigue crack shapes and crack-growth mechanisms, and these different mechanisms finally affect their fatigue life. In the following, the fatigue mechanisms of the four categories of HEAs will be introduced separately, based on their microstructures.

### *3.1.1. FCC HEAs*

In FCC HEAs, cracking is attributed to the irreversibility of cyclic slips caused by dislocation motion. The two main cracking mechanisms observed in FCC HEAs are slip-band (SB) and twin-boundary (TB) cracking, which are shown in Figure 3.1.1 [23]. Because fatigue failure is a dynamic process, there is a one-way transformation relationship between these two mechanisms. During the initial cyclic-load application process, the complex dislocation activity formed a slip zone. High-density dislocations and vacancies accumulate in the slip band to form a squeeze that yields stress concentration, causing fatigue cracks to nucleate and propagate along the slip-band first. Since the intersections between the crystal grains are not collinear with one another, dislocations and vacancies accumulate at the twin boundaries [1]. Although the twin boundary is regarded as a barrier to the movement of dislocations in many studies, when the accumulation of dislocations and vacancies reaches a critical value, the slip-band cracking turns into inter-granular fatigue and twin-boundary cracking. Figure 3.1.2 describes the process of this transformation in a schematic fashion.

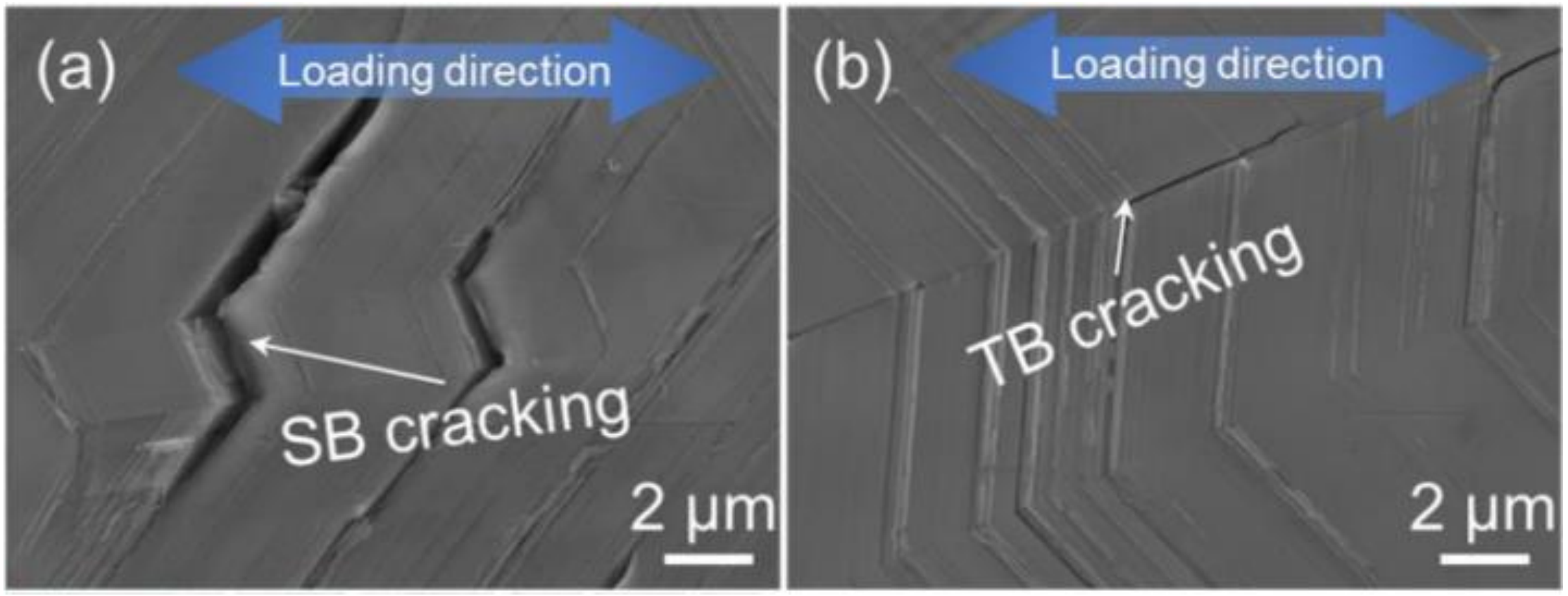

**Figure 3.1.1:** (a) Slip-band and (b) twin-boundary fatigue-crack initiation mechanisms in CoCrFeMnNi FCC HEAs [23].

In addition, the dominant mechanism of cracking is related to the grain size. For a coarse-grained specimen, the initial cracking will be dominated by a slip band along the {111} crystal plane [41], while twin-boundary cracking is more common to form in a fine-grained alloy earlier. As the difference in the Schmid factors between the matrix and the twin increases, and as the SFE decreases, twin-boundary cracking becomes easier [23].

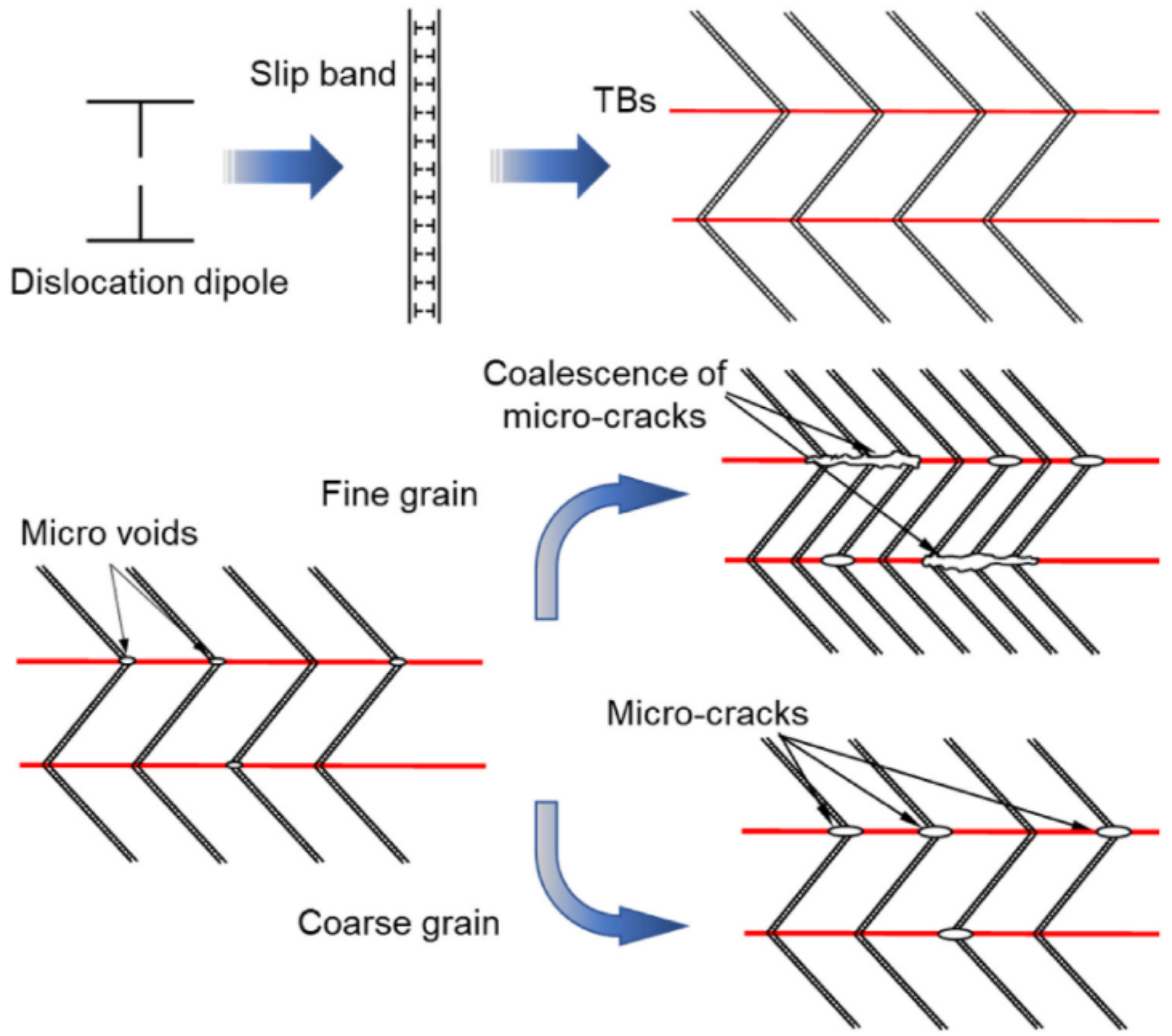

**Figure 3.1.2:** Dislocation-accumulation process in the CoCrFeMnNi FCC HEA [23].

### *3.1.2. BCC HEAs*

The particularity of BCC HEAs is their extremely high hardness and strength, but its ductility and fracture toughness are relatively low. This is reflected in their fatigue behavior in a high fatigue strength limit but fast crack-growth rate, as explained in Section 2.3 [17, 18]. Therefore, when explaining the fatigue mechanisms in BCC HEAs, it is necessary to explain these two features separately.

In materials with high stacking-fault energies, the recovery of dislocations is relatively easy, while twins and shear bands are more difficult to form. In the research of conventional alloys, materials with BCC structures usually feature a larger value for the unstable stacking fault energy than materials with FCC structures. This trend is considered to be one of the reasons for the higher strength of the BCC alloys [121]. Considering that research on BCC HEAs is still very limited, whether this theory is applicable to BCC HEAs remains to be further investigated.

Compared with the FCC structure, a BCC structure is a non-close-packed structure. Hence, the dislocations of the BCC phase do not strictly slip along the close-packed plane, which makes the dislocation recovery of the BCC HEA easier during the stretching process. It is not easy to form dislocation accumulation like in the FCC HEA. This feature can be manifested in mechanical behavior as BCC HEAs tend to feature higher tensile strength and the highest fatigue limit compared to other categories of HEAs. On the other hand, 18 slip planes were able to convey the crack propagation in the HfNbTaTiZr BCC HEA studied by Guennec et al., which according to the authors makes the fatigue cracks of the BCC HEA almost always start in the form of inter-granular cracking and expand in the form of trans-granular cracking [17]. Therefore, grain boundaries are far less obstructive to fatigue in the case of BCC HEAs than in the case of FCC HEAs. In the study by Guennec et al., this trend was demonstrated by measuring the changes in fringe spacing that occurred near the grain boundary. The striation spacing decrease (SSD) is described by Eq. (3.1.1) [43]:

$$SSD = \frac{ss_1 - ss_2}{ss_1} \tag{3.1.1}$$

Here, $ss_1$ and $ss_2$ represent the striation spacing before and after the grain boundary, respectively. As shown in Figure 3.1.3, the SSD value of the BCC HEA is very low, compared to the aluminum alloy material, which possesses FCC crystal structure at room temperature. The SSD value also increases with the inclination angle, α. Although the fatigue strength of the BCC HEA is high, the fatigue-crack propagation is also very easy, which is consistent with the extremely high d*a*/d*N* value of the HfNbTaTiZr BCC HEA, when in the stage of steady-state crack propagation [17]. The profusion of available slip planes in the HfNbTaTiZr BCC HEA, in comparison with the Al 2024 FCC alloy, provides a lot more possibilities to minimize the rotation angle and twist angles at the grain boundary. Therefore, the impact of the grain boundary on the crystallographic crack propagation mechanism in the HfNbTaTiZr BCC HEA alloy is less preeminent than in the Al 2024 FCC alloy, which is consistent with the experimental results in Figure 3.1.3 [17].

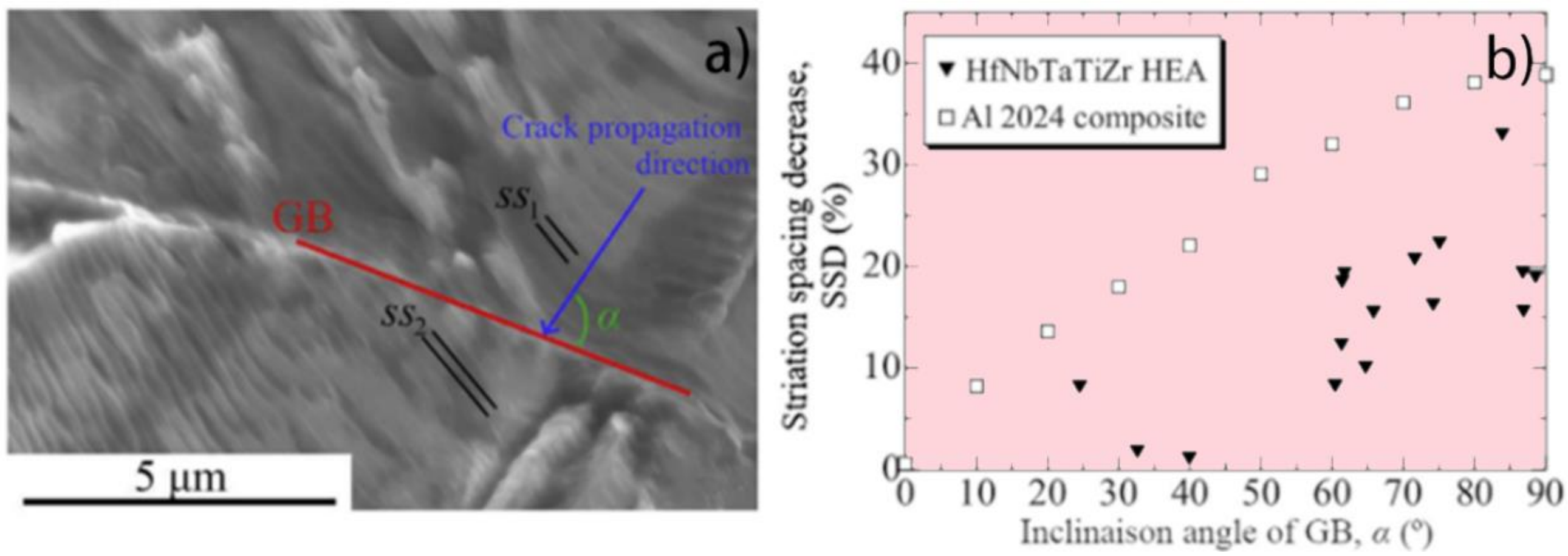

**Figure 3.1.3:** a) The striation spacing of BCC HEA fatigue cracks before and after the grain boundary. b) Comparison of the BCC HEA striation-spacing difference and aluminum alloy [17].

As shown in Figure 3.1.4, in the study of the HfNbTaTiZr BCC HEA, Chen et al. [44, 52] determined four external toughening mechanisms that can delay the fatigue-crack growth in this BCC HEA, namely:

1. *Crack deflection and meandering*, which cause part of the original Mode I stress to be decomposed into Mode II. As a result, the original Mode I stress intensity directly acting on the crack propagation is reduced.
2. *Roughness-induced crack closure*, which occurs mostly in areas with low stress intensity factors, is caused by the shear component in mixed-mode fractures and causes the zigzag crack closure to slow down the crack growth.
3. *Fracture-debris-induced crack closure,* which is the closure caused by insoluble corrosion products at the crack wake. It is similar to the closure caused by roughness.
4. *Crack branching*, where potential energy is shunted to a larger fracture-surface area. Hence, the growth rate of the main crack is correspondingly reduced.

Considering that the crack-growth rate of the HfNbTaTiZr BCC HEA is the fastest among the four categories of HEAs, we believe that these four toughening mechanisms are only auxiliary functions, not the main reason for maintaining the fatigue resistance of this BCC HEA.

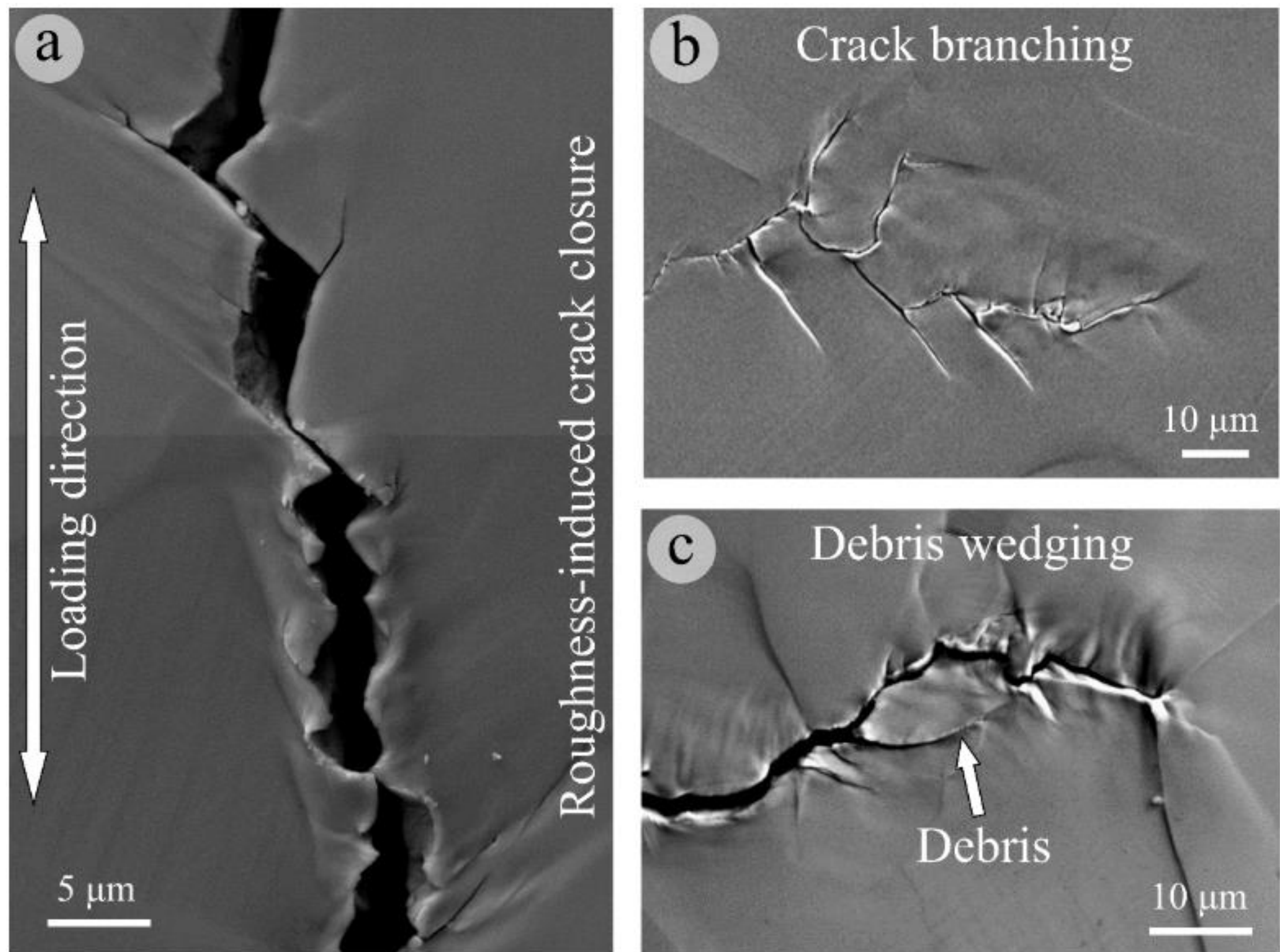

**Figure 3.1.4:** Roughening mechanisms that delay the fatigue-crack growth in the HfNbTaTiZr BCC HEA [44, 52].

### *3.1.3. Multiphase HEAs*

The multiphase HEAs currently studied refer, in general, to HEAs with the FCC + BCC dual-phase structure [22, 45, 122], or to HEAs with the FCC + BCC + sigma three-phase structure [22]. For multi-phase HEAs with a large fraction of FCC phase, the fatigue mechanism is similar to that of a single-phase FCC HEA with very fine grains. The $Al_{0.3}CoCrFeNi$ HEA has a large fraction of FCC phase. The microstructure shows an FCC matrix with ultrafine grains and a small amount of second-phase precipitates [22]. Although these hard second-phase particles are very strong, the resulting microstructural inhomogeneities allow dislocations to accumulate near them. These second-phase particles act as the source of crack initiation, and make the cracks propagate along the interface between them and the matrix. However, these second-phase precipitates also refine the grains of the FCC matrix. The low SFE of the FCC structure leads to easy formation of thin deformation twins (DTs) under cyclic stress loading, increasing the fraction of DT boundaries in the microstructure of the FCC. DTs increase the interaction between twin boundaries and dislocations and improve the dislocation storage rate of the ultra-fine grain (UFG) structure. This mechanism causes a work-hardening effect, thereby furnishing the multiphase HEA with quite high fatigue resistance, per Figure 2.1.15. The high fatigue strength is due to composite strengthening by the grain-boundary strengthening mechanism and the secondary-phase hardening mechanism. The higher ductility, on the other hand, is due to enhanced dislocation storage by the dynamic Hall-Petch effect [123].

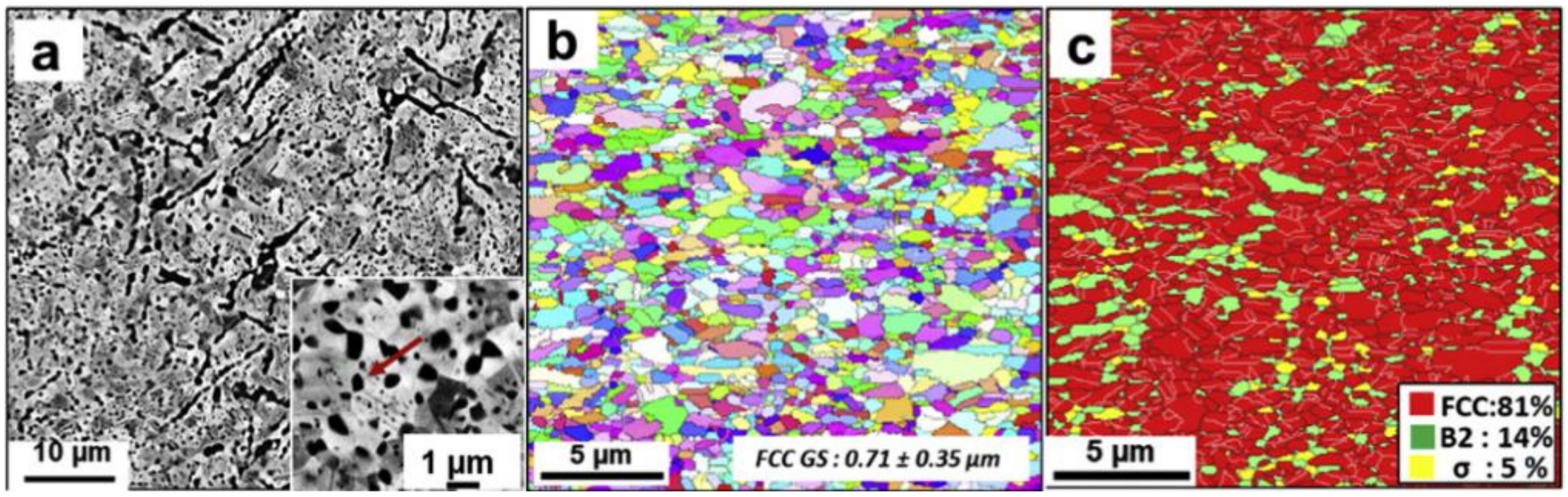

**Figure 3.1.5:** Microstructure of the $Al_{0.3}CoCrFeNi$ multiphase HEA [22].

As the fraction of the BCC phase increases, the microstructure of a multiphase HEA can gradually transform into a layered structure. Taking the $AlCoCrFeNi_{2.1}$ HEA, shown in Figure 3.1.6, as an example, its microstructure is composed of flakes of FCC and BCC phases. Studies have shown that for such structures, crack initiation and propagation mostly occurs in the FCC-matrix region rather than in the hard BCC region [45]. Loading causes the dislocations in the FCC region to nucleate and form persistent slip bands, through climbing and cross-slips. Furthermore, the interaction of slip bands with grain boundaries or twin boundaries leads to crack initiation and propagation in a trans-granular manner. The propagation of cracks generally follows a zigzag pattern, resulting from intersection of multiple slip systems. After the cracks propagated to the boundary of FCC and BCC phases, they began to propagate along the phase boundary. The cracks will enter the BCC region only when the stress intensity at the crack tip is large enough to penetrate the BCC phase. As analyzed in the previous two subsections, on the fatigue mechanism of FCC and BCC structures, the BCC HEA leads to more severe dislocation-initiation conditions, due to its higher stacking-fault energy and larger number of slip systems (48, compared to 12 slip systems for FCC structures and 3 slip systems for HCP structures). But once the crack source is formed, its propagation rate is higher than for the FCC phase [45]. Therefore, among the two multiphase HEAs mentioned in Section 3.4, the $Al_{0.2}CrFeNiTi_{0.2}$ HEA, with a higher BCC phase fraction, has a greater crack-growth rate and a more brittle fracture surface feature, than the $AlCrFeNi_2Cu$ HEA, as shown in Figure 3.1.7 [31].

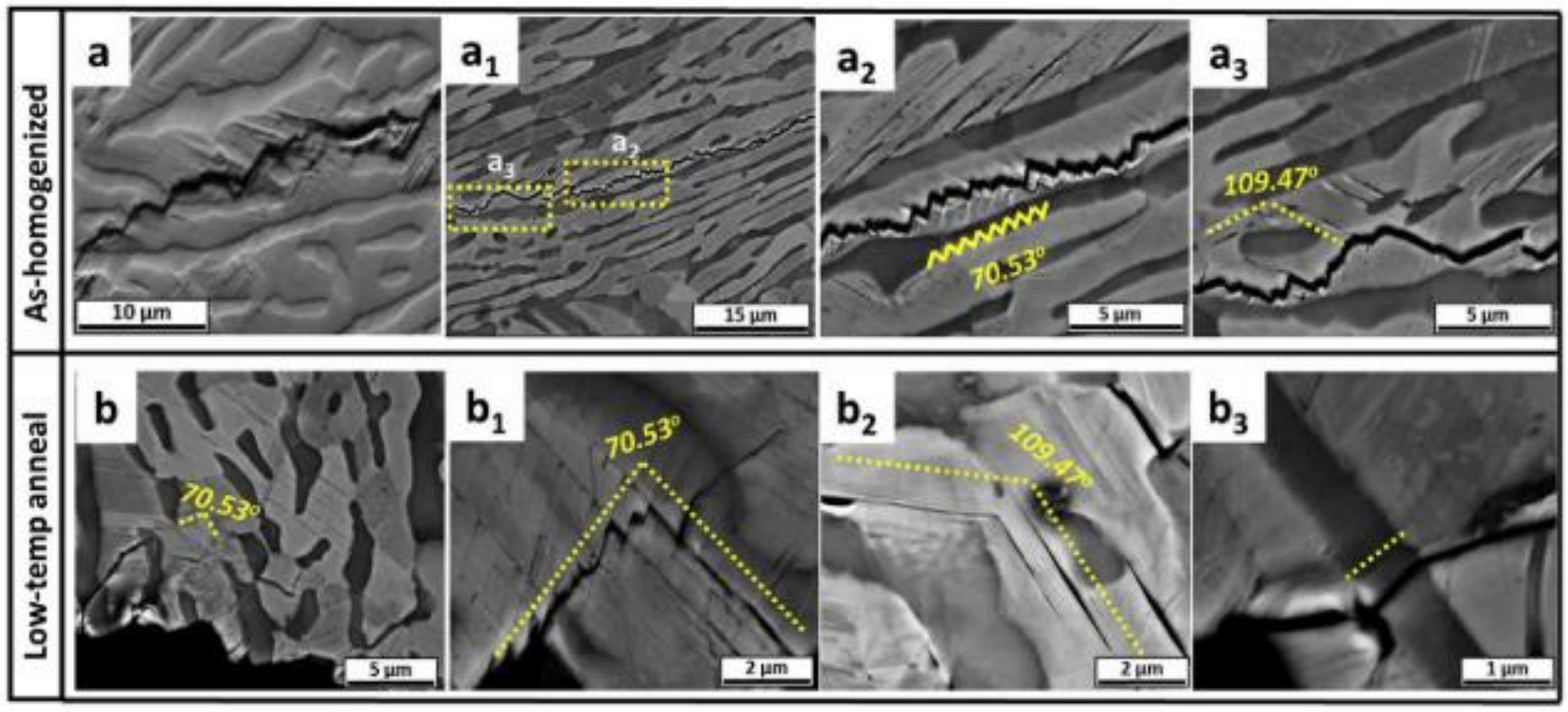

**Figure 3.1.6:** Typical cracking mechanism in the $Al_{0.7}CoCrFeNi$ dual-phase HEA [45].

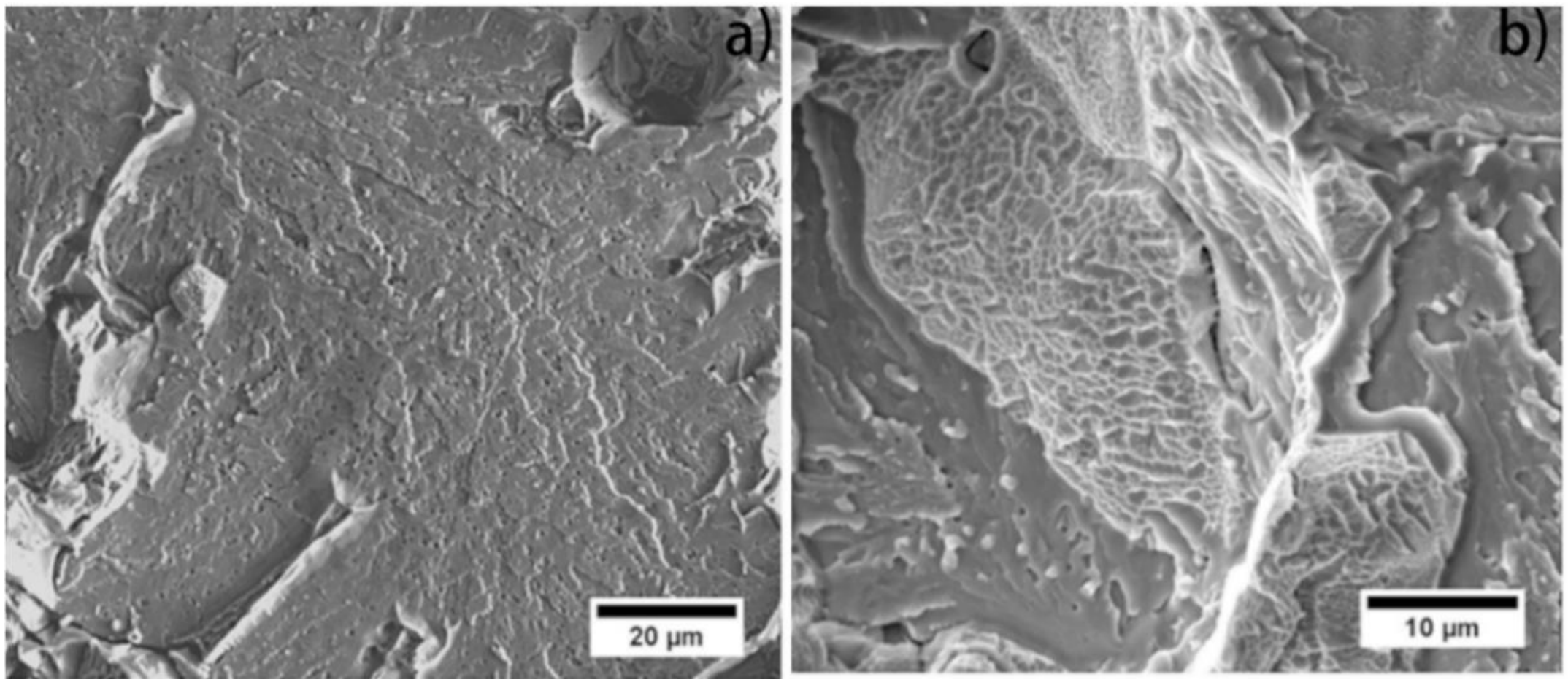

**Figure 3.1.7:** Fracture surfaces of multiphase HEAs: a) $Al_{0.2}CrFeNiTi_{0.2}$ and b) $AlCrFeNi_2Cu$ [31].

### *3.1.4. Metastable HEAs*

In metastable HEAs, the extremely strong fatigue resistance is the result of the TRIP effect [20]. The TRIP effect refers to a tensile stress-induced phase transformation phenomenon that has been widely used in conventional alloys. In metastable HEAs, stretching induces a unique phase transition from the FCC phase to the HCP phase near the crack. As illustrated in Figure 3.1.8, the content of the HCP phase is greatly increased in the plastic zone near the crack, while almost no phase transformation occurs in the region outside the plastic zone [46]. The evolution of the volume fraction of the martensite phase and its distribution, as the number of cycles increases, is shown in Figure 3.1.8(b) [19]. This unique work-hardening mechanism, caused by the phase transformation, results in local work hardening in the region near the crack tip. The finely divided phase fractions also passivate the crack tip to a large extent, which in turn plays a key role in slowing down the crack propagation. Figure 3.1.8 shows the shape of the fatigue crack in the metastable HEA. Compared to the single-phase HEA mentioned above, the crack in the metastable HEA exhibits more obvious deflection and branching. This situation is attributed to the formation of deformation twins in the HCP phase (in the plastic zone). The dispersion of cracks increases energy dissipation and improves the fatigue resistance of the material, and helps avoid traditional sudden failures [33].

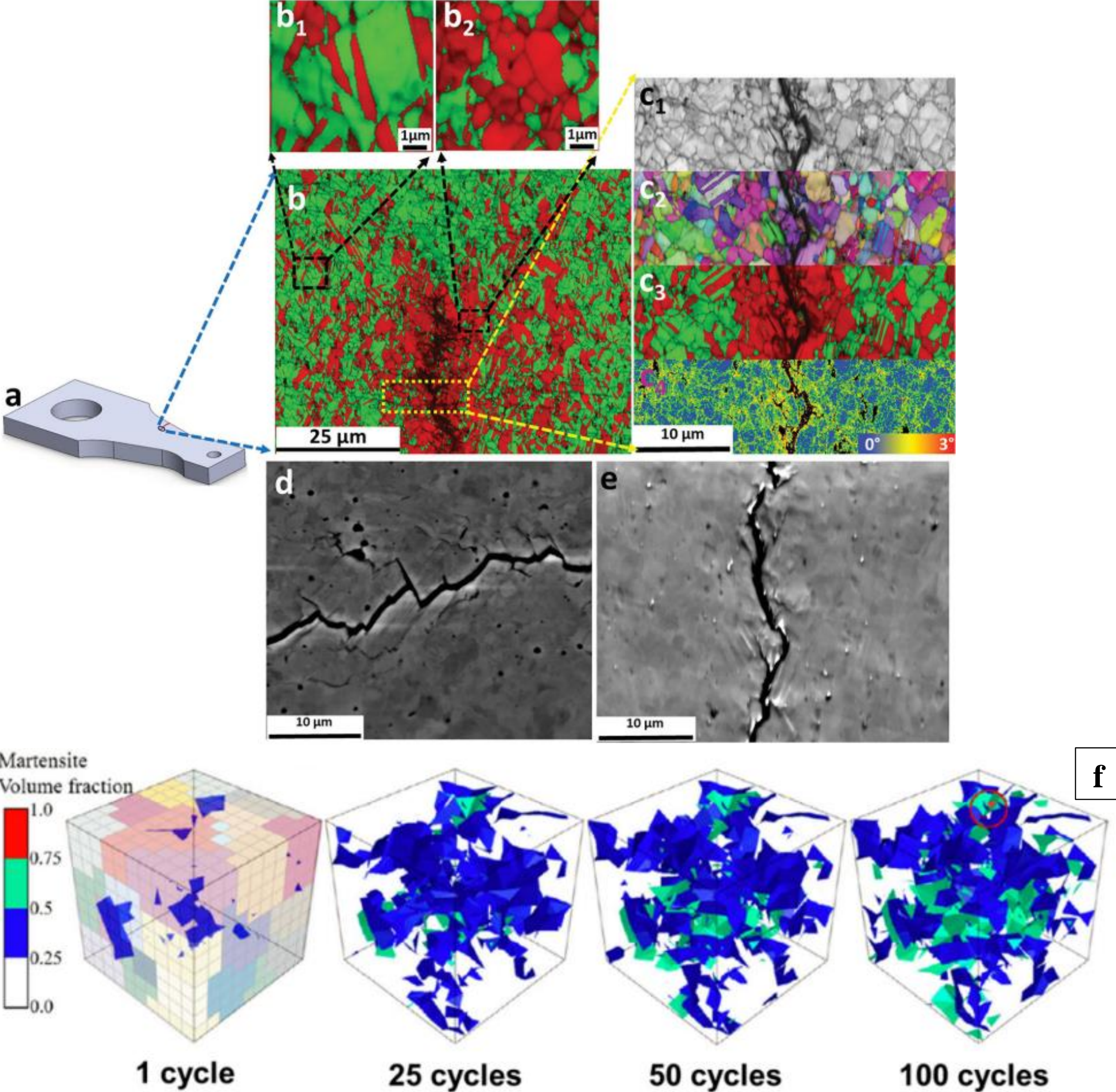

**Figure 3.1.8:** Microstructure of a metastable HEA. (a) Schematic representation of the actual fatigue sample. (b) Phase map at the crack tip on the fatigue sample, (c1– c4) higher magnification scan around the crack with an image quality map, IPF map, phase map, and KAM, respectively. (d) Back scattered electron (BSE) image of the crack. (e) image of the crack branching near the Paris region [46]. (f) fraction and distribution of hcp-martensite changes as the cycles increase [19].

### 3.2.Microstructure effects

Different microstructures have been proven to exert huge influence on the mechanical properties and deformation mechanisms of metallic materials. The impacts and effects of microstructures or other related features on the fatigue performance of HEAs are discussed in this section. These impacts and effects include the effect of different grain sizes, the existence of defects, impurities, segregations, and secondary phases, the phase-transformation behaviors, as well as the effects of dislocation structures and deformation twins that are generated during cyclic loading.

### *3.2.1. Grain size*

Due to the different levels of grain boundaries, different grain sizes are well known to exert great impact on the movement of dislocations, during plastic deformations. Similarly, under cyclic loading, the HEAs with finer grains tend to exhibit better fatigue resistance. Tian et al. [15] prepared coarse grained (CG) and ultra-fine-grained CoCrFeMnNi HEAs by altering the cold rolling and annealing processes. Their HCF performances were examined under the fully-reversed uniaxial loading condition. Their respective average grain sizes were measured to be 30 μm and 0.65 μm for the CG and UFG conditions [as shown in Figure 3.2.1(a)-(b)]. The HCF property of the UFG specimens was found to be much better, compared to the CG ones. Even considering their different ultimate tensile strengths (UTS), the fatigue ratio (the ratio of fatigue strength to UTS) of the UFG HEA is still higher than that of the CG HEA (0.32 versus 0.28). This superior fatigue resistance could be related to the recrystallized fine grains. It should be noted that the slopes of the two S-N curves presented in Figure 3.2.1(c) are different, and this phenomenon is attributed to the pre-existing cracks introduced from the cold rolling. The influence of such cracks will be further discussed in Section 3.2.2. Another study by Chlup et al. [124] on the same composition prepared by ball-milling (BM) and spark plasma sintering (SPS) suggests a similar trend. With two different sintering times, the specimens show slightly different average grain sizes. But their resulting HCF performance under a three-point bending (3PB) test was observed to differ significantly. The large grains were found to be the crack-initiation sites under both conditions. Thus, the fatigue-endurance limit of the specimens with the smaller average grain sizes is higher, due to the lack of large grains and crack initiations.

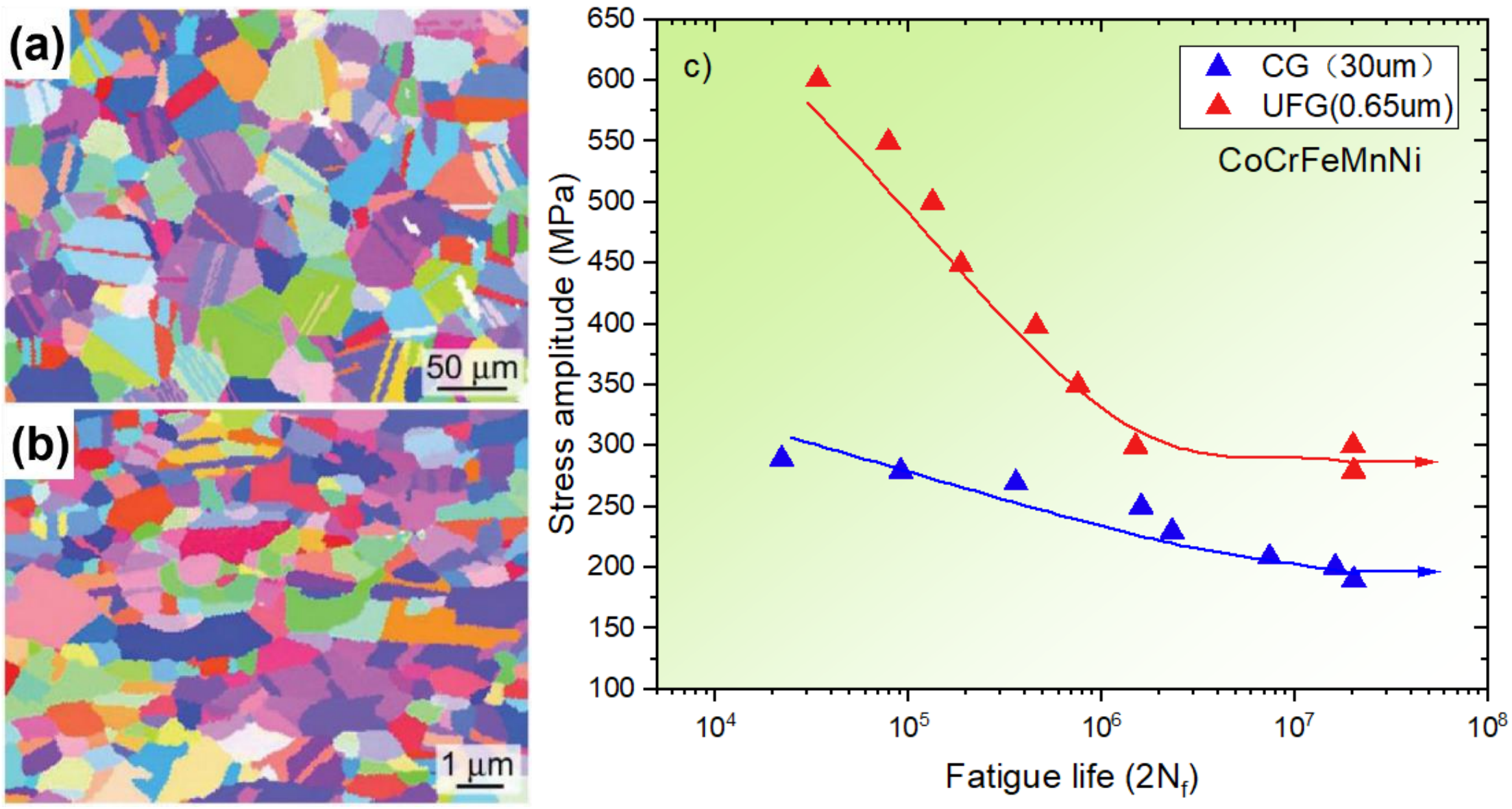

**Figure 3.2.1:** Inverse pole figures (IPF) of (a) CG and (b) UFG CoCrFeMnNi HEAs showing the average grain sizes, and (c) S-N curves under fully-reversed uniaxial HCF [15].

CoCrFeMnNi HEAs with smaller grain sizes also exhibit longer LCF life especially at lower total strain amplitudes, as confirmed by Shams et al. [67]. As can be seen from Figure 3.2.2(a), the number of reversals to failure increases with decreasing grain size, which are labeled in the figure legends. The plastic-strain-amplitude plots in Figure 3.2.2(b) suggest that the plastic resistance of these HEAs is not sensitive to the grain size, indicating that the higher LCF life can be mainly attributed to the elastic resistance in the specimens with finer grains. Lu et al. [65] investigated the LCF behaviors of similar materials with distinct grain sizes by altering the annealing temperature. The reduced grain size leads to an improvement in the LCF life at all of the strain amplitudes investigated. Another study by Picak et al. [24] on the same composition, comparing hot-extruded (HE) and equal channel angular pressed specimens, shows improved LCF life at the lowest strain amplitude for the ECAP specimens with refined grains. However, the fatigue lives of the ECAP specimens are shorter at higher strain amplitudes, due to their higher stress amplitudes and cyclic softening. Such a grain-refining strategy has been utilized in HEA studies to improve the fatigue properties of the HEAs, especially in the HCF regime [20, 46].

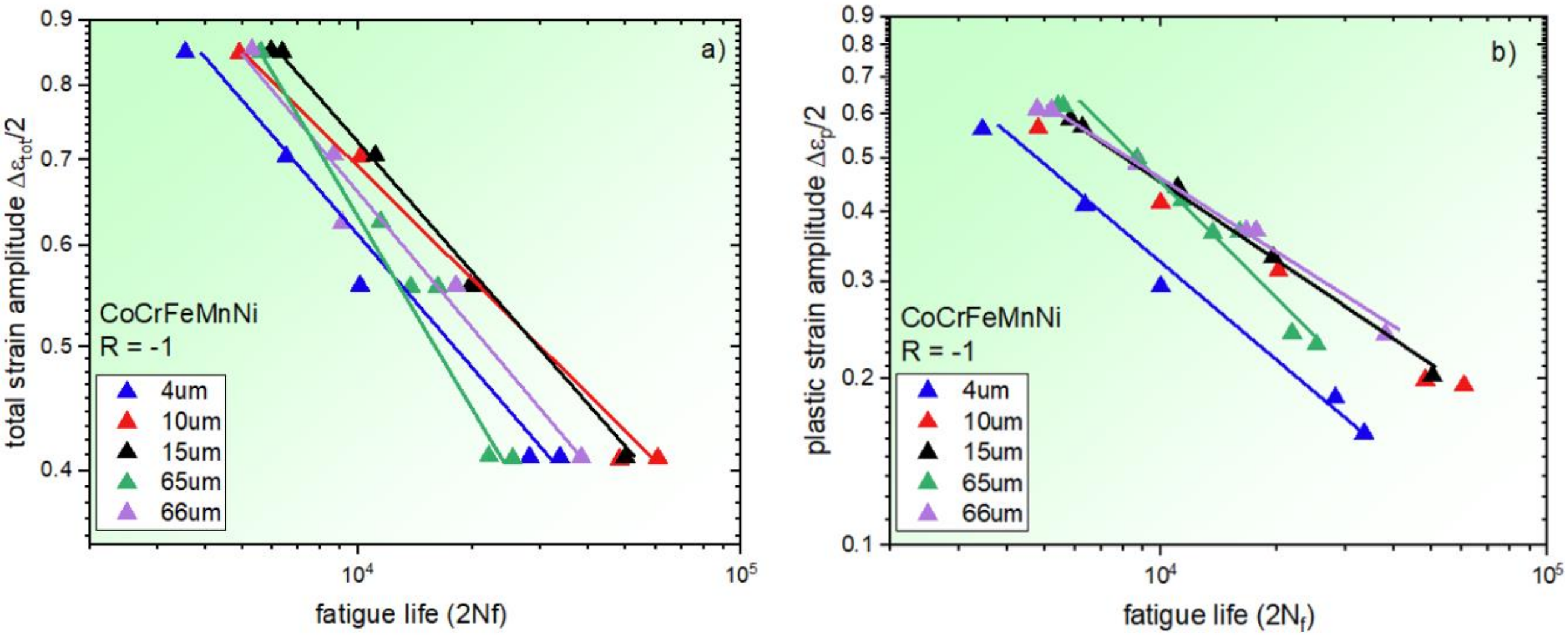

**Figure 3.2.2:** The LCF curves of (a) total strain amplitude, (b) plastic strain amplitude, and (c) stress amplitude versus the number of reversals to failure for the CoCrFeMnNi HEAs [67].

The single-edge notch tension (SENT) study by Sidharth et al. [109] on single-crystalline CoCrFeMnNi HEAs loaded along the <001> and <111> crystallographic directions [109] presents an extreme condition, since the grain size can here be assumed to be infinitely large, and the effect of the grain boundaries hence can be ignored. Based on the fatigue-crack propagation (FCP) results, the RT threshold of the stress intensity factor range, $\Delta K_{th}$, is found to be independent of the crystallographic orientation. Moreover, the specimens loaded parallel to the <001> direction present higher stress intensities and slower crack-growth rates. Although the fatigue lifetime of the single crystalline HEAs was not investigated here, this study on the FCP provides insight into how the crack grows within the grains and into how the intra-granular failure happens.

### *3.2.2. Defects, impurities, and segregations*

Defects, impurities, and chemical segregations are typically considered to be harmful to the fatigue lives of HEAs, as they usually act as crack-initiation sites or promote crack growth. Stress could be concentrated around the pre-existing defects, and dislocations could accumulate, where the chemical composition or microstructure changes. Lee et al. [38] pre-strained the CoCrFeMnNi HEAs at RT and cryogenic temperature, and thus introduced different levels of micro-voids. As presented in Figure 3.2.3(a)-(d), the voids formed in the specimens pre-strained at RT are fewer but larger in size. On the other hand, a large number of smaller voids were found in the specimen pre-strained at the cryogenic temperature. The HCF performance is better in the specimens pre-strained at RT (the RT specimen has fewer defects than the one pre-strained at the cryogenic temperature). It can be concluded that the number of micro-voids plays an important role in determining the HCF life of HEAs, since the micro-voids greatly promote the crack initiations as well as the subsequent crack growth.

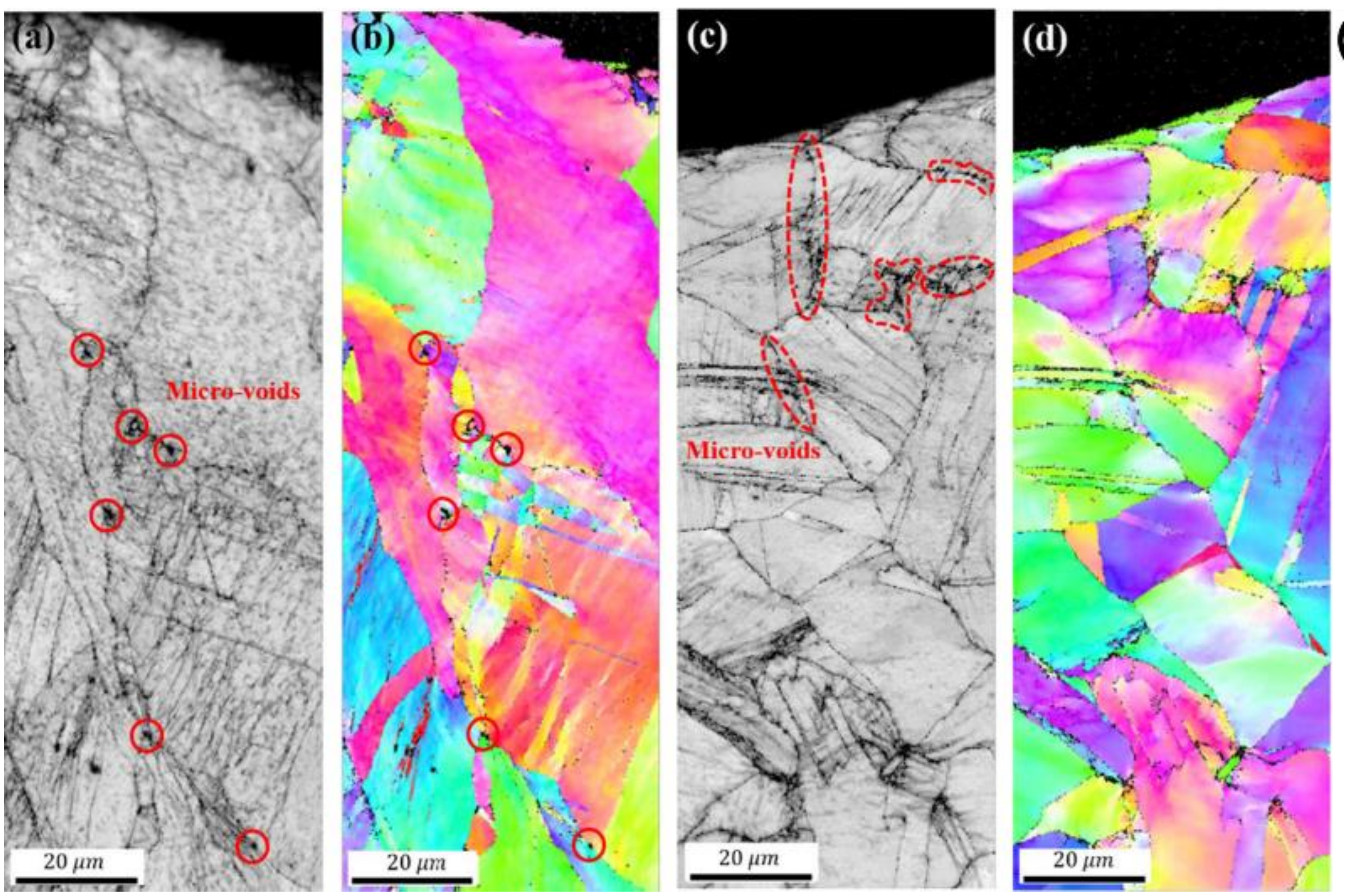

**Figure 3.2.3:** EBSD IQ and IPF maps of specimens pre-strained at (a) - (b) RT and (c) - (d) a cryogenic temperature [38].

To investigate the effect of the purity of raw materials and porosities, Tang et al. [49] prepared the $Al_{0.5}CoCrCuFeNi$ HEAs under three different conditions. Specimens in Condition-1 and Condition-2 were fabricated using raw metals of commercial purity, whereas samples in Condition-3 were fabricated using elements of high purity. In addition, shrinkage pores and macro-segregation on the ingot surfaces after fabrication were carefully removed for the Condition-2 and Condition-3 specimens. The HCF testing results using the 4PB method indicate that the specimens with Condition-3 (the ones with high-purity elements and defects removed) show the best fatigue resistance and less scattering in data. This phenomenon is believed to be related to the pre-existing defects and impurities. The research work of Tang et al. [49] suggests that the HCF properties of HEAs could be enhanced by utilizing high-purity raw materials and improved fabrication processes.

The existence of carbides [43, 67], oxides [21, 40], and other segregations [51] or clusters [69] has been observed in fatigue studies of HEAs. Such entities are viewed as crack-initiation sites, which degrade the fatigue resistance in LCF tests [21, 40, 51, 67], especially at high strain amplitudes. As illustrated in Figure 3.2.4, the clusters or other similar features could retard the reversible motion of planar dislocations [69]. Figure 3.2.4(a) depicts a case, where the dislocations are pinned by the nano-clusters, at a low strain amplitude. Figure 3.2.4(b), on the other hand, describes a case, where the copper rich nano-clusters get sheared, at higher strain amplitude or at high stress applied during cyclic loading. At high strain amplitudes, the accumulated high density of dislocations could lead to local failures, which could further initiate microcracks. However, carefully controlled formations of carbides or oxides could also be utilized to improve the HCF properties, under uniaxial loading, or to improve the LCF life, at low strain amplitudes [21, 67]. The introduction of uniformly distributed carbides or oxides typically increases the strength of HEAs, and thus also enhances the HCF endurance limit or improves their elastic resistance under LCF.

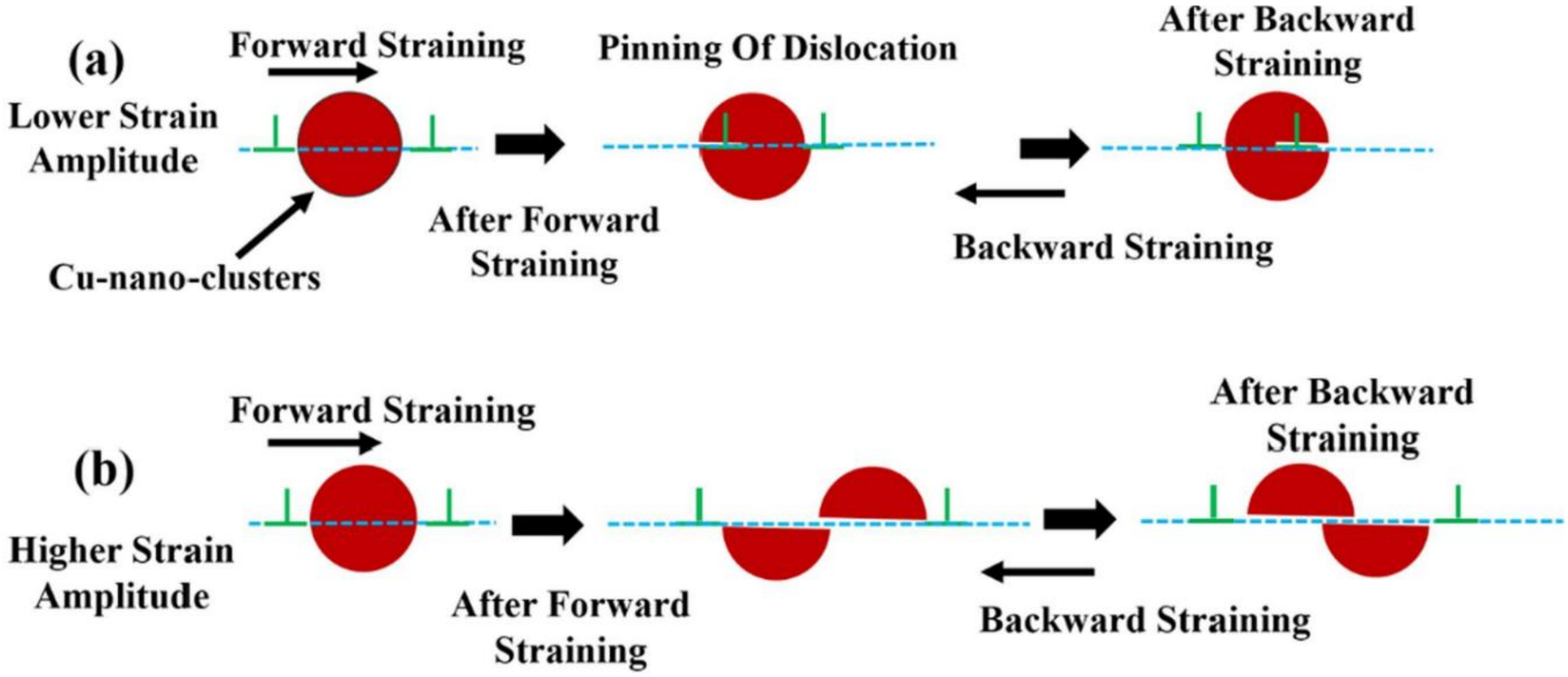

**Figure 3.2.4:** A schematic illustration of the interaction between dislocations and copper rich nano-clusters at (a) low strain amplitudes and (b) high strain amplitudes in a CoCuFeMnNi HEA under LCF [69].

#### *3.2.3. Secondary phases*

The formation of secondary phases typically improves the fatigue resistance, due to the interaction between secondary phases and dislocations. As illustrated in Figure 3.2.5, uniformly distributed particles with an incoherent secondary phase could hinder the movement of dislocations, and thus, impede the formation of persistent slip bands (PSBs) in HCF. Such a phenomenon would result in the reduction of cyclic strain localization and in the delay of fatigue-crack nucleation. The fatigue resistance could then be improved, since the PSBs play an important role in the initiation of microcracks and in the propagation of fatigue cracks. For instance, the BCC phase is found to form in $Al_x(CoCrFeMnNi)_{100-x}$ HEAs with high Al content. Furthermore, the HCF resistance under cantilever bending is enhanced in this HEA with BCC particles distributed in the FCC matrix [70]. However, if the precipitates formed are coherent with the matrix, then improvement in fatigue properties is not obvious, since these coherent precipitates tend to resist shearing during cyclic loading [45].

Another example of a eutectic HEA (EHEA) with the composition of $AlCoCrFeNi_{2.1}$ under HCF cantilever-bending tests is presented in Figure 3.2.6 [16]. Compared to a hierarchy structure containing

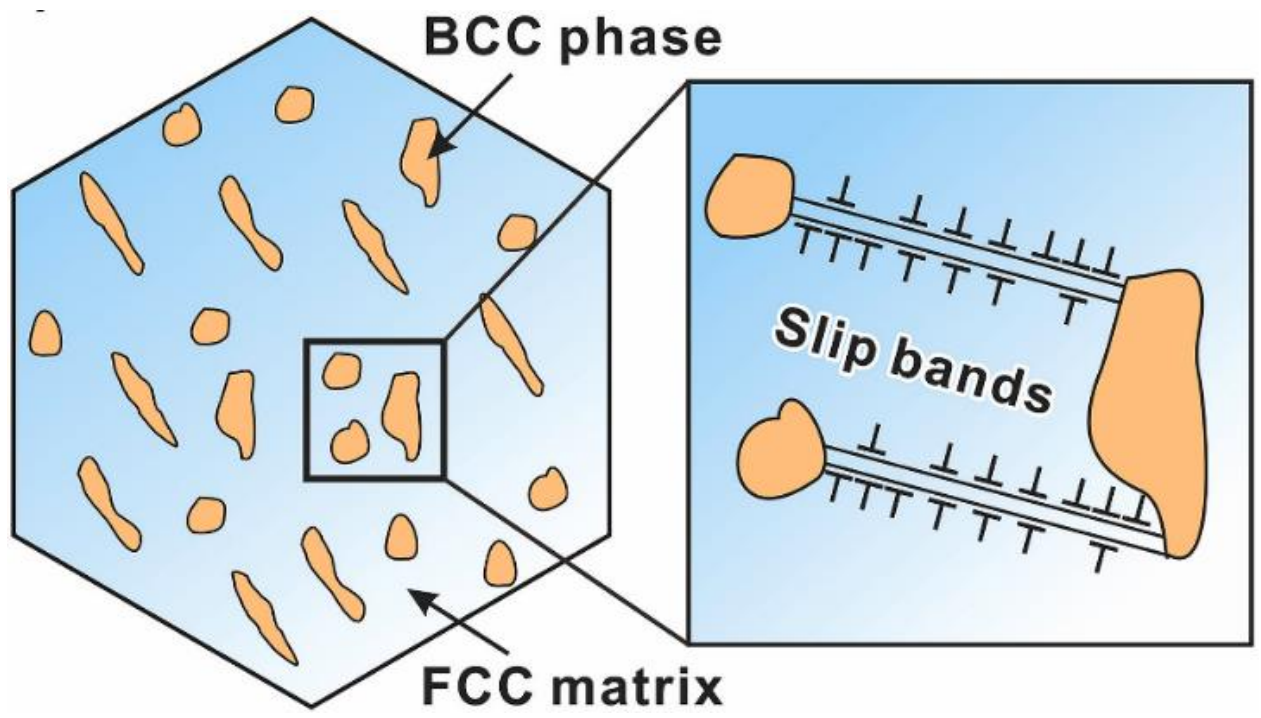

**Figure 3.2.5:** A schematic illustration of the influence of a secondary phase on the formation of slip bands [70].

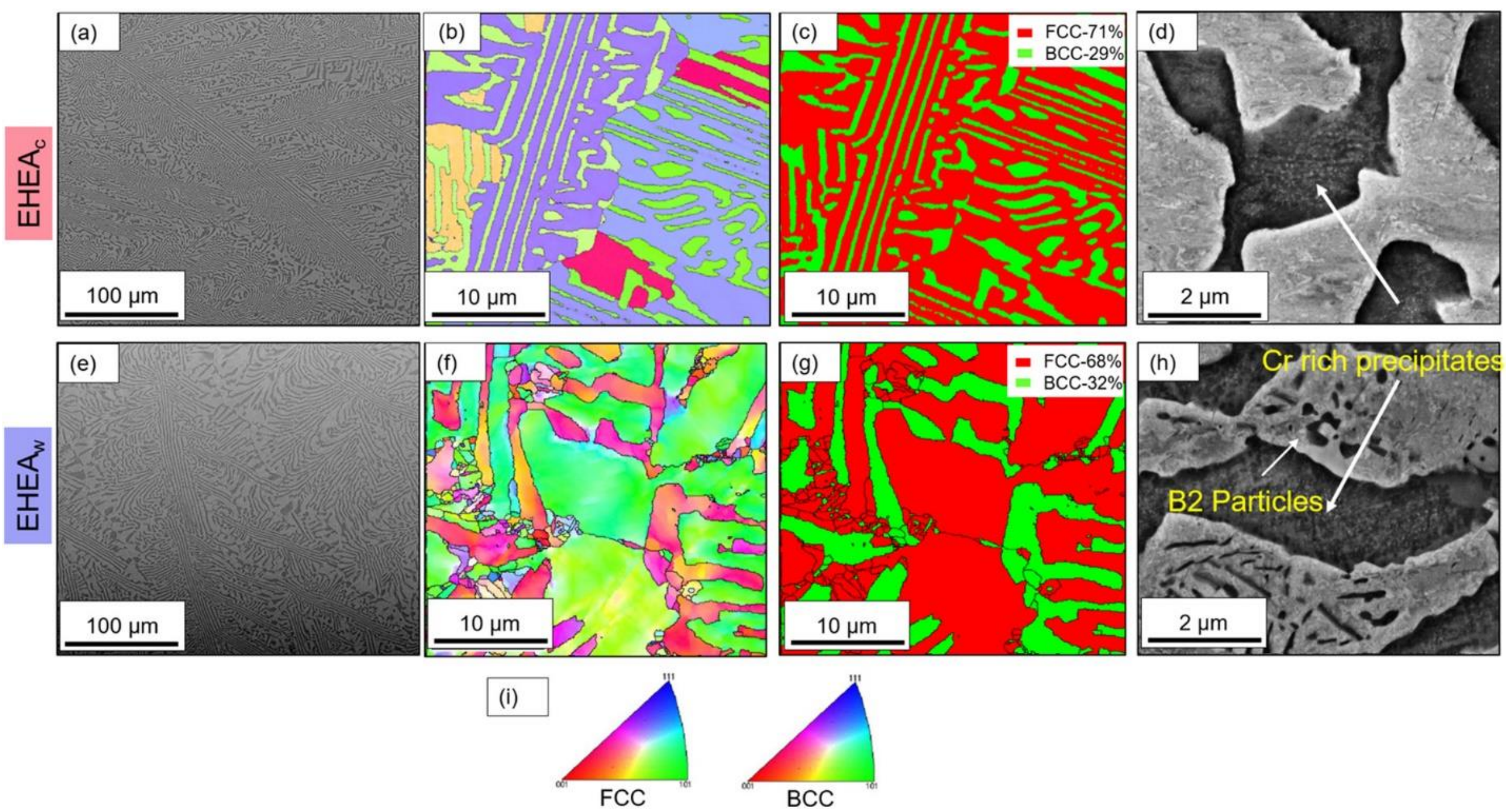

**Figure 3.2.6:** BSE images, EBSD-IPFs, phase maps, and high-magnification BSE images of (a) - (d) EHEA$_c$ and (e) - (h) EHEA$_w$ [16].

FCC lamellae and BCC phases in the as-cast specimens (named as EHEA$_c$), B2 particles could be observed inside the deformed and recrystallized FCC matrix in the cold-rolled and heat-treated specimens (named as EHEA$_w$). Such secondary phases, introduced by thermomechanical processes, could improve the HCF performance of HEAs by stopping fatigue-crack initiations, resulting from the hindered path of PSBs.

### *3.2.4. Phase transformations*

Phase-transformations during plastic deformation could retard the crack propagation under cyclic loading by introducing work hardening in the plastic regions near the crack tip. The metastability of the FCC phase has been well studied in some of the HEAs containing 3d-transition elements. Such HEA compositions, including $Fe_{50}Mn_{30}Co_{10}Cr_{10}$, $Fe_{42}Mn_{28}Cr_{15}Co_{10}Si_5$, and $Fe_{38.5}Mn_{20}Co_{20}Cr_{15}Si_5Cu_{1.5}$, have been proven to exhibit great fatigue performance under HCF, LCF, and FCP tests [20, 27, 33, 46, 110, 125]. By modifying chemical compositions and utilizing the friction stir process, Liu et al. [20, 46] developed metastable HEAs that show excellent HCF properties under cantilever bending. The crack initiation is delayed by applying an aforementioned grain-refining strategy. More importantly, the crack propagation is hindered, and the crack branching is promoted through the TRIP effect, by introducing FCC to HCP phase transformations. As presented in Figure 3.2.7, the phase transformations could be observed near the crack tip. The EBSD phase maps and phase-fraction-distribution curves in Figure 3.2.7($b_1$)-($b_3$), and (d) indicate that the HCP phases are mainly distributed in the plastically deformed regions behind the crack tip, while the microstructure is kept in a nearly pure FCC phase in front of the crack tip with no failures. Such a controlled TRIP effect, attributed to the metastability of the FCC matrix, could greatly improve the fatigue resistance of HEAs.

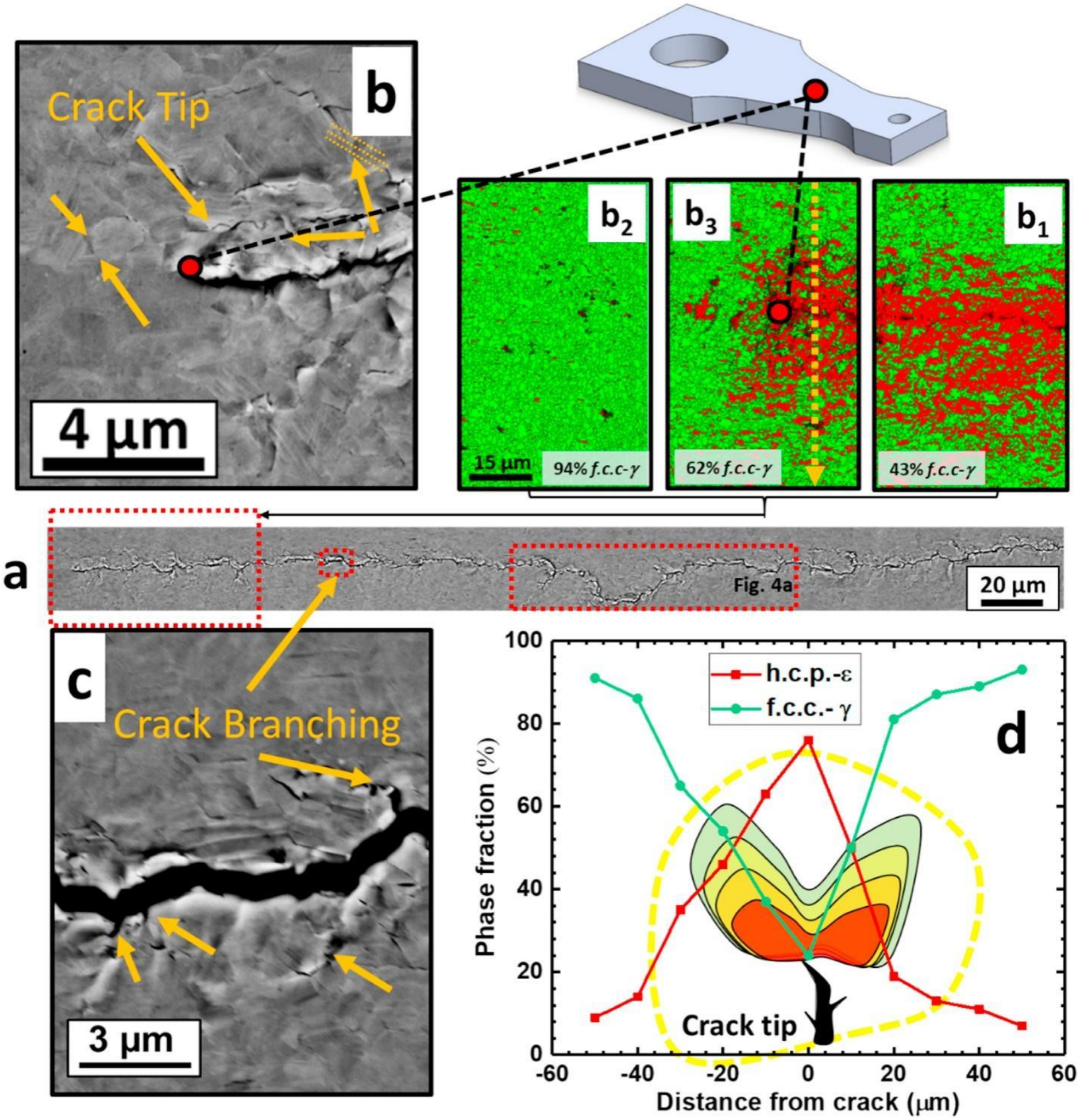

**Figure 3.2.7:** BSE images (a) overviewing the whole crack, (b) focusing on the crack tip with higher magnification, and (c) showing crack branching. ($b_1$) - ($b_3$) EBSD phase maps revealing the distribution of FCC and HCP phases. (d) Evolution of HCP and FCC phases in vicinity of the crack tip, due to plastic zone [20]. HCP dominates at the crack tip. But as one moves away from the crack tip, FCC becomes more prevalent.

The deformation-induced phase transformation in the metastable $Fe_{50}Mn_{30}Co_{10}Cr_{10}$ HEA is also the dominant deformation mechanism under LCF [27, 125], especially at lower strain amplitudes. FCP studies using CT specimens on the same material [33, 110] also confirm that a massive FCC to HCP phase transformation occurs near the propagated-crack tips. However, Bahadur and collaborators [125] noted that the bi-directional phase-transformation behavior between FCC and HCP phases could be observed close to the fracture tip at higher strain amplitudes in LCF. The higher cyclic irreversibility of the dislocations generates local stress-concentration fields, which could increase the local temperature. Thus, the reverse transformation from the HCP to the FCC phase could be promoted.
A different martensitic-transformation path originating from B2 precipitates to an orthorhombic structure is utilized by Feng et al. to improve the LCF life of a thermo-mechanically-processed $Al_{0.5}$CoCrFeNi HEA [25]. These intermetallic B2 precipitates could effectively retard the initiation of fatigue microcracks, due to their ductility and transformability. Unlike traditional intermetallic precipitates, that cannot bear too much plastic deformation, the ductile and transformable B2 structure, shown in Figure 3.2.8, could be plastically deformed and transformed in its microstructure, which actives additional dislocation-slip systems.

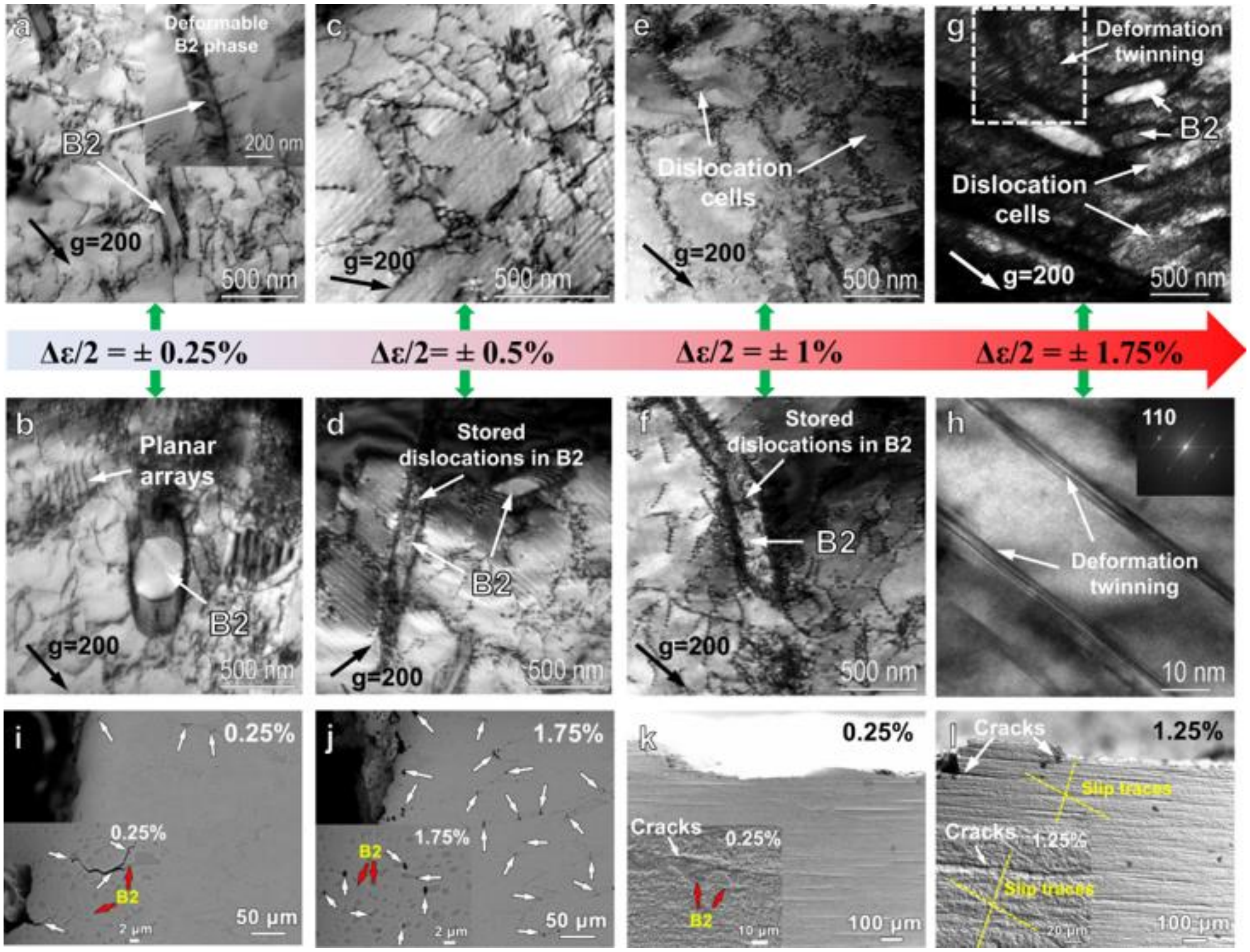

**Figure 3.2.8:** TEM and SEM characterizations of $Al_{0.5}$CoCrFeNi HEA with B2 precipitates and other related structures under LCF [25]. Bright-field TEM images with strain amplitudes ranging ±0.25%, ±0.5%, ±1%, and ±1.75% (a - h), the microstructural cyclic-deformation behavior is shown at several nm scales wherein the inset figure in Fig. K1a shows the plastically deformable B2 phase in a dark-field image. SEM images of fractured samples with strain amplitudes ranging from ± 0.25%, ± 1.25%, to ± 1.75% (i–l) [25].

### 3.2.5. *Dislocation structures*

Dislocations are commonly formed under cyclic loading because of the reversable planar slips of atoms. Discrete dislocation lines would tangle together and then form specific structures, which will further cause local stress or strain concentrations. Then the accumulated local plastic deformation would become the initiation sites for microcracks and further propagate to cause the final fatigue failure. Typical dislocation structures, that have been observed in HEAs under cyclic loading, are summarized in Figure 3.2.9 [21, 24, 25, 39, 40, 43-45, 51, 65-68, 70, 106, 108, 111, 125]. These dislocation structures mainly include random dislocation structures, such as discrete dislocations and tangled dislocations, planar dislocation structures, such as slip bands and stacking faults, and wavy dislocation structures, such as dislocation loops, veins, walls, and cells. It has been suggested that the dislocations mainly consist of planar dislocation structures at low strain amplitudes in LCF, while they generally consist of wavy structures at high strain amplitudes [65]. Moreover, the slip mode may change from planar to wavy structures, with increasing number of cycles, which can also be attributed to the cyclic-stress response [65]. Therefore, pre-existing dislocation structures could act as crack-initiation sites, and thus, might degrade the fatigue resistance of HEAs.

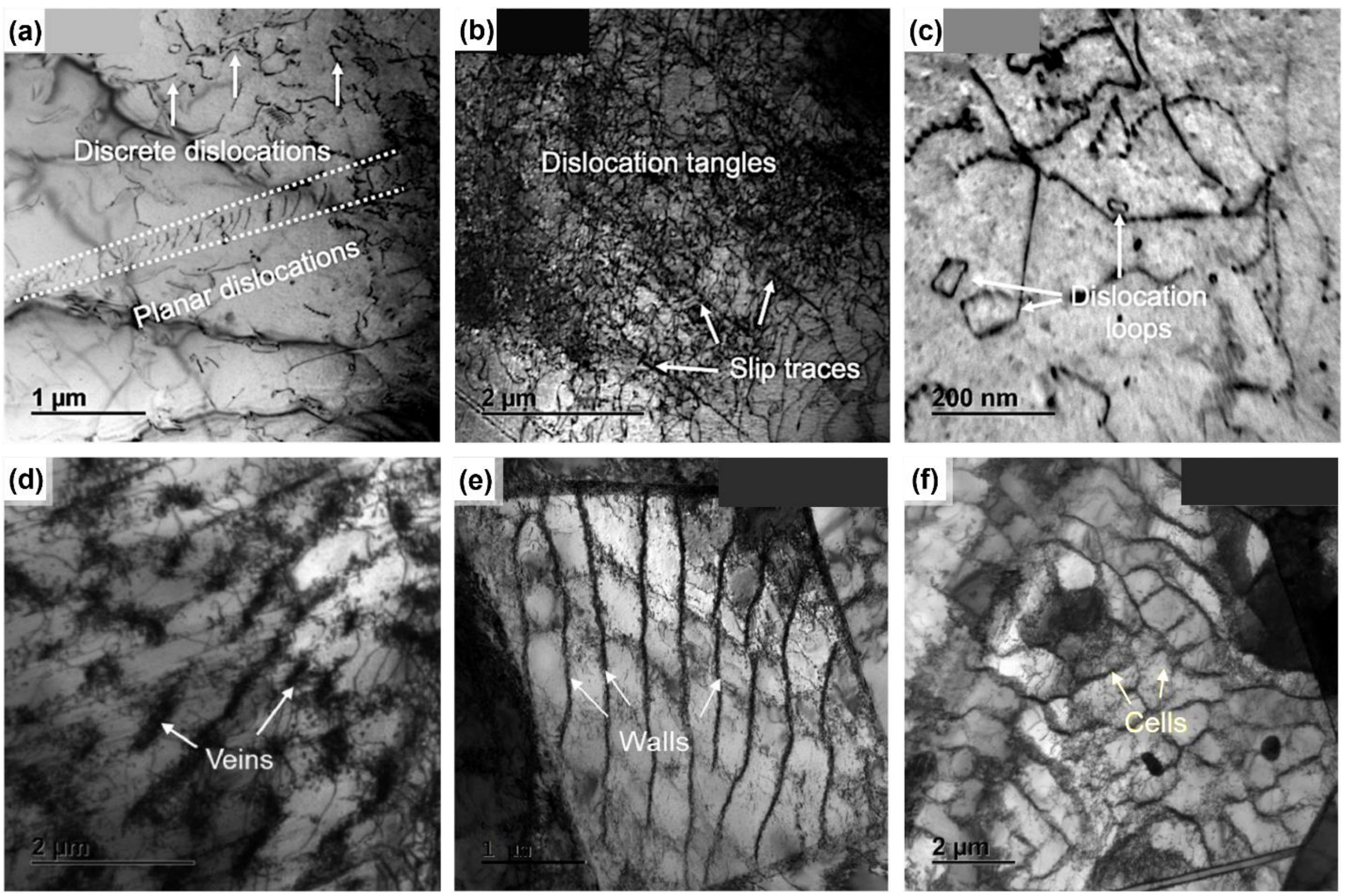

**Figure 3.2.9:** TEM images of typical dislocation structures in HEAs formed under cyclic loading [51, 65, 66].

### 3.2.6. *Deformation twins*

Besides the aforementioned dislocation structures, the formation of deformation twins or nanotwins is the other dominating deformation mechanism that is commonly observed in HEAs under cyclic loading [21, 22, 24, 25, 39, 40, 43, 49, 67, 68, 106, 108, 109, 111, 124, 125]. The deformation twins could form before crack initiation or during the crack-propagation process. Figure 3.2.10 presents a typical morphology of deformation twins and shows how they are intersected. The deformation twins can be induced at a relatively low stress during cyclic loading, compared with the critical stress for such twin formation in monotonic deformation [25, 39]. The increasing density of deformation twins could result in higher work hardening, and could give better resistance to the initiation of microcracks, since more twin boundaries tended to degrade the local stress concentration [49]. Hence, the formation of deformation twins could strengthen HEAs and improve their fatigue life. The deformation twins were also observed to be at the crack-propagation path under HCF [40]. However, the propagation mechanism, and the influence of deformation twins on the crack-growth rate, still needs to be further investigated.

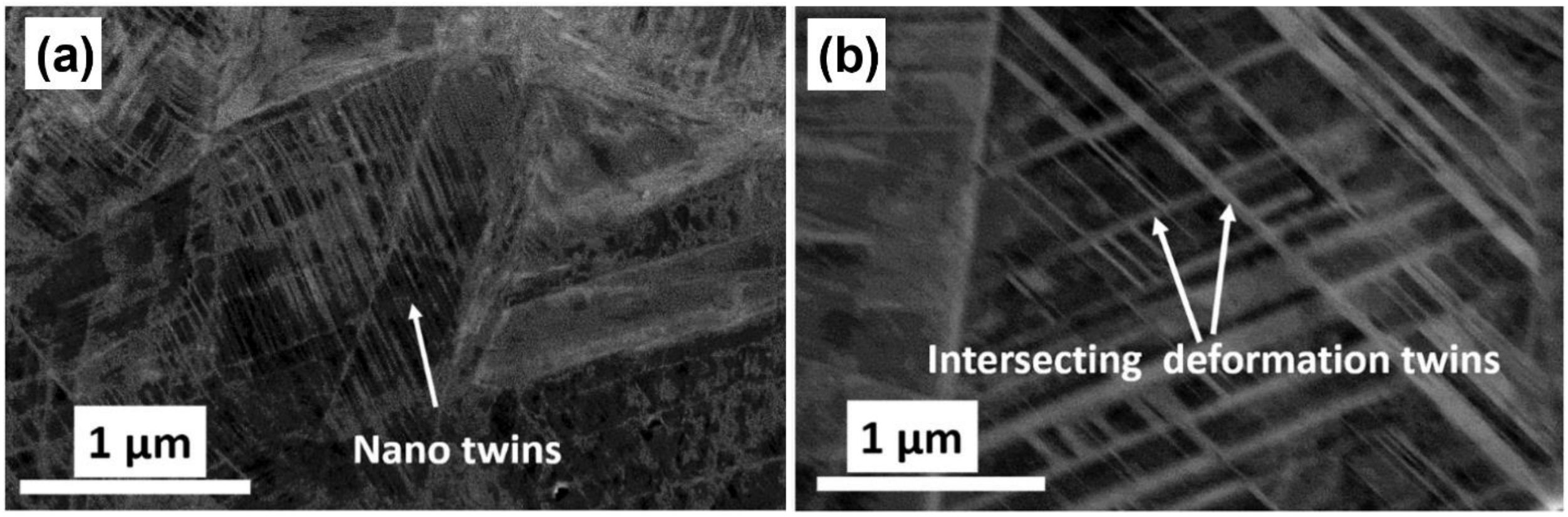

**Figure 3.2.10:** Typical TEM images of (a) nano-sized deformation twins and (b) the intersection between twins [125].

### 3.3.Temperature effects

Temperature has a significant effect on the mechanical properties of alloy materials. For most alloys, a higher environment temperature reduces the stiffness and strength, while the ductility increases with increasing temperature. Research by Thruston et al. demonstrated that this rule is also applicable to HEAs [30]. Just like other mechanical properties, the fatigue behavior of HEAs is also affected by ambient temperature, this effect is reflected in HCF, LCF, and crack propagation, as described below.

#### *3.3.1. Temperature effect on accumulated damage*

Figure 3.3.1 captures the fully reversed strain-controlled fatigue data of two sets of CoCrFeMnNi with similar grain sizes and the same strain rate ($\dot{\varepsilon} = 3 \times 10^{-3}\ s^{-1}$) a) at RT and b) at high temperature (550°C) [51]. From the perspective of both total strain amplitude ($\Delta\varepsilon_{total}/2$) and plastic-strain amplitude ($\Delta\varepsilon_{p}/2$), the fatigue behavior of CoCrFeMnNi under a high-temperature (HT) environment is inferior to that of a RT specimen, which indicates that high-temperature environment has a negative impact on the LCF behavior of the CoCrFeMnNi HEA. It is worth mentioning that when the strain amplitude is large, the number of cycles experienced by the HT specimen is significantly lower than that of the RT specimen. As the total strain amplitude gradually approaches 0.3, the gap in fatigue life also has decreased.

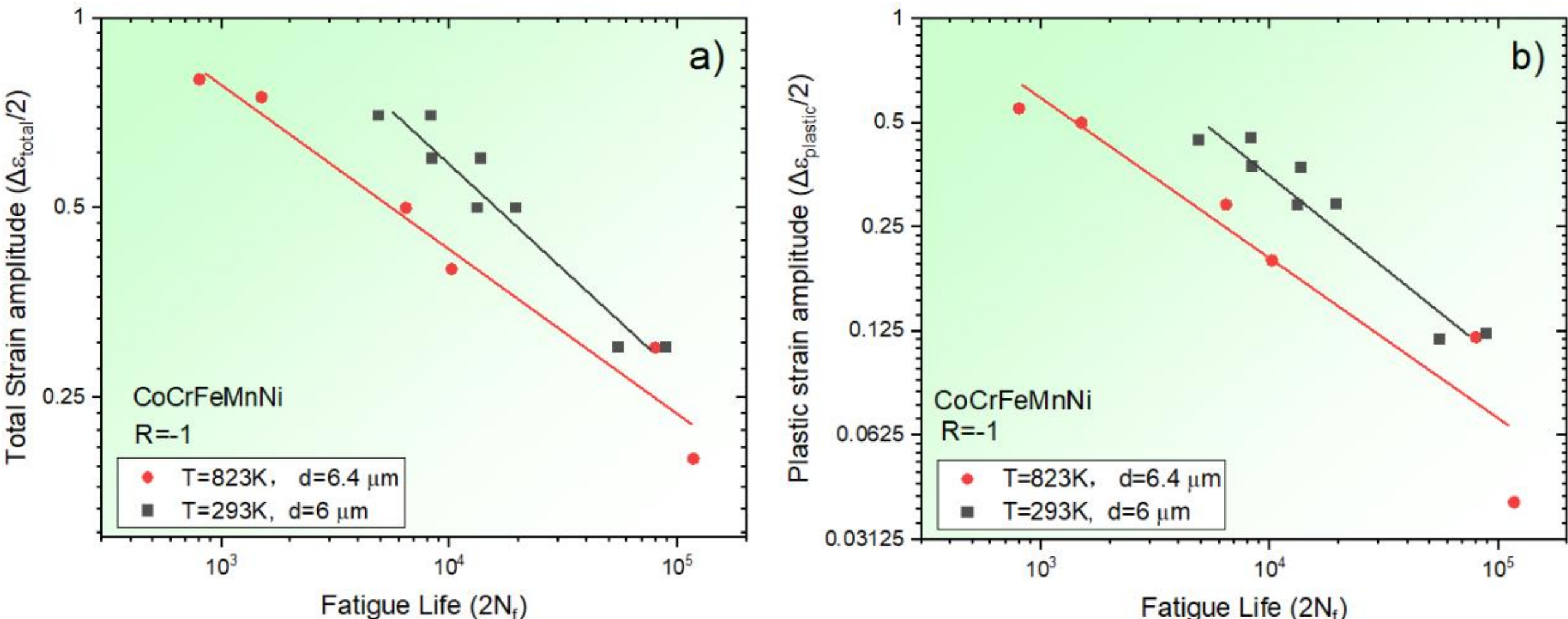

**Figure 3.3.1:** The a) total strain-life and b) plastic strain-life for CoCrFeMnNi under the room and high temperatures, respectively (adapted from [51]).

The increase in temperature not only shortens the life of the HEA, under the same $\Delta\varepsilon_{total}/2$, but also significantly affects the stress that it can withstand. In LCF, there is a power relation between the cyclic stress amplitude, $\Delta\sigma_t/2$, and plastic strain amplitude, $\Delta\varepsilon_p/2$, captured in Eq. (2.2.4). In the double logarithmic plot in Figure 3.3.2(a) of the stress amplitude, $\Delta\sigma_t/2$, and the plastic-strain amplitude, $\Delta\varepsilon_p/2$, the linear correlation confirms this relationship. The $K$ and $n$ values fitted through the HT experimental data are both lower than those of the RT specimen, which suggests that the CoCrFeMnNi HEA experiences a softening effect at high temperatures.

In addition, Figure 3.3.2(b) shows the changes in the tensile peak stress of the HT specimen and the RT specimen, for $\Delta\varepsilon_{total}/2$ of 0.3% and 0.5%, respectively [51]. During the entire fatigue test, the peak stress of the HT specimen is ~100 MPa lower than that of the RT specimen, under the same $\Delta\varepsilon_{total}/2$. But the RT specimen presents a stronger cyclic-stress sensitivity: It exhibits a cyclic-hardening phenomenon faster and enters a cyclic-softening stage after hardening reaches a certain level. In contrast, HT specimens show the cyclic-hardening phenomenon only when the $\Delta\varepsilon_{total}/2$ is high. But the peak stress remains stable in the post-hardening stage, until the fracture happens. In the low $\Delta\varepsilon_{total}/2$ experiment, the peak stress of the HT specimen has remained stable, and there is no obvious sign of cyclic hardening or softening.

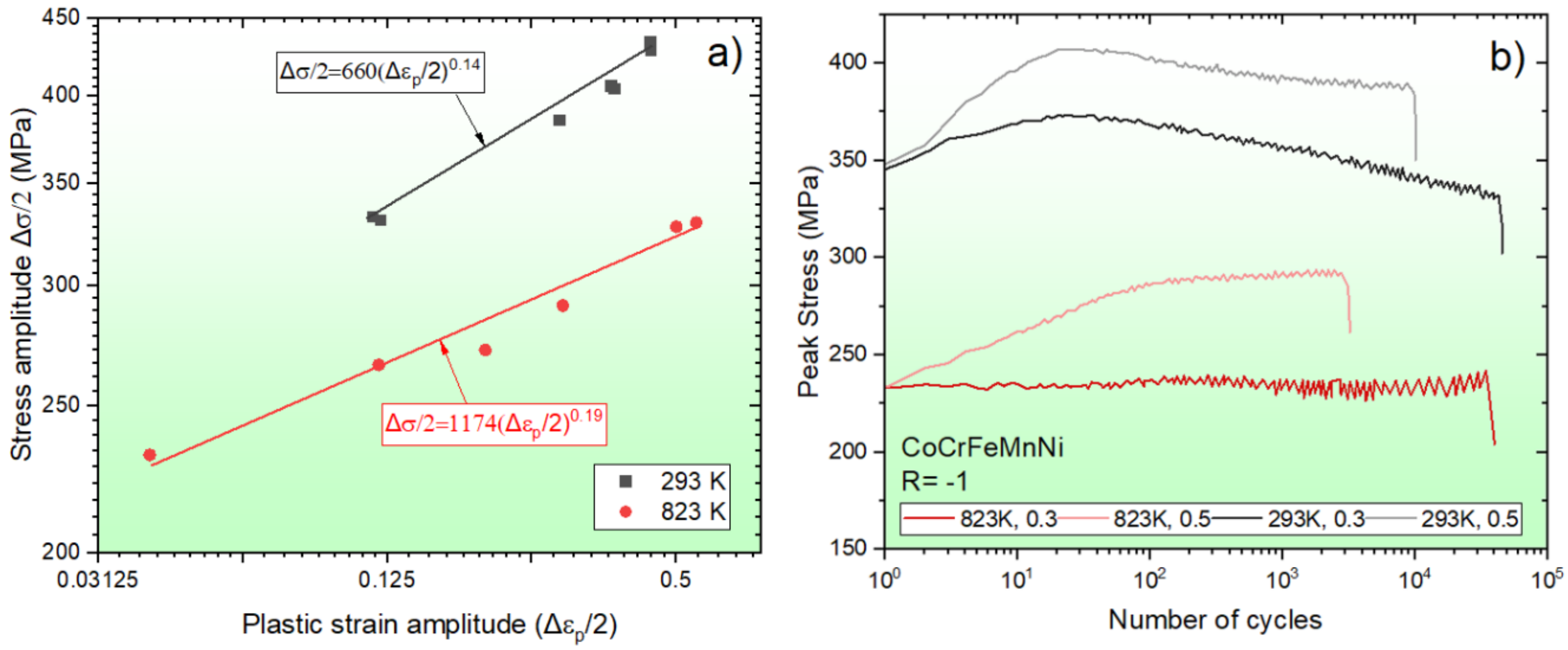

**Figure 3.3.2:** a) The relationship between the stress amplitude and strain amplitude under room and high temperatures. b) The cyclic-stress responses of specimens under room and high temperatures when the strain amplitudes are 0.3 and 0.5, respectively [51].

The cause of this high-temperature effect can be explained, to a certain extent, by looking at the microstructure of the CoCrFeMnNi HEA. In the HT samples after cycling, the content of LAGBs has been significantly increased, compared to RT, which is also an embodiment of the high-temperature softening effect on the HEA, shown in Table 3.3.1 [51]. The microstructure of the cross sections of RT and HT specimens prove that the fatigue mechanisms of the CoCrFeMnNi HEA at different temperatures are almost the same. But the high-temperature environment significantly speeds up the entire process of fatigue degradation. The same dislocation structure often appears earlier in the HT specimen. At the same time, additional types of dislocation structures are found in the HT specimen (Figure 3.3.3) than that in the RT specimen (Figure 3.3.4). Some structures that were originally produced only under high strain amplitude (0.5% ~ 0.7%) conditions at RT have also appeared in HT samples under lower strain amplitudes (0 ~ 0.3%). Most of the dislocations are arranged succinctly along the <111> plane, while a considerable amount of discrete dislocations, dislocation loops and tangles appear in the HT specimen [51].

**Table 3.3.1:** Increase of low-angle grain boundaries in HT samples, compared to RT, is an embodiment of a high-temperature softening effect [51].

| **Condition** | **Average grain size [μm]** | **Twins' area fraction** | **HAGBs fraction (>15°)** | **HAGBs fraction (<15°)** |
|---|---|---|---|---|
| Undeformed | 6.4 | 35% | 96% | 4% |
| 0.3% | 6.7 | 34% | 88% | 12% |
| 0.5% | 6.5 | 34% | 86% | 14% |
| 0.75% | 6.4 | 36% | 82% | 18% |

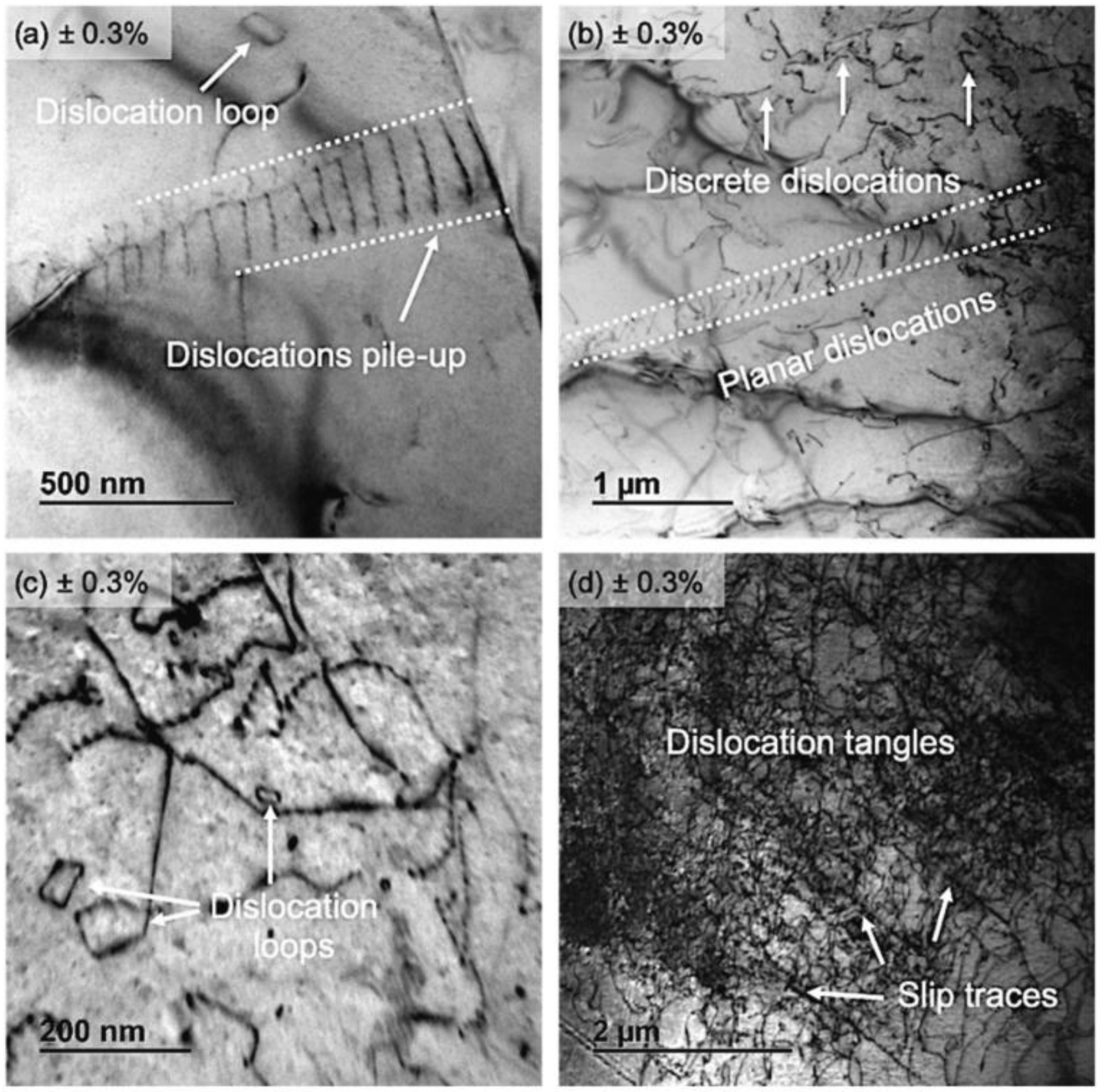

**Figure 3.3.3:** Dislocation structures in a high-temperature CoCrFeMnNi specimen. The Figure shows cyclic straining under a strain amplitude of $\pm$ 0.3% at 550°C. An inhomogeneous dislocation density was seen both between and within grains at lower strain amplitudes. (a) dislocation pile-up observed at grain boundaries, (b) planar slip bands, (c) discreet dislocations, after jerky motion, coupled with dislocation loops, and (d) slip traces and dislocation entanglements [51].

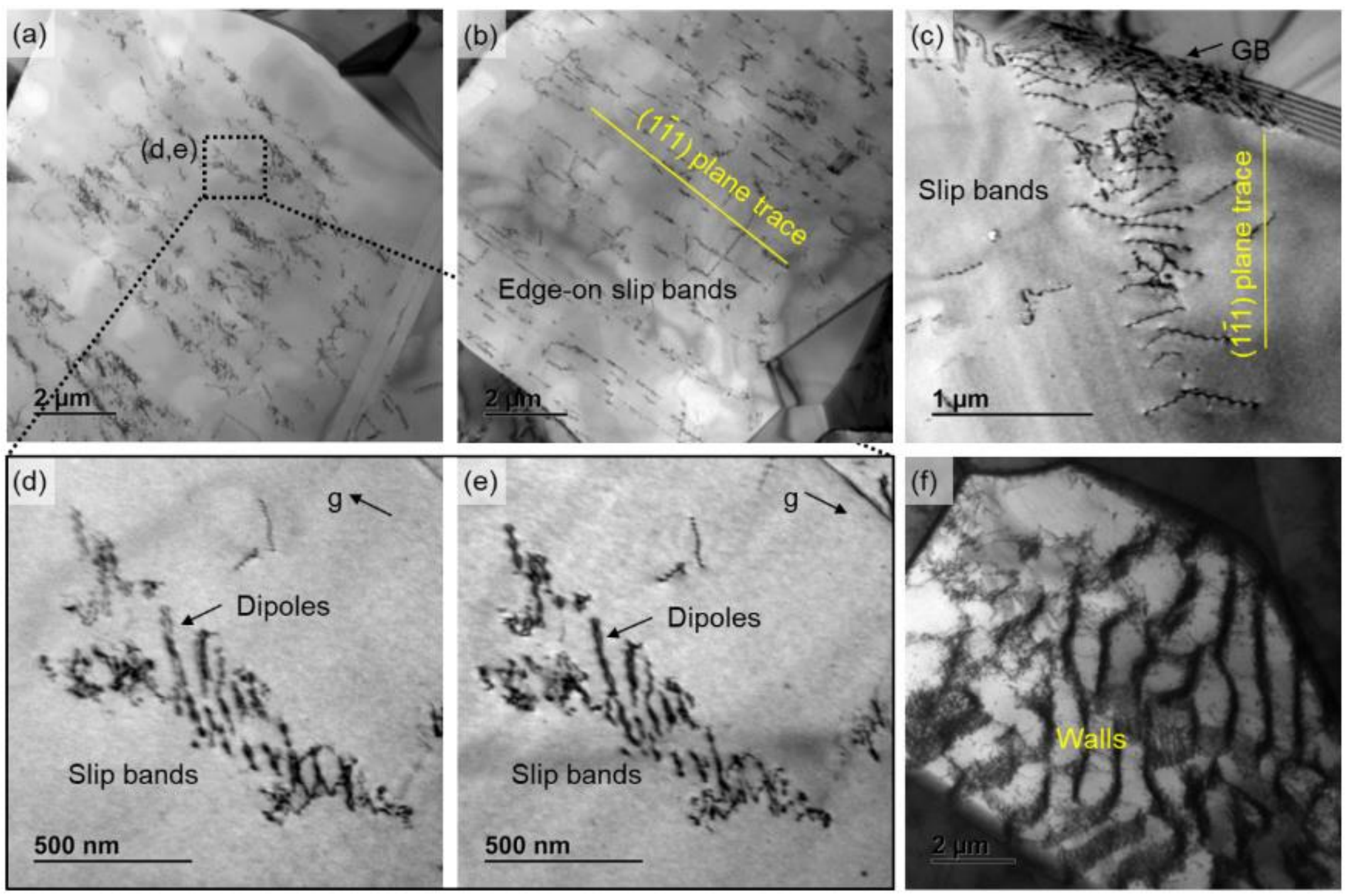

**Figure 3.3.4:** Dislocation structure in a typical room-temperature CoCrFeMnNi specimen [51].

In addition, another negative effect of high temperature on the fatigue behavior of the CoCrFeMnNi HEA relates to elemental segregation. In an environment below 800 °C, the phase stability is relatively fragile, and the grain boundary becomes the preferential nucleation point of the second phase. The main components are Cr-enriched and NiMn-enriched sub-micron-sized precipitates at the grain-boundary segregation. After reaching the critical size, these segregated particles will become a harmful structure that will reduce solid-solution strengthening and induce the embrittlement of grain boundaries. In this case, inter-granular propagation is prone to occur, and cracks are easier to form, which ultimately affects the fatigue resistance of the CoCrFeMnNi HEA [51].

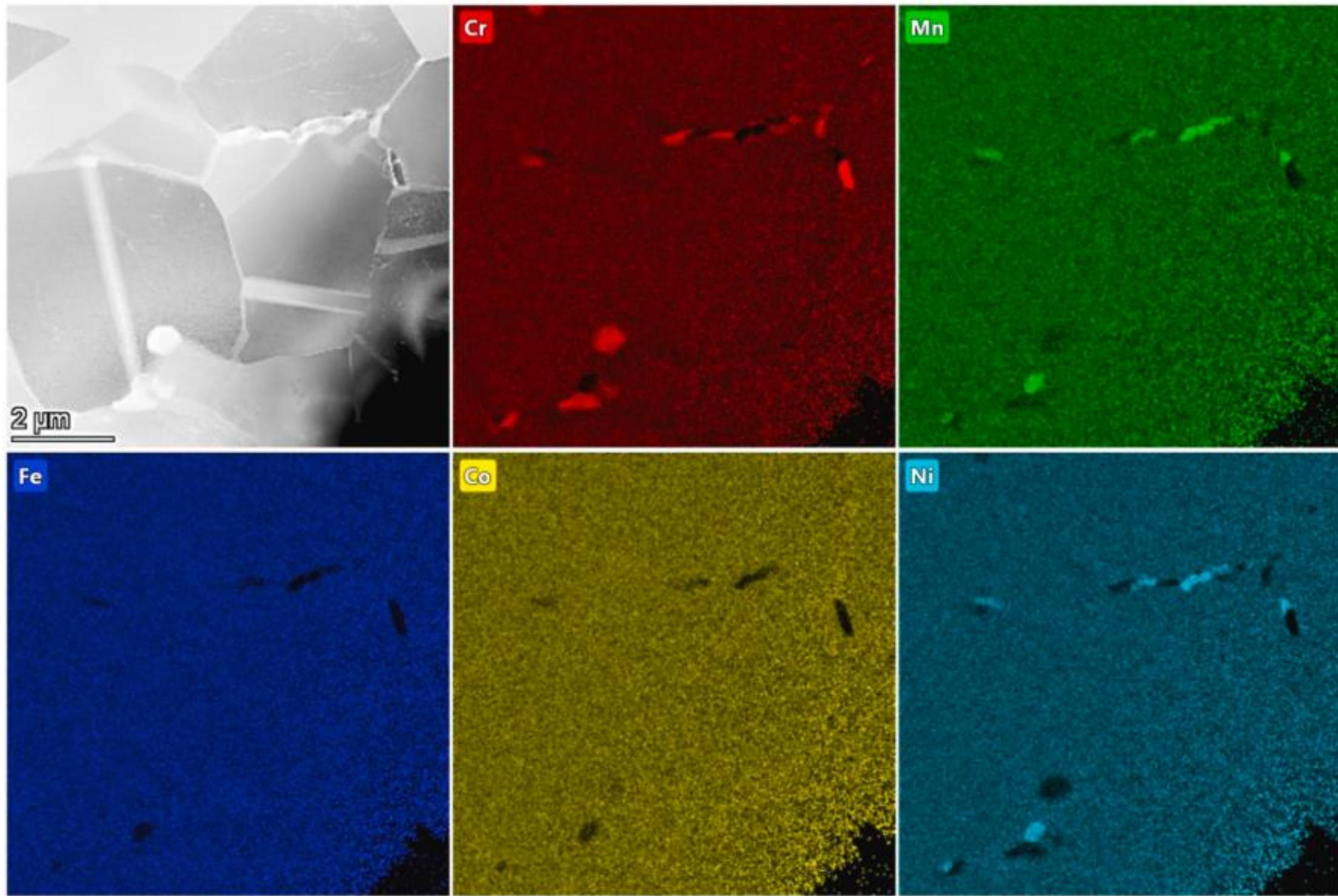

**Figure 3.3.5:** Cr-enriched and NiMn-enriched elemental segregation caused by the high-temperature environment is concentrated at grain boundaries or twin boundaries [51].

Considering that the tensile strength of materials tends to increase, as the temperature decreases, and that the fatigue endurance limit tends to increase with increasing UTS, one can expect that HEAs will exhibit higher fatigue resistance in a low-temperature environment than in a high-temperature environment. However, the S-N data and fatigue ratio vs. fatigue-life data in Figure 3.3.6, obtained by Lee et al. [38], shows that the HCF behavior of the CoCrFeMnNi HEA in the cryogenic-temperature (CRT) environment is lower than at RT, which is opposite to what one may expect. In order to minimize the impact of the UTS, Lee et al. used a pre-strain process (RT = 30%, CRT = 18%) to adjust the ultimate tensile strength of the CoCrFeMnNi specimen at RT (904 MPa) and at low temperature (849 MPa) to a fairly close level. Based on the difference of ~50 MPa in the UTS, the difference in the fatigue limit between the two specimens is estimated to be 20 - 30 MPa. However, in the final experimental results, the fatigue limit of the RT specimen (550 MPa) is nearly 80 MPa higher than that obtained for the CRT environment (474 MPa). Furthermore, the average fatigue-striation spacing of the CRT specimen (1.29 μm), under the same stress amplitude (606 MPa), is also wider than that of the RT specimen (0.24 μm). These results demonstrate that the fatigue resistance of the CRT specimens is actually worse in this case.

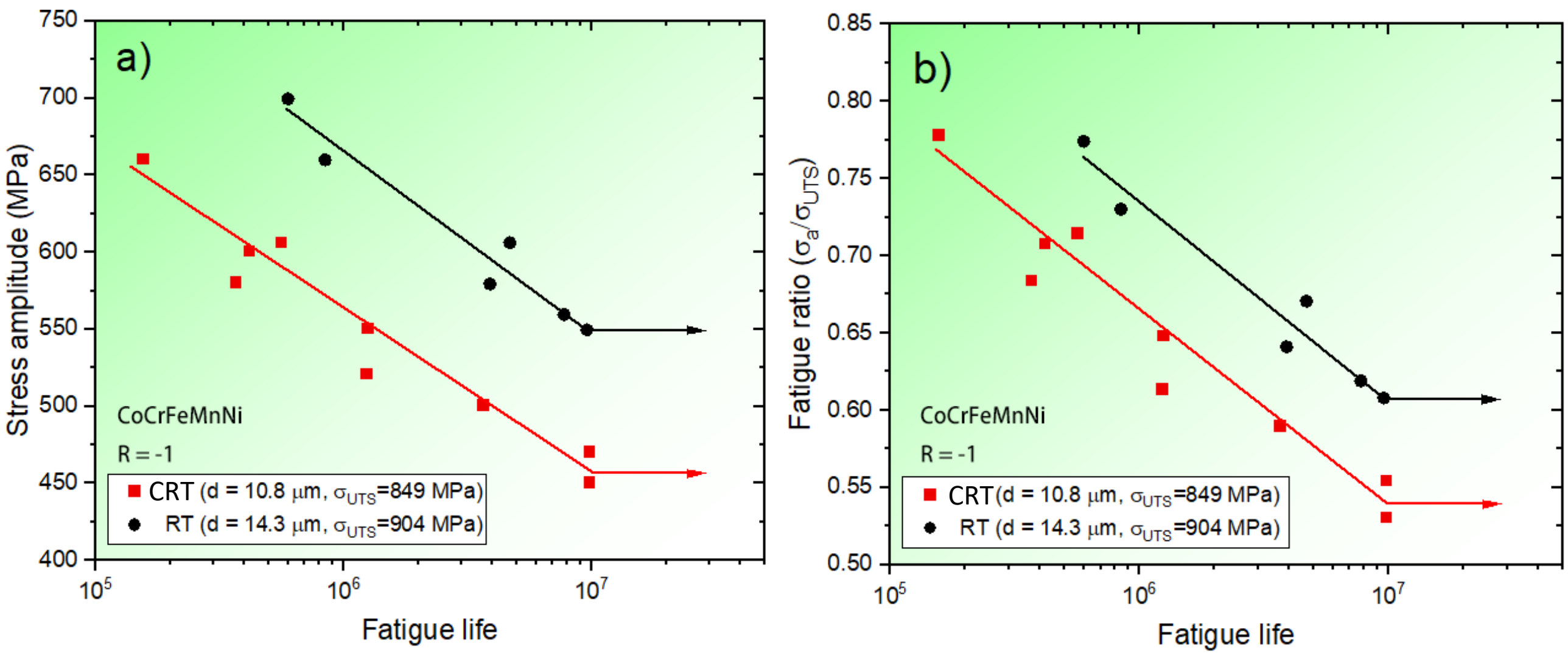

**Figure 3.3.6:** HCF behavior of RT and CRT CoCrFeMnNi under a) S-N model and b) fatigue ratio vs. fatigue life [38].

Unlike the influence of ambient temperature on fatigue behavior, this result can be primarily attributed to the co-action of the pre-strain process and the low-temperature environment. The EBSD results in Figure 3.2.3 show that more LAGBs with similar orientations are formed in the RT specimens after pre-straining than in the CRT specimens [38]. In the RT specimens, dislocations are more likely to move in the same direction and do not accumulate easily. On the contrary, in the CRT specimens, a large number of twin structures are observed in the microstructure. The number of micro-voids, that are observed at the grain boundary, or in the region where the twin boundary and the grain boundary intersect, is also larger in the CRT specimens than in the RT specimens. These comprehensive characteristics make it easy for the dislocations to accumulate and accelerate. The local stress concentration eventually causes cracks to nucleate and propagate more easily. At the same time, the results of the Kernel average misorientation analysis proved that, also in the case of $\Delta\sigma_a = 606$ MPa, even if the fatigue cracks of RT and CRT specimens were mainly formed at the grain boundary, triple junctions or micro-voids formed during the pre-strain. The RT specimen is also more evenly distributed inside the crystal grain, while the stress in the CRT specimen is more concentrated at the grain boundaries (see Figure 3.3.7), which is one of the reasons why the CRT specimen is more prone to failure [38].

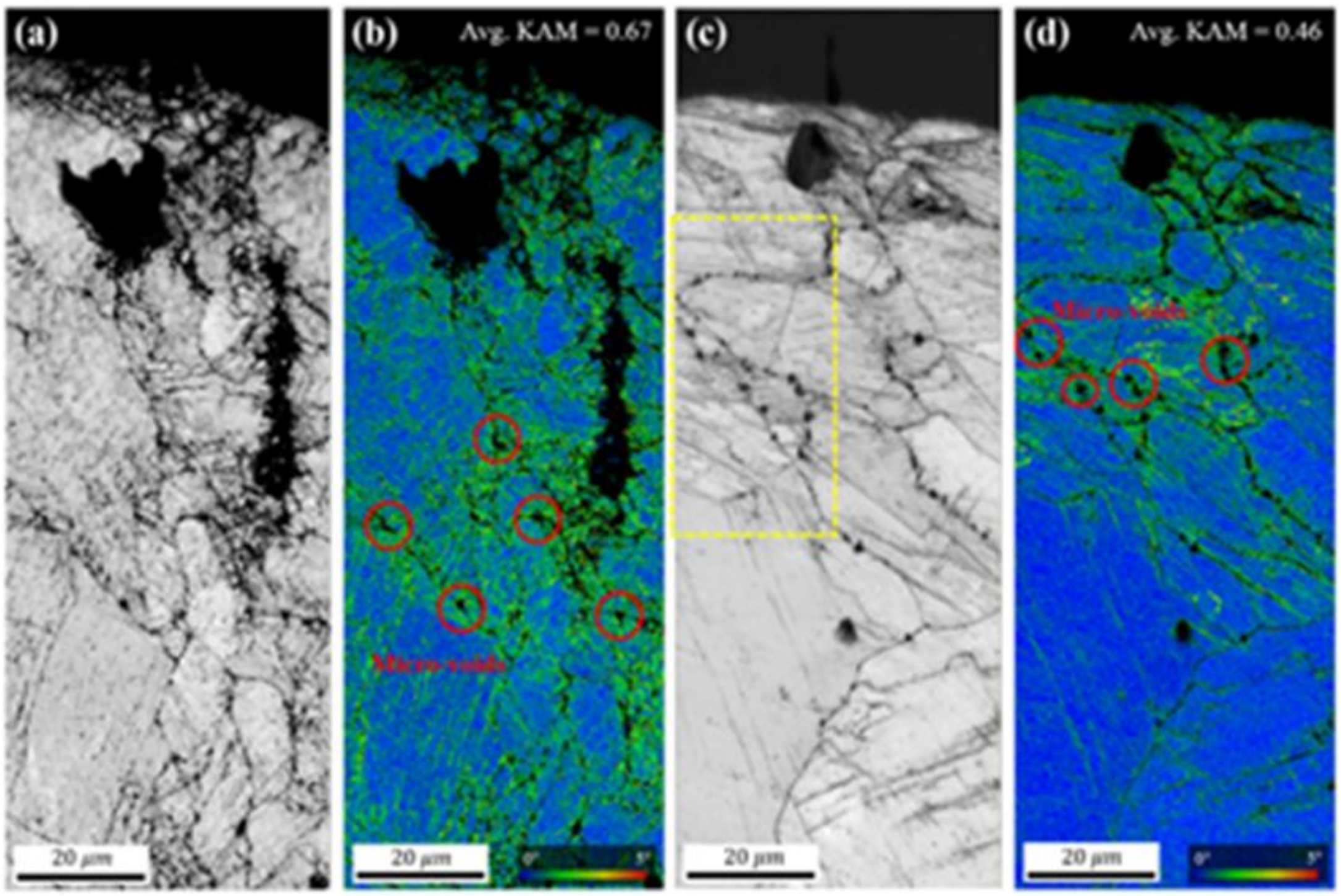

**Figure 3.3.7:** SEM and KAM of the pre-strained CoCrFeMnNi specimen under (a, b) RT and (c, d) CT [38].

### *3.3.2. Temperature effect on fatigue-crack propagation*

Thruston was the first one to study the crack growth of CoCrFeMnNi with an FCC structure at RT and at low temperature, respectively [30, 107]. The experimental crack-propagation rate vs. stress intensity factor range plot conclusively illustrated the effect of temperature on the fatigue resistance of the FCC CoCrFeMnNi (see Figure 3.3.8). In terms of the crack growth, the influence of the temperature is mainly reflected in the crack-initiation stage, that is, in the $\Delta K$ range close to the threshold region. The threshold $\Delta K_{th}$ usually increases as the temperature decreases. This means that cracks are more difficult to form at low temperatures. After the crack expansion is transferred to Stage II (the Paris region), the effect of temperature is not so obvious. But the temperature effect still exists: As the temperature decreases, the crack-growth rate under a given stress intensity factor range becomes lower. However, the Paris power law exponent, *m*, is basically not affected by temperature, during Stage II. The value of *m* at all temperatures is confined to relatively narrow range (to the range of 2.9 – 4.5).
Both at RT and at lower temperatures, the fatigue-failure mechanism of the CoCrFeMnNi follows the description of the FCC HEA outlined in Section 4.1 [107]. The SEM and EBSD results shown in Figure 3.3.9 and Figure 3.3.10 support the reason for higher crack resistance of the cryogenic specimen. Figure 3.3.9 shows the fracture surface of samples failing at different temperatures (from a region with similar stress intensity factor range) [107]. As the temperature increases, the fracture surface of the material gradually changes from sharp, multi-surface features to fine-jagged features. This trend means that cracks will propagate in the form of an inter-granular mode in a low-temperature environment. As the temperature gradually increases, the propagation mode gradually changes to a trans-granular mode [30]. The EBSD results in Figure 3.3.10 demonstrate this behavior more clearly. In case of the RT test, cracks from the specimen mainly penetrate through grains and twins in a trans-granular fashion, but there is only a small amount of inter-granular cracking observed. Furthermore, there is a large amount of plastic deformation observed in the grains near the two sides of the crack, apparently caused by a slight change in Mode II displacement of the adjacent sides of the crack. In contrast, at 198 K, the crack propagation is completely dominated by the inter-granular mechanism, which contributes to more crack-path deviations. Too much inter-granular mechanism also increases the roughness of the fracture surface, which increases the roughness-induced crack-closure level, but decreases the effective stress intensity factor range ($\Delta K_{eff}$). Hence, the CoCrFeMnNi HEA can achieve a higher value for $\Delta K_{th}$ at a lower temperature. In addition, the higher strength at the low temperature decreases the effective crack-tip opening displacements at a given $\Delta K$, which is one of the reasons for the decrease in the crack-growth rate in the Paris region [30].

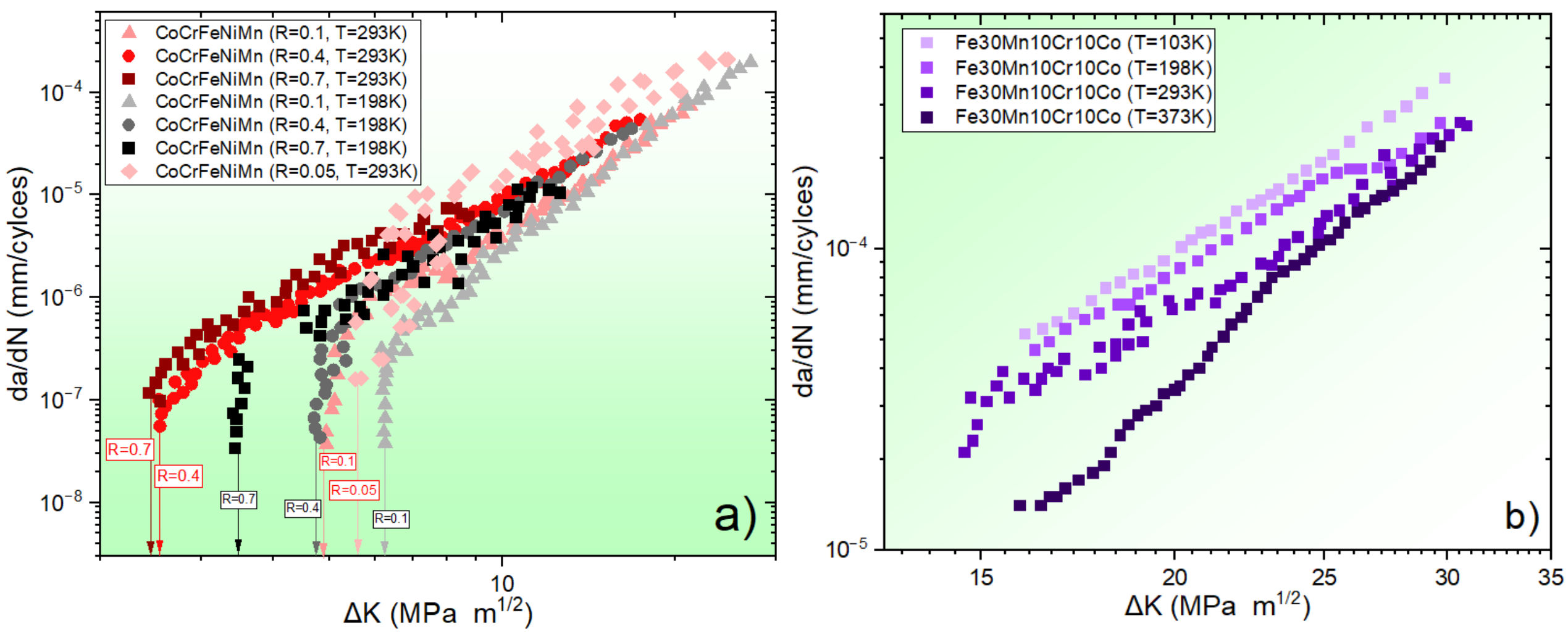

**Figure 3.3.8:** FCGR plot of a) CoCrFeMnNi and b) $Fe_{30}Mn_{10}Cr_{10}Co$ at different temperatures [107, 110].

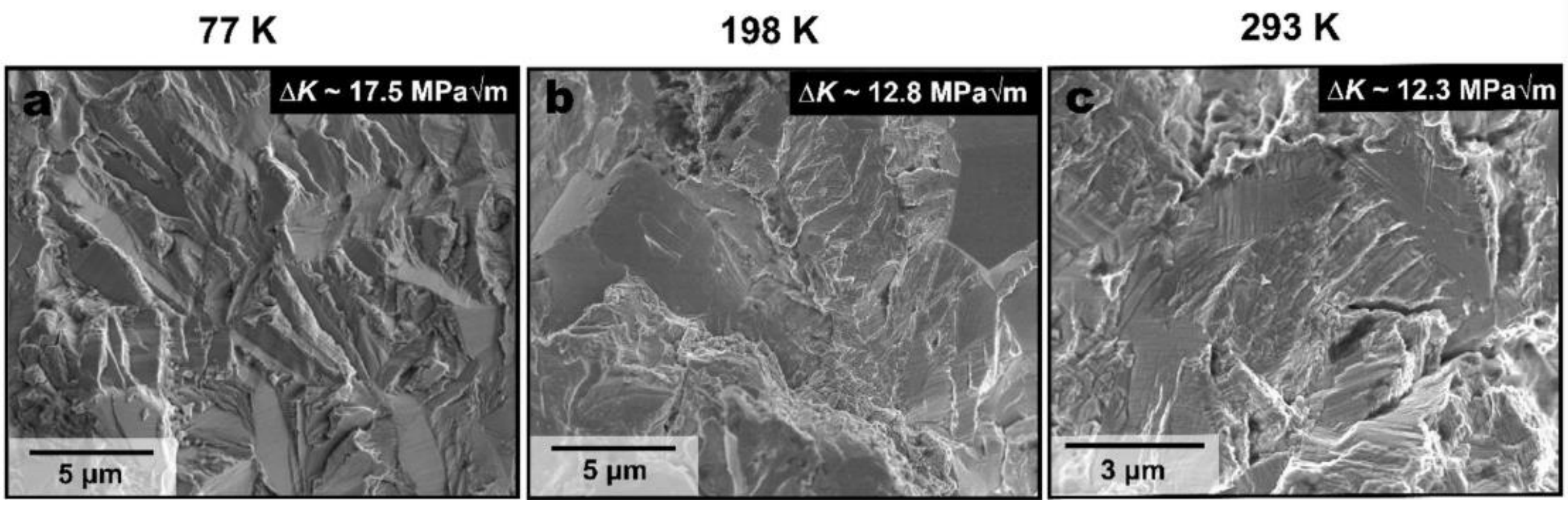

**Figure 3.3.9:** Fracture surface of CoCrFeMnNi under a) 77 K, b) 198 K, and c) 293 K (from a region with similar stress intensity factor range) [107].

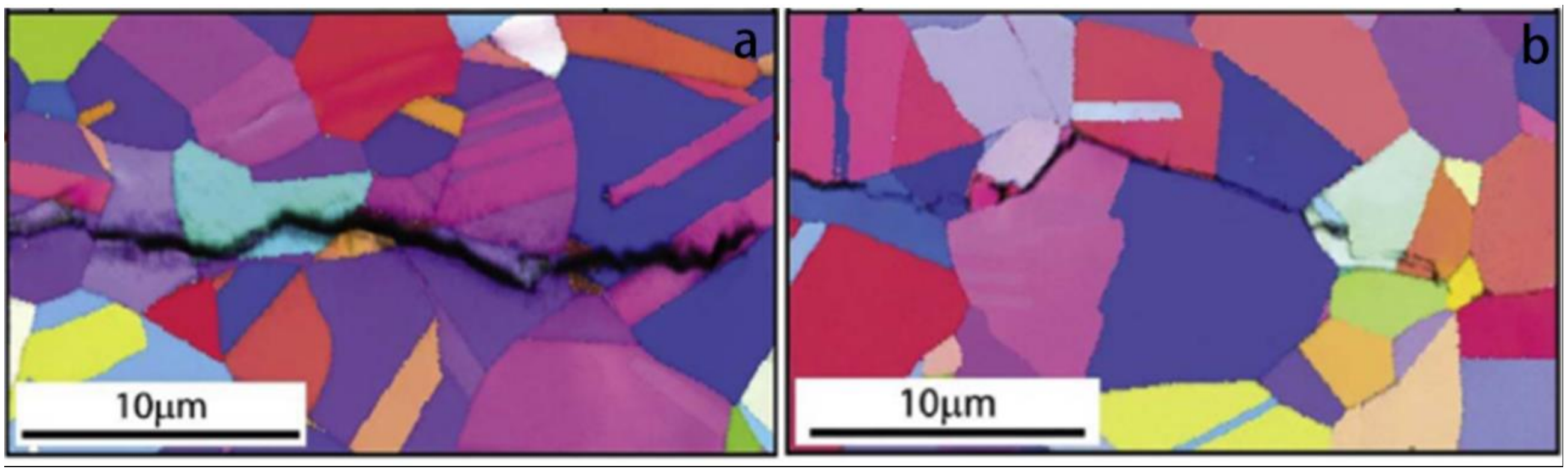

**Figure 3.3.10:** Crack shapes of the CoCrFeMnNi HEA at a) RT and b) low temperature. Trans-granular cracking is dominant at RT, while inter-granular cracking dominates at low temperature [107].

Fatigue crack propagation has also been studied for the $Fe_{30}Mn_{10}Cr_{10}Co$ HEA, which is a Fe-based metastable high-entropy alloy [110]. Figure 3.3.8(b) shows the fatigue-crack-propagation rate vs. $\Delta K$ for the $Fe_{30}Mn_{10}Cr_{10}Co$ HEA at 103 K, at 198 K, RT, and at 373 K, respectively. Since the data only account for the Paris region (Stage II), it is impossible to infer from the data how the temperature effect is related to the threshold $\Delta K_{th}$. But in terms of crack-growth rates, the $Fe_{30}Mn_{10}Cr_{10}Co$ HEA presents an opposite trend to that of the CoCrFeMnNi FCC HEA. In case of the $Fe_{30}Mn_{10}Cr_{10}Co$ HEA, the crack-growth rate at a given $\Delta K$ value in the Paris region gradually decreases, as the temperature increases. This trend shows that the fatigue resistance of the material at high temperatures tends to be higher.

The most important factor responsible for the change in the rate of fatigue crack growth in the $Fe_{30}Mn_{10}Cr_{10}Co$ metastable HEA at high temperatures is believed to be related to the FCC-to-HCP-martensite phase transformation [54]. The changes in the rate of fatigue crack growth with increasing temperature are mainly attributed to two factors: (1) Reduction in HCP martensite phase fraction and (2) ductility of the HCP martensite phase. As for the first factor, Figure 3.3.11 shows how the HCP-martensite phase fraction changes with temperature [110]. At low temperature, a part of the HCP-martensite phase transforms into an FCC phase, and more slip systems in the FCC structures reduce the resistance of the $Fe_{30}Mn_{10}Cr_{10}Co$ HEA to plastic deformations, thus affecting the fatigue-crack-growth rate. As for the second factor, the HCP-martensite phase observed in the microstructure of the low-temperature (103 K) specimen is sharper than for the other temperatures. In case of the low-temperature (103 K) specimen, more secondary cracks are observed on the crack surface, indicating that the low-temperature (103 K) environment reduces the ductility of the HCP martensite phase, which in turn affects the plastic-strain component inside the HEA, per Figure 3.3.12 [110]. Compared to the relatively uniform stress and strain distribution in higher-temperature specimens, the stress/strain in low-temperature specimens tends to be concentrated at grain or twin boundaries. The combined effect of these two factors counteracts the usual phenomenon of high temperatures accelerating fatigue degradation. As a matter of fact, the fatigue resistance of the $Fe_{30}Mn_{10}Cr_{10}Co$ metastable HEA increases with increasing temperature. Finally, as the temperature increases, the crack-growth rate of the $Fe_{30}Mn_{10}Cr_{10}Co$ metastable HEA exhibits a decreasing pattern, which is opposite to what one may expect from the FCC phase alone.

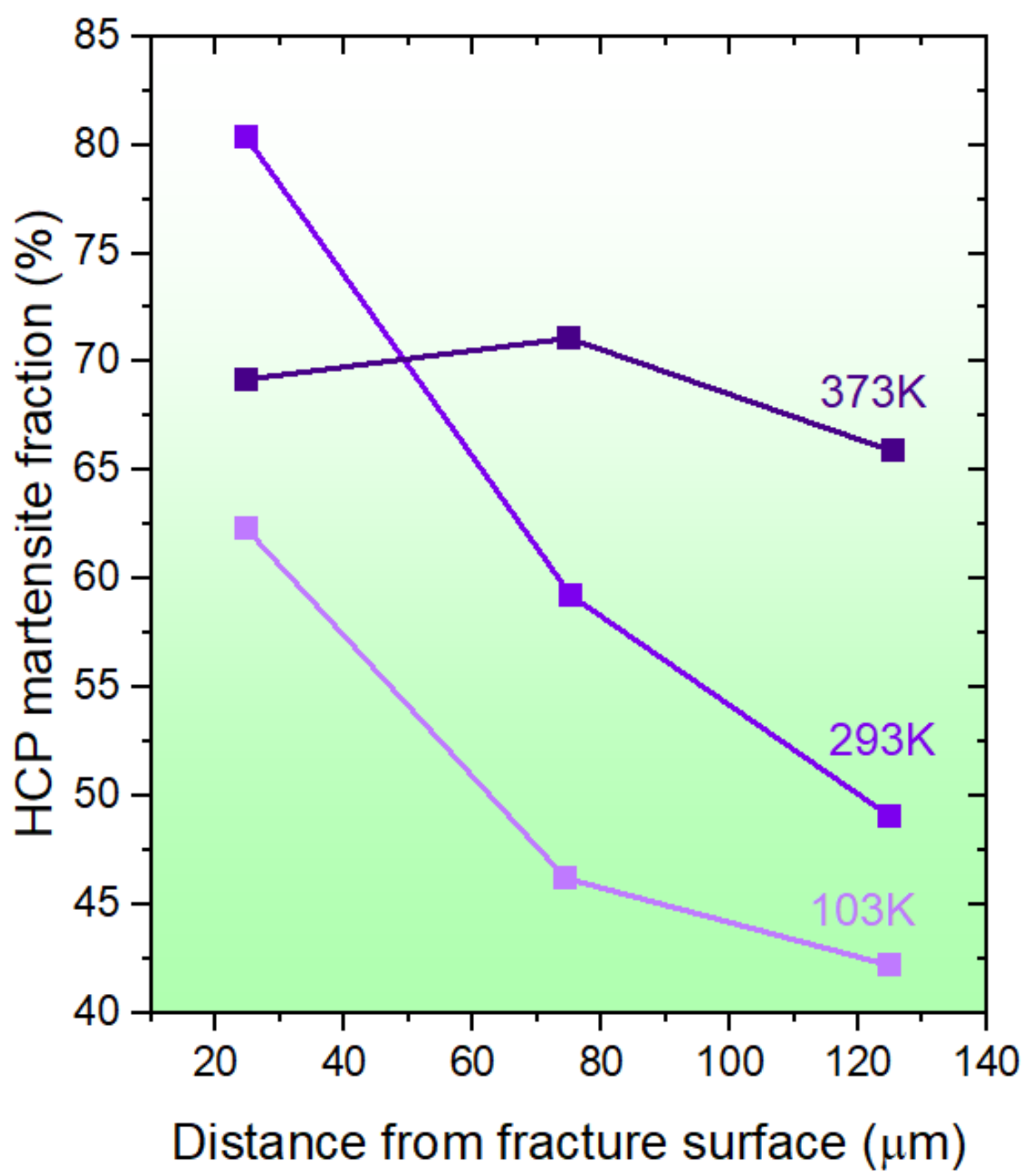

**Figure 3.3.11:** Change in HCP-martensite phase fraction in the $Fe_{30}Mn_{10}Cr_{10}Co$ HEA under different temperature environments [110].

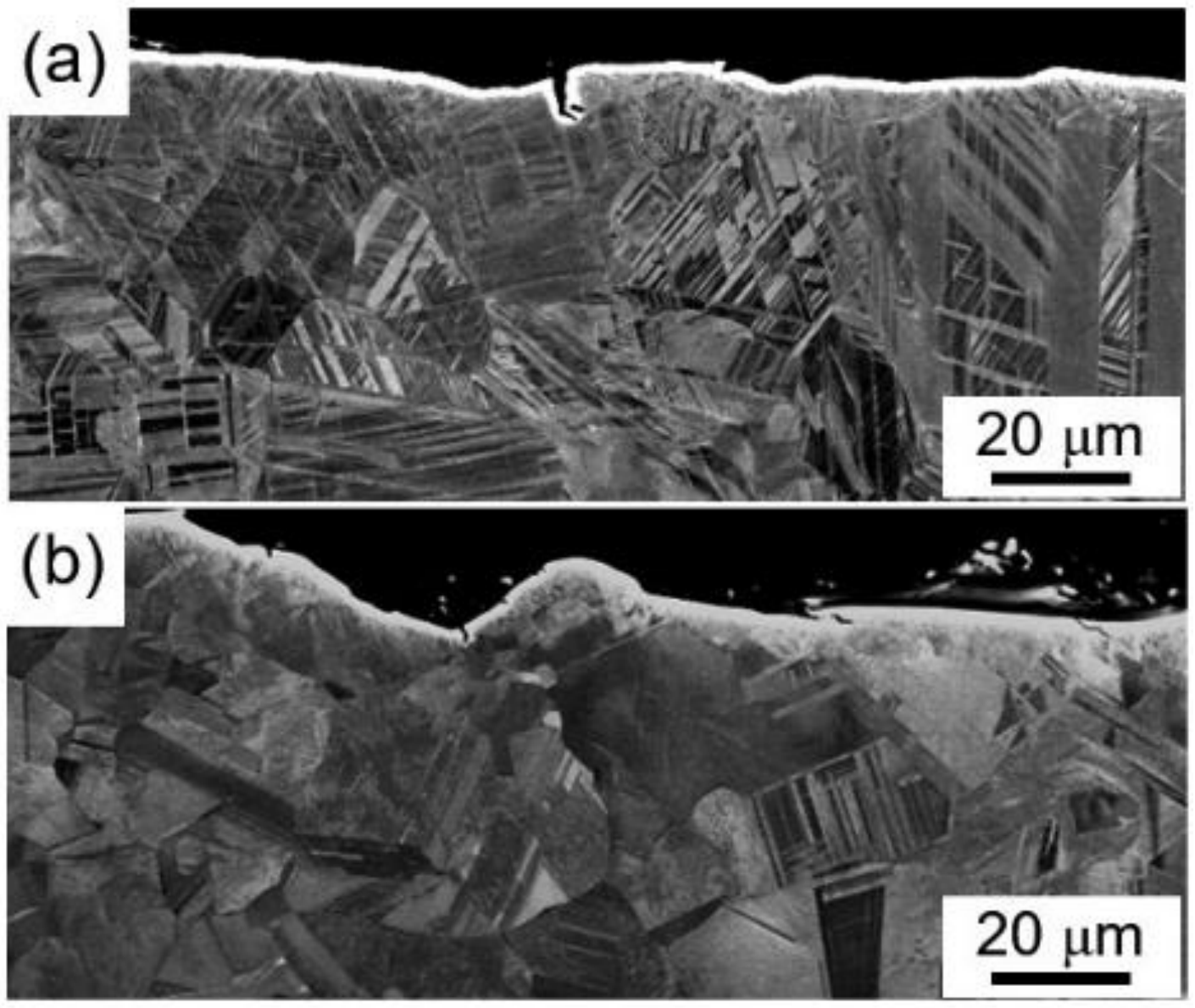

**Figure 3.3.12:** Microstructure of the $Fe_{30}Mn_{10}Cr_{10}Co$ metastable HEA at a) 103 K and b) 373 K (from regions with the same $\Delta K$). The HCP-martensite becomes sharper in the low-temperature environment [110].

Accurate description of the influence of temperature on the fatigue behavior of the $Fe_{30}Mn_{10}Cr_{10}Co$ metastable HEA makes for a rather complex story. Improvements in the fatigue behavior of the $Fe_{30}Mn_{10}Cr_{10}Co$ HEA results from a combined effect of many factors. In all temperature environments, the fatigue mechanism of this metastable HEA occurs in the form of trans-granular fracture. Similar slip lines and striation pattern also demonstrate that the failure mechanism of the $Fe_{30}Mn_{10}Cr_{10}Co$ HEA does not fundamentally change with temperature. For the high-temperature (373 K) specimen in Figure 3.3.12, the formation of slip lines and short cracks happens earlier than for the low-temperature (103 K) sample. Similar acceleration of fatigue degradation with increasing temperature was observed in case of the CoCrFeMnNi HEA. However, multiple slip systems tend to be activated, especially at elevated temperatures. Figure 3.3.13 shows uniformly oriented slip lines at RT, in the case of the $Fe_{44}Mn_{36}Co_{10}Cr_{10}$ HEA, but multiple slip systems leading to staggered slip bands at elevated temperature (300 $^{o}$C) [126]. Due to non-uniform dislocation direction, it is difficult for microcracks to form a main crack, in the early fatigue degradation process, which reduces stress localization during loading, thereby decreasing the crack-growth rate that is needed to improve the fatigue resistance at the higher temperatures [126].

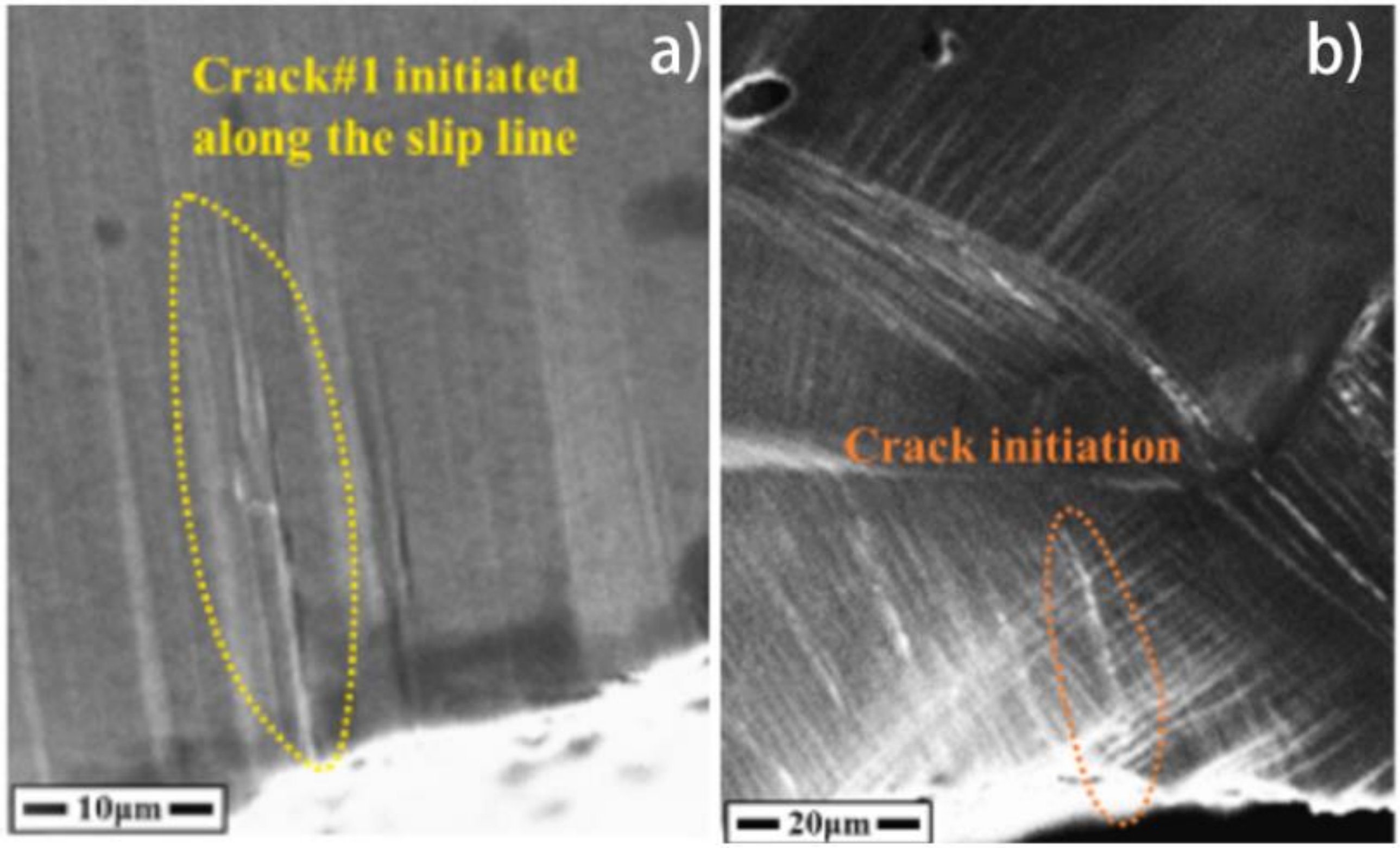

**Figure 3.3.13:** Temperature-dependent fatigue response of a $Fe_{44}Mn_{36}Co_{10}Cr_{10}$ HEA. a). The slip lines oriented uniformly under RT. b). Multiple slip systems at elevated temperature (300 $^{o}$C) led to staggered slip bands [126].

### 3.4.Stress-ratio effects

As mentioned in Section 2.1.1, only the data obtained under the same stress/strain ratio are comparable. Hence, the stress ratio ($R = \sigma_{min} \ / \sigma_{max}$) and the strain ratio ($R = \varepsilon_{min} \ / \varepsilon_{max}$) are factors that cannot be ignored in fatigue studies of HEAs. Fatigue data at different stress ratios can be converted, using the Smith-Watson-Topper equation [37]. Based on Eqs. (3.1.1) and (3.1.2), it can be known that under the same stress amplitude, the higher the stress ratio, the shorter the fatigue life of HEA. Figure 3.4.1 shows the high-cycle fatigue data of all the HEAs mentioned in Section 2.1 at the stress ratios of $R = -1$ and $R = 0.1$, respectively.

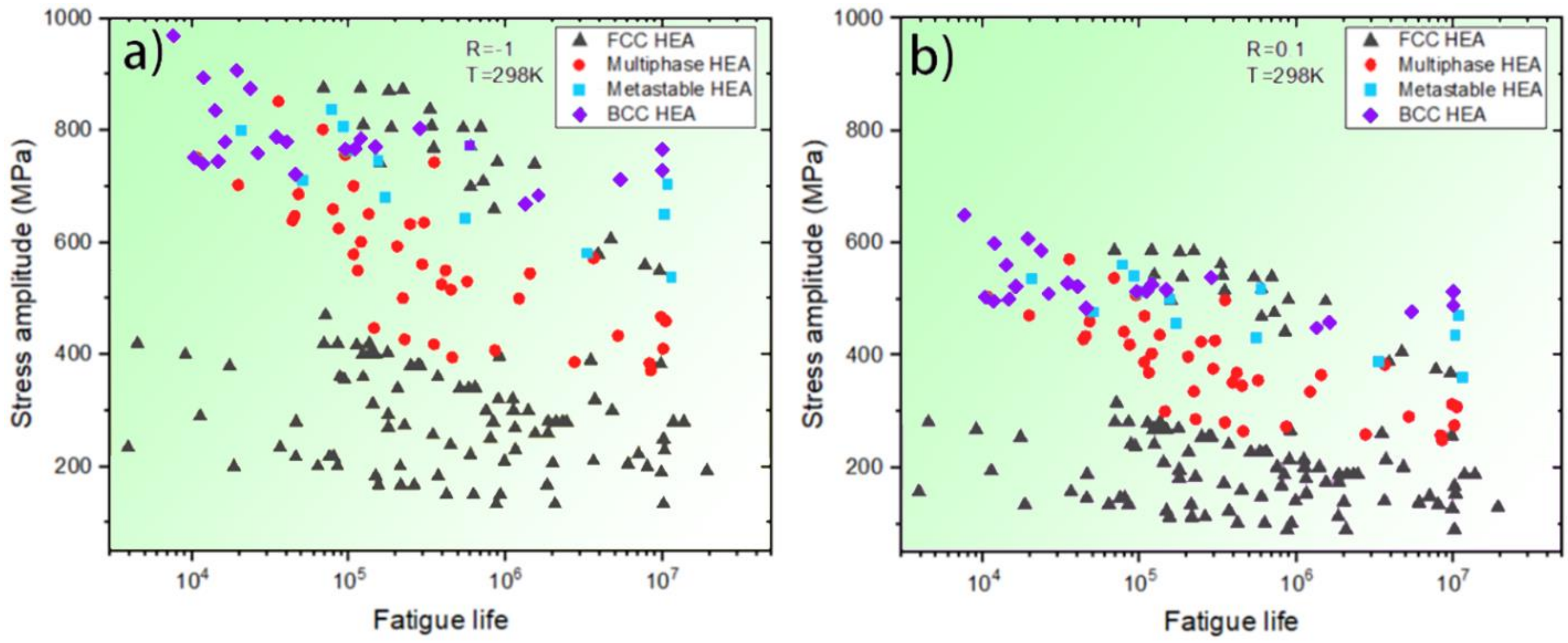

**Figure 3.4.1:** HCF data of the same set of HEAs at a) $R = -1$ and b) $R = 0.1$, respectively.

Figure 3.4.1 illustrates that a high stress ratio negatively affects the fatigue resistance of HEAs. Many studies have shown that this effect is also reflected in the fatigue-crack growth rate. There are currently three studies on stress-ratio effects on the fatigue-crack growth rates of HEAs. The materials studied include two FCC HEAs (CoCrFeMnNi and CoCrFeNi) [106, 107, 111] and two multiphase HEAs ($AlCrFeNi_2Cu$ and $Al_{0.2}CrFeNiTi_{0.2}$) [31]. The crack-growth-rate data are shown in Figure 3.4.2. Table 3.4.1 records the corresponding $\Delta K_{th}$ values as well as the corresponding exponent values fitted by the Paris law.

**Table 3.4.1:** Detailed information of HEAs under different stress amplitudes.

| | ***R*** **(stress ratio)** | **Paris slope [*m*]** | $\Delta K_{th}$ | **Ref** |
|---|---|---|---|---|
| AlCrFeNi2Cu | 0.1 | 3.4 | 17 | [31] |
| | 0.3 | 6.5 | 5 | |
| | 0.7 | 14.5 | 7 | |
| $Al_{0.2}CrFeNiTi_{0.2}$ | 0.1 | 4.9 | 16 | [31] |
| | 0.3 | 5.3 | 7 | |
| | 0.7 | 25.8 | 5 | |
| CoCrFeMnNi <001> | 0.05 | 2.51 | 5.4 | [109] |
| CoCrFeMnNi <111> | 0.05 | 3.21 | 5.5 | [109] |
| CoCrFeMnNi | 0.1 | 3.5 | 4.8 | [107] |
| | 0.4 | 2.8 | 2.6 | |
| | 0.7 | 2.6 | 2.5 | |
| CoCrFeNi | 0.05 | 2.12 | N/A | [111] |
| | 0.2 | 2.02 | N/A | |

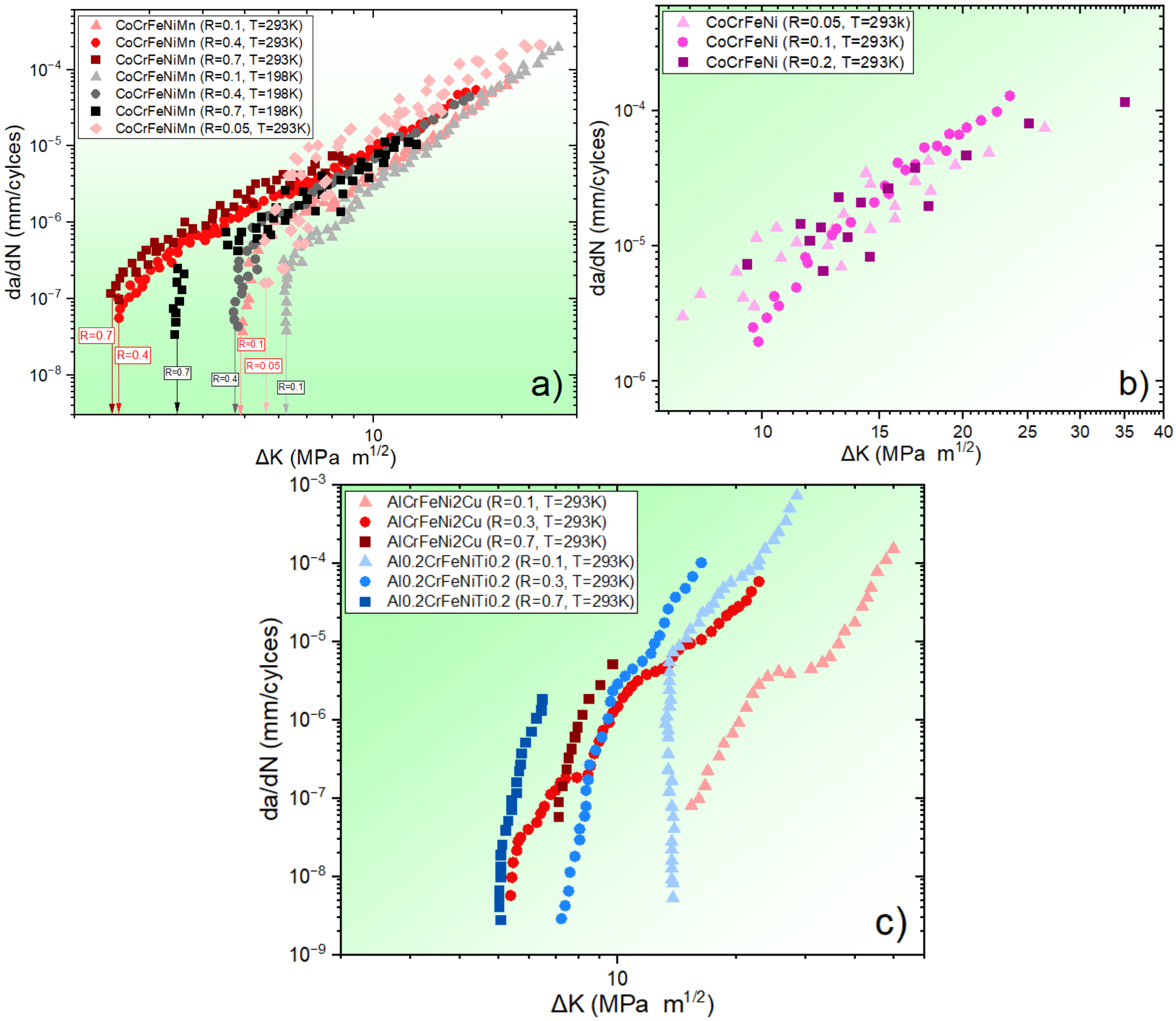

**Figure 3.4.2:** FCGR plot of a) CoCrFeMnNi [107], b) CoCrFeNi [111] and c) the two multiphase HEAs [31], under different stress ratios.

The first conclusion that can be drawn from combining the data in Figure 3.4.1, Figure 3.4.2 and Table 3.4.1 is that within a certain critical value, when the stress ratio is relatively high, the $\Delta K_{th}$ threshold of the alloy is relatively low. This phenomenon is primarily attributed to the crack-closure level. Figure 3.4.3 shows the crack-closure level of the CoCrFeNi HEA, when the stress ratio equals 0.05 and 0.2, respectively [111]. In the low stress ratio test (at $R = 0.05$), a certain level of crack closure ($K_{cl}$) was observed, which resulted in a shielding effect and reduced the $\Delta K_{eff}$ at the crack tip. Hence, the $\Delta K_{th}$ threshold becomes relatively high at low stress ratios. But as the stress ratio increases, the crack-closure level gradually decreases, as shown in Figure 3.4.3(b), and the relative threshold stress intensity factor range also decreases. Once the stress ratio exceeds a certain critical point, crack closure no longer occurs, and the threshold stress intensity factor range no longer decreases. This trend explains why the $\Delta K_{th}$ difference in the CoCrFeMnNi HEA studied by Thruston et al. at stress ratios of 0.4 and 0.7 is not very large [107].

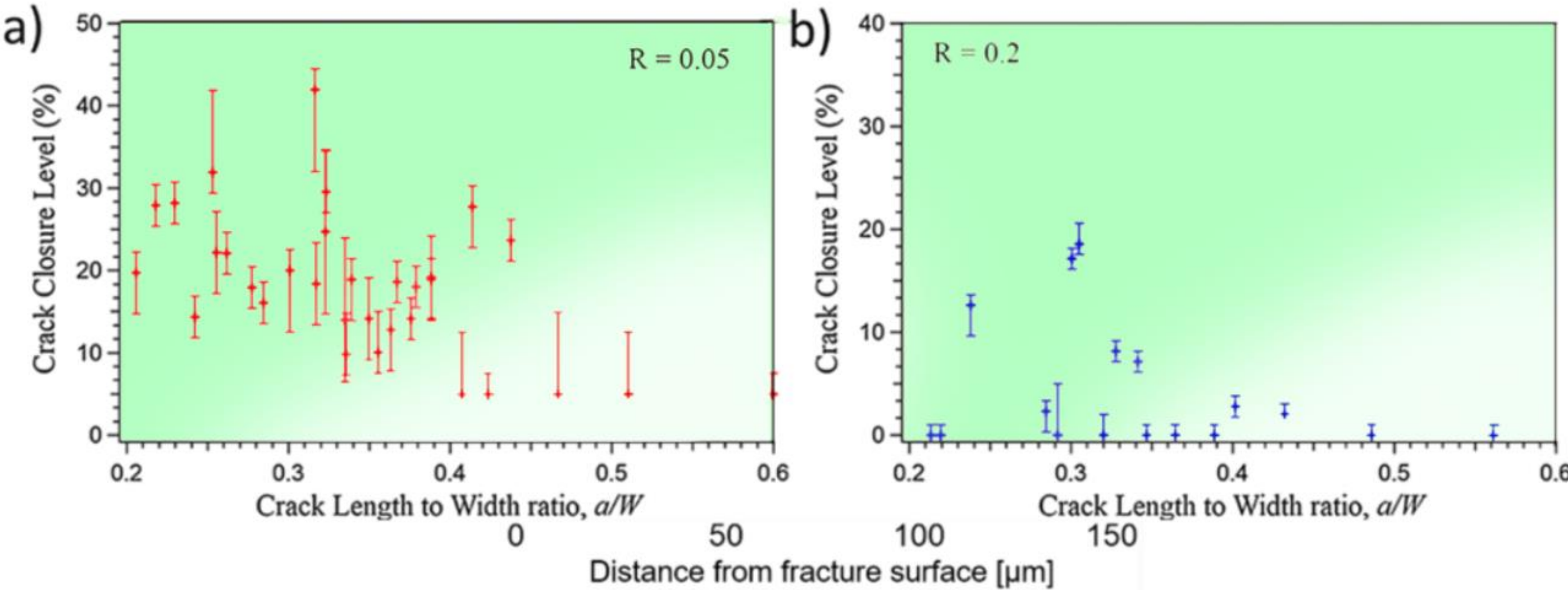

**Figure 3.4.3:** Crack closure of CoCrFeNi at different *R* values. The higher the *R* value, the lower the crack-closure rate [111].

In the Paris region, the stress-ratio effect showed a different trend in FCC and multiphase HEAs. The studies by Thruston et al. [107] and by Williams et al. [111] showed that an increase in the stress ratio slightly increases the d*a*/d*N*, at a given Δ*K*, but reduces the Paris slope, in case of FCC the HEAs. However, most of the data still show overlapping trends. Nevertheless, the multiphase HEA studied by Seifi et al. [31] exhibits an obvious stress-ratio dependence in the Paris region (Stage II), that is, a higher crack-growth rate at the same Δ*K*, and a Paris slope that increases together with *R*. The authors believe that this phenomenon is related to the fatigue mechanism of the multiphase HEAs outlined in Section 3.1. In general, cracks often initiate and propagate in the FCC phase with higher toughness. When the propagation of cracks is blocked by the BCC phase, they tend to continue to grow along the FCC/BCC boundary rather than penetrate into the BCC phase. Under the same stress amplitude, a high *R* means a greater stress, which increases the possibility of cracks penetrating into the BCC phase, resulting in more brittle failures, and finally resulting in an increase in the crack-growth rate and in the Paris slope [122].

Since the transformation of fatigue data obtained under different *R* needs to be performed, based on stress rather than strain, most fatigue studies on the stress-ratio effect focus on HCF and FCGR. In terms of the LCF, there is currently no research related to the effect of the strain-ratio on the fatigue behavior of high-entropy alloys. But some studies on conventional alloys have demonstrated the same importance of the strain-ratio effect. Based on related research data for conventional alloys, we offer the following two speculations about the effect of strain-ratio on the behavior of HEAs:

1. When the strain ratio increases, the cyclic-hardening stage of HEAs will end earlier and enter the cyclic-softening stage earlier; when the strain ratio is high to a certain extent, the HEAs will not exhibit hardening, but will directly enter the softening stage.
2. As the strain ratio increases, stress relaxation occurs, and the cyclic strength coefficient, *K*', and the cyclic work-hardening exponent, *n*', of the HEAs will decrease.

## 3.5.Strain effects

In this section, we provide insight into the various effects of strain on the fatigue behavior of HEAs.

Typically, the range of the total strain amplitude vs. the number of cycles to failure is used to evaluate the LCF performance of various materials [67]. The typical expression of the total strain amplitude, the sum of elastic and plastic contributions, is seen as:

$$\frac{\Delta\epsilon_t}{2} = \frac{\sigma_f'}{E}\left(2N_f\right)^b + \epsilon_f'\left(2N_f\right)^c \tag{3.5.1}$$

Here, the total strain amplitude is defined as $\frac{\Delta\epsilon_t}{2}$, the elastic portion is the first term in the right-hand side of the equation, and the plastic portion is the second term in the right-hand side of the equation. $c$ is the fatigue-ductility exponent, $\epsilon_f'$ is the fatigue-ductility coefficient, $2N_f$ is the number of reversals to failure, $b$ is the fatigue-strength exponent, $E$ is the elastic modulus, which is typically measured in GPa, and $\sigma_f'$ is the fatigue-strength coefficient. The plastic-strain amplitude ($\frac{\Delta\epsilon_p}{2}$), elastic-strain amplitude ($\frac{\Delta\epsilon_e}{2}$), and stress amplitude ($\frac{\Delta\sigma}{2}$) can be observed via half-life stress-strain hysteresis loops [67].

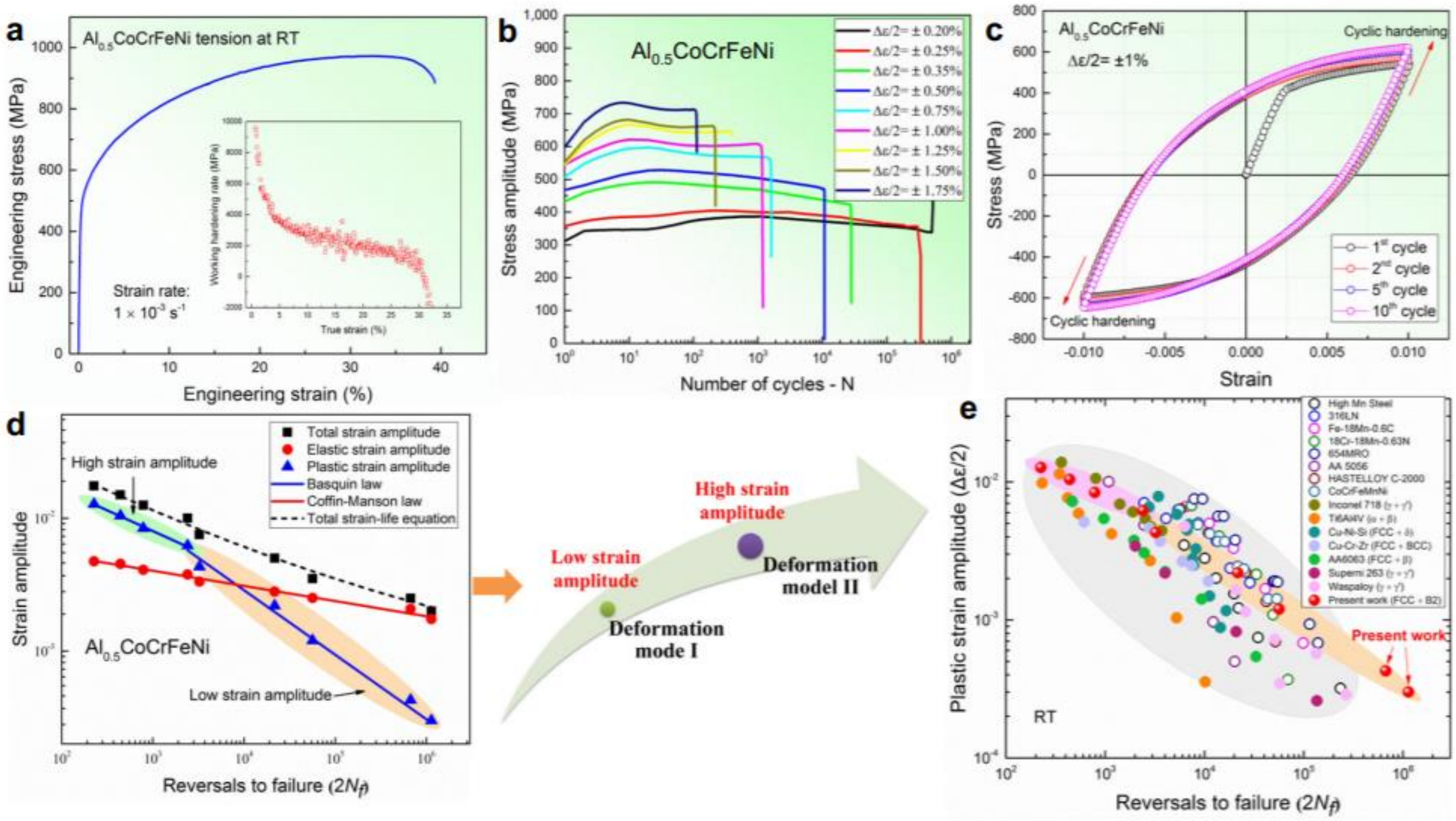

**Figure 3.5.1:** LCF and tensile results for the $Al_{0.5}CoCrFeNi$ HEA. a) Tensile results at room temperature (inset demonstrated the work-hardening rate against true strain), b) Stress amplitude vs. lifetime cycles using varying strain amplitudes, c) Stress-strain hysteresis loops at the strain amplitude of ±1%, d) plastic-strain, elastic-strain, and total-strain amplitudes vs. reversals to failure ($2N_f$), e) comparison of LCF performance of conventional alloy vs. the work of Feng et al. [25] demonstrates the enhanced LCF properties of $Al_{0.5}CoCrFeNi$ at low strain amplitudes compared to the traditional alloys (solid-circle and open circle) designate precipitation-strengthened and single-phase alloys, respectively [25].

Drawing upon Figure 3.5.1(a), the tensile results for the $Al_{0.5}CoCrFeNi$ demonstrated the following: The ultimate tensile stress of 973 ± 25 MPa, yield strength of 493 ± 4 MPa, ductility of 40 ± 2%, and work hardening > 1,000 MPa throughout deformation. From Figure 3.5.1(b), the evolution of stress amplitudes was observed against the number of cycles under varying strain amplitudes to inquire about the cyclical performance, and by extension, the primary driver of fatigue performance is the cyclic-stress response (CSR). The observation here was that the fatigue life decreased as the stress and strain amplitudes increased; and that as the number of fatigue cycles increased, cyclic hardening to softening and saturation were recorded. Cyclic softening and saturation occurred after initial cycling, wherein the stress amplitude increased sharply. Initial cyclic hardening can be seen from Figure 3.5.1(c), where the hysteresis loops were of strain amplitude of ± 1%. At strain amplitudes < ± 1%, cyclic softening occurred [see Figure 3.5.1(b)], whereas at strain amplitudes ≥ ± 1%, minimal cyclic softening was observed post initial hardening. This trend, in turn, was followed by the cyclic saturation till fracture. From Figure 3.5.1(d), the bilinear Coffin-Manson behavior of the plastic-strain amplitude as a function of the fatigue life in the form of reversals to failure ($2N_f$) was observed. The implication is that a change in the cyclic-deformation mode occurs as the applied strain amplitudes change (from deformation Mode I to deformation Mode II). In addition, after excluding the contribution of the elastic strain from different material stiffness for both the conventional alloys and for the $Al_{0.5}CoCrFeNi$ HEA, the plastic-strain amplitude vs. reversals to failure ($2N_f$) was demonstrated in Figure 3.5.1(e). The $Al_{0.5}CoCrFeNi$ HEA performed as well as (at plastic-strain amplitudes $< 10^{-3}$) or superior to (at plastic-strain amplitudes $> 10^{-3}$) conventional alloys in terms of LCF, which was confirmed by the smaller slope of between reversals to failure ($2N_f$) and the plastic strain amplitude. Lastly, the multicomponent B2 precipitates-strengthened HEA at low strain amplitudes outperformed the conventional alloys, in terms of LCF. One illustration is at a plastic-strain amplitude of ~0.03%, the $Al_{0.5}CoCrFeNi$ HEA exuded a fatigue life four times larger than that of the conventional alloys [25].

### ***3.5.1. Strain amplitude***

One of the most important strain effects to investigate is the applied strain amplitude during the fatigue tests. Below, we inquire about various ranges of strain amplitudes with their subsequent effects.

At *low total strain amplitudes* (generally 0.3% to 0.6% unless otherwise noted): In HEAs, deformation-induced slip bands are one of the most common failure mechanisms due to the accrual of localized cyclic deformation at low strain amplitudes [51, 65 {Lu, 2020 #33]. However, other factors affecting fatigue at low strain amplitudes include phenomena, such as martensitic transformation and cyclic-strain hardening. To list some examples, deformation-induced martensitic transformation in the FCC phase is the dominating mechanism in the $Fe_{50}Mn_{30}Co_{10}Cr_{10}$ metastable dual phase (FCC + HCP) HEA [125]. Dislocation structures primarily consisted of planar slip bands [65]. Equal channel angular pressing played a significant role in increasing the LCF life of the CoCrNiFeMn in comparison to the same without ECAP, wherein the applied total strain amplitude ($\frac{\epsilon_t}{2}$) was $\pm 0.2\%$ strain amplitude. However, at a total strain amplitude of $\pm$ 0.6%, the fatigue life of an hot-extruded CoCrNiFeMn HEA demonstrated a greater fatigue life than its ECAP counterpart, as a result of higher stress amplitudes as well as cyclic softening due to the accelerated dislocation annihilation [24]. Large contributions to initial cyclic-strain hardening at elevated temperatures (550°C) was attributed to dislocation-solute atom interactions, dislocation-dislocation, and dislocation multiplication, which in turn, drove the elevation of the dislocation density, while simultaneously becoming more homogenous, as the applied strain amplitude increased [51]. For the $AlCoCrFeNi_{2.1}$ eutectic HEA, the predicted fatigue limit was acquired, wherein the fish-bone type fracture surface was brought about by the nucleation-controlled process, which also drove the formation of persistent slip bands in the subsurface zones at strain amplitudes of 0.01% to 0.015% [127].

At *high total strain amplitudes* (generally 0.9% to 2.0% unless otherwise noted): At high strain amplitudes within HEAs, twinning and dislocation substructures are commonly observed [19, 38]. For instance, deformation-induced twinning in the HCP and FCC phases of an $Fe_{50}Mn_{30}Co_{10}Cr_{10}$ system were observed to be the dominating deformation mechanism that drove an $\epsilon$-martensite phase transformation [125]. Dislocation substructures (i.e., labyrinths, walls, and veins) were dominant at strain amplitudes between 0.5% and 0.7% for a CoCrFeMnNi quintenary system [65]. For $AlCoCrFeNi_{2.1}$ under higher strain amplitudes, the fatigue cracks were trapped between the dendritic branch despite the ease of initiation and propagation of the crack inside the B2 ordered phase. The crack-growth stage was limited until the stress level raised, and the crack sheared and passed through the dendrite, which in turn, resulted in a two-stage fatigue-hardening response [127]. The prevalence of twinning and dislocation substructures at high strain amplitudes is believed to be governed by the enhanced levels of cyclic plastic deformation, that accrue over a larger length scale than the localized deformation mechanisms observed at the low strain amplitudes.
It, therefore, can be concluded that the length scale of cyclic plastic deformation accrual is the primary ingredient that governs the deformation mechanisms observed at low vs. high strain amplitudes. It is not yet known, however, how differing deformation mechanisms can interfere with one another in HEA systems as a result of non-symmetric stress or strain applications in fatigue. As additional research work is done in the area of fatigue, random loading effects, which could more closely represent real world effects on fatigue performance in HEAs, need to be studied.

### ***3.5.2. Microstructural effects***

Here, we provide a brief overview of the effects of strain on the microstructure of the alloys under study, which is another important aspect.

As demonstrated by Figure 3.5.2, crack nucleation carries higher priority in comparison to the propagation stage within the low total strain amplitude regime (HCF). Crack initiation can be observed in the inter-dendritic regions (in case of B2 phase) as well as the dendritic regions (in case of $L1_2$ phase). If the red and blue arrows in Figure 3.5.2(a) are followed (the highest maximum shear stress), the propagation along the slip planes meet at the interface and continue along an energetically favorable path. This trend drives the creation of the fish-bone shape behavior [Figure 3.5.2(a)], perpetuating through the subsurface zone underneath the surface of the fracture. The creation of persistent slip bands is common in the HCF. Thus, following the indicated slip planes is thought to be a clue for the aforementioned. With higher strain amplitudes (LCF), two-stage fatigue-hardening behavior was observed. This characteristic was governed by the crack-propagation behavior in the controlling stage. Despite the simple crack propagation/initiation in the B2 phase, crack trapping was observed within the dendritic branch, during its growth [Figure 3.5.2(b)]. As demonstrated by following the red and yellow rectangles (Figure 3.5.2b), it could be observed that the growth stage stayed limited until the stress level increased to sufficient extent to allow the crack to cross the dendrite via the shear stress. During cyclic compressive and tensile deformation, dendrites were crushed and dispersed within the structure, as seen in Figure 3.5.2c. As shown by the green circles in Figure 3.5.2(c), a decrease in the crack-growth rate occurs because of dendrite fragmentation resisting crack propagation [127]. This process provides a clear example of how varying strain amplitudes generate fatigue-hardening mechanisms, and as a result, differing forms of microstructural morphologies.

Figure 3.2.8 provides a set of microstructural insights into the $Al_{0.5}CoCrFeNi$ HEA over a wide variety of strain amplitudes. Some observations were drawn in the study, which included that, as a result of severely stored dislocations in the B2 phase (see Figure 3.2.8), the plastic deformation incrementally propagated into the hard B2 phase as the cyclic strain grew. It must also be noted, however, that the martensitic transformation also occurred at a low strain amplitude (0.25%), because the B2 phase experienced plastic deformation at those strain amplitudes, as shown in Figure 3.2.8(a)-(b). The microcracks were caused by the localized stress concentration between the boundaries of the soft FCC and the hard B2 phases. The accommodation between B2 and FCC phases is given via the increased plastic deformation and martensitic transformation, that comes from the multicomponent NiAl-rich B2 phase, which, as a result, causes a decrease of the stress concentration and slows microcrack initiation. In Figure 3.2.8(e), longer fatigue life was observed at low strain amplitudes (within the $Al_{0.5}CoCrFeNi$

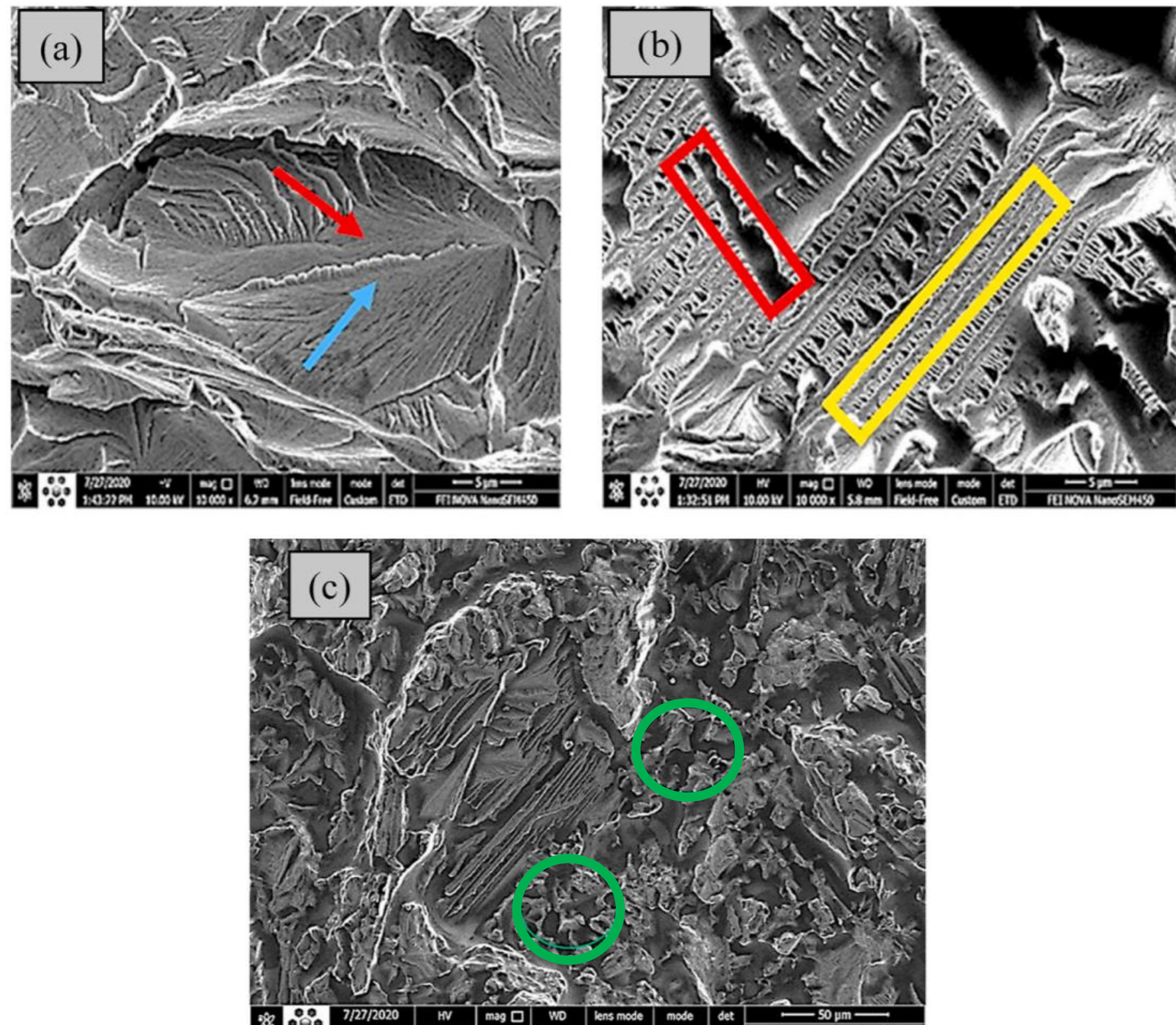

**Figure 3.5.2:** Fracture morphology of the $AlCoCrFeNi_{2.1}$ EHEA and their deformation under LCF loading for the strain amplitudes of (a) 0.01%, (b) 0.02%, and (c) 0.04%, respectively. (a) The red and blue arrows depict patterns of crack growth along slip planes with the largest maximum shear stress in the hard and soft phases, respectively. (b) The red crack rectangle shows that the crack passes through the dendrites at higher stress levels. The yellow rectangle highlights crack trapping, which can be observed inside the dendritic structure. (c) The green circles highlight crushed dendrites during cyclic deformation [127].

HEA studied). This feature was achieved by high fatigue-crack-initiation resistance, which is evidenced by the near absence of microcracks, as seen in Figure 3.2.8(i). Stress concentration could not be accommodated well close to the band-like B2 precipitates with incoherent interfaces, which is the reason why an elevated number of cracks initiated and propagated at higher strain amplitudes (1.75%), as seen in Figure 3.2.8(j). Moreover, the B2 phase demonstrated that they performed well as crack arresters, that inhibit crack coalescence and propagation, as a result of forming blunted crack ends, during cyclic deformation. All these factors contributed to the B2 phase, enhancing the prevention of fatigue life and crack initiation [25]. From the above, it can be concluded that the presence of intermetallic precipitates, such as the B2 phase precipitates, can significantly enhance the fatigue life of systems, such as the $Al_{0.5}CoCrFeNi$ HEA, at low strain amplitudes, while simultaneously retaining comparable fatigue life, at high strain amplitudes, via the retardation of microcrack initiation and propagation.

In sum, as stated in Section 3.5.1, the governing component in the observed deformation mechanism is the length scale of the localized accrual of cyclic deformation. However, the resulting phase transformations, as demonstrated from the microstructure above, can significantly influence the location and manner of crack nucleation, in systems such as $Al_{0.5}CoCrFeNi$. It can further be concluded, that energetic barriers, such as phase boundaries or interfaces, significantly influence the crack path. By taking the above three points into account, a key area in need of investigative work, would include the high-throughput theoretical modeling wherein all three aspects are optimized via compositional fine tuning, in order to further enhance the fatigue life of HEAs. It is important to note that possible interference between deformation mechanisms, resulting in a new deformation mechanism in certain HEA systems, is not yet well established.

### *3.5.3. Strain hardening*

Another important strain effect relates to strain hardening due to an intrinsic ability to increase the number of dislocations in HEAs, and thereby increase the mechanical strength of the HEAs under study. Strain hardening can occur due to inter-phase incompatibility within HEAs, which can also contribute to dislocation pile-ups. Additionally, strain hardening can occur during cyclic fatigue deformation, which in some cases can promote nano-twinning, which would also enhance the mechanical properties of the HEAs, and by extension, their fatigue lives.

Here, we provide a representative list of HEAs with observed strain hardening, as a result of fatigue, and their respective causes, as a mean of identifying primary mechanisms: In the as-cast and HE CoCrNiFeMn, significant strain hardening was observed. However, for ECAP samples, a distinct absence of strain hardening was seen because of the large pre-existing defect density being the dominant reason for the limited work hardening response (also known as strain hardening) [24]. In this case, the authors described dislocation density hardening as one of the three main processes by which the HEA can linearly enhance its strength in an additive manner [24]. Given the direct comparison of fatigue lives between the two processes, the HE samples performed half as well as the ECAP samples did in fatigue due to the ultra-fine grain material demonstrating a better postponing of crack nucleation at low strain amplitudes. But at high strain amplitudes, the HE samples performed better [24]. In other work, however, when CoCrFeMnNi with an isotropic microstructure produced by a variety of homogenization, water quenching, cold rotary swaging and recrystallization processes were studied, an initial hardening of 55 MPa and 30 MPa was observed at strain amplitudes of 0.5% or 0.4%, respectively. The authors, however, elaborate that of the three stages undergone by the sample (fast cyclic hardening, cyclic quasi-stable behavior and fatigue crack nucleation/growth, respectively), the governing aspect, that dictated the presence or absence of strain hardening, was the applied strain amplitude used in the fatigue tests [51].

In the $Al_{0.5}CoCrFeNi$ HEA, dense dislocation pile-ups were observed, because of the presence of B2-phase boundaries, during LCF testing. This trend occurred due to strain incompatibility between the FCC and B2 phases in the material. Moreover, the contribution to cyclic strain hardening occurred, due to the development of the inter-phase stress and the hindrance of dislocation movement. The dislocation movement was arrested by obstacles in the form of tough B2 precipitates, which also contributed to the resistance to plastic deformation, which in turn, elevated the fatigue performance of the HEA [25].

Lastly, the primary deformation mechanism, in dual-phase $Al_{0.5}CoCrCuFeNi$, that had been cold rolled to 84% thickness reduction in preparation for fatigue testing, was observed through TEM to be nano-twinning. Moreover, severe cyclic stress action at the sites of the maximum tensile stress was identified to be the primary crack-initiation mechanism. This mechanism drove the accelerated formation of nano-twins, prior to crack initiation. Furthermore, the increase in twin boundaries allow for the suppression of localized stress concentrations throughout the alloy, which in turn, result in an enhancement of resistance to crack initiation, due to work hardening (which also contributed to the fatigue performance) [49].

From the above, it can be concluded that the strain hardening is typically governed by increased strain incompatibility typically by dense dislocation pile-ups, phase incompatibility, and – last but not least - the applied strain amplitude. A dedicated study is recommended to explicitly observe if the accumulation of localized damage accrual from cyclic plastic deformation is the primary contributor or detractor of the presence of strain-hardening phenomena within HEAs.

#### *3.5.4. Strain rate*

Cui et al. present a review of dynamic behavior of HEAs, in context with data-driven design of HEAs for lightweight and dynamic applications [128]. The review covers the dynamic-compression behavior of HEAs, the dynamic-tension behavior of HEAs, the ballistic-impact response, as well as the shock-compression response and spall behavior [128]. Despite a somewhat limited number of investigations, some studies have inquired into the dependence of HEA fatigue properties on the strain rate applied. One such study was conducted by Jin et. al [129], wherein the CrMnFeCoNi HEA was selected for a series of investigations into the thermo-mechanical behavior of an additively manufactured laser powder bed fusion (LPBF) HEA. When the strain rate dependency was examined in relation to fatigue within the envelope of $1 \times 10^{-2}\ s^{-1}$ and $1 \times 10^{-3}\ s^{-1}$, the primary cyclic softening and hardening mechanism in the LPBF HEA demonstrated little to no dependence on the strain rate [129]. It must be noted, however, that the reversed strain sensitivity observed between 200 °C - 400 °C and the secondary cyclic hardening observed at 400 °C were driven by dynamic strain ageing (DSA). A decrease in fatigue resistance was observed, as a result of a change in the fracture mode. Namely, because of decohesion at grain boundaries, inter-granular-like fracture was observed at 600 °C. But between 22 - 400°C, the fracture mode observed was trans-granular [129].

For conventional FCC and BCC metals, the effects of strain rates are elaborated by Mayer et al. [130]. A key observation by Mayer et al., however, relates to the presence of a minimal to moderate fatigue crack-growth response generally to cycling frequency and stress amplitude within conventional FCC cell dislocation structures [130]. This observation was confirmed by Lukás et al. by measuring the same stress amplitude for varying cyclic frequencies ranging between 100 Hz to 20 kHz [131]. The BCC metals, however, demonstrate an enhanced plastic-deformation strain-rate dependence in comparison to the FCC metals. The high effective stresses in BCC metals come from a limited number of thermally activated screw dislocations due to lattice friction stresses at elevated strain rates [130]. A theoretical inquiry by Shao et. al [132] into the cyclic-strain-rate dependency (otherwise known as the frequency effect) of metals is elaborated on within the 3.7 "Frequency Effects" section of this review.

Mechanisms in HEAs, that are known to exhibit heavy dependence on the strain rate, include the serration behavior [133, 134] and the creep characteristics [135-137]. For the purposes of this review, we suggest other works, such as [138, 139], for a review of creep properties, and [140, 141], for a comprehensive guide on the serrated flow behavior in HEAs.

In short, the effects of strain rate on the LCF fatigue performance of HEAs are not yet a well understood. Given the limited data so far, no major discovery on the effects of variable strain rates on fatigue performance of HEAs has yet been made.

#### *3.5.5. Grain size*

Next, we present the effects of changes in grain size coupled with varying strain amplitudes and their subsequent effects on fatigue performance.

From Figure 3.5.3(a), it can be observed that the inelastic strain amplitude in fine-grained (FG, ~ 6 μm) material is lower than that of coarse grained (CG, ~ 60 μm), which is indicative of higher cyclic/yield strength and elastic strain in comparison to CG. The fundamental relationship can be expressed as:

$$\frac{\Delta\sigma}{2} = K'\left(\frac{\Delta\epsilon_{in}}{2}\right)^{n'} \tag{3.5.2}$$

where $K'$ represents the cyclic strength coefficient, $n'$ denotes the work hardening exponent, $\frac{\Delta\sigma}{2}$ is the saturated stress amplitude, and $\frac{\Delta\epsilon_{in}}{2}$ is the inelastic strain amplitude. Figure 3.5.4(b), demonstrates how the behavior of the two grain sizes at half-life fit well with recorded data as a linear relationship within the log-log plot of saturation stress and inelastic strain amplitude.

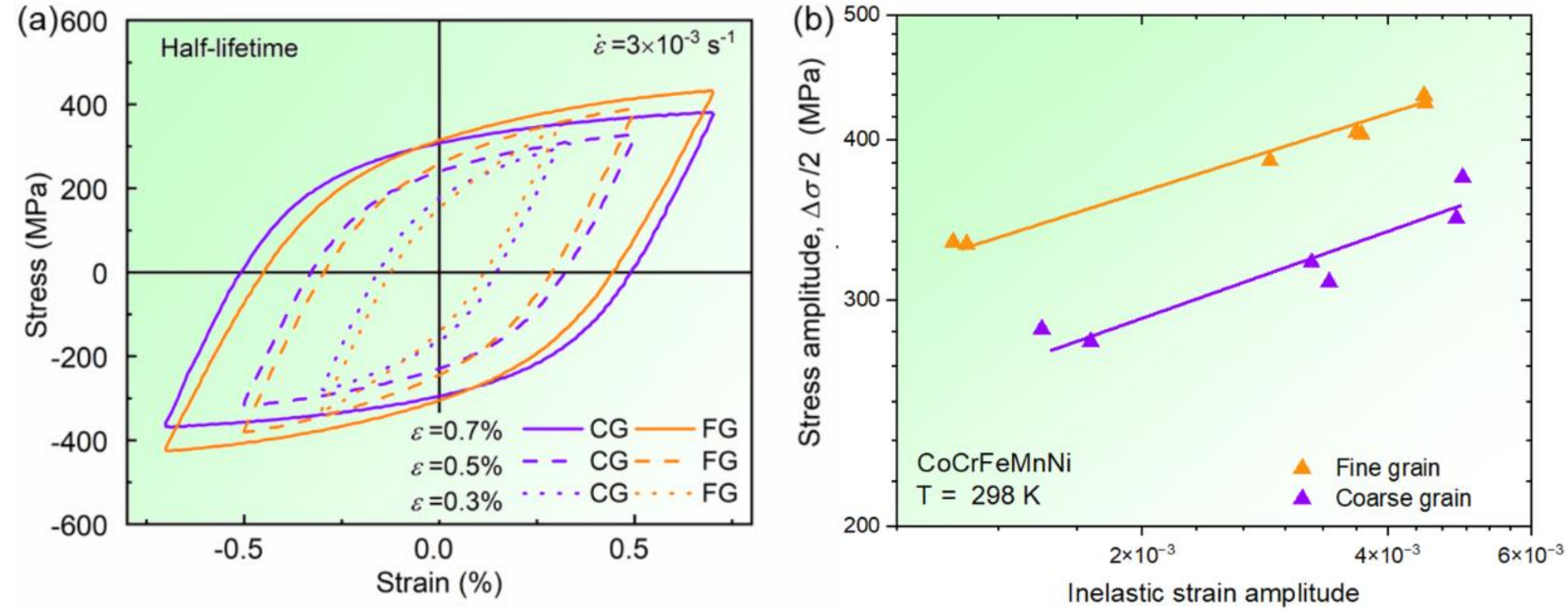

**Figure 3.5.3:** Hysteresis loops at half-life. (a) Stress amplitude vs. inelastic-strain-amplitude plot of coarse grained (CG, ~ 60 μm) and fine grained (FG, ~ 6 μm) CoCrFeMnNi at room temperature. (b) Image modified from [65].

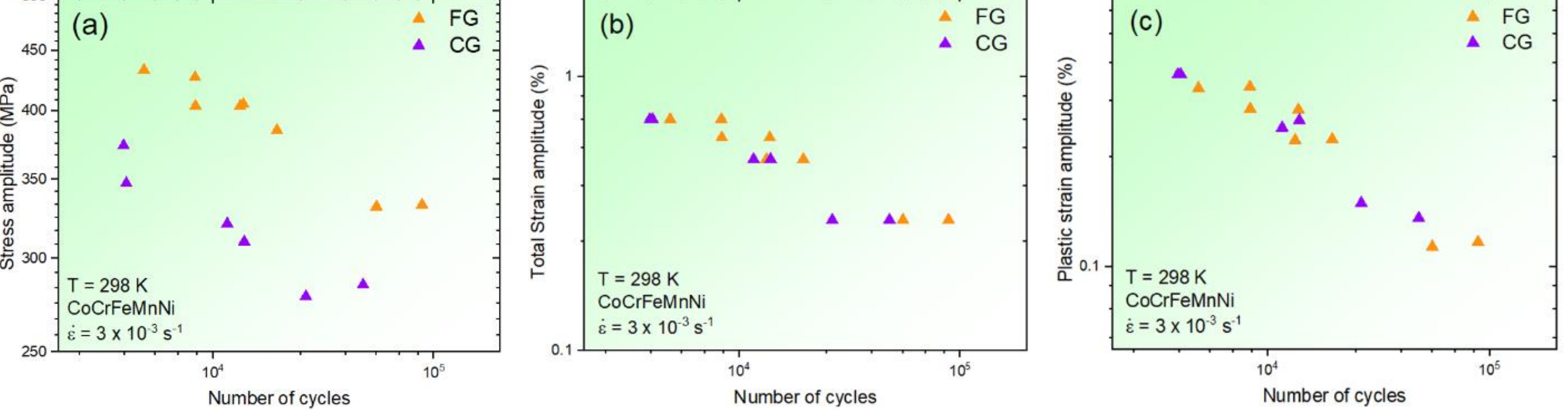

**Figure 3.5.4:** (a) Stress amplitude ($\frac{\Delta\sigma}{2}$) vs. number of cycles to failure. (b) Total strain amplitude ($\frac{\Delta\epsilon_t}{2}$) vs. reversals to failure ($2N_f$). (c) Inelastic strain amplitude ($\frac{\Delta\epsilon_{in}}{2}$) vs. reversals to failure ($2N_f$) for FG and CG CoCrFeMnNi HEAs. The image has been modified from [65].

In Figure 3.5.4(a), the stress amplitude is plotted against cycles to failure for FG and CG CoCrFeMnNi HEAs. When CG and FG materials are compared at similar stress amplitudes, a better cyclic-stress resistance, resulting from the grain refinement, can be observed for the FG CoCrFeMnNi HEA. This trend, in turn, produces a better lifetime performance in FG than in CG CoCrFeMnNi HEAs. In Figure 3.5.4(b), the total strain amplitude is plotted against reversals to failure. Here, it was concluded, after careful inspection, that fatigue life of the FG CoCrFeMnNi HEA was still higher than that of the CG CoCrFeMnNi HEA, for the same total strain amplitude. In contrast to Figure 3.5.4(c), however, when the inelastic strain amplitude is plotted against the number of reversals to failure, the conclusion made was that the grain size was independent in the micrometer regime. because the fitted Coffin Manson curves [$\frac{\Delta\epsilon_{in}}{2} = \epsilon_f'\left(2N_f\right)^c$] lay virtually on top of each other. It was believed that this phenomenon was related to the lower induced inelastic strain observed in the FG CoCrFeMnNi HEA in comparison to the CG CoCrFeMnNi HEA at the same strain amplitude in Figure 3.5.3(a) [65].

From Figure 3.2.2. above, which shows total fatigue life curves for CoCrFeMnNi HEA samples, with different grain size, and with processing conditions and fatigue parameters specified in Table 3.5.1, it can be seen, that in the warm-rolled samples (the WR samples), the fatigue performance increased over a wide range of strain amplitudes, as the grain-size diameter decreased [see Figure 3.2.2(b)] [67].

**Table 3.5.1:** Fatigue parameters of CoCrFeMnNi HEA samples with varying processing conditions [67].

| Material | Specifications | $\boldsymbol{\epsilon_f'}$ | $\boldsymbol{C}$ | $\boldsymbol{b}$ | $\boldsymbol{\sigma_f'}$ [MPa] |
|---|---|---|---|---|---|
| Cantor+C | CR (cold-rolled) - A800 (4 $\mu$m) | 0.49 | -0.549 | -0.099 | 1180 |
| | WR-A1000 (10 $\mu$m) | 0.24 | -0.441 | -0.124 | 1457 |
| | WR-A1030 (15 $\mu$m) | 0.46 | -0.502 | -0.128 | 1384 |
| | WR-A1125 (66 $\mu$m) | 0.42 | -0.494 | -0.13 | 1220 |
| Cantor | C-free (65 $\mu$m) | 0.61 | -0.532 | -0.133 | 1202 |

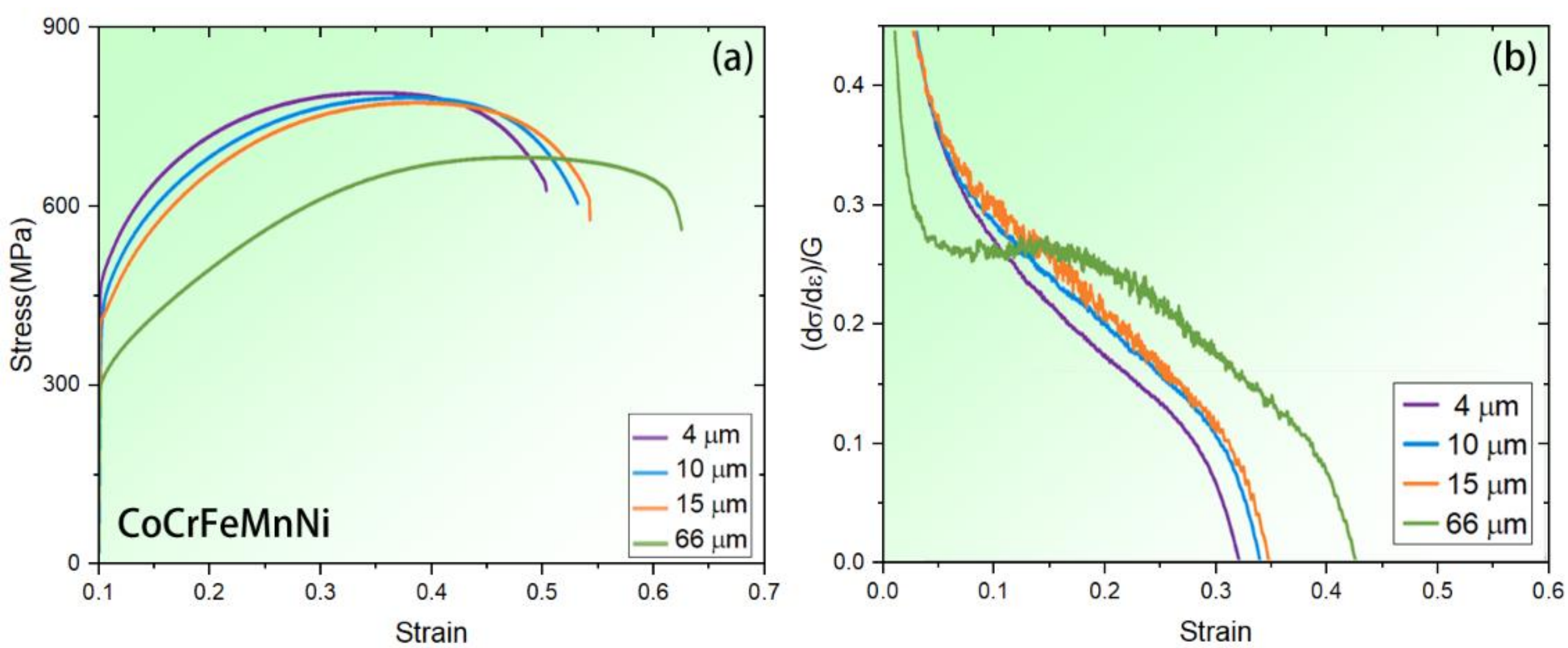

**Figure 3.5.5:** Stress-strain curves of the CoCrFeMnNi HEAs listed in Table 3.5.1. (a) Strain-hardening behavior of samples with varying heat treatment parameters, listed in Table 3.5.2 (b). Image modified from [67].

**Table 3.5.2:** Tensile properties of CoCrFeMnNi HEA specimens with different processing conditions [67].

| Material | Specification | YS [MPa] | UTS [MPa] | Elongation [%] | Grain Size [mm] |
|---|---|---|---|---|---|
| Cantor+ | CR* - A***800 | 474 | 791 | 47 | 4 ± 1 |
| C | WR***- A1000 | 410 | 783 | 50 | 10 ± 3 |
| | WR-A1030 | 409 | 775 | 51 | 15 ± 4 |
| | WR-A1125 | 300 | 683 | 60 | 66 ± 24 |

*CR = cold-rolled **A = annealed ***WR = warm-rolled

From Table 3.5.2, the tensile stress-strain data presented in Figure 3.5.5(a) were obtained. A subsequent conversion to strain-hardening behavior, shown in Figure 3.5.5b, followed. When the grain sizes were refined, the ultimate tensile strength increased in the warm-rolled samples. Moreover, the yield strength (YS) increased, due to a reduction in the heat-treatment temperature. According to Figure 3.5.5(b), the formation of mechanical twin boundaries governed the different strain-hardening behavior in the CoCrFeMnNi HEA samples, especially in the samples with finer grain diameter than WR-A1125 [67].

**Table 3.5.3:** Summary of LCF results for an Fe-Mn-C twinning-induced plasticity (TWIP) steel and CoCrFeMnNi HEAs, governed by varying grain sizes and strain amplitudes [73].

| HEAs & Reference Systems | Grain size [μm] | Strain amplitude [%] | Fatigue lifetime [cycles] | Mechanisms |
|---|---|---|---|---|
| CoCrFeMnNi | 18 | 1 | 570 | A transition from planar to wavy slip-driven sub-grain structures in the fine-grained alloy was found, while prevailing twin structures in the coarse-grained were obtained. |
| | 184 | 1 | 603 | The combined effects of grain refinement and twinning-induced cyclic deformation were beneficial for fatigue lifetime. |
| CoCrFeMnNi | 5 | 0.3 & 0.4 | $2 \sim 2.3 \times 10^4$ | Decreasing grain size promoted twin-boundary cracking and slightly extended fatigue lifetime. |
| | 30 | 0.3 & 0.4 | $2 \sim 2.3 \times 10^4$ | |
| | 165 | 0.3 & 0.4 | $2 \sim 2.3 \times 10^4$ | |
| Carbon-containing CoCrFeMnNi | 4 | 0.4 / 0.7 / 0.85 | 16,694/3,191/1,743 | Reducing the grain size to 10 mm prolonged fatigue lifetime at low strain amplitudes due to the increased elastic resistance. |
| | 10 | 0.4 / 0.7 / 0.85 | 30,048/5,012/2,403 | |
| | 15 | 0.4 / 0.7 / 0.85 | 24,993/5,490/2,899 | The precipitation of coarse carbides reduced fatigue lifetime at the strain amplitude of 0.4% in the finest grain size (4 μm). |
| | 66 | 0.4 / 0.7 / 0.85 | 18,990/4,328/2,574 | |
| CoCrFeMnNi | < 1 | 0.2 / 0.4 / 0.6 | 80,000/4,000/1,000 | Superior fatigue lifetime at low strain amplitudes in the refined grain size (< 1 mm) was found but better fatigue lifetime at high strain amplitudes in the grain size (12 mm) were obtained, driven by high-density dislocation walls and dislocation annihilation. |
| | 12 | 0.2 / 0.4 / 0.6 | 37,000/10,000/4,000 | |
| CoCrFeMnNi | 6 | 0.3 / 0.5 / 0.7 | 55,970/13,584/5,015 | Decreasing grain size enhanced fatigue lifetime. |
| | 60 | 0.3 / 0.5 / 0.7 | 26,487/11,566/3,995 | Increasing strain amplitude caused a transition from a planar-slip to wavy-slip mode. |
| TWIP steel | 6 | 0.3 / 0.6 / 1.0 | 199,144/13,433/4,332 | Reducing the grain size delayed the crack initiation by the enhanced deformation homogeneity. |
| | Gradient (6-50) | 0.3 / 0.6 / 1.0 | 90,106/15,211/4,699 | |
| | 50 | 0.3 / 0.6 / 1.0 | 49,455/9,781/3,222 | |

Upon reviewing Table 3.5.3, one notices that the fatigue lifetime of the various CoCrFeMnNi HEAs listed diminishes, as the strain amplitude applied is increased. However, this fatigue lifetime, which is driven by the strain amplitude, fluctuates with the grain size of the specimen. The finding that a decrease in the grain size of the specimens increases the expected fatigue lifetime is consistent with other literature wherein the same conclusion has been made [65, 67, 73].

Given the above body of literature, it can be concluded that a strong correlation has been observed where a decrease in grain size typically enhances both the yields strength as well as the HCF lifespan (per Section 2.1.3) of the material, by delaying the crack initiation, whereas the opposite holds for the LCF lifespan (per Section 2.2.3). Very little work has been done as of yet to explicitly measure the effects of a change in grain size within HEAs to the crack propagation pattern, which are indicative of the localized length scale of cyclic damage accrual due to fatigue.

### *3.5.6. Strain effects at elevated temperatures*

In this section, we provide an overview of various strain effects at elevated temperatures. LCF tests for various HEAs were carried out at elevated temperatures. One such example is the CoCrFeMnNi HEA, wherein the fatigue test was conducted at 550°C. The behavior of the hysteresis loops, after a set number of cycles, demonstrates that the inelastic strain is the primary cause for the cyclic-stress response observed in Figure 3.5.6(a). Upon further investigation of the hysteresis loops, the progressive width decrease is indicative of shrinking inelastic strain, until a constant loop width is achieved, and hence, a state of saturation, wherein each is associated with strain hardening and quasi-stable cyclic performance, respectively. Additionally, a serrated flow, a second phenomenon observed in the hysteresis loops in Figure 3.5.6(a), happened both under the tensile and compressive loading. As noted in Figure 3.5.6(b), the serration amplitude shrinks iteratively as a function of loop cycles. In the case of a 0.75% strain amplitude, the original 10 MPa serration amplitude shrunk iteratively after approximately 100 cycles to a serration amplitude of nearly zero [51]. An overview of the CoCrFeMnNi LCF results is presented in Table 3.5.4. The saturated hysteresis loops at half-life were used to measure the elastic strain amplitudes ($\frac{\Delta\epsilon_e}{2}$), the inelastic strain amplitude ($\frac{\Delta\epsilon_{in}}{2}$), and the cyclic stress amplitude ($\frac{\Delta\sigma_t}{2}$). The elastic strain amplitude was defined as the subtraction of the inelastic strain amplitude from the total strain amplitude. Conversely, the number of cycles at which 90% of the saturated value for the peak stress was used to define the number of cycles to failure ($N_f$). The stress amplitude (denoted as $\frac{\Delta\sigma_t}{2}$) followed a power-type relationship throughout LCF testing. As a function of the inelastic strain amplitude (denoted as $\frac{\Delta\epsilon_{in}}{2}$), the behavior of the stress amplitude is governed by the following [51]:

$$\frac{\Delta\sigma_{\mathrm{t}}}{2} = K\left(\frac{\Delta\epsilon_{\mathrm{in}}}{2}\right)^n; \tag{3.5.3}$$

$$\frac{\Delta\epsilon_{\mathrm{in}}}{2} = \epsilon_f'\left(2N_f\right)^c. \tag{3.5.4}$$

Here, $n$ is defined as the strain-hardening exponent, and $K$ is defined as the strain-hardening coefficient. Upon further analysis of the LCF data collected for the CoCrFeMnNi HEA, the elastic modulus, $E$, was 155 GPa, the strain-hardening coefficient, $K$, was 660 MPa, the strain exponent, $n$, was 0.14, the fatigue ductility coefficient, $\epsilon_f'$, was 0.14, and the fatigue-ductility exponent, $c$, was -0.46 [51].

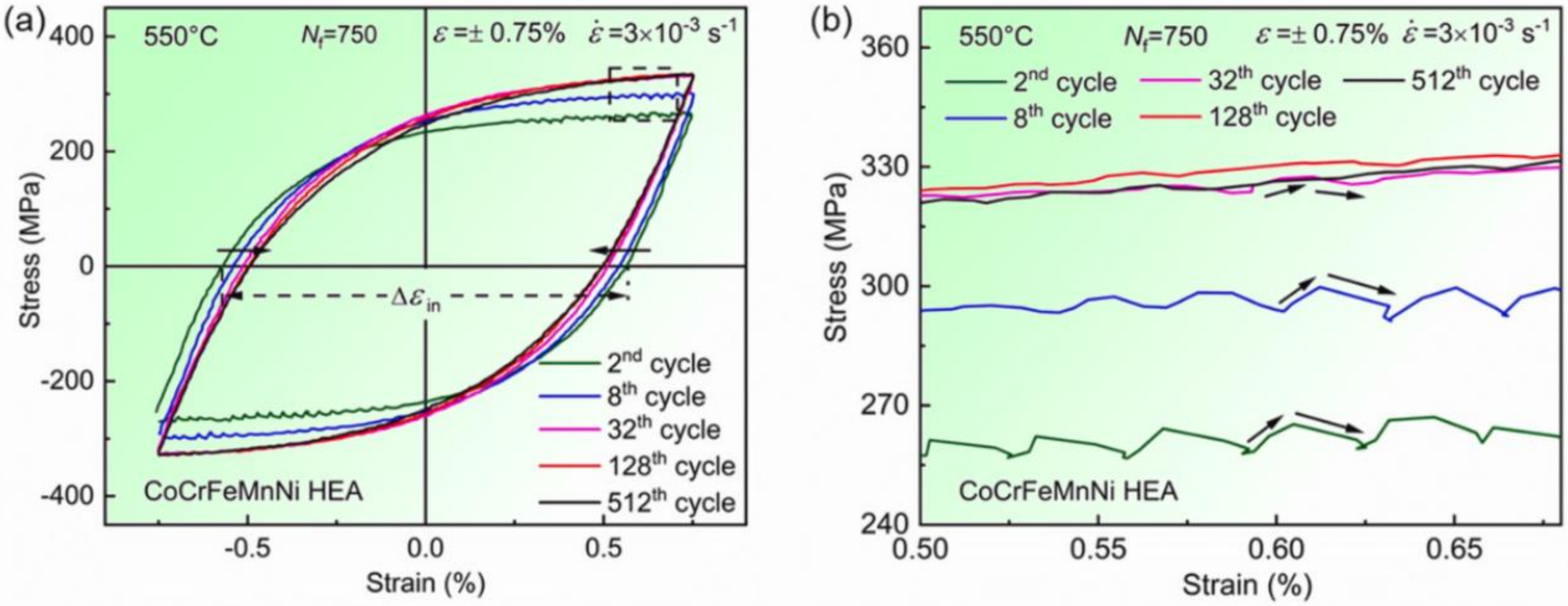

**Figure 3.5.6:** Low-cycle fatigue behavior of CoCrFeMnNi HEA. (a) A CoCrFeMnNi HEA sample tested under a 0.75% strain amplitude demonstrates a decrease in the inelastic strain by the reduction of the loop width within the hysteresis loops at various cycles till saturation as well as serrated flow both in the tensile and compression loading regimes. (b) Enlarged view of the inset in (a) demonstrates the iterative decrease of the serrated flow as the number of cycles increases. Image modified from [51].

**Table 3.5.4:** Summary of results from LCF tests for CoCrFeMnNi conducted at 550°C [51]. *= Measured at half-life.

| **Total strain amplitude, $\Delta\varepsilon_t/2$ [%]** | **Stress amplitude*, $\Delta\sigma_t/2$ [MPa]** | **Inelastic strain amplitude*, $\Delta\varepsilon_{in}/2$ [%]** | **Elastic strain amplitude*, $\Delta\varepsilon_e/2$ [%]** | **Fraction of inelastic strain* [%]** | **Number of cycles to failure, $N_f$** |
|---|---|---|---|---|---|
| 0.2% | 232 | 0.04 | 0.16 | 21 | 58,460 |
| 0.3% | 266 | 0.12 | 0.19 | 38 | 39,862 |
| 0.4% | 272 | 0.20 | 0.20 | 50 | 5,104 |
| 0.5% | 291 | 0.29 | 0.21 | 58 | 3,217 |
| 0.75% | 328 | 0.50 | 0.26 | 66 | 750 |
| 0.8% | 330 | 0.55 | 0.25 | 68 | 400 |

When Table 3.5.4 is analyzed, the concluding observation, is that, as the applied strain amplitude increases, the corresponding fatigue lives decrease. The cyclic lifetime, however, is primarily governed by the inelastic strain within the LCF domain, whose fraction ranges between 68% - 21% of the highest to lowest tested strain amplitudes, respectively. Table 3.5.4 is modified from [51].
Through comparison of Figure 3.5.7 and Figure 3.3.3(a)-(c), one can tell that an increased density of dislocations was observed, at the elevated strain amplitudes of $\pm$ 0.5%, in comparison to those observed, at the lower strain amplitude of $\pm$ 0.3%. In turn, this would drive increased cyclic hardening, during elevated strain amplitudes, whose initial hardening behavior mimics that of the austenitic steels. Through further comparison of Figure 3.3.3 and Figure 3.5.7, it can be seen that entanglements, dislocation loops, planar slip-bands, and dislocation pile-ups are among the family of dislocation structures that result from cycling. The most typical causes for formation of dislocation structures entail low SFE, which can contribute to planar movement, co-planar slip activity, and super-jogs. Lastly, according to Figure 3.3.3(d), complicated structures, such as tangled structures, are formed from wavy slips. These structures are developed as a result of cross-slip and glide along secondary slip systems [51].

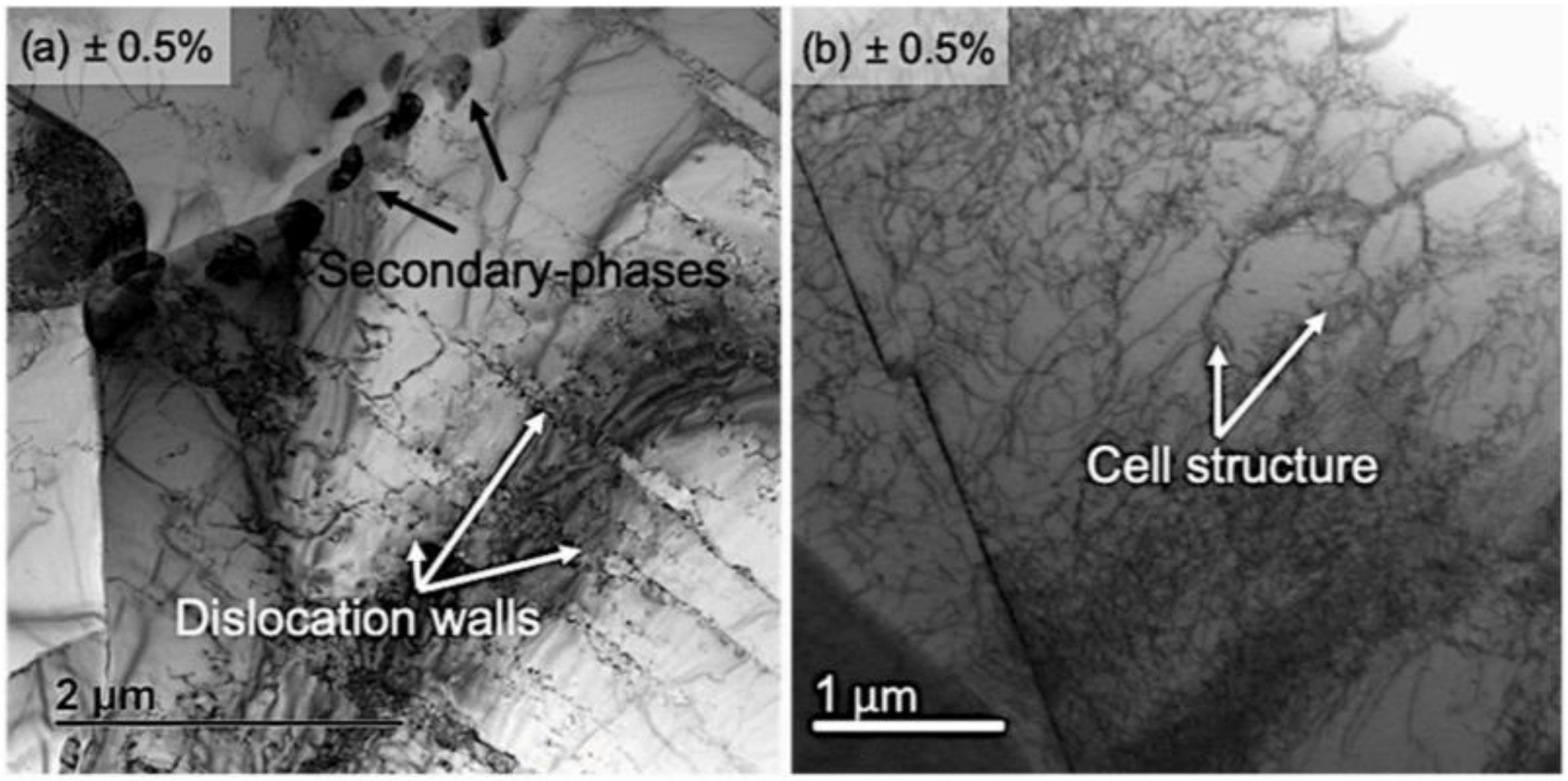

**Figure 3.5.7:** (a) TEM image of cyclic strain at strain amplitude of ±0.5% at 550 °C using secondary phase from segregation at grain boundaries. (b) Demonstration of badly defined cell structures via HAADF-STEM [51].

With the increase of dislocation structures due to iterative dislocation density/regions enhancement [such as dislocation walls, see Figure 3.5.7(a)], the promotion of cross-slip and secondary-slip systems is encouraged via dislocation barriers arresting dislocation motion. Annihilation processes and dynamic recovery have been known to form, at elevated temperatures, because of cross-slip and climb from screw and edge dislocations, respectively. A quasi-stable cyclic response is caused by a quasi-stable density of dislocations, which itself is governed by the dynamic equilibrium between the dislocation annihilation and multiplication. According to Figure 3.5.7(b), badly defined cell structures, that are considered to be among the more stable lower-energy configurations, occur due to self-arranged dislocations [51].

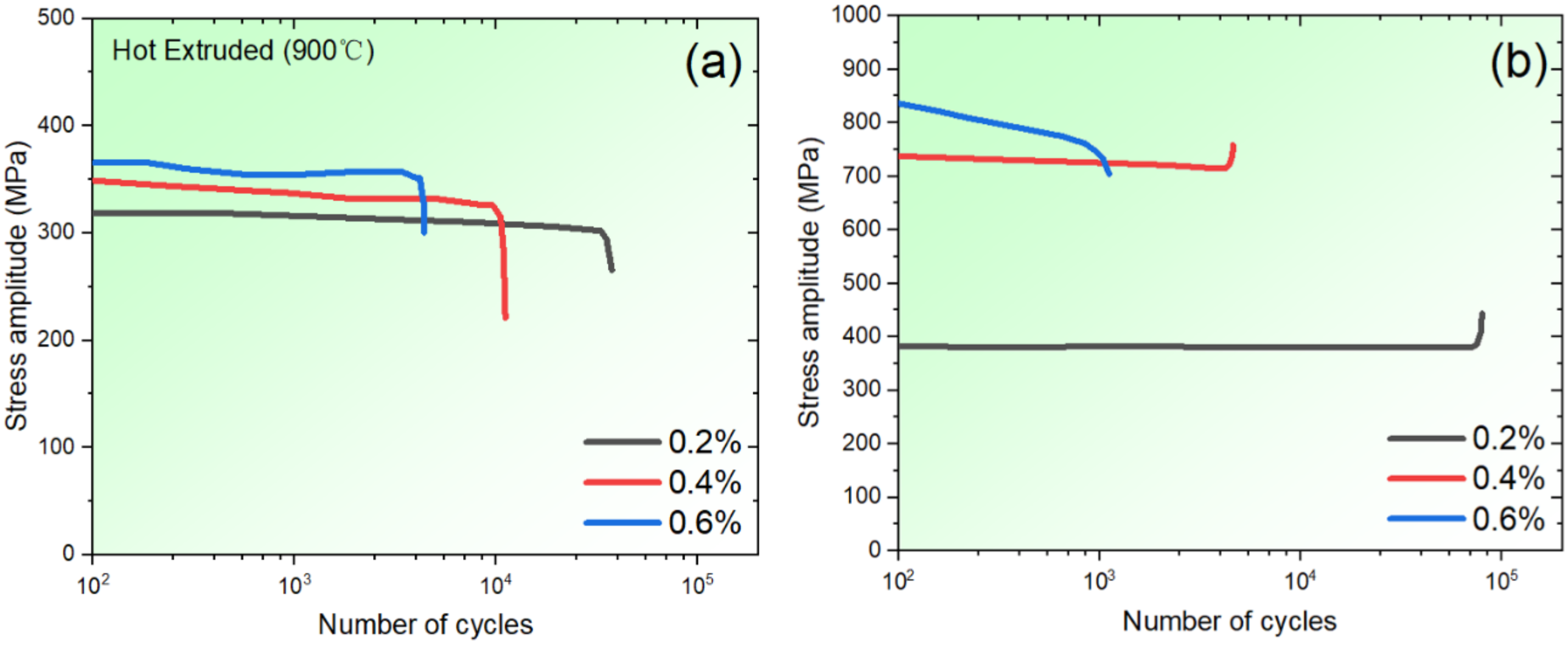

**Figure 3.5.8:** Cyclic deformation response (CDR) of the CoCrFeMnNi (a) hot extruded at 900°C, and (b) equal channel angular processing at 300°C. Total strain amplitudes used: $\frac{\Delta\epsilon_t}{2} = \pm 0.2\%, \pm 0.4\%,$ and $\pm 0.6\%$. Inset within (b) demonstrates half-life plastic strain amplitude ($\Delta\epsilon_{pl,amp}$, %) vs. half-life stress ($\frac{\Delta\sigma}{2}$, MPa) behavior. Only high levels of plastic strain were considered. Image modified from [24].

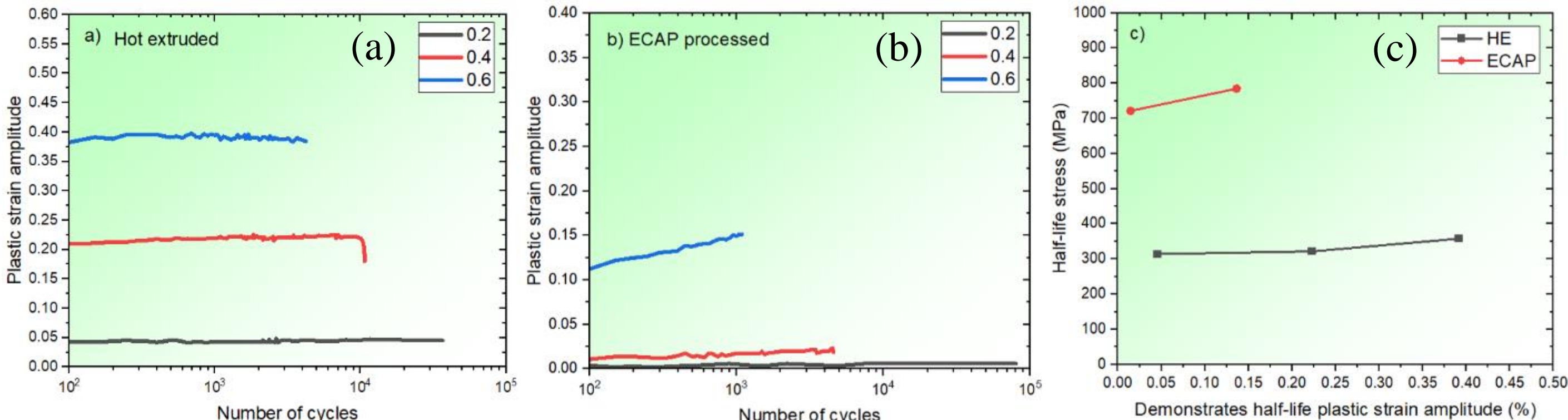

**Figure 3.5.9:** Cyclic-plastic-strain behavior of the CoCrFeMnNi HEA. (a) Hot extruded at 900°C. (b) ECAP at 300°C. Total strain amplitudes used: $\frac{\Delta\epsilon_t}{2} = \pm 0.2\%, \pm 0.4\%,$ and $\pm 0.6\%$. Inset within (b) demonstrates half-life plastic-strain amplitude ($\Delta\epsilon_{pl,amp}$, %) vs. half-life stress ($\frac{\Delta\sigma}{2}$, MPa) behavior. Only high levels of plastic strain were considered. Image modified from [24].

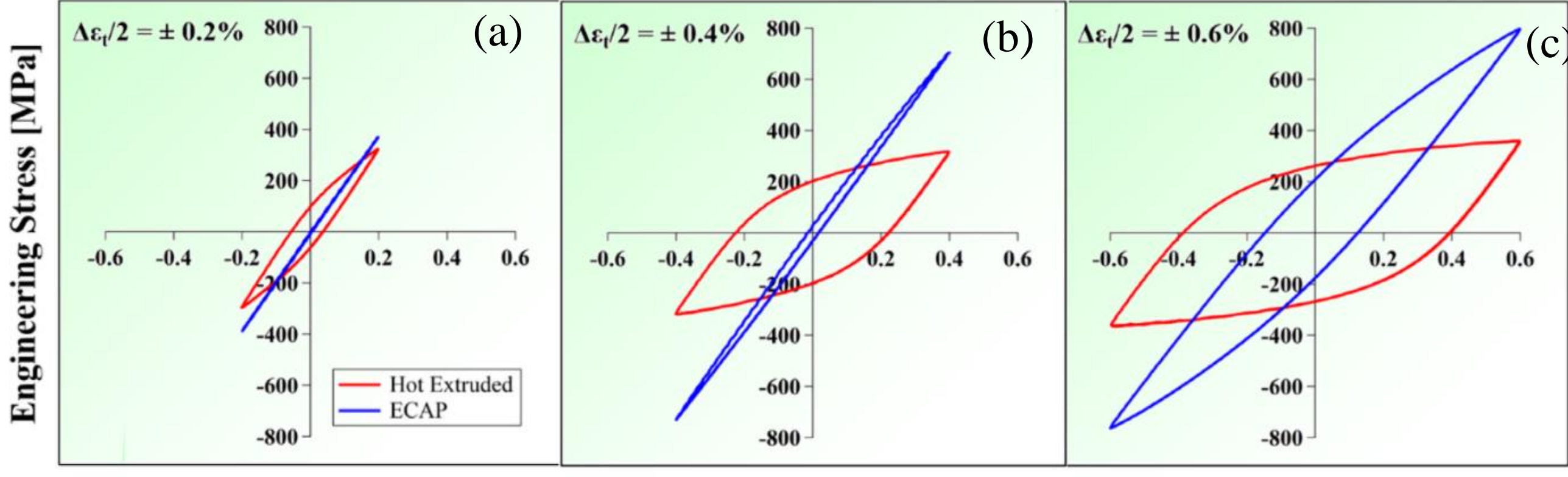

**Figure 3.5.10:** CoCrFeMnNi half-life hysteresis loops for HE and ECAP processes. Total applied strain amplitudes of (a) $\frac{\Delta\epsilon_t}{2} = \pm 0.2\%$, (b) $\frac{\Delta\epsilon_t}{2} = \pm 0.4\%$, and (c) $\frac{\Delta\epsilon_t}{2} = 0.6\%$. Image modified from [24].

According to Figure 3.5.8, the hot-extruded CoCrFeMnNi samples experienced a lower stress level than those samples processed with ECAP. The differences can be seen most clearly at the elevated strain amplitudes. It must be noted, however, that the cyclic-deformation response of both the HE and ECAP processes demonstrated a lack of hardening, i.e., of a stress amplitude saturation state, at the strain amplitudes of $\frac{\epsilon_t}{2} = 0.2\%$ and $0.4\%$. At the strain amplitude of $\frac{\epsilon_t}{2} = 0.2\%$, the fatigue life of the ECAP-processed samples performed almost twice as well as the HE processed samples at ≈80,000 and ≈37,000 cycles, respectively. This is not surprising, since ultra-fine-grained materials generally offer good resistance to crack initiation in both cyclic and monotonic domains. However, cyclic softening is observed in the ECAP-processed samples, when the applied strain amplitude is increased to $\frac{\epsilon_t}{2} = 0.6\%$, while the HE processed samples stayed stable at the same strain amplitude. This behavior points to a re-arrangement of dislocation structures, because of cyclically induced microstructural instability, plasticity, and strain path changes. In case of the applied strain amplitudes of 0.4% and 0.6%, the ECAP processed samples were noted to fail sooner, when the cycles to failure were compared, than those that were HE processed.

According to Figure 3.5.9, the hot-extruded CoCrFeMnNi HEA samples demonstrated absence of either cyclic softening or hardening, which was indicative of a stable stress response as the CDR. From the same figure, this conclusion can be further verified by the absence of change in the plastic-strain behavior over the range of the total strain amplitudes. In Figure 3.5.10, the hysteresis loop for the ECAP CoCrFeMnNi HEA samples made a complete loop, which pointed towards the presence of pure elastic strain in the macro regime, which was further demonstrated by the virtual absence of plastic strain, as indicated in Figure 3.5.9. However, the wide-opened half-life hysteresis loops from Figure 3.5.10 are indicative of cyclic plasticity within the HE samples. In essence, cyclic softening is clearly observed in all CDRs within the ECAP-processed CoCrFeMnNi HEA samples at elevated applied strain amplitudes, which is further validated by the plastic-strain-range enhancement stress response observed in Figure 3.5.8. In sum, regardless of the processing method, an increase in the strain amplitude applied causes a decrease in the fatigue life. Cyclic plasticity is the primary mechanism that drives softening and shortening of fatigue life, because of the high strain amplitudes. Due to higher energy dissipation per cycle, a smaller number of cycles occurs, which, in turn, dictates the deterioration of the fatigue performance [24].

In summary, we see a variety of fatigue responses due to strain behavior over a wide range of applied strain amplitudes, cyclical deformation, post-processing conditions, microstructural changes, and even changes in grain sizes. It is safe to conclude that varying the range of strain amplitudes can drive different deformation modes [25], strain hardening [24, 25, 51], crack initiation in microstructural effects [127], and can increase the density of dislocations [51]. When other aspects are factored in, such as a ultra-fine grained processing, fatigue life can be enhanced at low strain amplitudes [24, 73]. An increase in the applied strain amplitude causes a decrease in the fatigue life, as a direct result of cyclic plasticity, wherein it is governed by the elevated energy dissipation per cycle, resulting in a shorter cyclic lifespan [24]. Changes in grain-size diameter can also assist in the cyclic-stress resistance and varying mechanisms, which contribute to the enhancement of fatigue performance in HEAs with smaller grain size [65, 67, 73].

### 3.6.Overload effects

In the process of cyclic-load applications, sudden overload or underload (usually one or more times) will affect the fatigue behavior of the material under study. At present, fatigue studies of many alloys have shown that overloading will increase the fatigue life of the alloys, while underloading will cause the fatigue life of the alloys to decrease [108]. This phenomenon is mainly reflected in the crack-growth rate. As shown in Figure 3.6.1 [142], when overloading happens, the crack-growth rate curve exhibits a very short initial acceleration, followed by a brief delay stage, which is referred to as retardation. In the retardation region, the acceleration of the crack-propagation rate will decrease. But after the delay period, the crack-growth rate will return back to a steady state.

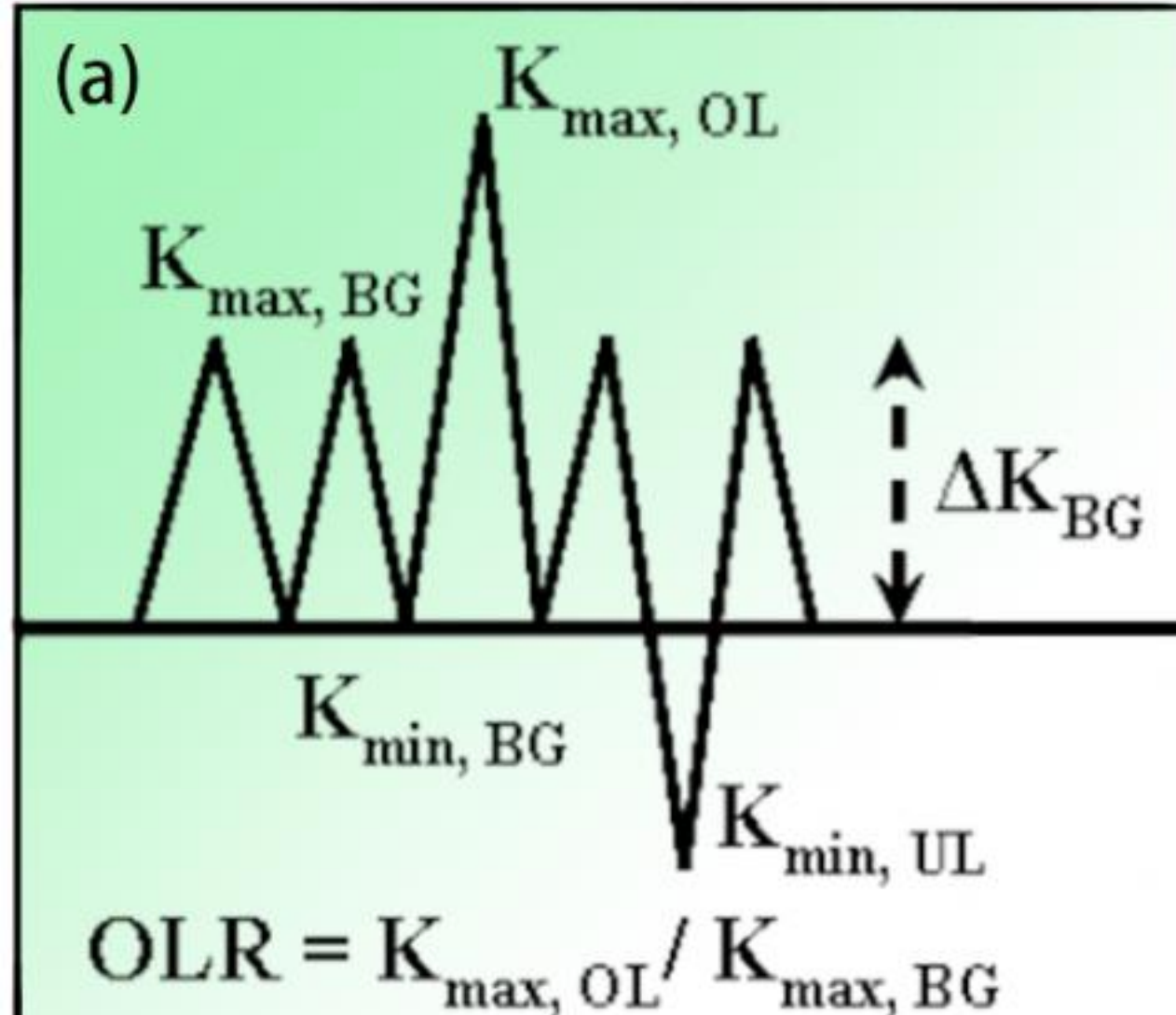

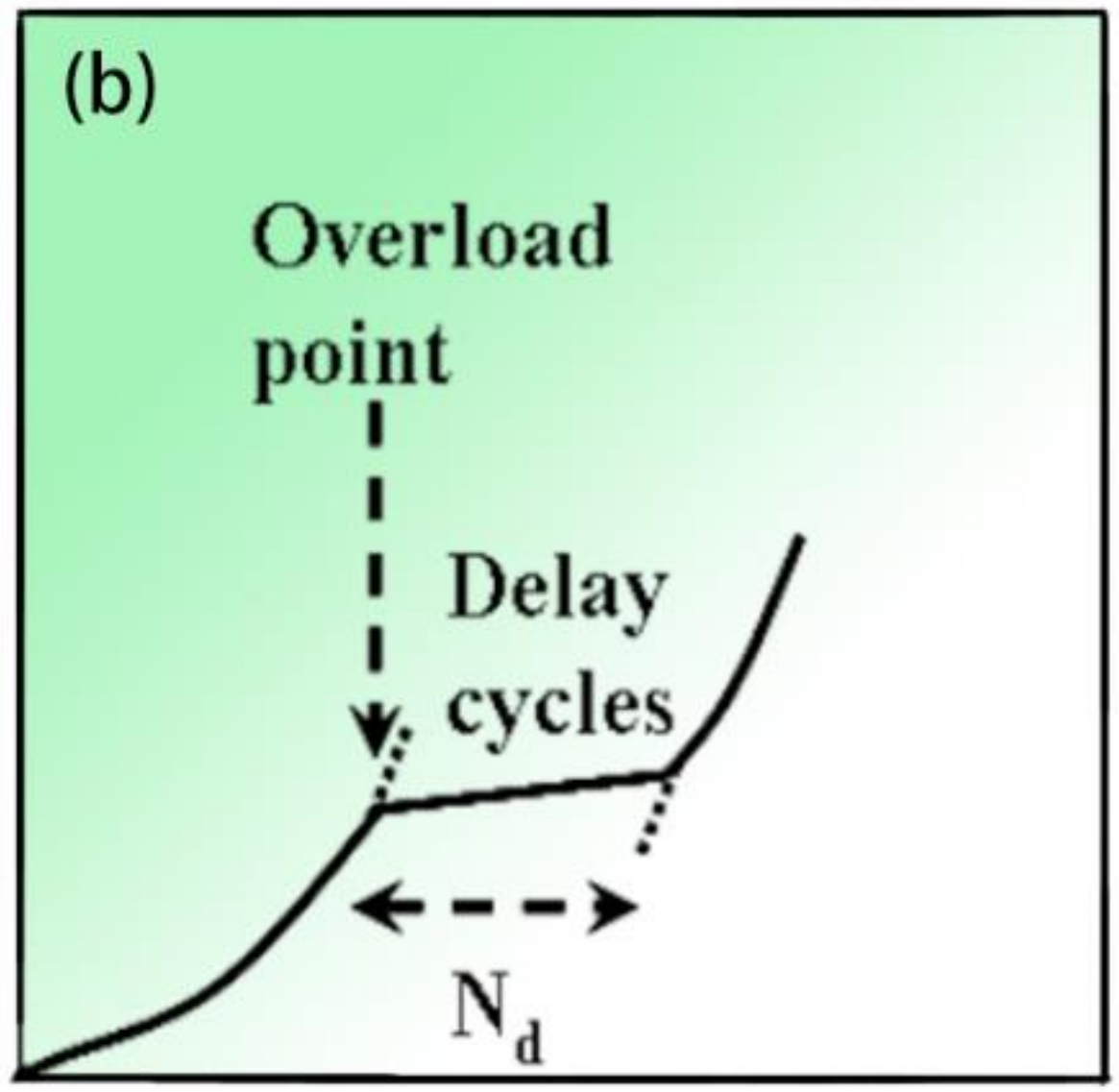

**Figure 3.6.1:** (a) Stress intensity factor range under overloading and underloading. (b) Representation of the overload effect in a diagram depicting the crack-growth rate [142].

The occurrence of the overload effect can be explained by the bulk elastic-lattice strain in the crack-tip region. In the fatigue test, the larger the applied load, the greater the compressive residual stress/strain near the crack tip, and correspondingly, the greater the plastic-zone size. The plastic-zone sizes under as-fatigue and overload conditions can be expressed by Irwin's equation [143]:

$$R_{y(as-fatigue)} = \frac{1}{\beta\pi}\left(\frac{K_{max,BG}}{\sigma_y}\right)^2 \quad (3.6.1)$$

and

$$R_{y(overload)} = \frac{1}{\beta\pi}\left(\frac{K_{max,OL}}{\sigma_y}\right)^2 \quad (3.6.2)$$

where $\beta$ = 1 or 3 under plane-stress or plane-strain conditions, $K_{max,BG}$ and $K_{max,OL}$ are stress-intensity factors for normal conditions and overload, as shown in Figure 3.6.1. It can be estimated from the above equations that the overload causes the crack tip to form a large plastic region, and in the next few cycles, the current plastic region still stays within the overload-induced plastic region. The crack-growth rate does not return to the normal state until the plastic region under the current cycle leaves the overload-induced plastic region. When the retardation stage ends, the crack-growth rate returns to the original trend again.

The retardation effect is related to the overload ratio. During loading, the stress-intensity factor in the normal stage can be marked as $K_{max,BG}$, and the K value, when overloading happens, can be marked as $K_{max,OL}$. The overload ratio is calculated as $\frac{K_{BG}}{K_{max}}$, and the range of the overload ratio is usually around 2 ~ 3. The number of cycles experienced in the retardation stage is $N_d$, The higher the value of the overload ratio, the longer the retardation period.

Lam et al. [108] prove that the overload effect is also effective for HEAs. In the study by Lam et al. [108], the crack-propagation-rate data of CoCrFeMnNi under normal and overload conditions was plotted, as shown in Figure 3.6.2, and a small retardation region was found on the plot at the overload point.

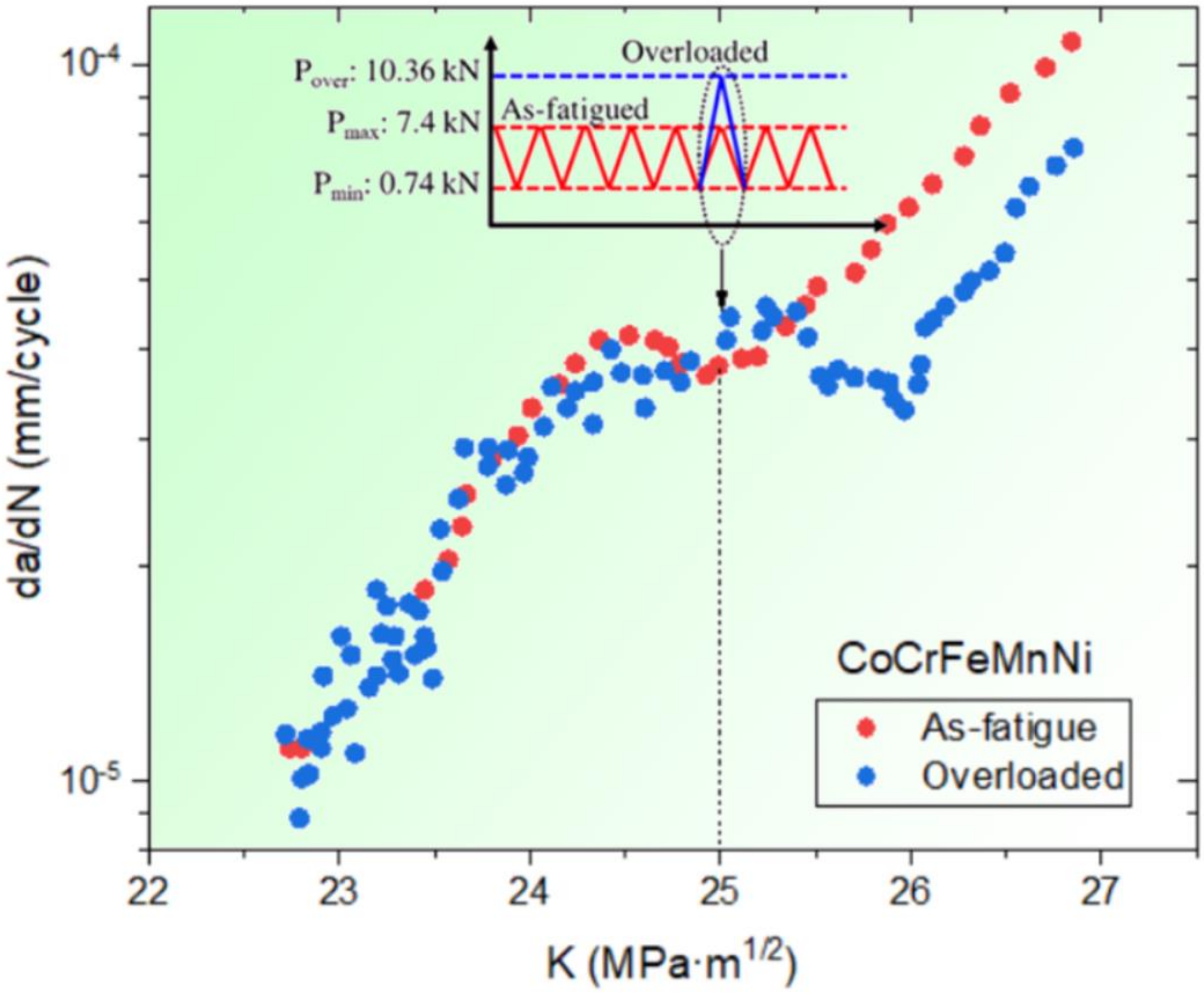

**Figure 3.6.2:** Reflection of the overload effect in the CoCrFeMnNi fatigue-crack-propagation experiment [108].

In addition to the difference in the crack-growth rate, the study by Lam et al. also compared the residual stress at the crack tip of the specimens under normal fatigue and overload tests [108]. The results show that the residual stress at the crack tip of the as-fatigue specimen is always at the current maximum value. But the residual stress at the crack tip of the specimen after overloading was slightly relaxed.

According to the estimation of Irwin's equation [Eq. (3.6.1)], the size of the plastic zone under an overload condition is about twice that under an as-fatigue condition. Furthermore, the study also shows that the large plasticity induced under overloading generates a critical analytical shear stress beyond the onset of twinning, resulting in the generation of a large number of deformation twins near the crack tip. As shown in Figure 3.6.3, twinning was not observed in the microstructure of the as-fatigued CoCrFeMnNi specimens, whereas in the overloaded CoCrFeMnNi specimens, strip-like twinning structures appeared around the cracks under RT.

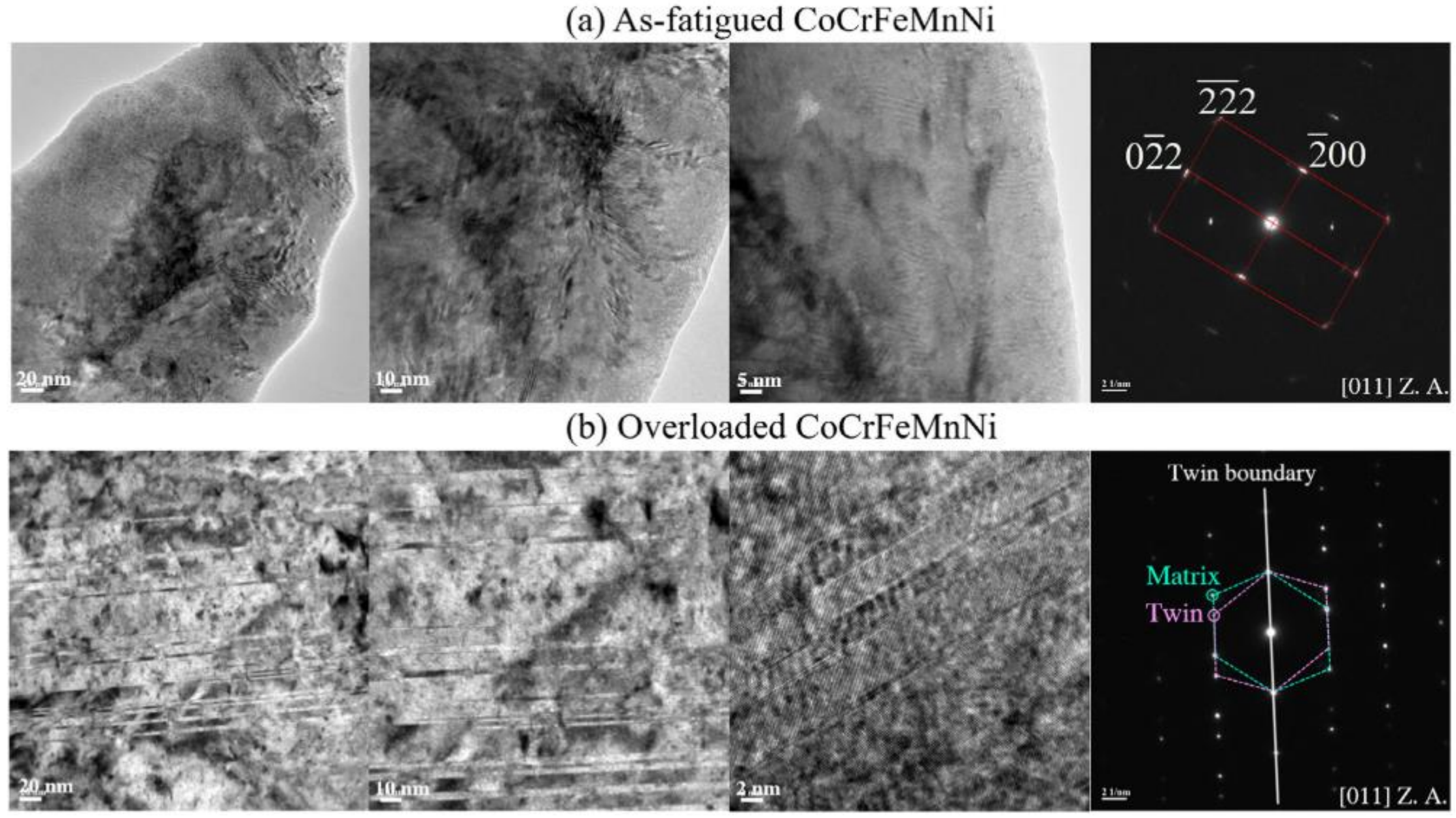

**Figure 3.6.3:** Microstructure of the CoCrFeMnNi specimen under (a) a normal fatigue-crack-growth-rate test and (b) an overload fatigue-crack-growth-rate test [108].

## 3.7. Frequency effects

Not much investigation has conducted addressing the question of specifically measuring a change in the frequency during fatigue tests. However, this section will elaborate on the effects of frequency on the fatigue behavior of HEAs, based on the limited research work identified. Figure 3.7.1 shows the effect of frequency on CoCrFeMnNi fatigue properties, in particular on the fatigue limit, the UTS and the fatigue ratio. All the data is obtained from Table 2.1.1. The frequencies used in all experiments range from 10 Hz to 20,000 Hz. As shown in the results, the effect of the change in testing frequency on the HCF behavior of CoCrFeMnNi was not obvious, in regards to microstructure, temperature, strain and overload. Therefore, frequency is identified as a sub-important factor in the evaluation of fatigue performance.

### *3.7.1. Metals and conventional technical alloys*

In case of the conventional technical alloys, ultrasonic testing frequencies had no major impact on the HCF resistance at diminished stress levels. Moreover, the testing frequency range of 20 kHz – 200 Hz was reported for various technical alloys and other metals with the lifetime ranging between $\sim 10^4$ to $\sim 10^8$ cycles [144]. No noticeable change in the lifespan of X3CrNi134 steels and 316 steels was observed at $\sim 10^8$ cycles within the ultrasonic frequency range reported [39].

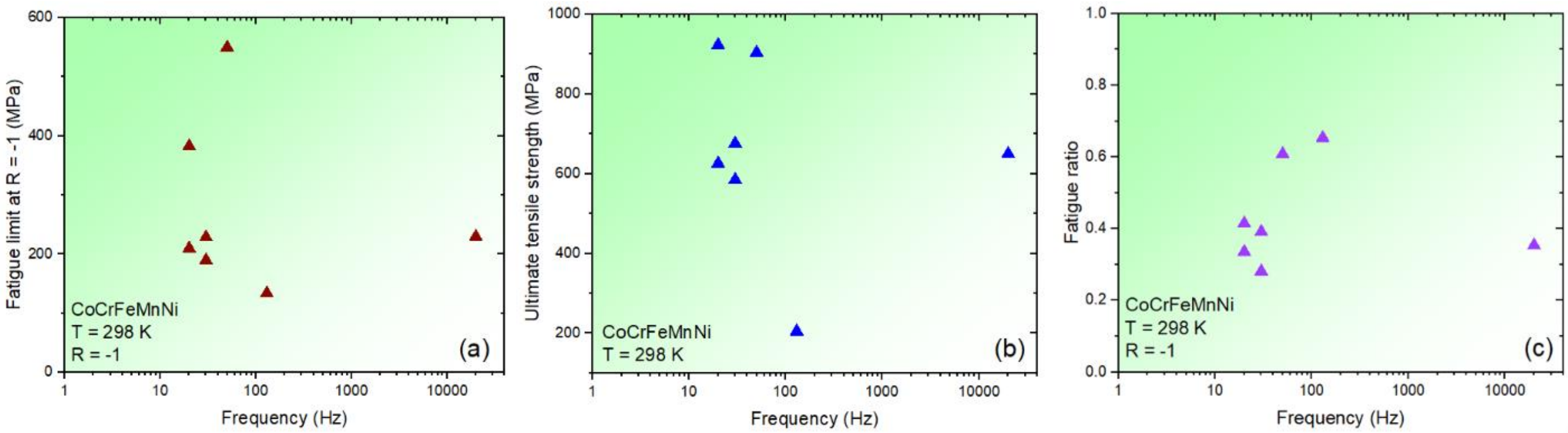

**Figure 3.7.1**: The relationship between frequency and (a) fatigue limit, (b) UTS and (c) fatigue ratio of CoCrFeMnNi under $R$ = -1 and $T$ = 298 K [21, 38, 42]. The data is obtained from Table 2.1.1.

Moreover, Shao et al. [132] conducted a theoretical inquiry into the high-frequency behavior of metals, using discrete dislocation dynamics, in an effort to provide insights into the effects of frequency on the dislocation behavior of the metals. The conclusion provided was that, due to a combination of 1) a large degree of reversibility in plastic deformation, caused by a decrease in cross-slip events, and 2) an enhanced contribution of anelastic/elastic deformation as a result of the minimal dislocation mobility, a large percent of cyclically induced reversible plastic deformation via high-frequency testing would be formed [132]. Here, the high-frequency testing involved the range of multiple tens or hundreds of Megahertz and as high as 20 kHz for fatigue at ultrasonic frequencies.

### *3.7.2. Pure metals or alloys with FCC and BCC phases*

In the case of pure FCC metals, the high-cycle fatigue resistance of pure Al increased by ~50%, upon increasing the testing frequency from 200 Hz to 20 kHz (and estimating the high-cycle fatigue resistance after $10^4$ to ~$10^8$ cycles) [39]. For pure Cu, only ~10% increase in fatigue resistance has been observed, upon increasing the testing frequency from 200 Hz to 20 kHz [39]. Moreover, the damping capability of FCC-heavy alloys exhibit similar, if not increased, loss tangent value, and by extension, a decrease in the elastic storage modulus, in comparison to Fe-Al alloys, within the applied frequency range of 1 - 16 Hz [145]. This phenomenon, however, is thought to be primarily temperature dependent and not a frequency-dependent. As a result of the absence of an observed change, when the testing frequency was varied, it was discovered that the storage modulus of the $Al_x$CoCrFeNi HEA ($x$ = 0, 0.25, 0.5, 0.75, and 1) exhibited excellent kinetic stability [145].

Thermally activated dislocation movement in alloys and crystal structures have been thought to be driven principally by cyclic strain-rate dependency (otherwise known as the frequency effect) [39]. Additionally, a large strain-rate dependency was observed in BCC alloys, because of the association of dislocation movement to large activation energy and large lattice friction [39]. This phenomenon of strain-rate dependency, however, did not hold the same amount of significance for FCC alloys [132]. From Figure 3.7.1, it can be seen that the frequency effect carries moderate-to-low significance within HEAs, such as the CrMnFeCoNi HEA. However, it must be noted that, in the work of Ghomsheh et al. [39], a direct study of the effects of ultrasonic vs conventional testing frequency testing on the fatigue performance of the CrMnFeCoNi HEA was not available.

### *3.7.3. Fatigue-crack propagation*

Another topic of immense importance pertains to crack propagation due to fatigue. In this section, we provide an overview of some prominent crack-propagation behavior in HEAs and of how the behavior relates to changes in the applied frequency and temperature.

From Figure 3.7.2, it can be noted that little to no significant changes have been observed in the crack-growth rate of the $Fe_{30}Mn_{10}Cr_{10}Co$ HEA upon varying the testing frequency from 1 Hz to 25 Hz. However, the presence of slight changes in the rate of fatigue-crack growth can be observed in the low $\Delta K$ regime in Figure 3.7.2. Moreover, the possibility of crack surface contact was established based on evidence left behind in the form of chipped parts. As a result of the aforementioned, it was deduced that the measure of crack-surface roughness was the dominant driver for the observed change in the lower $\Delta K$ regime at the testing frequency of 25 Hz [110]. As a result, it can be concluded that, in case of the HEAs, the crack-growth rate as a function of the stress-intensity factor might not be very sensitive even at different frequencies and temperatures.

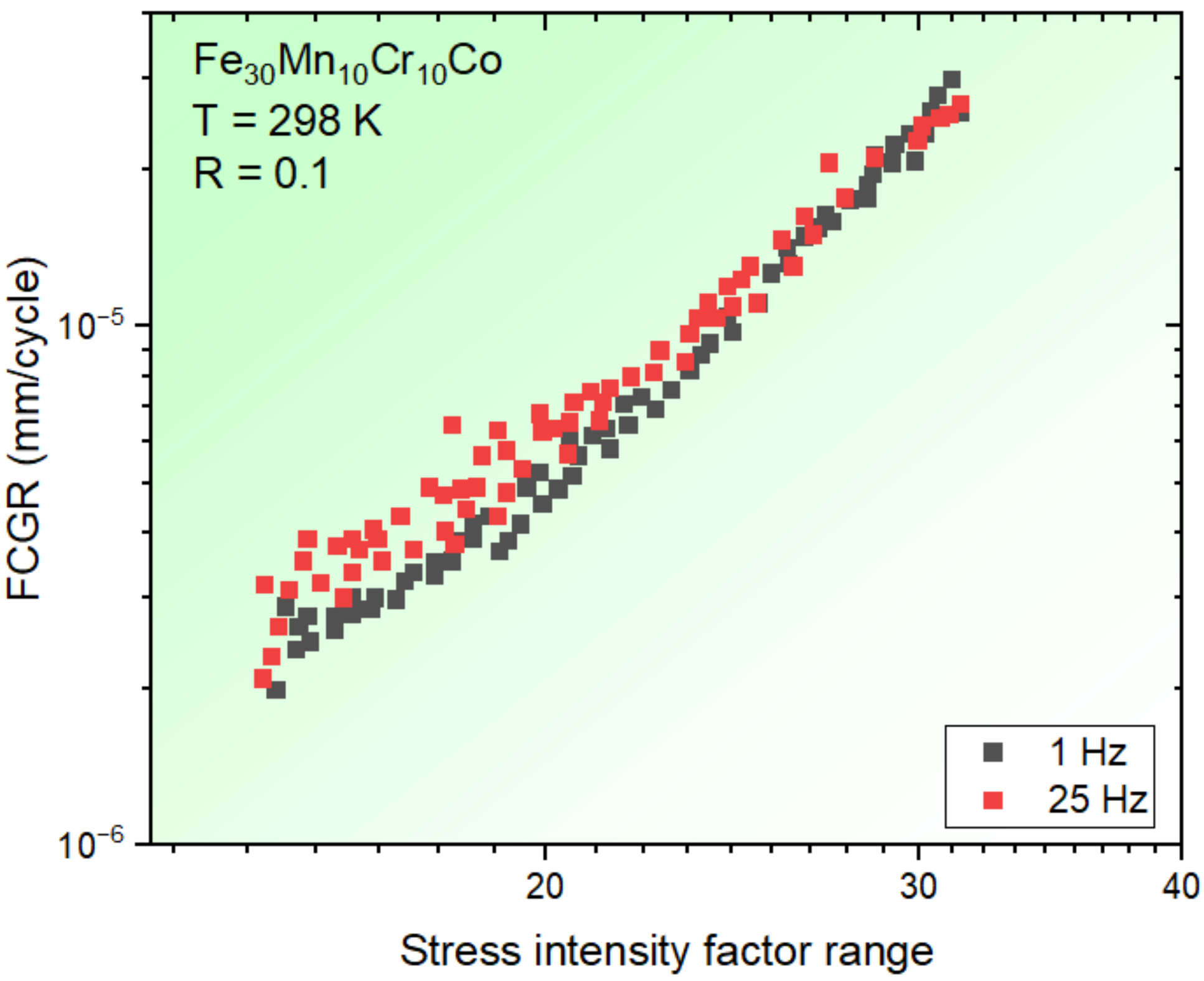

**Figure 3.7.2:** Fatigue crack growth in the $Fe_{30}Mn_{10}Cr_{10}Co$ HEA at testing frequencies of 1 Hz and 25 Hz. Image modified from [110].

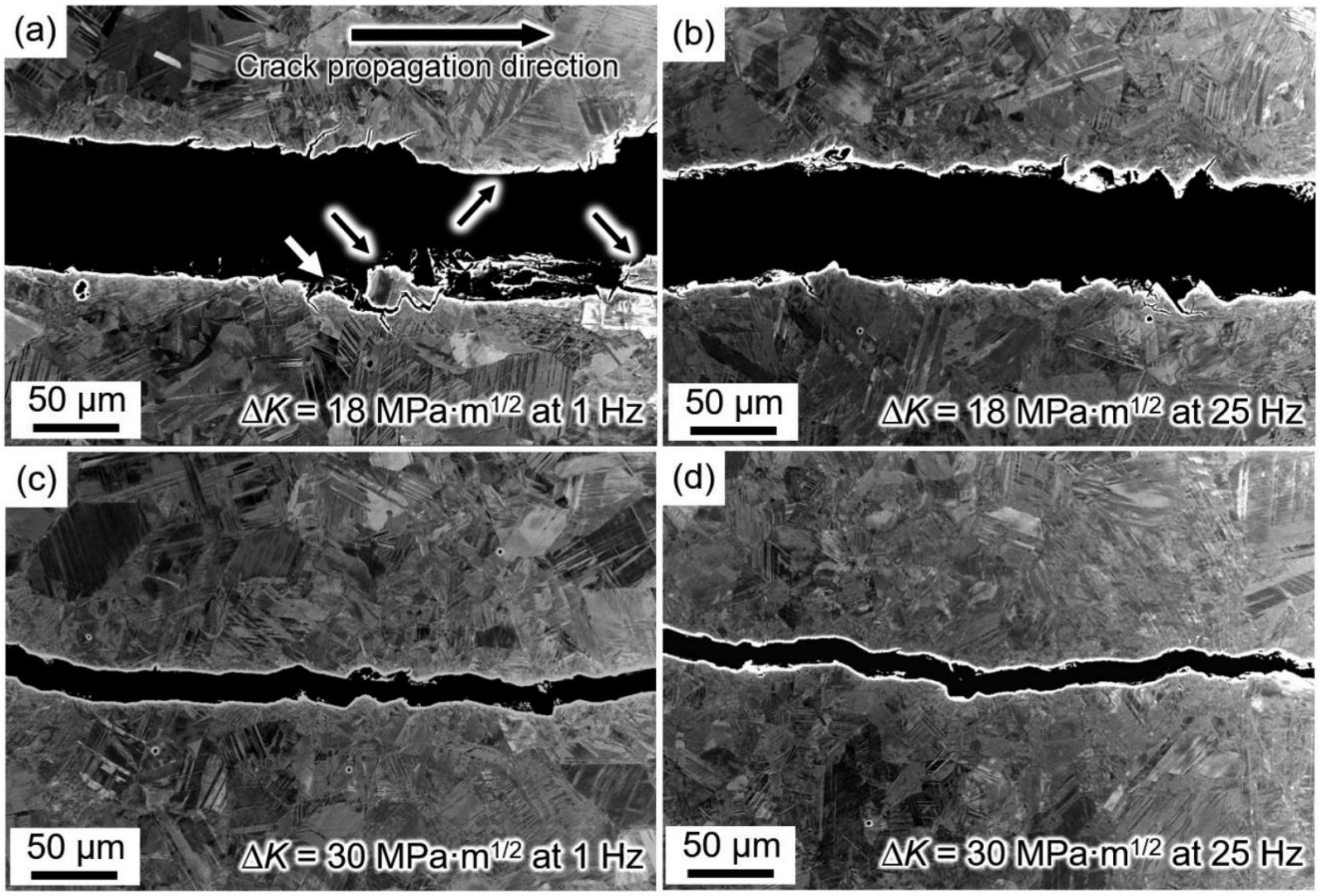

**Figure 3.7.3:** Electron channeling contrast (ECC) cross-sectional images of $Fe_{30}Mn_{10}Cr_{10}Co$ surface fracture areas obtained at 1 Hz (a, c) and 25 Hz (b, d). Change in the stress intensity factor range ($\Delta K$) is 18 MPa $m^{1/2}$ (a, b) and 30 MPa $m^{1/2}$ (c, d). Unusually large crack-roughness areas are marked with black arrows [110].

The presence of chipping within the fracture surface of the $Fe_{30}Mn_{10}Cr_{10}Co$ HEA specimen has been observed [110]. The white arrow in Figure 3.7.3(a) highlights this phenomenon. Here, the large crack roughness was observed for the $Fe_{30}Mn_{10}Cr_{10}Co$ HEA specimen that was tested at 1 Hz. This chipping (crack roughness) phenomenon is indicative of contact between crack surfaces throughout cyclic loading. In turn, this chipping phenomenon could mitigate, to some extent, the propagation of cracks during fatigue testing. This observation, however, did not hold, when the applied frequency was elevated to 25 Hz. As the frequency of the cyclic loading was increased to 25 Hz, the extent of crack roughness decreased, as demonstrated in Figure 3.7.3(b). Nevertheless, it must also be noted that an increase in the change in the stress intensity factor range ($\Delta K$) also played a role in mitigation of the crack roughness. The stress intensity factor range was increased from 18 to 30 MPa $m^{1/2}$, as noted in Figure 3.7.3(c)-(d). As a follow-on observation, it is worth mentioning that the parameters describing the roughness of the crack surface roughness (namely its sharpness, frequency and size) were noted to be higher, in the case of laminated steels (e.g., in the case of pearlitic steels), than in the case of the $Fe_{30}Mn_{10}Cr_{10}Co$ HEA in the low stress intensity factor regime at 1 Hz. From the aforementioned, it can be established that for the $Fe_{30}Mn_{10}Cr_{10}Co$ alloy, the crack-growth was relatively smooth on a macroscopic scale, but the crack-surface roughness caused minimal crack-growth deceleration [110].

Prior work has demonstrated some key mechanisms through which the fatigue-crack growth in the low $\Delta K$ regime interact. Some relevant research work in this area include the work of Liaw et al. on the evolution of fatigue crack growth in metals [146], the work of Suresh et al. on the effects of surface roughness on fatigue-crack closure [147], and even the demonstration by Suresh et al. of strain rate sensitivity to certain deformation mechanisms [137]. Of note, it is known that the Young's modulus is proportional to the stress intensity factor range for the effective threshold [146]. As a result, this phenomenon would dictate that, in order to enhance the propagation performance of near-threshold cracks, an increase in the Young's modulus would generally be necessitated [146]. Another important relationship is the Elber equation, which is defined as $\frac{da}{dN} = C\left(\Delta K_{eff}\right)^n$ [148], where $\boldsymbol{C}$ denotes a coefficient, $n$ represents an exponent, and $K_{eff}$ is the effective stress intensity [148]. An increase in the fracture surface roughness (i.e., asperity height) is known to increase the crack-closure level ($K_{cl}$), thereby decreasing $\Delta K_{eff}$ [147] which, in turn, would result in a decreased crack growth rate, as governed by the Elber equation [147, 148]. One possibility for an increase in the roughness of the fracture surface at 1 Hz, in comparison to 25 Hz, could relate to the rate of localized material deformation being lower at lower frequencies, which would allow for a longer length scale of localized deformation. As the testing frequency increases, the length, over which plastic deformation (i.e., permanent separation of material) occurs, would decrease, due to the more rapid accumulation of localized fatigue damage, thereby decreasing the roughness of the fracture surface in vicinity of the fatigue crack.

From the above, it can be concluded that the tests of fatigue-crack growth in compact tension (CT) specimens, with applied testing frequency of 1 Hz, and conducted at RT, have demonstrated the following [110, 149]:

1. High concentrations of HCP martensites were observed on the fatigue crack.
2. Crack propagation of Mode-I type was thought to be observed because of the presence of a smooth crack-growth track, much like for stable austenitic steels.
3. For FCC and HCP phases, plastic deformation drove fatigue-crack growth of Mode-I type.
4. In spite of the presence of secondary cracks, no acceleration of fatigue-crack growth was observed.

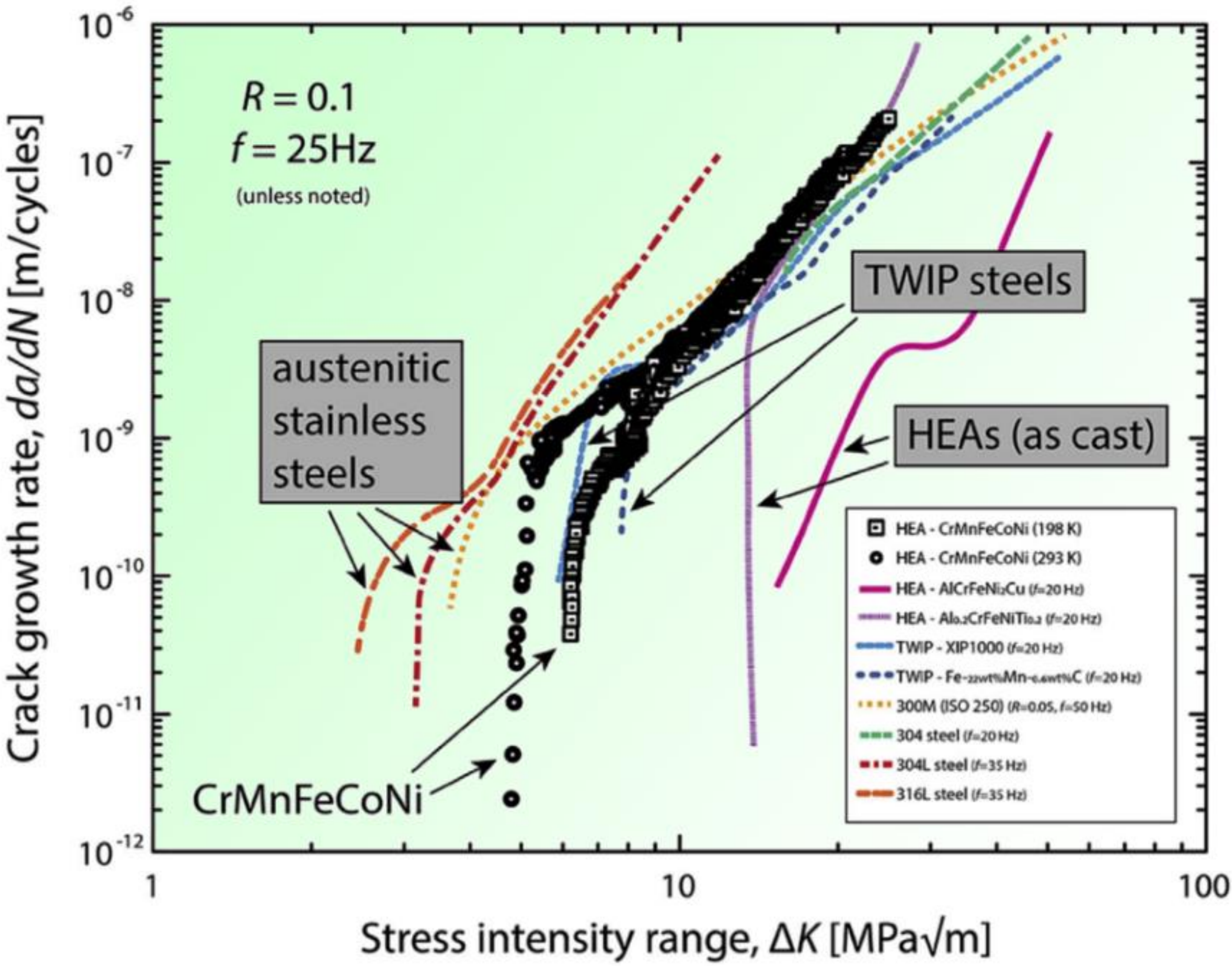

**Figure 3.7.4:** Fatigue crack growth behavior of TWIP steels (298K unless stated otherwise), austenitic stainless steels, and CrMnFeCoNi HEA (Cantor alloy). The image has been modified from [30]. HEAs with the Al content exhibited enhanced fatigue thresholds and slopes in the Paris regime [31]. This was thought to be due to smaller sample sizes used during testing as well as the as-cast microstructure involved [30]. Moreover, comparable Paris slopes of 3.5 - 4.5 were observed in austenitic steels with similar microstructures [150]. Quenched and tempered low-alloy steels [151], and twinning-induced plasticity [152, 153] resemble most closely to the Cantor alloy, from perspective of the fatigue resistance.

Details on the particular alloys and testing conditions presented in Figure 3.7.4 are listed in Table 3.7.1. The testing frequencies applied range between 20 - 50 Hz. After the close inspection of Figure 3.7.4, it was concluded that the Cantor alloy (CrMnFeCoNi) demonstrated unusually good fatigue-crack-growth resistance, at the temperatures of 198 K and 298 K, in comparison to austenitic steels, TWIP steels and other HEAs, given similar microstructure and testing conditions. TWIP steels demonstrated a distinct absence of twinning, under fatigue loading conducted at room temperature. This was replicated closely in the Cantor alloy [152, 153]. Moreover, upon further inspection of Figure 3.7.4, and upon comparing the austenitic and TWIP steels to the Cantor alloy, it can be noted that the Cantor alloy performs similarly to the other materials, in terms of the fatigue threshold and the Paris slope [30].

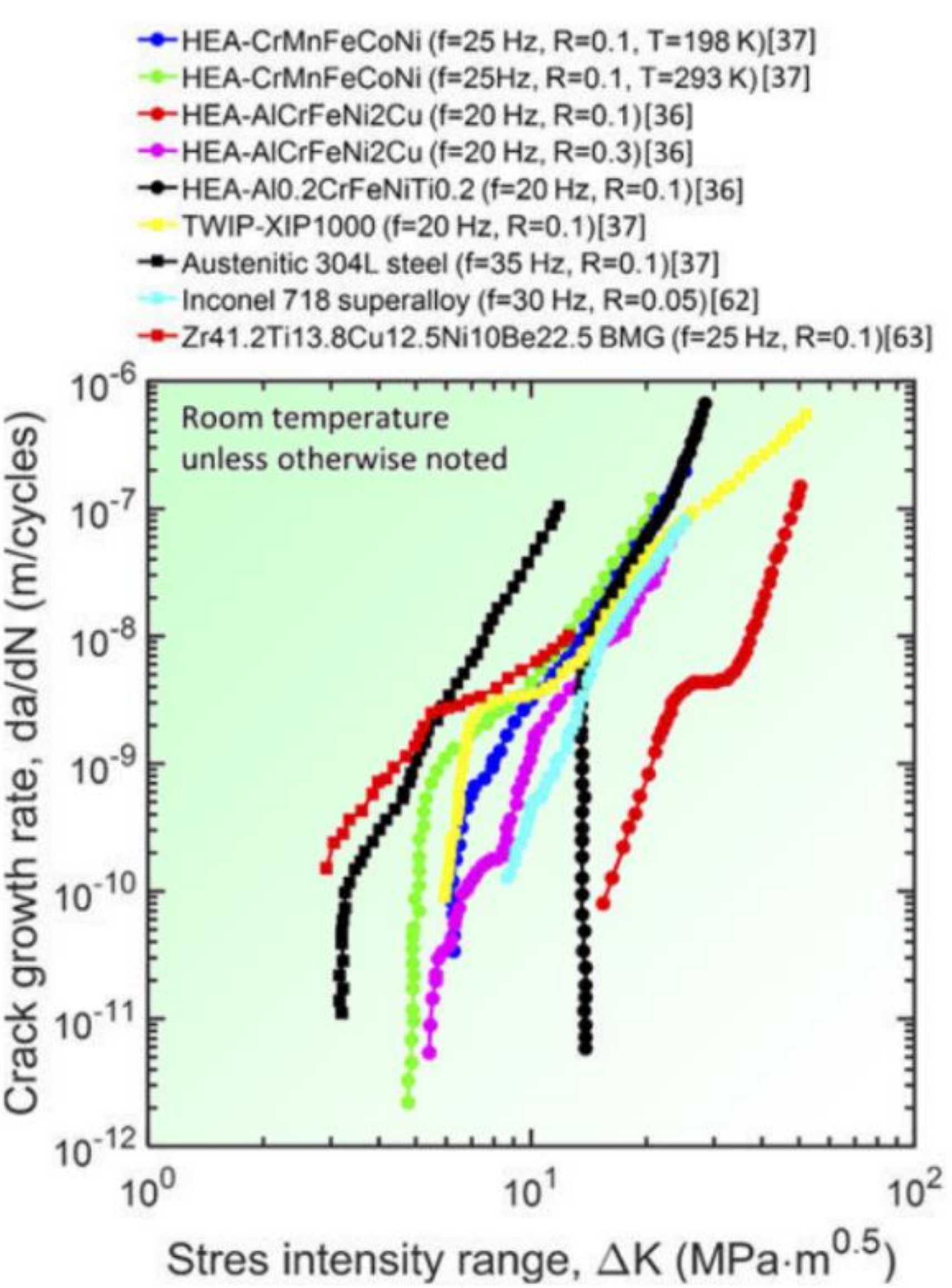

**Figure 3.7.5:** Fatigue-crack-growth rate of $Al_{0.2}CrFeNiTi$, $AlCrFeNi_2Cu$, and CrMnFeCoNi HEAs compared to the fatigue-crack-growth rate of a number of conventional alloys from previous studies [30, 31, 154, 155]. The image has been modified from [138].

Regarding the effect of the stress ratio ($R$) on the fatigue-crack-growth behavior of as-cast $Al_{0.2}CrFeNiTi_{0.2}$ and $AlCrFeNi_2Cu$ HEAs at RT, the purple and red curves in Figure 3.7.5 (denoted by circles) provided the necessary insights to establish that there was a tendency towards the low $\Delta K$ regime for fatigue-crack-growth curve. In case of the as-cast $Al_{0.2}CrFeNiTi_{0.2}$ HEA, the calculated Paris slope ($m$) was computed to be 4.9, 5.3, and 25.8, at the stress ratios of $R$ = 0.1, 0.3, and 0.7, respectively [31]. Given this positive trend in the Paris slope with increasing $R$, the resulting implication could be that brittle failure should be expected within the Paris regime for $Al_{0.2}CrFeNiTi$, $AlCrFeNi_2Cu$, and other alloys [31, 138, 156, 157]. Based on the aforementioned, lower magnitudes of fatigue-crack closure is considered to be the driver of increased fatigue-crack growth as the stress ratios applied are increased [138]. It must also be noted that the environmental temperature also had some effect on fatigue-crack-propagation properties for some of the HEAs. In particular, the CrMnFeCoNi HEA demonstrated elevated crack-propagation resistance at cryogenic temperatures. This was demonstrated, when the testing temperature was diminished to 198 K from 293 K and an increase in $\Delta K_{th}$ by ~ 30% was observed, coupled with a minor shift to 4.5 from 3.5 in the Paris slope [138]. Other literature references also confirm similar reports of enhanced mechanical properties at cryogenic temperatures [158, 159].

From Table 3.7.1, the observation can be made that amongst the structural alloy candidates listed, the Al and Mn containing HEAs provide good fatigue-crack-growth behavior. In particular, the CrMnFeCoNi HEA (the Cantor alloy) demonstrated excellent mechanical properties under harsh environments. These properties include, but are not be limited to, immense fatigue-crack-growth resistance, good fracture toughness, excellent ductility, and high strength. The properties may make the Cantor alloy a competitive candidate for applications involving structural materials [138]. This conclusion was obtained by comparing the Paris slope and fatigue thresholds of the Cantor alloy to those of several major structural materials, under comparable fatigue test conditions, wherein the testing frequency applied ranged between 20 - 50 Hz (with minimal observable changes from the frequency alone).

**Table 3.7.1:** Fatigue-crack-growth properties with relevant testing parameters of HEAs vs. conventional alloys [30, 138].

| **Alloy** | ***R* (stress ratio)** | **Frequency [Hz]** | **Grain size [μm]** | **Young's Modulus [GPa]** | **Paris Slope [m]** | **Threshold fatigue stress-intensity factor range [$\Delta K_{th}$, MPa $m^{1/2}$]** | **$\Delta K_{th}$ / *E* [$\sqrt{m}$]** | **Temperature [K]** |
|---|---|---|---|---|---|---|---|---|
| CrMnFeCoNi | 0.1 | 25 | 7 | 202 | 3.5 | 4.8 | 0.024 | 293 |
| CrMnFeCoNi | 0.1 | 25 | 7 | 210 | 4.5 | 6.3 | 0.03 | 198 |
| $AlCrFeNi_2Cu$ | 0.1 | 20 | As-cast | - | 3.4 | 17 | - | RT |
| $Al_{0.2}CrFeNiTi_{0.2}$ | 0.1 | 20 | As-cast | - | 4.9 | 16 | - | - |
| XIP 1000-TWIP steel | 0.1 | 20 | 2 | 188 | 2.7 | 5.9 | 0.031 | RT |
| Fe-22 wt%Mn-0.6 wt%C TWIP steel | 0.1 | 20 | 5 | - | 3.8 | 7.6 | - | RT |
| 300-M steel (ISO250) | 0.05 | 50 | 20 | 205 | 2.5 | 3.6 | 0.018 | - |
| 304 Steel | 0.1 | 20 | - | 200 | 3.8 | 15.8 | 0.079 | RT |
| 304L Steel | 0.1 | 35 | 40 | 193 | 5.8 | 4.8 | - | RT |
| 316L Steel | 0.1 | 35 | 38 | 193 | 4.9 | 3.3 | - | RT |
| $AlCrFeNi_2Cu$ | 0.3 | 20 | - | - | 6.4 | 5 | - | RT |
| $AlCrFeNi_2Cu$ | 0.7 | 20 | - | - | 14.5 | 7 | - | RT |
| $Al_{0.2}CrFeNiCu_{0.2}$ | 0.1 | 20 | - | - | 4.9 | 16 | - | RT |
| $Al_{0.2}CrFeNiCu_{0.2}$ | 0.2 | 20 | - | - | 5.3 | 7 | - | RT |
| $Al_{0.2}CrFeNiCu_{0.2}$ | 0.7 | 20 | - | - | 25.8 | 5 | - | RT |
| Ti-6Al-4V | 0.1 | 50 | - | - | 5.1 | 4.7 | - | RT |
| Inconel 718 superalloy | 0.05 | 30 | - | - | 5.62 | 7.8 | - | RT |
| $Zr_{41.2}Ti_{13.8}Cu_{12.5}Ni_{10}Be_{22.5}$ | 0.1 | 25 | - | - | 2.7 | 3 | - | RT |

In sum, the properties identified (i.e., the $R$ stress ratio, grain size, $\Delta K_{th}$, etc.) tend to affect the overall fatigue performance of HEAs in a complementary fashion. However, a close cross-examination is recommended of how the length scale of the cyclic localized damage accrual affects crack initiation and propagation as a result of fatigue in HEAs. It is hypothesized by the authors of this review that the length scale of the cyclic damage accrual has a major influence on the crack initiation and propagation pattern, which would be further complicated by a heterogenous energetic landscape throughout the solid solution of a given HEA.

#### 3.7.4. *Conventional alloys*

In the case of certain conventional alloys, behavior somewhat different from the HEAs is observed, because of change in the testing frequency applied. One such example is the fatigue response of pure polycrystalline copper. In that case, variations in the testing frequency applied ranging from 100 Hz to 20 kHz contributed to a shorter fatigue life, compared to the conventional frequency range of 100 Hz [160, 161]. Another example is seen in the case of a high-carbon chromium steel (GCr15, which is equivalent to SUJ2 or SAE 52100 steels). Within the very high-cycle regime (defined as $\geq 10^7$ cycles), electromagnetic resonance axial loading (EA, 120 Hz) resulted in lower fatigue strength than observed under ultrasonic axial loading with cooling (UA, 20 kHz). Moreover, EA and UA fatigue specimens both exhibited significantly larger fatigue strength than observed under ultrasonic axial loading with no cooling (UA-NC, 20 kHz) [162].

In sum, low-to-moderate changes in fatigue life, and in fatigue-crack propagation, have been observed up to this point in HEAs as a result of a wide range of testing frequencies applied. The testing frequencies applied have ranged from 1 – 25 - ≤ 200 Hz, for conventional frequencies, to 20 kHz, for ultrasonic frequencies [39, 110]. This conclusion, however, does not necessarily hold valid for conventional metals and alloys, such as pure polycrystalline copper, and high-carbon chromium steel [160-162]. However, with a distinct absence in change in certain properties in FCC heavy HEAs, good kinetic stability can be verified [145]. Nevertheless, a definite possibility of Al and Mn containing HEAs being primary candidates of future work for fatigue inquiries ranging from fatigue-crack propagation, to fatigue resistance in unusual environments is highly promising [30, 138].

## 4. Theoretical Modeling

### 4.1. Stochastic models – for enabling prediction of fatigue life

The mechanical-fatigue phenomenon is stochastic in nature. Stochastic (probabilistic) modeling is, therefore, frequently employed for the prediction of fatigue life [8, 44, 52, 163]. Stochastic modeling is needed to account for the scatter in the fatigue life, which is observed under constant amplitude loading, and which tends to be smaller for lower fatigue life, but higher for lower stress amplitudes, as exhibited in Figure 4.1.2 and S2.4.1.3. In [35], Hemphill et al. present a study of the connection between defects and fatigue-endurance limits, defined as the stress levels below which no fatigue failures occur, of HEAs by means of statistical fatigue-life modeling. The defects can be broadly considered as a part of the microstructural effects listed in Figure 4.1.1 (per the "Microstructure" label). Reference [123] addresses the intrinsic role of microstructures on persistent slip bands. The authors noted that although the nano-sized $L1_2$ precipitates enhanced tensile strength, no improvement in fatigue properties was observed [123].

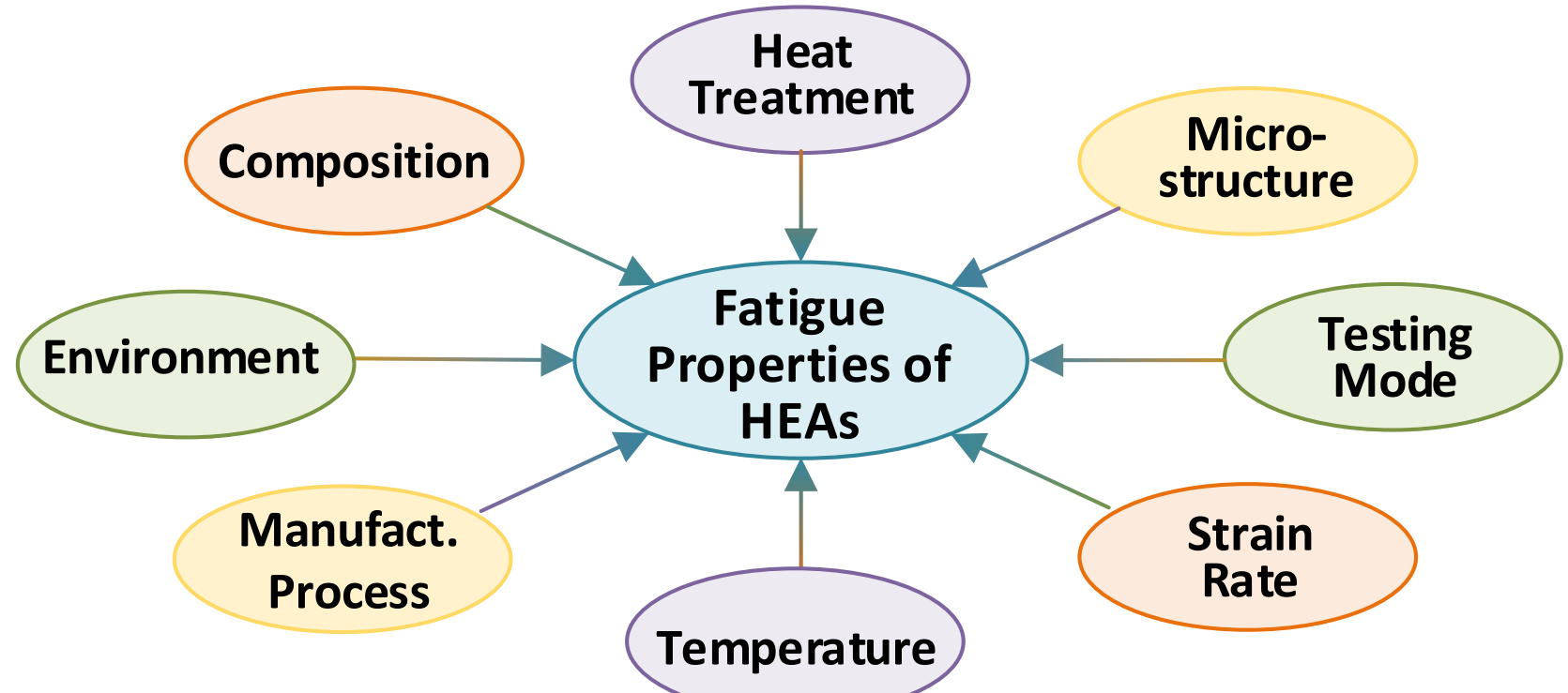

**Figure 4.1.1:** Overview of the sources impacting the fatigue properties of HEAs (adapted from [164, 165]). We are here primarily looking at the mechanical fatigue properties of HEAs.

The stochastic fatigue lifespan models under the consideration here consist of
1. Weibull predictive models;
2. Weibull mixture predictive models;
3. General log-linear models;
4. Random endurance limit fatigue life models.

For a general background on these stochastic fatigue lifespan models, refer to the supplementary manuscript.

#### *4.1.1. Weibull predictive models*

The Weibull family of distributions has been used to model properties of materials pushed to the brink of failure or fracture, such as fatigue lives (also referred to as failure times). In the context of the extreme value theory, a subfield within the probability theory and statistics, the Weibull family is considered a type-III extreme-value distribution, along with the Gumbel (type I) and the Fréchet (type II) families of distributions. The Weibull family is a system of versatile distributions, which have been widely used for describing lifetime distributions in reliability engineering, fatigue, and fracture [59, 166, 167]. The Weibull family of distributions is the most widely used family of lifetime distributions in reliability engineering, and has been employed for modeling fatigue behavior of a variety of materials, such as steels [168, 169], aluminum alloys [170], and metallic glasses [171, 172]. Popularity of the Weibull distributions relates in part to the fact that all the model parameters have clear physical meaning in the fatigue-deformation model [173].

The probability density function (PDF) and the cumulative distribution function (CDF) for the fatigue life at a given stress level can be described, using the Weibull distribution, as Eqs. (4.1.1) and (4.1.2) illustrate, respectively [35]:

$$f(N(S)|\alpha(S),\beta) = \begin{cases} \dfrac{\beta}{\alpha(S)}\left(\dfrac{N(S)}{\alpha(S)}\right)^{\beta-1} \exp\left(-\left(\dfrac{N(S)}{\alpha(S)}\right)^{\beta}\right) & \text{for } N(S) \geq 0 \\ 0 & \text{for } N(S) < 0 \end{cases} \quad (4.1.1)$$

$$F(N(S)|\alpha(S),\beta) = \begin{cases} 1-\exp\left(-\left(\dfrac{N(S)}{\alpha(S)}\right)^{\beta}\right) & \text{for } N(S) \geq 0 \\ 0 & \text{for } N(S) < 0 \end{cases} \quad (4.1.2)$$

Here, the second factor, $\beta$, refers to the Weibull shape factor. Motivated by the structure of the S-N curves, the first factor, the Weibull scale factor, $\alpha(S)$, may be tied to the applied stress, $S$, as follows [35]:

$$\log(\alpha(S)) = \gamma_0 + \gamma_1 \log(S). \quad (4.1.3)$$

The two Weibull factors, $\alpha(S)$ and $\beta$, are subject to the limits listed below:

$$\alpha(S) > 0 \text{ and } \beta > 0. \quad (4.1.4)$$

The authors of [44, 52] demonstrate that the two-parameter Weibull model can describe the fatigue life distribution of the TiZrNbHfTa HEA quite well over a wide range of applied stress amplitudes. The Weibull model is capable of capturing the probabilistic nature of the HCF behavior of the TiZrNbHfTa and of predicting the fatigue life in a statistical sense [44, 52]. Further to such an effect, the authors of [174] and [175] report that all possible basic shapes of load spectra can be approximated well by the two-parameter Weibull distribution.

For additional background information on the Weibull predictive models, in context with reliability engineering (but not specific to HEAs), refer to the supplementary manuscript.

#### *4.1.2. Weibull mixture predictive models*

A single Weibull predictive model may not be able to adequately characterize variability in the observed fatigue data, such as in Figure 4.1.2, especially if operating conditions vary during the service life. If the operating conditions vary during the service life, the shape of the load spectra (S-N curves) may become multi-modal, i.e., form more than one group of concentration, and may thus not be accurately described by a unimodal distribution function [174]. Quite a few researchers have confirmed such a multi-modal nature of the load spectra [176-178]. The mixed Weibull distribution has the ability to accurately model load spectra whose data exhibits concentration around more than one center [179]. A mixed PDF, used in a mixture predictive model, is defined as

$$f(s) = \sum_{l=1}^{m} w_l \, f_l(s), \quad (4.1.5)$$

where $w_l > 0$ ($l = 1, \ldots, m$), and $\sum_{l=1}^{m} w_l = 1$ [174]. The constants, $w_l$, represent weighting factors, whereas $f_l(s)$ denotes an arbitrary PDF. In general, the PDF $f(s)$ can be composed of $m$ component distributions, each having a different type [174]. The identification and unwrapping of a model, meaning the estimation of $f_l(s)$, $w_l$ and $m$ from $f(s)$, can be a difficult undertaking. Significant simplification may be attained, if all $f_l(s)$ have the same type [174]. Moreover, the estimation of the total number of component distributions, $m$, can also be influenced by the choice of $f_l(s)$ [174]**.**

By inserting Eq. (4.1.1) into the definition in Eq. (4.1.5), one obtains the PDF for the Weibull mixture models:

$$f(s) = \sum_{l=1}^{m} w_l \frac{\beta_l}{\theta_l} \left(\frac{s}{\theta_l}\right)^{\beta_l - 1} \exp\left(-\left(\frac{s}{\theta_l}\right)^{\beta_l}\right) \quad (4.1.6)$$

The cdf for the Weibull mixture model can be obtained, in a similar fashion, through analogy to Eqs. (4.1.1) and (4.1.2):

$$F(s) = 1 - \sum_{l=1}^{m} w_l \exp\left(-\left(\frac{s}{\theta_l}\right)^{\beta_l}\right) \quad (4.1.7)$$

In the special case of only two groups, such as the weak and the strong groups in Figure 4.1.2, the probability density and the cumulative density functions can be written as [35]

$$f(N(S)|p, \alpha_w(S), \beta_w, \alpha_S(S), \beta_S) = p \frac{\beta_w}{\alpha_w(S)} \left(\frac{N(S)}{\alpha_w(S)}\right)^{\beta_w - 1} \exp\left(-\left(\frac{N(S)}{\alpha_w(S)}\right)^{\beta_w}\right) + (1-p) \frac{\beta_S}{\alpha_S(S)} \left(\frac{N(S)}{\alpha_S(S)}\right)^{\beta_S - 1} \exp\left(-\left(\frac{N(S)}{\alpha_S(S)}\right)^{\beta_S}\right) \quad (4.1.8)$$

$$F(N(S)|p, \alpha_w(S), \beta_w, \alpha_S(S), \beta_S) = p\left[1 - \exp\left(-\left(\frac{N(S)}{\alpha_w(S)}\right)^{\beta_w}\right)\right] + (1-p)\left[1 - \exp\left(-\left(\frac{N(S)}{\alpha_S(S)}\right)^{\beta_S}\right)\right] \quad (4.1.9)$$

Here, the subscripts, $w$ and $s$, reference the weak and the strong group, respectively, the parameter, $p$, represents the fraction of samples belonging to the weak group, and the parameter, $(1-p)$, denotes the fraction of samples belonging to the strong group. The Weibull-scale parameters for the weak and the strong groups, $\alpha_w(S)$ and $\alpha_s(S)$, can further be assumed to depend on the stress level, $S$, in accordance with Eq. (4.1.3) [35]:

$$\log(\alpha_w(S)) = \gamma_{w,0} + \gamma_{w,1}\log(S) \quad (4.1.10)$$

$$\log(\alpha_s(S)) = \gamma_{s,0} + \gamma_{s,1}\log(S). \quad (4.1.11)$$

In Figure 4.1.2, the fatigue lives observed for the $Al_{0.5}CoCrCuFeNi$ HEA appear to form two groups, a strong group and a weak group, with the fatigue lives in the weak group being much shorter than in the strong group, especially when the applied stress is less than 1,000 MPa, as predicted by the above Weibull mixture model [35]. Such variability in the fatigue lives may be caused by varying defect levels in the test specimens under a study [35].

For additional background information on Weibull mixture predictive models, in context with reliability engineering (but not necessarily specific to HEAs), refer to the supplementary manuscript.

#### ***4.1.3. General log-linear model***

For the purpose of relating to general log-linear models, it helps first acquaint oneself with microstructure (or phase) morphologies. A parallel morphology for fatigue testing of $Al_{0.5}CoCrCuFeNi$ HEA samples refers to the microstructure morphology for the case, where the HEA fatigue samples are machined parallel to the rolling direction [35]. A vertical morphology for fatigue testing of the $Al_{0.5}CoCrCuFeNi$ HEA samples refers to the microstructure morphology for the case, where the HEA fatigue samples are machined perpendicular to the rolling direction [35]. The $Al_{0.5}CoCrCuFeNi$ HEA samples consist of an $\alpha$ matrix phase together with a Cu-rich $\beta$ phase.

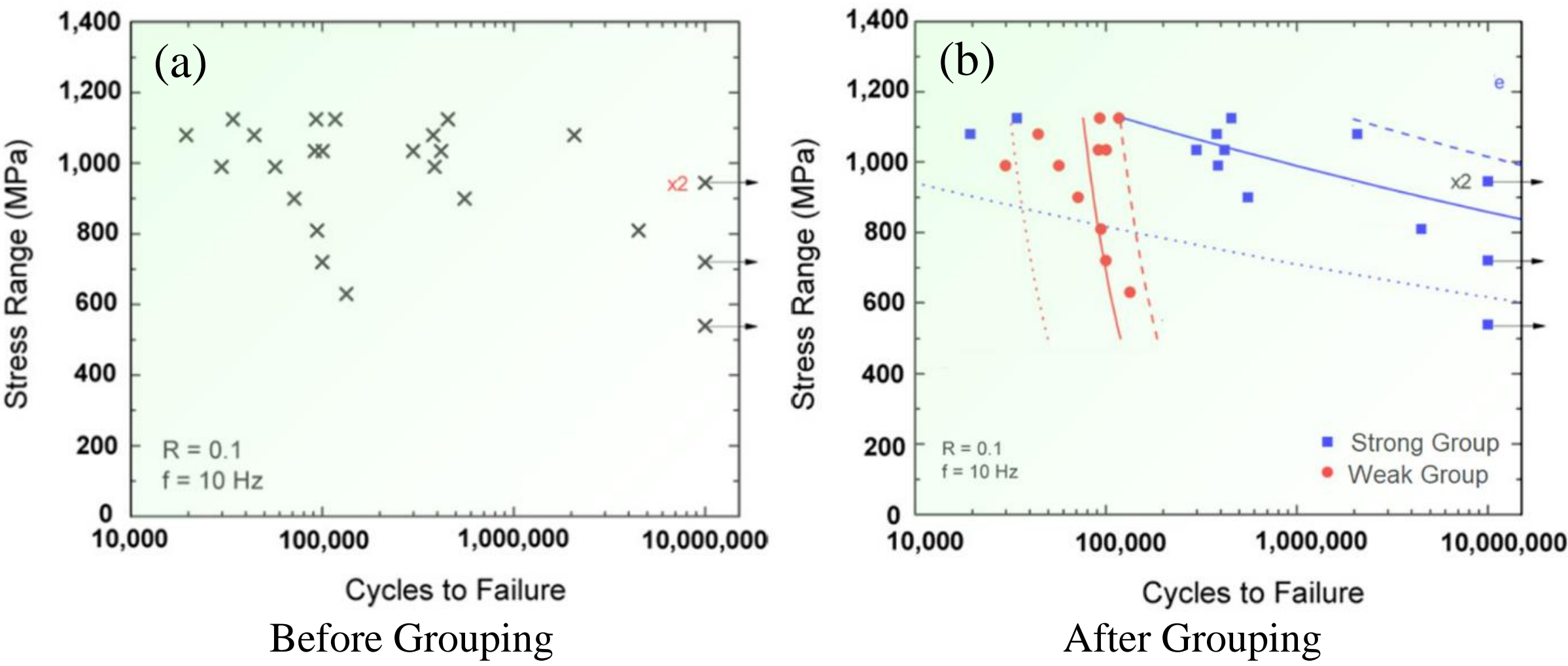

**Figure 4.1.2:** Fatigue lives for the $Al_{0.5}CoCrCuFeNi$ HEA plotted as the stress range vs. the number of cycles to failure (adapted from [35]). Left: Multi-modal load spectrum before grouping into strong and weak groups. Right: Multi-modal load spectrum after such grouping.

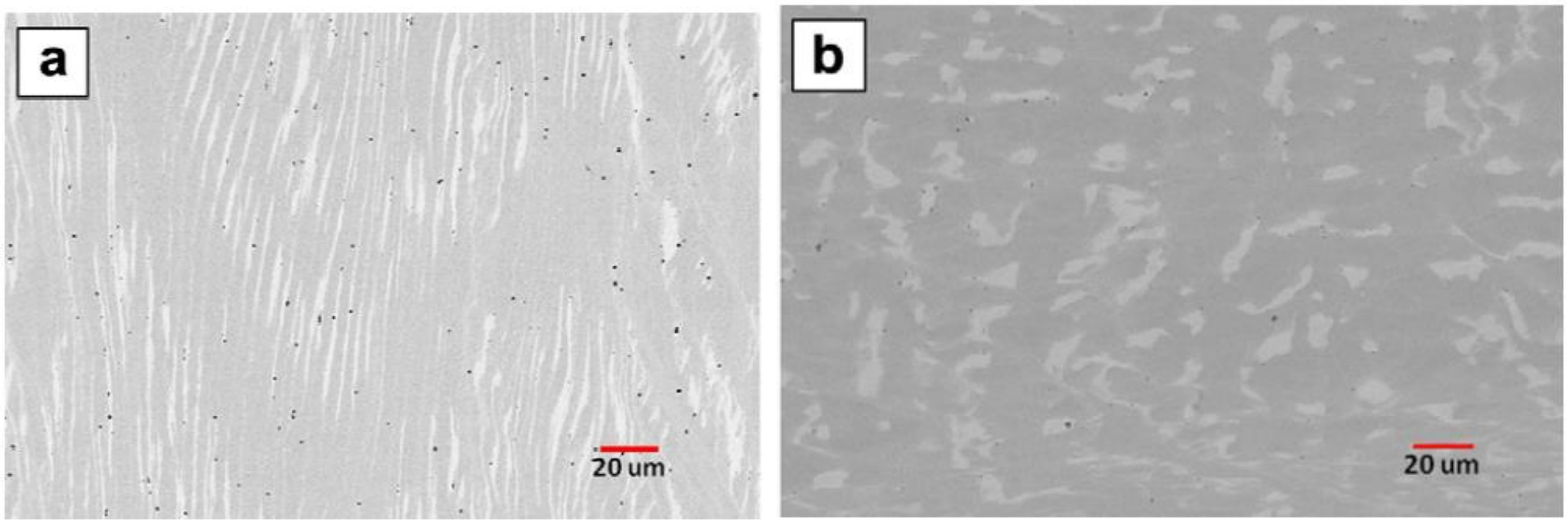

**Figure 4.1.3:** SEM images showing two different types of phase morphologies for the $Al_{0.5}CoCrCuFeNi$ HEA specimens subjected to fatigue testing: (a) Parallel morphology with a lamellar flow pattern of alternating $\alpha$ and $\beta$ phases. (b) Vertical morphology with random orientations of $\alpha$ and $\beta$ phases. The figure has been adapted from [35].

In order to account for correlations between the fatigue life and the parallel or the vertical morphology, a binary variable, $X_i$, can be introduced into Eq. (4.1.3), in order to indicate the type of the morphology for the $i$th experimental unit [35]. A general log-linear model describes the inference of stress and morphology on the Weibull scale parameter [35]:

$$\ln(\alpha_i) = \gamma_0 + \gamma_1 \log(S_i) + \gamma_2 X_i. \quad (4.1.12)$$

Here, we assume that $X_i = 0$ in case of the parallel morphology, but $X_i = 1$ in case of the vertical morphology [35].

For additional background information on general log-linear models, in context with reliability engineering (but not necessarily specific to HEAs), refer to the supplementary manuscript. Section 1.2.2 addresses the capability of the weak and the strong groups to elucidate the impact of defects on the fatigue life of the $Al_{0.5}CoCrCuFeNi$ HEA.

#### *4.1.4. Random endurance limit fatigue life models*

For background information on general log-linear models, in context with reliability engineering (but not specific to HEAs), refer to the supplementary manuscript.

### 4.2. Machine-learning models

In [164], the authors establish that higher UTS contributes to greater fatigue-endurance limits. Physics-based models can be introduced to facilitate prediction of the UTS or the fatigue resistance. In [164], the authors devise models capable of accounting for physics-based dependencies. The authors factor such dependencies into the models as a priori information [164].

The mathematical model presented {Eq. (4.79) of [164]} has the following structure:

$$\text{Endurance limit} = f[\text{UTS}, \text{process}, \text{defect (process)}, \text{grain (process)}, \text{microstructure (process)}, T. \ldots)] \quad (4.2.1)$$

where, according to Eq. (4.80) of [164]),

$$\text{UTS} = h[\text{composition}, \text{heat treatment process}, \text{defect level(process)}, \text{grain size}, T]. \quad (4.2.2)$$

Here, defect (process), grain (process), and microstructure (process) are taken to represent the defect level, the distribution in the grain size, and the microstructure, resulting from the application of a given heat-treatment process, respectively [180]. The term, microstructure (process), may be interpreted such as to broadly account for microstructures, both at nano- and micro-scales, as well as the phase properties [180]. The term, "process," broadly captures both manufacturing processes and post-processing [180]. Similarly, "defects" are defined broadly, such as to include inhomogeneities, impurities, dislocations, or unwanted features [180]. In case of $TiAl_6V_4$, the fatigue properties may further depend on the application of hot iso-static pressing (HIP), which – again through a broad interpretation - may be captured as a part of the heat-treatment process, and on the type of machining applied, which may be broadly combined with the defect level (process) [181].
Figure 4.2.1 presents a graphical overview of the sources impacting the fatigue properties of HEAs, including inter-dependencies. Figure 4.2.2 outlines an overall methodology for predicting the fatigue-endurance limits of HEAs from the contributing sources [164, 165].

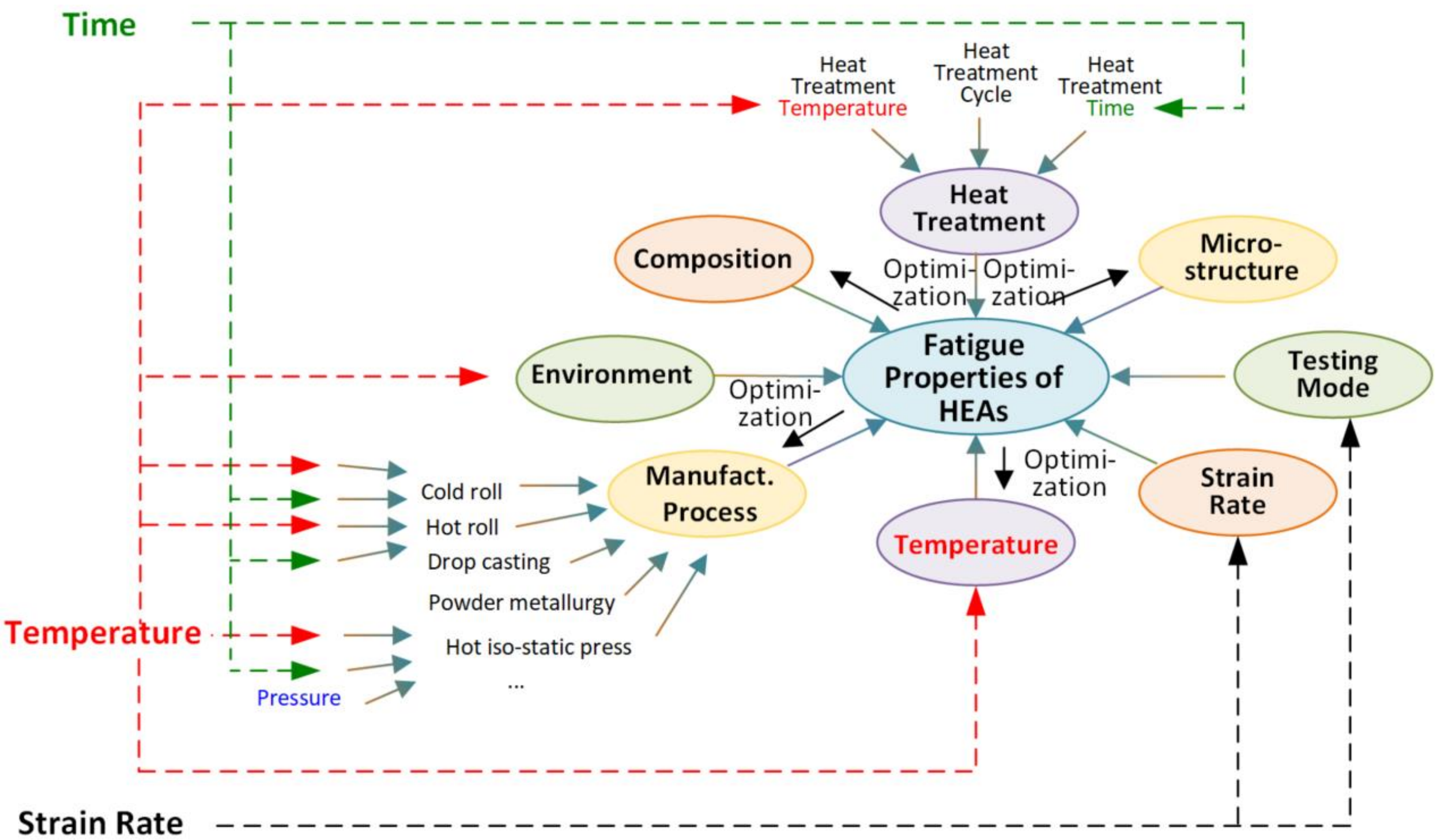

**Figure 4.2.1:** Overview of the sources impacting the fatigue properties of HEAs and their inter-dependence, e.g., as is the case for the temperature, relative to Figure 4.1.1 (adapted from [164, 165]).

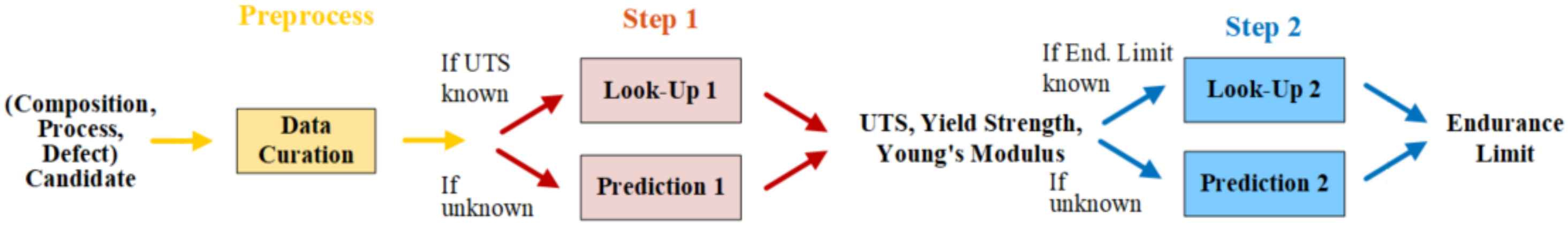

**Figure 4.2.2:** Overall methodology for deriving the output quantity of interest (the endurance limit) from the input sources (compositions, processes, and defects) [164, 165].

## 4.3. Other non-stochastic models

### *4.3.1. Macroscopic models*

In [44, 52], Chen et al. applied classical beam theory to calculate the maximum stress. $\sigma$, on a tensile surface within the span of two outer pins for TiZrNbHfTa samples subjected to the four-point bending fatigue experiment. In [17], Guennec et al. similarly applied the classical beam theory to four-point bending tests and single-edge notch-bending tests employed for the analysis of the fatigue-crack-growth mechanisms in an equimolar BCC HfNbTaTiZr HEA.

The classic phenomenological model by Laird and Smith [182] prescribes that fatigue-crack growth in stage II is accomplished through a repetitive process of plastic blunting and sharpening of crack tips under the action of alternate tensile and compressive loads [44, 52]. Large plasticity over a wide range of strain levels can retard both crack initiation and crack propagation. In [44, 52], Chen et al. describe an intrinsic toughening mechanism affecting the HfNbTaTiZr HEA, a mechanism that involves large-scale dislocation activities, which remain active irrespective of crack size and location. In [52], Chen et al. present similar descriptions in the context of stress-controlled fatigue deformation and fracture mechanisms for the HfNbTaTiZr HEA.

Finite-element modeling (FEM) enables modeling in smaller parts and assembles these smaller parts into a larger mesh of the objects and equations that model the entire system. In [108], Lam et al. conducted FEM simulation with the ANSYS software [ANSYS 8.0 Manual Set, ANSYS Inc., Canonsburg, PA (1998)], [ANSYS Theory Reference, Seventh Edition, Swanson Analysis Systems (1998)], [ANSYS – Engineering Analysis System. Theoretical Manual (for ANSYS Revision 8.04), Swanson Analysis Systems (1998)] to analyze the distribution of the principal stress and shear stress in the vicinity of the crack tip of CoCrFeMnNi test specimens upon unloading at the crack length of 16 mm under constant- and tensile overloaded-fatigue conditions. The CoCrFeMnNi specimens simulated complied with the ASTM Standard E647-99 [97, 108]. The configuration of the CoCrFeMnNi specimens assumed isotropic elasticity/plasticity, which enabled the occurrence of large plastic deformations by taking geometric nonlinearity into account [108].

### *4.3.2. Mesoscopic or microscopic models*

In 2020, Gao et al. took a step towards applying artificial neural networks (ANNs) used in conjunction with a crystal-plasticity finite-element method (CPFEM), to the NiCoCrFe HEA system [183]. In a fashion analogous to Ali et al. [184], Gao et al. employed the CPFEM method, based on their experimental data and physical mechanisms, to provide the data set for training the ANNs. Here, CPFEM provides a huge amount of data for the ANN models, compared to the experimental data available. The ANN models of Gao et al. were first trained with benchmark data and then validated through experiments and physics simulations. With the physics-based CPFEM simulations used in conjunction with the ANNs, Gao et al. demonstrated the benefits of ANNs (or machine learning) in investigating the plasticity of the NiCoCrFe HEA [183].

In 2021, Feng et al. reported on an $Al_{0.5}CoCrFeNi$ HEA with the enhanced fatigue life through ductile-transformable multicomponent B2 precipitates [25]. The cyclic-deformation mechanisms were revealed through real-time in-situ neutron diffraction, transmission-electron microscopy, crystal-plasticity modeling, and Monte-Carlo simulations. An Elastic-Visco-Plastic Self-Consistent (EVPSC) model incorporated with the martensitic transformation was able to predict the internal strain evolution, macroscopic stress–strain curves, and phase-transformation behavior quite well both under monotonic and cyclic deformations [25]. For detailed description of the EVPSC model, refer to [185]. In the EVPSC model, the polycrystalline aggregate is regarded as a homogeneous effective medium (HEM) and each grain is treated as an ellipsoidal inclusion [25]. The Eshelby's solution is an analytical method that can be used to derive the elastic field within and around an ellipsoidal inclusion embedded in a matrix. Through the analysis of linear elasticity, Eshelby derived the elastic field inside the elliptic inclusion using the biharmonic potential and Green's tensor [186]. More specifically, the Eshelby theory predicts that the dependence of the displacement field, $u(r)$, on the distance away from an inclusion, $r$, is given by [186, 187]

$$u(r) \propto \frac{1}{8\pi}(1-\upsilon)r^2, \tag{4.3.1}$$

where $\nu$ represents the Poisson's ratio. As applied to [25], the Eshelby's solution describes their interaction between each grain and the HEM. The EVPSC model simultaneously solves a series of nonlinear differential equations consisting of the elastic-visco-plastic single-crystal constitutive relations, the Eshelby's solution for the interaction between each grain and the HEM as well as of a self-consistency criterion [25].

In terms of specific contributions to single-crystal CoCrFeMnNi HEAs, Qi et al. revealed a linear relationship between the square root of the dislocation density and the alloy hardness, by analyzing the evolution of the dislocation density and hardness, in good agreement with the classic theory of Taylor hardening [188]. Conventional continuum mechanics work-hardening theory, i.e., the Taylor hardening model, was employed for the exploration of the effect of crystallographic orientation on the mechanics properties of the CoCrFeMnNi HEA material during nanoindentation [188].

In [189], Li et al. applied the density-functional theory (DFT) and continuum mechanics to systematically investigate the competition between the cleavage decohesion at a crack tip and dislocation emission from a crack tip in the body-centered-cubic refractory HEAs of HfNbTiZr, MoNbTaVW, MoNbTaW, MoNbTiV, and NbTiVZr. Such a crack-tip competition was evaluated for tensile loading and an entirety of 15 crack configurations and slip systems. They presented continuum analysis that assumed a sharp crack tip. Their theoretical predictions utilized analytical solutions from the Rice theory [190]. In [189], Li et al. applied the linear-elastic fracture mechanics for anisotropic media to determine the displacement and stress fields (in plain strain) for a sharp crack in an infinite medium. In their analysis, they introduced the unstable stacking fault energy as a key material parameter measuring the resistance to dislocation nucleation. Their results predicted that dislocation plasticity at the crack tip was generally unfavorable, for all 15 crack configurations and slip system considered, although borderline cases ($K_{Ie} \approx K_{Ic}$, with $K_{Ie}$ representing the critical stress intensity factor for dislocation emission under pure mode-I loading and $K_{Ic}$ denoting the critical stress intensity factor for crack propagation under pure mode-I loading) existed for HfNbTiZr and MoNbTaVW. This process pointed towards intrinsic brittleness and low crack-tip fracture toughness in these five HEAs (HfNbTiZr, MoNbTaVW, MoNbTaW, MoNbTiV, and NbTiVZr) at zero temperature [189].

In [191], Tong et al. experimentally investigated the continuum assumption for the local strain field induced by a relatively large size mismatch in Mn-and Pd-HEAs free from the external perturbation through an experimental approach. Figure 1.2.5 of [191] compares the local strain field for the Mn-HEA and the Pd-HEA with the Eshelby theory. The authors conclude that the strain field caused by atomic-size mismatch scales in proportion with $1/r^2$, with $r$ representing the distance away from an inclusion, in accordance with Eq. (4.3.1). The authors, further, conclude that the local lattice distortion complies with the assumptions of continuum mechanics.

In [192], Xiao et al. have studied the homogeneous and heterogeneous nucleation of dislocations in a representative FCC HEA, CoNiCrFeMn, by combining molecular dynamics simulations and continuum mechanics modeling. The dislocation theory based on continuum mechanics is here applied to explore the strain rate, temperature, and diameter-dependent yield strength of the FCC CoNiCrFeMn HEA nanowires, by directly fitting the activation energy and athermal stress for the surface dislocation nucleation from the atomistic simulations, as further explained in the section titled Modeling of Dislocation Dynamics in the supplementary manuscript.

### *4.3.3. Nanoscopic models*

Material degradation, due to fatigue, during crack initiation and growth, is a multi-scale process, where atomic bonds rupture, defects, and dislocations nucleate, and build up from the microscopic to the macroscopic scale, leading ultimately to a macroscopic fracture. Starting with processes at the atomic scale, the toughness and overall fracture behavior of materials, therefore, can be affected. Reference [193] presents a nice review of atomistic aspects of fracture, largely aimed at brittle crystalline materials, which typically cleave to produce atomically sharp surfaces at low fracture energies and in the absence of defects [194]. Fracture processes, and in particular brittle fracture processes, are clear cases, where the macroscopic properties of materials are largely determined by events at the atomistic scale. In the context of Ref. [193], atomistic-simulation methods are taken to include large-scale molecular dynamics simulations with classical potentials, density functional theory calculations, and advanced concurrent multiscale methods. Such methods have led to new insights, e.g., on the role of bond trapping, dynamic effects, crack/microstructure interactions, and chemical aspects on the fracture toughness and crack-propagation patterns in metals and ceramics [193].

With the exception of [195], not much has been reported, regarding the atomistic mechanisms of dislocation nucleation in HEAs, despite the importance of atomistic mechanisms in determining the mechanical behavior in HEA materials [192], In [195], Patriarca et al. studied slip nucleation in single-crystal FeNiCoCrMn HEAs, and presented an advanced atomistic modified Peierls–Nabarro modeling formalism. Patriarca et al. undertook atomistic DFT calculations to obtain a lattice constant, $a_0$, the shear modulus, $G$, the slip stress, $t$, and important energy terms, such as the unstable stacking fault energy, $\gamma_{us}$, and the intrinsic stacking fault energy, $\gamma_{isf}$, associated with a generalized stacking fault energy (GSFE) curve. The GSFE curve captures free-energy differences between a crystal fault and the bulk lattice with various degrees of shear displacements [195, 196]. The value of $\gamma_{us}$, represents the fault energy per unit area required to nucleate a slip. The $\gamma_{isf}$ is equivalent to the differential of the HCP and FCC free energy per unit area [195]. By utilizing a well-developed modified Peierls–Nabarro formalism for cubic metals [197-202], Patricia et al. calculated the critically resolved shear stress theoretically, producing close agreement with the experimentally observed value [195]. Patriarca et al. determined the equilibrium lattice constant of FeNiCoCrMn as $a_0$ = 3.59 Å, using the first-principle-based DFT calculations [195]. This step became a starting point for simulations of slip nucleation in the single-crystal FeNiCoCrMn HEA [195]. The DFT calculations were carried out, using the Vienna Ab initio Simulation Package (VASP) software package with a Projector Augmented Wave method and a Generalized Gradient Approximation [195].

Table 4.3.1 and Table 4.3.2 compare the estimates for the intrinsic stacking fault energy, $\gamma_{isf}$, and the critical resolved shear stress (CRSS) for FeNiCoCrMn, obtained from the DFT calculations to the corresponding experimental values [195]. Patriarca et al. noticed, after many simulations, that the presence of Co atoms was favored near the stacking fault and reduced the intrinsic stacking fault energy by almost 55% (lowered from 38 mJm$^{-2}$ to 17 mJm$^{-2}$). For that reason, the stacking-fault energies presented in Table 4.3.1 accurately reflect the lowest energy values possible for the {111}<110> slip system [195].

**Table 4.3.1:** The equilibrium lattice constant, $a_0$, the {111}<110> shear modulus, the unstable and intrinsic stacking fault energies, $\gamma_{us}$ and $\gamma_{isf}$, for the FeNiCoCrMn estimated from DFT and comparison to an experimental value for $\gamma_{isf}$ [195].

| $a_0$ [Å] | Shear modulus [GPa] | $\gamma_{us}$ [mJm$^{-2}$] | $\gamma_{isf}$ [mJm$^{-2}$] | $\gamma_{isf}$ [mJm$^{-2}$] |
|---|---|---|---|---|
| DFT calculations [195] | | | | Experimental [203] |
| 3.59 | 84 | 192 | 17 | 20 – 25 |

**Table 4.3.2:** Comparison of the DFT estimate for the CRSS for slip, $\tau$, for the FeNiCoCrMn to the experimental value [195].

| Temperature [K] | Critical resolved shear stress for slip $\tau$ [MPa] | |
|---|---|---|
| | Experimental | DFT calculations [195] |
| 77 | 175 ± 5 | 178 |

Further towards the application of the ab initio DFT to the analysis of fatigue properties of HEAs, Li et al. have applied DFT simulations and continuum mechanics analysis to systematically investigate the competition between the cleavage decohesion and dislocation emission from a crack tip in five, previously synthesized, equi-molar BCC HEAs, HfNbTiZr, MoNbTaVW, MoNbTaW, and NbTiVZr, under tensile loading and for 15 crack configurations and slip systems [189]. Li et al. present the systematic theoretical assessment of these BCC refractory HEAs by studying the competition between the dislocation emission and cleavage at the crack tip. Experimental fracture results by Zou et al. from 2017 [204] for pre-cracked, small-sized, single-crystal MoNbTaW specimens motivated the theoretical assessment by Li et al. These experimental results suggested a quasi-cleavage fracture mode with a limited amount of crack-tip plasticity. The theoretical predictions by Li et al. account for analytical solutions from Rice theory [190], [205] for the nucleation of two-dimensional dislocation geometries. The theoretical analysis by Li et al. draws upon continuum elasticity to determine the stress field for a crack in an infinite, anisotropic linear-elastic medium, similar to the previous work for Mg [206].
Important to efforts to develop or employ nano-scale modeling methods to predict meso- or macro-scale fatigue or fracture behavior, Bitzek et al. noted that many otherwise reliable empirical potentials fail to correctly describe the fracture behavior of specific materials, particularly brittle semiconductors and ceramics [193], [207], [208]. This trend could be attributed, in part, to the atomic environment near a crack tip deviating strongly from the equilibrium-bonding situation [193]. Nano-scale ab initio methods like DFT may traditionally not be well suited for modeling of meso- or macro-scale systems like the crack tip. According to [193], atomistic models, which are able to model such systems, may deploy

1. Potentials constructed automatically from quantum mechanical calculations, either by coarse-graining the electronic structure to produce bond order potentials [209, 210] or approximating the feature using machine learning (ML) [211].
2. Screened potentials, which recover the correct bond-breaking behavior by extending the interaction range, as proposed by Pastewka et al. [212, 213].
3. Novel multi-scale methods that combine DFT-calculations with classical potentials [214].

Further towards such an end, one can use ML methodologies, along with the previous data from the literature, to reduce the configurational and compositional space, for material systems and surface conditions analyzed using DFT, but without speeding up the core DFT calculations themselves. Hence, one can reduce the number of DFT calculations needed, or simulate larger material systems, for given computational resources. By configurational space, we are here referring to different structures and chemical environments, for a single chemical composition and a perfect crystal. This process may include different surface atoms exposed to various surface structures or terminations. The configurational space would increase, once defects were included. For a single HEA composition, subject to the external exposure, one can loosely define the configurational space as follows:

**X** = Configurational space $_{\text{HEA, perfect crystal}}$ = [ (parameters related to different structures and chemical environment), (parameters related to the type and strength of environmental exposure), (parameters characterizing different surface atoms exposure to various surface structures / terminations), (parameters related to defect properties) ]. (4.3.2)

The optimization problem addressed can be loosely formulated as follows:

$$\min_{\mathbf{X}} \ [\text{number_of_DFT_computations}(\mathbf{X})]. \quad (4.3.3)$$

Figure 4.3.1 offers the conceptual depiction of different options for combining ML with MD and DFT [215]. For the information on the use of ML for joint optimization of alloy properties, refer to [216]. For additional background information on nanoscopic models, in context with HEAs or fatigue properties, refer to the supplementary manuscript.

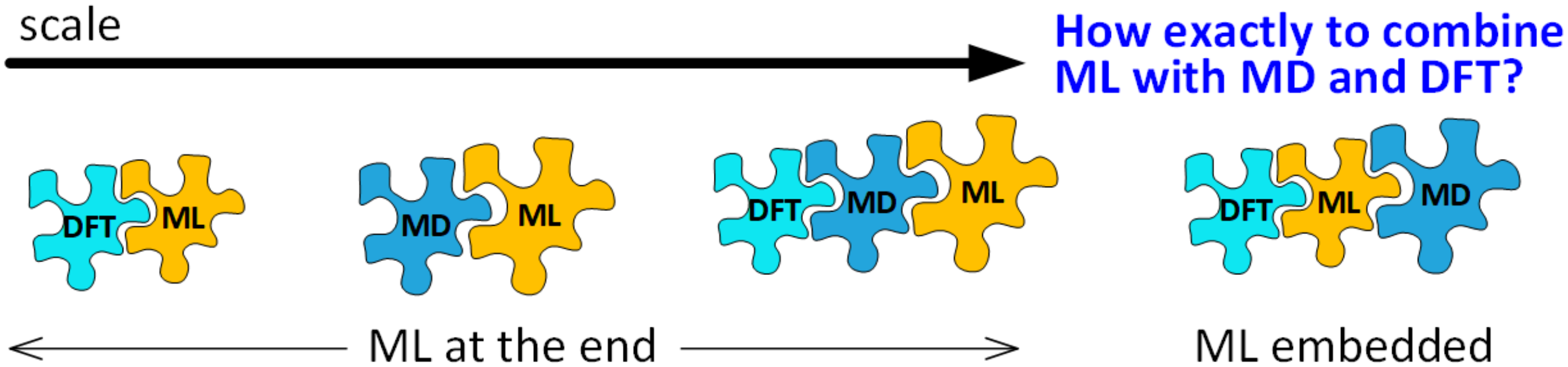

**Figure 4.3.1:** Conceptual depiction of different options for combining ML with MD and DFT (motivated by [215]).

# 5. Future Research

HEAs are currently one of the hottest fields of research in physical metallurgy. And fatigue is a crucial property that dictates the usability and applicability of materials in structural-engineering applications. Given these two actualities, fatigue of HEAs warrants thorough and extensive investigation. Nevertheless, according to Figure 1.4.1, this sub-area is way less explored than warranted, considering the rapid yearly growth of publications in the field of HEAs as a whole [217]. Besides, the vast majority of these investigations focus on the classical CoCrFeMnNi alloy, while other HEAs are not receiving much attention. A part of the reason for insufficient investigations of the fatigue behaviors of HEAs pertains to the fact that fatigue tests, either high-cycle fatigue tests, low-cycle fatigue tests, or fatigue-crack growth tests, are invariably time-consuming, costly and complex, when compared to other forms of mechanical tests.

We can see the promise of HEAs as structural materials with good fatigue resistance, even from the limited publications presently available. But the current findings are still far from enough to provide an all-round appraisal of the practicality of differently-structured HEAs in fatigue-critical applications. In addition, within the small pool of research publications presently available, the fatigue data are inevitably scattered due to the variations in microstructures and testing conditions, and some findings in fatigue mechanisms are contradictory to one other.

Given these limitations in investigations of the fatigue properties of HEAs to date, the following, broad directions are suggested for future research in this area:

1. Fatigue investigations warrant expansion well beyond the CoCrFeMnNi HEA, to many more HEAs with intriguing microstructures. Refractory HEAs (e.g., HfNbTaTiZr [52]), eutectic HEAs, and multiphase HEAs are interesting candidates for further explorations.
2. Fatigue tests should be accompanied by in-depth, delicate analysis of underlying mechanisms. Fatigue data obtained from tests are useful for us to determine if a material can fit into certain applications, while a reliable appreciation of fatigue mechanisms allows us to design new HEAs with microstructures more resistant to fatigue failures.
3. The present studies are mostly limited to in-air and room-temperature conditions. Considering that some HEAs (e.g., refractory HEAs) have great potential for high-temperature applications, future investigations of thermal fatigue or fatigue at elevated temperatures of these alloys will likely prove worthwhile.
4. It is fruitful to study fatigue mechanisms using in-situ techniques, such as neutron/synchrotron diffraction, electron microscopy, acoustic emission, ultrasonics, eddy currents, etc.
5. Some HEAs exhibit remarkable properties in regard to corrosion resistance [218-221] and may prove promising for applications in corrosive environments (e.g., bio-applications in human bodies). In view of this potential, it is also sensible to evaluate the corrosion fatigue of HEAs with potential applications in corrosive as well as bio environments.
6. Dedicated work concerning random-loading effects on HCF performance as well as the cyclic damage accrual behavior of HEAs at low and high strain amplitudes for LCF performance with hold times, especially at elevated temperatures, would be valuable.

Fatigue is usually the most time-consuming and expensive experiment among all mechanical tests. The generation of each single data point in either the $S - N$ plot, or the $\varepsilon - N$ plot, or the $\mathrm{d}a/\mathrm{d}N - \Delta K$ plot often requires days of testing. The enormously large compositional space of HEAs makes fatigue investigations even more frightening. Therefore, accelerating fatigue investigations becomes important when it comes to searching for fatigue-resistant HEAs. To facilitate the assessment of the fatigue properties of HEAs, the following four, general routes are suggested [120]:

1. Acquire insights on the promise of HEAs based on their uniaxial mechanical properties (e.g., strength and ductility) to narrow down the search space;
2. Take advantage of the classic models to complement time-consuming experiments;
3. Devise and utilize high-efficiency or high-throughput fatigue tests;
4. Proactively tailor microstructures to pinpoint fatigue-resistant alloys.

A schematic showing the design strategy for developing advanced fatigue-resistant HEAs is provided in Figure 5.1.1. The themes in Figure 5.1.1 resonate with the high-level observations from Table 1.3.1.

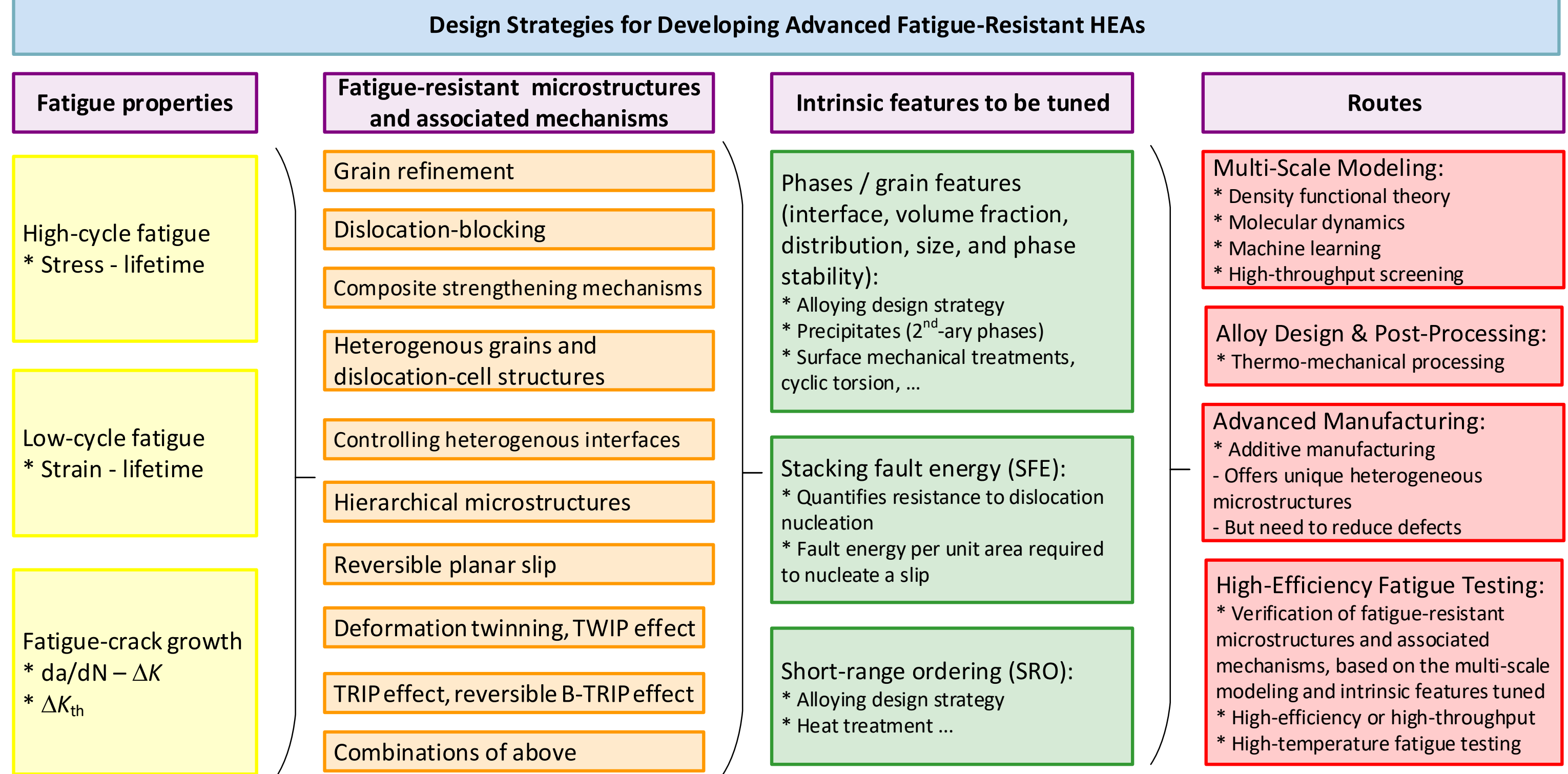

**Figure 5.1.1:** Towards design strategies for developing advanced fatigue-resistant HEAs (extended from [19]).

More specifically,

1. There is currently lack of studies on taking advantage of heterogeneous grains and dislocation cells, to enhance HCF resistance of HEAs through use of surface mechanical treatments (SMTs) or cyclic torsion, despite their efficacy in improving yield strengths of HEAs [19, 222, 223].
2. More careful experiments may be needed in the future to validate results obtained for Al-CrFeNi-based HEAs ($AlCrFeNi_2Cu$ and $Al_{0.2}CrFeNiTi_{0.2}$). The as-cast multiphase Al-CrFeNi-based HEAs outperform other materials, in terms of the fatigue-crack-growth resistance [31]. Nevertheless, very high fatigue thresholds, with $\Delta K_{th}$ values above 20 $MPa{\cdot}m^{1/2}$, came across as somewhat doubtful. Such results may be caused by their very coarse as-cast structures, the small size of the specimens tested, and the correspondingly excessive plasticity throughout their cross-section, but may need to be further validated [19, 30, 32].
3. While the unique heterogeneous microstructures generated by AM have great potential in improving HCF resistance, reducing AM-induced defects as much as possible is crucial. Extraordinary HCF resistance of the hetero-structured in-situ formed oxides reinforced

CoCrFeMnNi HEA by Kim et al. is mainly attributed to the unique heterogeneous microstructure in the AM-built samples, including heterogeneous grain structures, dislocation networks, and in-situ formed oxides, in addition to fatigue deformation induced deformation twinning [19, 21].

4. Motivated by the work of Kim et al., the design of a favorable microstructural hierarchy that can effectively enhance fatigue-crack initiation and propagation is a ripe area for future research on HEAs with exceptional fatigue resistance [19, 21]. Microstructural hierarchy can be gained by the alloy-design strategy, thermo-mechanical processing, surface mechanical treatments, cyclic torsion, additive manufacturing, etc. Tuning the interface structure, controlling the volume fraction of phases, along with their distribution and size, tailoring the phase stability, and engineering the grain boundary characteristics so that multiple micro-mechanisms that resist crack initiation and propagation can be activated simultaneous comprise attractive routes towards improving the fatigue performance of HEAs.
5. As noted above, a dedicated study is recommended to explicitly observe if the accumulation of localized damage accrual from cyclic plastic deformation is the primary contributor or detractor of the presence of strain-hardening phenomena within HEAs.
6. Also as noted above, very little work has been done as of yet to explicitly measure the effects of a change in grain size within HEAs to the crack propagation pattern, which are indicative of the localized length scale of cyclic damage accrual due to fatigue.
7. Hence, a close cross-examination is recommended of how the length scale of the cyclic localized damage accrual affects crack initiation and propagation as a result of fatigue in HEAs. It is hypothesized by the authors of this review that the length scale of the cyclic damage accrual has a major influence on the crack initiation and propagation pattern, which would be further complicated by a heterogenous energetic landscape throughout the solid solution of a given HEA.

## 6. Conclusions

For metallic structures in service, fatigue properties are of central importance. Yet, fatigue might be the most complex property among the various mechanical behavior of metallic materials, and normally requires a greater degree of efforts to evaluate. This statement especially holds true for HEAs, an emerging family of metallic materials. As the lab research of HEAs slowly progresses towards potential industrial applications, the assessment of the fatigue behavior of the HEAs gradually comes into focus and grows in importance. Ever since the first HEA fatigue publication in 2012, there have been tens of publications devoted to the study of the fatigue characteristics of HEAs from different perspectives. Broadly speaking, the fatigue studies of HEAs fall into three categories, namely, low-cycle fatigue, high-cycle fatigue, and fatigue-crack growth. These three types of fatigue behavior have all been reviewed in this paper.

Among the various investigative techniques, experimentation is still the dominant technique for investigating the fatigue behavior of HEAs, as well as that of many other structural metals. Through experiments, the low-cycle fatigue, high-cycle fatigue, and fatigue-crack growth of various structural HEAs have been studied, and their respective fatigue mechanisms disclosed. In addition, the effects of grain sizes, microstructural defects (e.g., dislocation and twin), phase structures and transitions on the fatigue behaviors of HEAs have been examined to varying degrees of detail.

Besides, a range of analytical and numerical computational techniques have been employed to predict certain fatigue behaviors and to deepen mechanistic understanding. These computer simulations cover stochastic models, ML models, empirical models, and microstructure-based models. The theoretical simulations have played a critical, complementary role in portraying the full picture of the fatigue behavior of HEAs.

From the literature survey, the following conclusions have been drawn:

1. Four types of HEAs have been studied, namely FCC, BCC, multiphase, and metastable structures. Among these, the research on the FCC HEA is the most comprehensive. Among these four types of HEAs, metastable and multiphase HEAs have exhibited excellent performance in all aspects and can be primarily considered for industrial applications in the future. In terms of the HCF, the fatigue behavior of BCC and metastable HEA seems superior. In terms of the LCF, the performance seems similar for the FCC, multiphase, and metastable HEAs. In terms of resistance to crack growth, metastable and multiphase HEAs tend to perform better, while BCC HEAs seem to offer insufficient fatigue resistance.
2. We note there are no LCF results reported in Table 2.2.1 on BCC HEAs, presumably because the LCF performance of BCC HEAs is inferior to that of FCC or multi-phase FCC with precipitates HEAs.
3. From a series of studies comparing CoCrFeMnNi HEA systems, we have come to learn about many microstructures that affect the fatigue mechanism of the CoCrFeMnNi HEA. The extremely fine grains obtained by processing help improve both the high-cycle and low-cycle fatigue resistance of the CoCrFeMnNi HEA, as indicated by Figure 2.1.4 - Figure 2.1.5 and Figure 3.2.1 - Figure 3.2.2. Grain-refined FCC HEAs can exhibit high-cycle fatigue resistance close to that of the metastable HEA, according to Figure 2.1.15, while some special microstructures, such as dislocation networks and low-angle grain boundaries, can also improve the high-cycle fatigue resistance, from the perspective of dislocation confinement, according to Figure 2.1.7. Grain refinement is more efficient for improving LCF resistance at low strain amplitudes, due to the enhanced fatigue crack-initiation resistance resulting from the grain refinement. The grain refinement does not work well at high strain amplitudes, because of accelerated fatigue-crack propagation along high-density grain boundaries. Furthermore, in Sections 2.1 and 2.2, we mentioned the manufacturing and processing methods that can furnish these special structures, which are instructive for the production of HEAs, with high fatigue resistance.
4. In Section 4, we summarize a range of mathematical models that may be used, from stochastic models (simple Weibull models, Weibull mixture models, general log-linear models and random endurance limit fatigue-life models), to machine-learning models and other mesoscale models. Considering the time and expense required for fatigue testing, effective prediction of fatigue behavior can greatly reduce the overall cost.
5. The impact of extreme environments on the fatigue behavior of HEAs deserves further study. The current investigation has demonstrated the influence of external factors, such as stress ratio and temperature, on the fatigue behavior. Especially with respect to the temperature, the present study shows that the low-cycle fatigue resistance of FCC HEAs tends to decrease with increasing temperature, according to Figure 3.3.1 - Figure 3.3.2, whereas the resistance of metastable HEAs to fatigue crack growth tends to increase with increasing temperature, according to Figure 3.3.8. It can be seen that the fatigue behavior of HEAs in extreme environments warrants further exploration.

6. In the high-level roadmap from the Introduction (Table 1.3.1), we noted that the outstanding LCF resistance of the multi-component B2 precipitate-strengthened $Al_{0.5}$CoCrFeNi HEA at low plastic strain amplitudes ($< 10^{-3}$), as reflected by its smaller slope of plastic-strain amplitude versus 2 $N_f$ at low plastic-strain amplitudes, demonstrates that the HEA-design concept provides a feasible way to design fatigue-resistant structural materials. At the plastic strain amplitude of ~0.03%, the multi-component B2 precipitate-strengthened $Al_{0.5}$CoCrFeNi HEA offers at least four times longer fatigue life than conventional alloys, as noted in Figure 3.5.1.
7. Consistent with the high-level roadmap from the Introduction (Table 1.3.1), the combination of UFG and TRIP effects suggests a reliable route to design HEAs with excellent fatigue resistance.

## 7. Acknowledgements

XF and PKL very much appreciate the support of (1) the National Science Foundation (DMR-1611180, 1809640, and 2226508), (2) the Army Research Office Project (W911NF-13-1-0438 and W911NF-19-2-0049), (3) the Department of Energy (DOE DE-EE0011185), and (4) the Air Force Office of Scientific Research (AF AFOSR-FA9550-23-1-0503). XF and PKL also appreciate the support from the Bunch Fellowship. XF and PKL would like to acknowledge funding from the State of Tennessee and Tennessee Higher Education Commission (THEC) through their support of the Center for Materials Processing (CMP). BS very much appreciates the support from the National Science Foundation (IIP-1447395 and IIP-1632408) with the program directors, Dr. G. Larsen and R. Mehta, from the U.S. Air Force (FA864921P0754), with J. Evans as the program manager, and from the U.S. Navy (N6833521C0420), with Drs. D. Shifler and J. Wolk as the program managers.
The authors also want to thank Dr. T. Yuan for helping us complete the identification of the references addressing the random endurance limit fatigue life model to high-entropy alloys. In addition, B.S. thanks Dr. L. Arnadottir for assistance with overall planning and coordination.

## 8. Author Contributions

X.F. was primarily responsible for the sections on microstructural effects (3.2). H.S. was primarily responsible for the sections on strain effects (3.5) and frequency effects (3.7). B.S. was primarily responsible for the sections on BCC HEA (2.1.4), on the cumulative damage (2.2.1 and 2.3.1), on the stochastic models (4.1), on the ML models (4.2), other models (4.3) as well as for corresponding sections in the supplementary manuscript. B.S. also helped craft Section 1 on the Introduction, Section 5 on Future Research and Section 6 on Conclusion, such as to address overall storytelling aspects. W.L. was primarily responsible for the sections on theoretical modeling (2.1.1) and future directions (section 4). S.C. was primarily responsible for the other sections (1, 2.1, 2.2, 2.3 3.1, 3.3, 3.4 and section 6.). S.C. was also primarily responsible for the overall organization and revisions of the paper together with B.S. and P.K.L.

# Supplementary Notes on Theoretical Modeling

## 1. More on Stochastic Models – for Enabling Prediction of Fatigue Life

As for the dependence on the temperature, we expect that the fatigue-endurance limit generally to degrade with increasing temperature [1], similar to the alloy strength. However, there may be temperature ranges, where the increased temperature allows the grains to expand, decreases porosity, improves the alloy strength as well as the fatigue life. Liu et al. report that the fatigue life of the Al-12Si-CuNiMg alloy increases, and the microstructure improves with the temperature increase at the same strain amplitude [2]. At given loading temperatures and strain amplitudes, the microstructure can be refined [2]. But in [3], Polák et al. note that the number of cycles decreases with increasing temperature, in case of the LCF behavior of Sanicro25 steel at room and elevated temperatures.

Concerning the dependence on the strain rate, Equation 2.4.1 captures an empirical relationship between true stress and strain rate, a relationship that forms the basis for research on high strain rate applications [4]:

$$\sigma = K\,\dot{\varepsilon}^{m} \qquad (2.4.1)$$

Here, $m$ represents a strain-rate sensitivity factor, $\dot{\varepsilon}$ the strain rate, $K$ a material constant, and □ true stress.

Concerning the other factors, listed in Figure 4.1.1 of the main manuscript, refer to the Smith-Watson-Topper equivalency relation Equation 2.6.1.1 listed in Section 3.1, for additional analytical description of the stress ratio, $R$, which is defined as $R = \sigma_{min} / \sigma_{max}$, where $\sigma_{min}$ represents a minimum stress amplitude, applied during testing, but where $\sigma_{max}$ denotes a maximum stress amplitude. As for the dependence on the testing frequency, the authors of [5] demonstrate that increasing the frequency of the load applied from 20 Hz to 1,000 Hz generally results in lowering of the fatigue life.

The Weibull models comprise a family of versatile distributions, which are the most widely used lifetime distributions in reliability engineering, and have been employed to model fatigue behavior for a variety of materials. The Weibull mixture predictive models are typically used, when the shape of a measured or design load spectra varies considerably during service life (exhibits significant scatter) and therefore, frequently cannot be approximated by simple unimodal [(1)] distribution functions. Here, the shape of representative load spectra tends to be multi-modal [(2)] [7]. The general log-linear models are introduced for the purpose of describing the effect of stress and morphology on the Weibull scale parameter. The term, “morphology,” refers here to the “microstructural morphology,” also referred to as the “phase morphology” or “grain morphology,” i.e., to distributions of the grains/phases within the sample relative to a build or a rolling direction. Pascual and Meeker introduced the random endurance limit fatigue life models to describe first the dependence of fatigue life on stress levels but also scatter in S–N data [8]. The models feature a fatigue-limit parameter and non-constant standard deviation of log fatigue life and can be used to describe the curvature and non-constant variance in S-N relationships [8]. The authors note that randomness in

[(1)] A unimodal distribution in statistics is a distribution with only one clear peak or most frequent value [6].

[(2)] A multi-modal distribution in statistics is a distribution with more than one clear peak or most frequent value [6].

the endurance limit is in part caused by the location, orientation, size, and number of defects (e.g., cracks) in the material, which themselves are random.

1.1 More on Weibull Predictive Models

*1.1.1 More on the General Structure of a Two-Factor Weibull Distribution*

The Weibull distribution can approximate the normal or the lognormal distributions [7]. The Weibull distribution is identical to the exponential or the Rayleigh distribution, when $\beta = 1$ or 2, respectively [7]**.** Figure S1 presents Weibull distributions for representative values for $\alpha(S)$ and $\beta$. As noted in [9], favorable properties of the Weibull model include:

(a) the model can be written in terms of elementary functions; and

(b) the model exhibits a continuous univariate distribution supported on a semi-infinite or finite interval.

By virtue of the first property, the model is usually considered to be simple and fairly easy to use. But the second property allows the model to satisfy the pertinent boundary conditions[9].

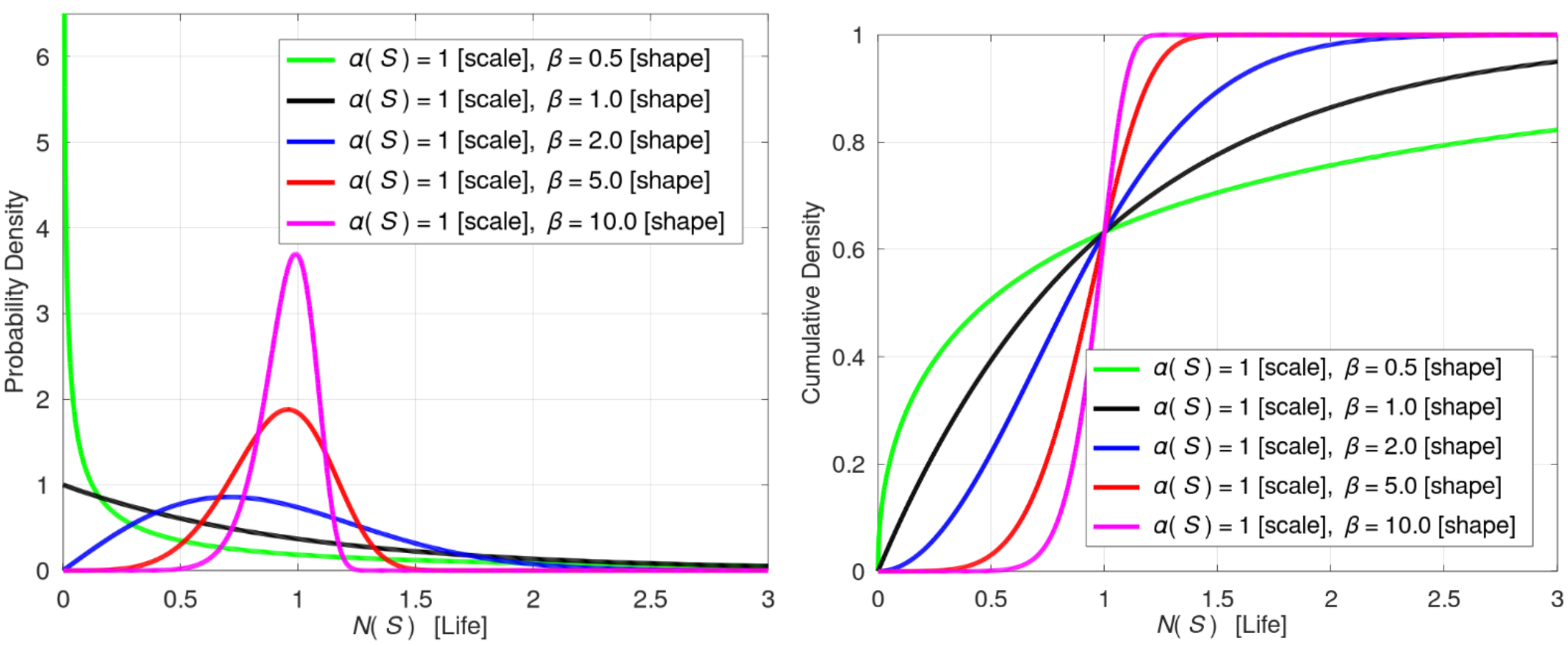

**Figure S1**: The probability density function (left) and the cumulative distribution function (right) of two-parameter Weibull distributions for representative values for $\alpha(S)$ and $\beta$.

*1.1.2 Standard Structure of a Two-Parameter Weibull Distribution*

The distributions presented in Figure S1 are considered two-parameter distributions. Similar to the general case, the pdf of the fatigue life, for a standard two-parameter Weibull distribution, can be represented as [10]

$$f(N;\lambda,\kappa) = \begin{cases} \dfrac{\kappa}{\lambda}\left(\dfrac{N}{\lambda}\right)^{\kappa-1} \exp\left[-\left(\dfrac{N}{\lambda}\right)^{\kappa}\right] & N \geq 0 \\ 0 & N < 0 \end{cases} \tag{2.4.1.5}$$

Here, $\kappa > 0$ denotes the *shape* parameter, and $\lambda > 0$, the *scale* parameter of the distribution. The corresponding cdf is given by [10]

$$F(N;\lambda,\kappa) = \begin{cases} 1 - \exp\left[-\left(\dfrac{N}{\lambda}\right)^{\kappa}\right] & N \geq 0 \\ 0 & N < 0 \end{cases} \tag{2.4.1.6}$$

$F(N;\lambda,\kappa)$ captures the probability that the fatigue life is equal to or less than $N$, given the shape parameter, $\kappa$, and the scale parameter, $\lambda$. With $N_c = 10^7$ cycles denoting the censor time, the probability of obtaining a censored observation at the stress level, $S$, can be expressed as

$$P(N \geq N_c) = 1 - F(N_c|\lambda,\kappa) = \exp\left[-\left(\frac{N}{\lambda}\right)^{\kappa}\right]. \tag{2.4.1.7}$$

*1.1.3 Three-Parameter Weibull Distributions*

Analogous to Eqs. (2.4.1.8) and (2.4.1.9), the pdf of the fatigue life, for a standard three-parameter Weibull distribution, can be written as

$$f(N;\lambda,\kappa,N_0) = \begin{cases} \frac{\kappa}{\lambda}\left(\frac{N-N_0}{\lambda}\right)^{\kappa-1} \exp\left[-\left(\frac{N-N_0}{\lambda}\right)^{\kappa}\right] & N \geq N_0 \\ 0 & N < N_0 \end{cases} \tag{2.4.1.8}$$

and the cdf as [9]

$$F(N;k,\lambda,N_0) = \begin{cases} 1 - \exp\left[-\left(\frac{N-N_0}{\lambda}\right)^{\kappa}\right] & N \geq N_0 \\ 0 & N < N_0 \end{cases} \tag{2.4.1.9}$$

As for the two-parameter case, $\kappa > 0$ denotes the shape parameter, and $\lambda > 0$ the scale parameter of the distribution. But $N_0$ is a new location parameter.

The authors of [9] employ a novel model based on the three-parameter Weibull model to describe the three-stage fatigue deformation behavior of plain and fiber-reinforced concrete. Alternatively, the two-factor model in Eqs. (4.1.1) – (4.1.3) from the main manuscript can be viewed as a three-parameter model with the unknown model parameters being $\beta$, $\gamma_0$, and $\gamma_1$.

*1.1.4 Maximum Likelihood Estimation of Parameters of a Generic Two-Factor Weibull Model*

Given $n$ observations, $\boldsymbol{N}(S) = [N_1(S), N_2(S), \ldots, N_n(S)]$, at a specific stress level, $S$, the maximum likelihood method estimates the mode parameters

$$\boldsymbol{\theta} = [\alpha(S),\beta] \tag{2.4.1.10}$$

by applying the principle [11]

$$\widehat{\boldsymbol{\theta}} = \underset{\boldsymbol{\theta}\in\boldsymbol{\Theta}}{\arg\max}\, L_n(\boldsymbol{\theta};\boldsymbol{N}(S)). \tag{2.4.1.11}$$

Here $L_n(\boldsymbol{\theta};\boldsymbol{N}(S)) \equiv L(\boldsymbol{\theta})$, referred to as the likelihood function, is obtained as [11]

$$L_n(\boldsymbol{\theta};\boldsymbol{N}(S)) = \boldsymbol{f}_n(\boldsymbol{N}(S);\boldsymbol{\theta}). \tag{2.4.1.12}$$

where $\boldsymbol{f}_n(\boldsymbol{N}(S);\boldsymbol{\theta})$ represents the joint probability density function from which $n$ observations comprising $\boldsymbol{N}(S)$ are drawn. In case of independent and identically distributed observations, $N_i(S)$, $\boldsymbol{f}_n(\boldsymbol{N}(S);\boldsymbol{\theta})$ can be computed as the product of probability density functions, $f(N_i(S)|\alpha(S),\beta)$:

$$\boldsymbol{f}_n(\boldsymbol{N}(S);\boldsymbol{\theta}) = \prod_{i=1}^{n} f(N_i(S)|\alpha(S),\beta)\ . \tag{2.4.1.13}$$

In other words,

$$\widehat{\boldsymbol{\theta}} = [\hat{\alpha}(S),\hat{\beta}] = \underset{\boldsymbol{\theta}\in\boldsymbol{\Theta}}{\arg\max}\, L_n(\boldsymbol{\theta};\boldsymbol{N}(S)) = \underset{\boldsymbol{\theta}=[\alpha(S),\beta]}{\arg\max} \prod_{i=1}^{n} f(N_i(S)|\alpha(S),\beta). \tag{2.4.1.14}$$

In case of the two-factor Weibull pdf in Eq. (4.1.1) of the main manuscript, we obtain

$$\widehat{\boldsymbol{\theta}} = [\hat{\alpha}(S),\hat{\beta}] = \underset{\boldsymbol{\theta}=[\alpha(S),\beta]}{\arg\max}\, L_n(\boldsymbol{\theta};\boldsymbol{N}(S)) = \underset{[\alpha(S),\beta]}{\arg\max} \prod_{i=1}^{n} \frac{\beta}{\alpha(S)}\left(\frac{N_i(S)}{\alpha(S)}\right)^{\beta-1} \exp\left(-\left(\frac{N_i(S)}{\alpha(S)}\right)^{\beta}\right). \tag{2.4.1.15}$$

The maximum likelihood estimator, $\widehat{\boldsymbol{\theta}}$, tends to possess favorable asymptotic (large-sample) properties [8]. In case of a "large" sample size, and for certain conditions on the fatigue-life distribution, the distribution of the maximum likelihood estimator, $\hat{\boldsymbol{\theta}}$, is approximately multivariate normal with the mean equal to the true values being estimated and standard deviation no larger than that of any alternative estimation technique considered[8].

*1.1.5 Formulation of a Weibull Regression Model*

The Weibull distribution can be associated with a common analytical expression for the S-N curve, given by [12]

$$N(S) = \begin{cases} c\,S^{-d} & S > S_0 \\ \infty & S \leq S_0 \end{cases} \tag{2.4.1.16}$$

Here, $S$ denotes the applied stress range, $S_0$ represents an endurance limit, $N(S)$ is the expected number of fatigue-life cycles at the stress level, $S$, but $c$ and $d$ represent positive material parameters. By applying the natural logarithm to both sides of Eq. (2.4.1.16), we obtain a relation resembling the one for the Weibull scale factor [listed in Eq. (4.1.1) of the main manuscript]:

$$\log(N(S)) = \gamma_0 + \gamma_1 \log(S), \quad S > S_0. \tag{2.4.1.17}$$

Here $\gamma_0 = \log(c)$ and $\gamma_1 = -d$ characterize the fatigue quality of the specimen under study. On a log-log chart, this model gives rise to a bilinear curve with the upper curve having a slope of $1/\gamma_1$ but with the lower curve being horizontal.

While the S-N relation, expressed through Eqs. (2.4.1.16) and (2.4.1.17), provides a simple way to relate the effect of a stress applied to the test item to the number of cycles to fatigue failure, *it does not capture the variability in the fatigue-lifespan data observed* [12, 13]. The Weibull regression model is obtained, by introducing an error term, $\varepsilon$, into Eq. (2.4.1.17), for purpose of accounting for such variability (inherent scatter):

$$\log(N(S)) = \gamma_0 + \gamma_1 \log(S) + \varepsilon, \quad S > S_0. \tag{2.4.1.18}$$

The fatigue–lifespan model in Eq. (2.4.1.18) clearly has the structure of a regression model, where $\log(N(S))$ is the dependent variable (output), where $\log(S)$ is an independent variable (input, also referred to as the covariate), and where $\varepsilon$ is an error term that follows a particular probability distribution [12, 14].

If the error term, $\varepsilon$, in Eq. (2.4.1.18) was assumed to follow a standardized normal distribution, then the fatigue life $N(S)$ at a given stress level, $S$, would follow a log–normal distribution[12]. A log-normal probability distribution can be expressed as

$$f(t \mid \mu, \sigma) = \frac{1}{t\,\sigma\sqrt{2\pi}} \exp\left[-\frac{1}{2}\left(\frac{\log(t)-\mu}{\sigma}\right)^2\right]. \tag{2.4.1.19}$$

But more commonly, it is assumed that the error term, $\varepsilon$, follows the standardized smallest extreme value distribution [12]. The association of the smallest extreme-value distribution to the Weibull distribution resembles that of the normal distribution to the log-normal distribution. The smallest extreme value distribution can be viewed as a log-Weibull distribution. In other words, if a random variable, $X$, conforms to a Weibull distribution, then $\log(X)$ will conform to the smallest extreme value distribution. In this sense, the smallest extreme value distribution can be regarded as a reparameterization of the Weibull distribution[15].

The Weibull regression model from Eq. (2.4.1.18) can be cast into the form of an equivalent Weibull-accelerated lifetime testing model, such as widely-used in reliability engineering and lifetime data analyses[12, 14]. When testing highly reliable components at normal stress levels, it may be difficult to obtain a reasonable amount of failure data in a relatively short period of time. For such reasons, accelerated lifetime tests are sometimes conducted at higher than expected stress levels. Such models, which predict failure rates at normal stress levels from test data obtained from items that fail at higher stress levels, are referred to as acceleration models. The fundamental assumption of the acceleration models is that the failure mechanism remains the same at the normal stress levels as well as at the higher stress levels[16].

*1.1.6 Formulation of S-N Curves Capturing P-Quantile Life*

In statistics, quantiles are defined as cut points that divide the range of a probability distribution into continuous intervals with probabilities consistent with the *p*-values specified. Figure S2 presents selected Weibull distributions from Figure S1 with the 0.05 and 0.95-quantile lives appended. Quantile life is defined mathematically in terms of the condition [8]

$$F(N_p(S)|\ \alpha(S),\beta) = p \qquad (2.4.1.20)$$

with $F(N_p(S)\mid\alpha(S),\beta)$ representing a cdf, such as in Eq. (2.4.1.6).

Figure S3 provides further graphical insight into S-N curves accounting for the *p*-quantile life. Here, each curve represents a constant probability of failure, *p*, as a function of $S$ [8]. Note that the fatigue life at a given stress level is considered to be stochastic in nature (subjected to a probability distribution). But this trend gives rise to the *p*-quantile S-N curves observed. Collins listed factors, such as the material composition, grain size, and direction, heat treatment, and surface conditions, affecting the *p*-quantile S-N curves [17].

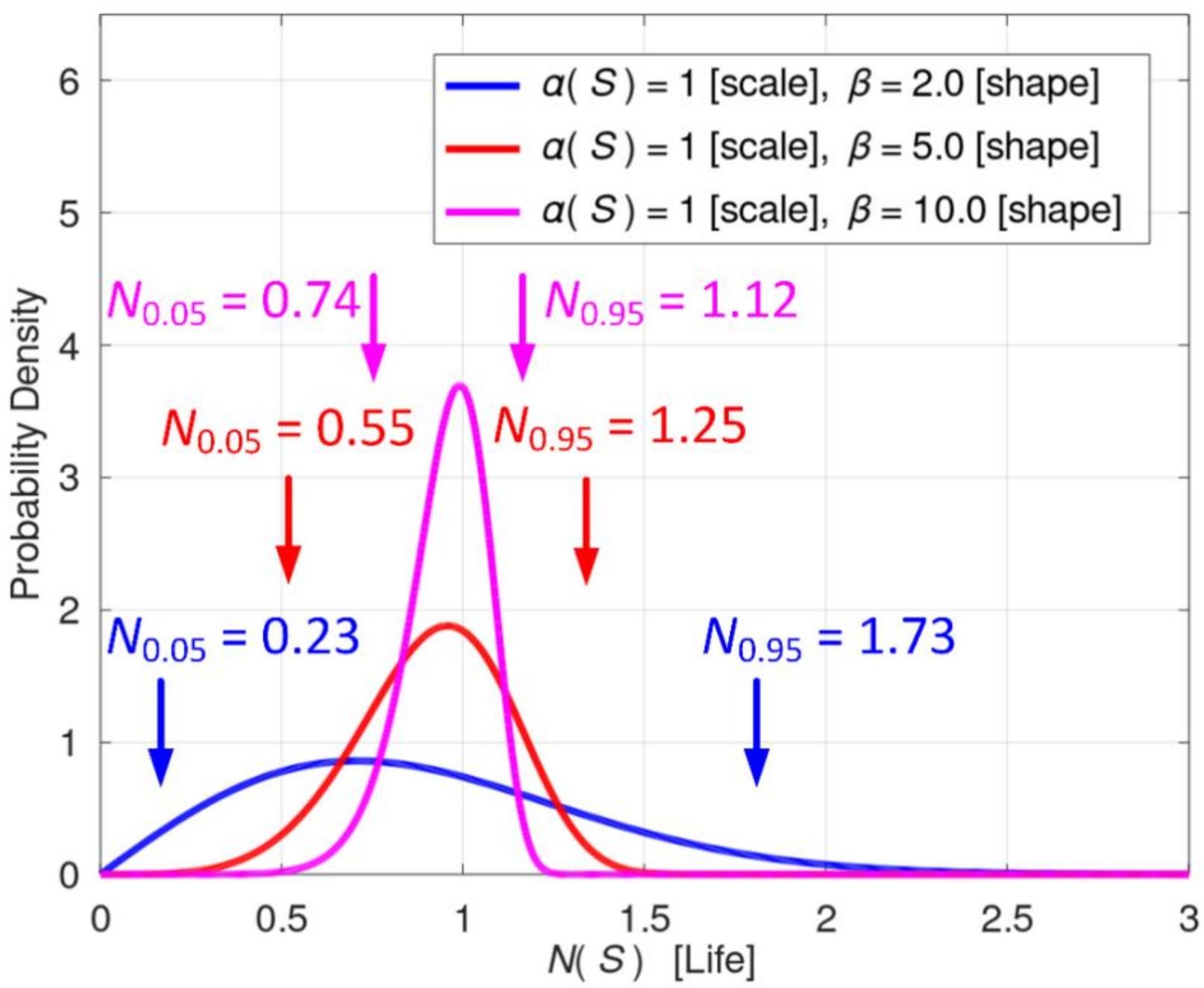

**Figure S2**: Selected Weibull distributions from Figure S1 with the 0.05 and 0.095-quantile lives appended.

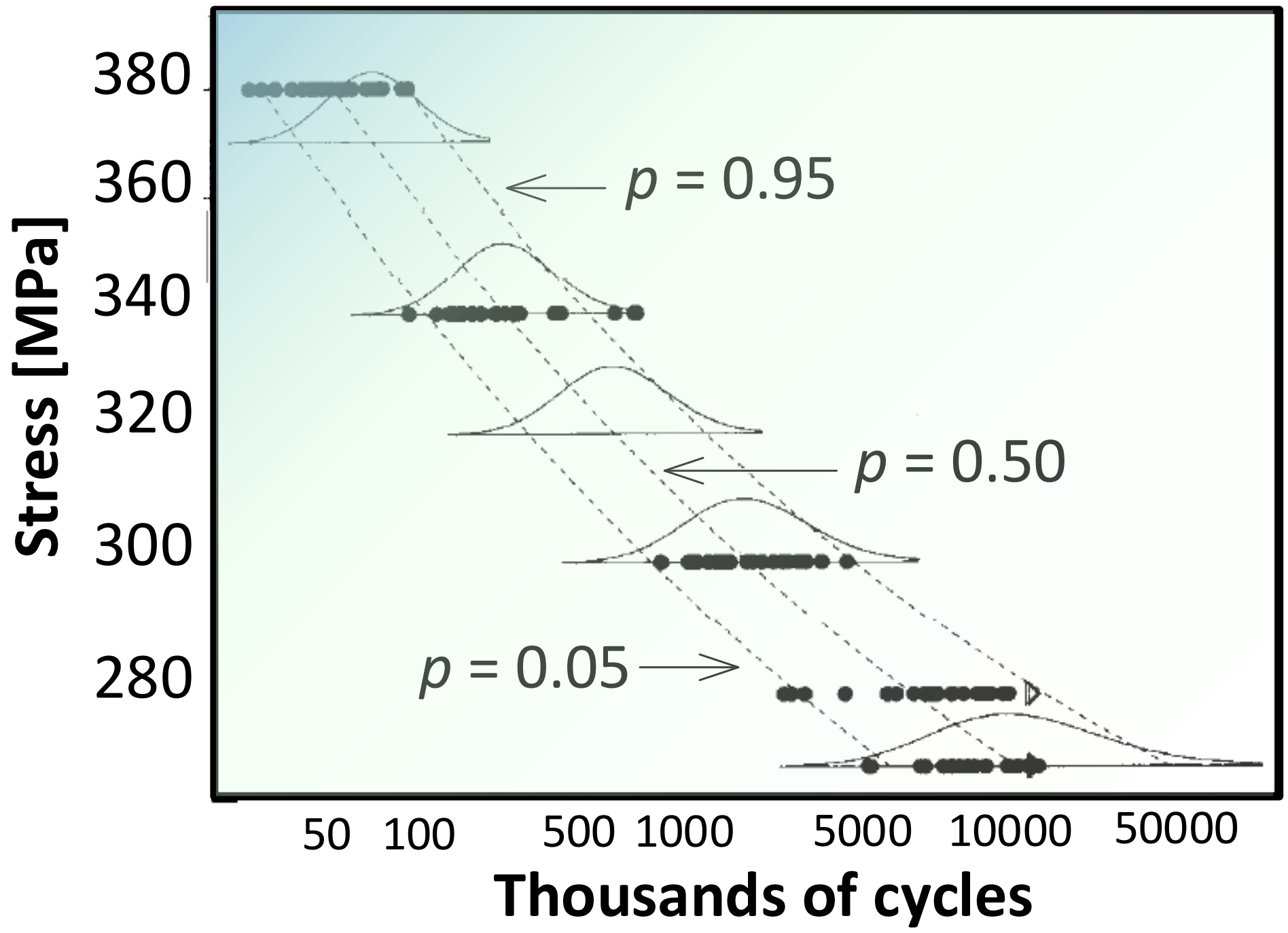

**Figure S3**: S-N plot of laminate panel data with probability density functions estimated, using the maximum likelihood method and $p = 0.05$, 0.50, and 0.95 quantile estimates added (adapted from Figure 1.2.8 of [8]).

When accounting for censoring, we represent the observed fatigue–lifespan data as $\{(N_i, S_i, \delta_i), i = 1, 2, \ldots, n\}$[18], where $n$ denotes – as before - the total number of samples tested, but $N_i$ and $S_i$ the fatigue–lifespan cycles and the stress applied to the $i$ th sample, respectively [12]. The binary indicator, $\delta_i$, is defined as.

$$\delta_i = \begin{cases} 1 & \text{in case sample failure is observed} \\ 0 & \text{in case of censored observation (in case that the sample failure is not obseredrved)} \end{cases} \quad (2.4.1.21)$$

and is given (not estimated). In the literature on reliability and survival analysis, this type of censoring is referred to as Type-I Censoring[19].

As shown in Chapter 6 of [19], the likelihood function for Type-I Censoring can be obtained as

$$L(\alpha(S),\beta) = \prod_{i=1}^{n} L_i(\alpha(S),\beta) = \prod_{i=1}^{n} \left[ f(N_i(S))^{\delta_i} \left(1 - F(N_i(S))\right)^{1-\delta_i} \right]. \quad (2.4.1.22)$$

When inserting the pdf and the cdf from Eq. (4.1.1) and (4.1.2) of the main manuscript, we obtain

$$L(\alpha(S),\beta) = \prod_{i=1}^{n} \left[ \left( \frac{\beta}{\alpha(S)} \left( \frac{N(S)}{\alpha(S)} \right)^{\beta-1} \exp\left( -\left( \frac{N(S)}{\alpha(S)} \right)^{\beta} \right) \right)^{\delta_i} \left( \exp\left( -\left( \frac{N(S)}{\alpha(S)} \right)^{\beta} \right) \right)^{1-\delta_i} \right]. \quad (2.4.1.23)$$

$$L(\alpha(S),\beta) = \prod_{i=1}^{n} \left( \frac{\beta}{\alpha(S)} \left( \frac{N(S)}{\alpha(S)} \right)^{\beta-1} \right)^{\delta_i} \exp\left( -\left( \frac{N(S)}{\alpha(S)} \right)^{\beta} \right). \quad (2.4.1.24)$$

Assuming

$$\log(\alpha(S)) = \gamma_0 + \gamma_1 \log(S). \quad (2.4.1.25)$$

the fatigue-lifespan behavior at a given stress, $S$, can be predicted by estimating the $p$-quantile life,

$$N_p(s) = \exp(\gamma_0 + \gamma_1 \log(S))\,(-\log(1-p))^{1/\beta}, \tag{2.4.1.26}$$

according to Eq. (2.4.1.17), once the model parameters, $\beta$, $\gamma_0$, and $\gamma_1$, have been estimated, for example using the maximum likelihood method. One can obtain Eq. (2.4.1.26) by combining Eqs. (4.1.2) and (4.1.3) from the main manuscript with Eq. (2.4.1.25):

$$F(N_p(S)|\,\alpha(S),\beta) = 1 - \exp\left(-\left(\frac{N_p(S)}{\alpha(S)}\right)^{\beta}\right) = p, \tag{2.4.1.27}$$

$$\exp\left(-\left(\frac{N_p(S)}{\alpha(S)}\right)^{\beta}\right) = 1 - p, \tag{2.4.1.28}$$

$$\left(\frac{N_p(S)}{\alpha(S)}\right)^{\beta} = -\log(1-p), \tag{2.4.1.29}$$

$$N_p(S) = \alpha(S)\,(-\log(1-p))^{1/\beta}, \tag{2.4.1.30}$$

$$N_p(S) = \exp(\gamma_0 + \gamma_1 \log(S))\,(-\log(1-p))^{1/\beta}. \tag{2.4.1.31}$$

## 1.2 More on the Weibull Mixture Predictive Models

### *1.2.1 Extending Maximum Likelihood Methods to Estimation of Parameters Comprising the Weibull Mixture Model – Case of the Weak and the Strong Groups*

A number of different methods can be applied to estimate the parameters of a Weibull mixture model [7]. In addition to the maximum likelihood estimation, one can consider applying pseudo-likelihood estimation, piece-wise likelihood estimation, or discriminative-based learning. In case of only two component distributions, the weak and the strong groups, the Weibull mixture predictive model consists of 7 unknown parameters, $p$, $\gamma_{w,0}$, $\gamma_{w,1}$, $\beta_w$, $\gamma_{s,0}$, $\gamma_{s,1}$, and $\beta_s$ [12]. But in case of the machine learning (ML) estimation, the likelihood function of the model parameters for Type-I Censoring can be obtained by Eq. (2.4.1.22), as shown in Chapter 6 of [19]. In case of the weak and the strong groups, the likelihood function is given by [12]:

$$L(p,\beta_w,\gamma_{w,0},\gamma_{w,1},\beta_s,\gamma_{s,0},\gamma_{s,1}) = \prod_{i=1}^{m} f(N_i)^{\delta_i}(1-F(N_i))^{1-\delta_i} \tag{2.4.2.8}$$

Where $f(N_i)$ and $F(N_i)$ are given by Eqs. (4.1.8) and (4.1.9) in the main manuscript, respectively. Once the maximum likelihood estimates of the seven model parameters have been obtained, the observed fatigue data can be clustered into the weak and the strong two groups [12]. If $N_i$ represents a failure observation, the likelihoods of the $i$th sample, $N_i$, belonging to the weak and the strong groups, are given by [12]

$$\frac{\beta_w}{e^{\gamma_{w,0}+\gamma_{w,1}+\log(S_i)}}\left(\frac{N_i}{e^{\gamma_{w,0}+\gamma_{w,1}+\log(S_i)}}\right)^{\beta_w-1}\exp\left(-\left(\frac{N_i}{e^{\gamma_{w,0}+\gamma_{w,1}+\log(S_i)}}\right)^{\beta_w}\right) \tag{2.4.2.9}$$

and

$$\frac{\beta_s}{e^{\gamma_{s,0}+\gamma_{s,1}+\log(S_i)}}\left(\frac{N_i}{e^{\gamma_{s,0}+\gamma_{s,1}+\log(S_i)}}\right)^{\beta_s-1}\exp\left(-\left(\frac{N_i}{e^{\gamma_{s,0}+\gamma_{s,1}+\log(S_i)}}\right)^{\beta_s}\right) \tag{2.4.2.10}$$

respectively. The $i$th sample, $N_i$, is then assigned to the group with a higher likelihood value[12]. Correspondingly, if $N_i$ represents a censored observation, the likelihoods of the $i$th sample, $N_i$, belonging to the weak and the strong group are given by [12]

$$\exp\left(-\left(\frac{N_i}{e^{\gamma_{w,0}+\gamma_{w,1}+\log(S_i)}}\right)^{\beta_w}\right) \tag{2.4.2.11}$$

and

$$\exp\left(-\left(\frac{N_i}{e^{\gamma_{s,0}+\gamma_{s,1}+\log(S_i)}}\right)^{\beta_s}\right) \quad (2.4.2.12)$$

respectively. The $p$ quantile fatigue lives for the strong group and for the weak group are given by Eqs. (2.4.2.13) and (2.4.2.14) [12]:

$$N_{p,w}(S) = \exp\big(\gamma_{w,0} + \gamma_{w,1}\log(S)\big)\,(-\log(1-p))^{1/\beta_w} \quad (2.4.2.13)$$

$$N_{p,s}(S) = \exp\big(\gamma_{s,0} + \gamma_{s,1}\log(S)\big)\,(-\log(1-p))^{1/\beta_s} \quad (2.4.2.14)$$

*1.2.2 Capability of the Weak and the Strong Groups to Elucidate the Impact of Defects on Fatigue Life*

The weak and the strong groups are capable of elucidating the impact of defects on the fatigue life of alloys. Fig. 8 and Fig. 12 of [12] confirm that the fatigue life of the $Al_{0.5}CoCrCuFeNi$ HEA. Figure S4 suggests that the strong group tends to exhibit fewer defects on average than the weak group. We believe the control of defects, during the fabrication process is of paramount importance, for future advancement and application of HEAs.

*1.2.3 Purported Critique of the Method for Estimating Mixed Weibull Distributions – Challenges in Estimating the Model Parameters*

A major drawback of the mixed Weibull distributions relates to difficulties with regards to the estimation of the unknown model parameters [7]. The efficiency of known methods for deriving the model parameters depends on the number of component distributions, *m*, and on the selection of the constituent pdf's, $f_l(s)$ [7]. Solving the system of equations for the unknown model parameters can be very difficult and only possible, when the spectrum consists of two Weibull distributions (when $m = 2$), or when other simplifications are made [20]. Problem simplifications are often required, in order to make the estimation of the unknown model parameters possible [7].

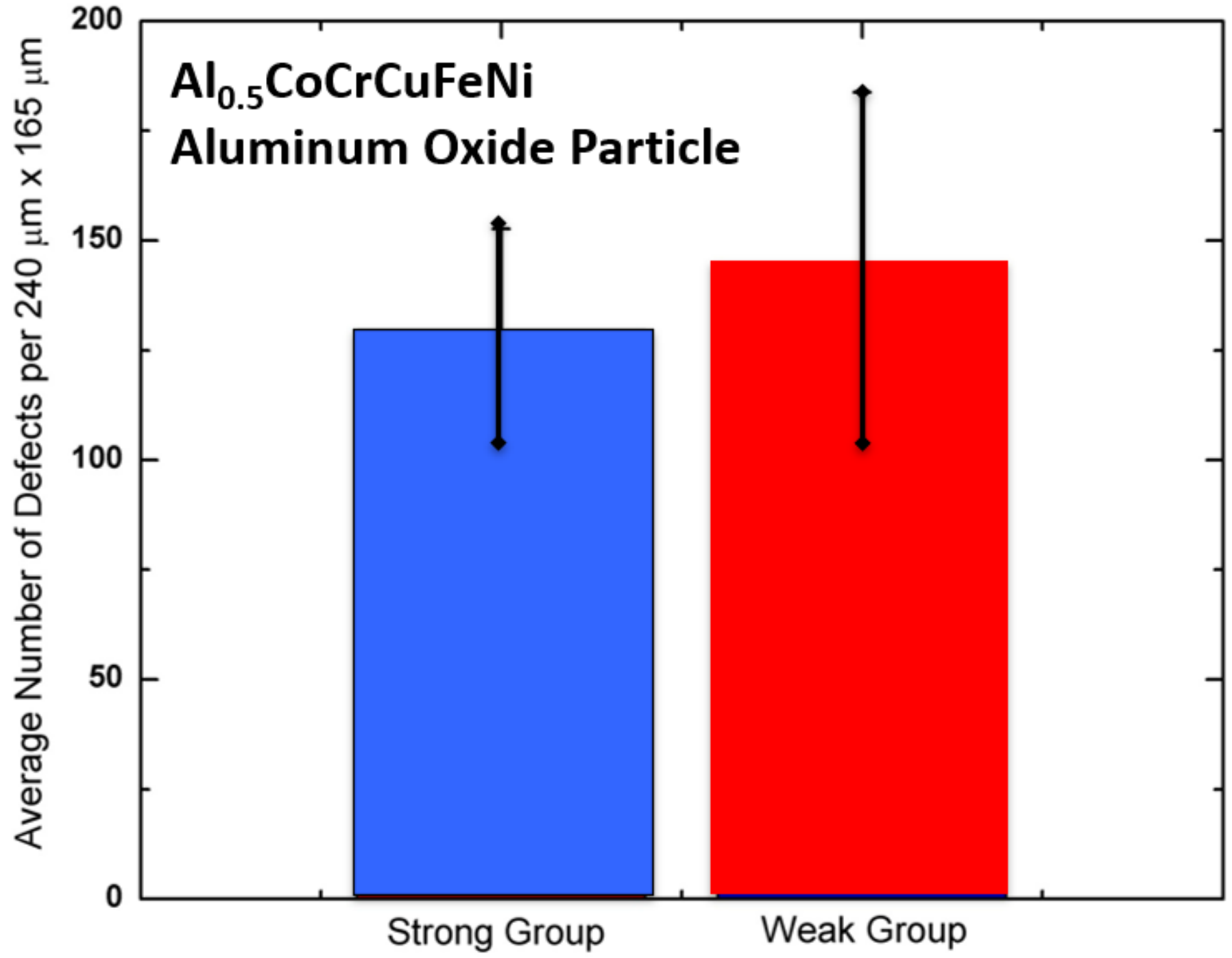

**Figure S4**: Average number of defects observed for the weak and the strong group, for the case of the Weibull mixture predictive model in Figure 4.1.2 from the main manuscript.

### *1.2.4 An Alternative Method for Rough Estimation of the Model Parameters of Mixed Weibull Distributions*

Because of the stated challenges of estimating the model parameters of mixed Weibull distributions, Nagode et al. present an alternative method [7]. The authors present an alternative algorithm, an iterative procedure, for rough parameter estimation and optimization, one suitable for the estimation of unknown parameters irrespective of the type of the component pdf's. The algorithm does not depend on the number of component pdf's, *m*, at all, and any other component pdf's can be used instead of the Weibull distribution[7]. Each component pdf is calculated separately, and the solving of complicated systems of equations is claimed not to be necessary[7].

## 1.3 More on the General Log-Linear Model

To establish whether the parallel or vertical morphology affects the fatigue life, a hypothesis test can be carried out on the regression coefficient, $\gamma_2$. There are two hypotheses to consider:

$$H_0\text{: } \gamma_2 = 0 \text{ vs. } H_1\text{: } \gamma_2 \neq 0 \qquad (2.4.3.2)$$

If the hypothesis, $H_0$, is rejected, there is evidence of the morphology type affecting the fatigue life. But if the hypothesis test does not lead to $H_0$ being rejected, there is no evidence against the hypothesis that the morphology type does not affect the fatigue life [12].

## 1.4 Random Endurance Limit Fatigue Life Models

In the random endurance limit fatigue-life models (RELFLMs), the fatigue endurance limit is modeled as a stochastic, as opposed to a deterministic, quantity. The RELFLMs simultaneously model the fatigue limit and the fatigue life as a stochastic process. The RELFLMs benefits from being able to (1) account for the variation in fatigue limits found from specimen to specimen, (2) easily include run-out results, (3) obtain the median S-N curve from the model coinciding with the conventional S-N model [Eqs. (2.4.1.16) – (2.4.1.17)], and (4) predict longer fatigue lives in the vicinity below the high-stress regime [21].

### *1.4.1 Origin of / Motivation for the Random Endurance Limit Fatigue-Life Models*

Dieter may have first discussed the stochastic (random) nature of fatigue-endurance limits [22]. Back in 1976, Dieter drew attention to the fact that 95% of fatigue-endurance limits fall between 40,000 and 52,00 psi, in the case of the heat-treated alloy of a forging steel [8]. Back in 1990, Nelson, further, suggested modeling the fatigue endurance limit as a random parameter [23]. He noted that test specimens could possess different fatigue-endurance limits, according to some probability distributions (referred to as the "strength distributions") [8]. In terms of these early efforts, Hirose, similarly, used ML methods back in 1993 to estimate the fatigue limit and mean life of polyethylene terephthalate films (used in electrical insulation) at the service stress [24]. Hirose fitted a Weibull inverse power relationship, but assumed a fixed fatigue-limit parameter [8].

Pascal et al. originally introduced a random endurance-limit fatigue-life model to describe (1) the dependence of fatigue life on the stress level and (2) scatter in the S-N data [8]. In their model, the scatter in the fatigue life and fatigue-endurance limit is combined in a single, joint model and treated simultaneously. Pascal et al. used their model to analyze four-point-bending fatigue data of carbon eight harness-satin/epoxy-laminate panels and smooth specimens with the metal-base material [25]. Their results provided evidence suggesting a random nature to the fatigue-endurance limit [25]. The authors believed that the randomness in the endurance limit was partially due to the location, orientation, size and number of defects (e.g., cracks), which themselves were subjected to randomness.
Lassen et al. further extended the random endurance-limit fatigue-life models of Pascal et al. and applied to the prediction of the fatigue life of fillet-welded steel joints, where cracks emanated from the weld toe and through the plate thickness [26]. Lassen et al. note that the experimental data in this stress regime are sparse and do not fit the knee point of the conventional bi-linear S-N curve. The authors indicate rightfully that few data exist to corroborate the abrupt shift in the slope of the conventional S-N curves in the log-log S-N diagram. But nonetheless, small alterations in the position of the knee point can have a strong bearing on fatigue-life predictions, fatigue design, and the final dimensions of welded details. The authors conclude that bilinear S-N curves may be excessively pessimistic in the stress regime, where service stresses frequently occur. Wallin has also examined statistical aspects of the fatigue-endurance limit, and similarly noted that the variation in the endurance limit had marked effects on the scatter in the fatigue life close to the endurance limit [12, 27]. To address this matter, Lassen et al. present a more accurate S-N curves, based on a random endurance-limit fatigue-life model of Pascal et al., one that features a smooth transition regime in the log-log scale, as opposed to the abrupt shift.
In terms of specific models, lognormal, normal, and Weibull distributions have been additionally proposed to describe randomness in the fatigue-endurance limit [25, 27].

*1.4.2 Structure of the Random Endurance-Limit Fatigue-Life Models*
Here, the S-N curve of the random fatigue-life model will not have an abrupt change from an inclined straight line to a horizontal line, but a gradual change in the slope as stress ranges become very low [21]. Lassen et al. observe that a nonlinear curve in a log-log scale is more consistent with the observed fatigue-life data for welded details at low stress levels than the bilinear relation, based on Eq. (2.4.1.18) [21]. Lassen et al. model the fatigue life as

$$\log(N(S)) = \beta_0 + \beta_1 \log(S - \gamma) + \varepsilon, \qquad S > \gamma, \tag{2.4.4.1}$$

where $\beta_0$ and $\beta_1$ represent coefficients of the S-N curve, $\gamma$ denotes the fatigue limit of the specimen, but $\varepsilon$ is an error term similar to Eq. (2.4.1.18). Let us now define [13, 21]

$$V = \log(\gamma) \tag{2.4.4.2}$$

and let us assume that $V$ can be described, using the pdf listed below:

$$f_V(v;\mu_\gamma,\sigma_\gamma) = \frac{1}{\sigma_\gamma}\phi_V\left(\frac{v-\mu_\gamma}{\sigma_\gamma}\right), \tag{2.4.4.3}$$

where $\mu_\gamma$ describes a location parameter and $\sigma_\gamma$ a scale parameter. The function, $\phi_V(\cdot)$, can represent the normal or the Weibull pdf [13, 21].
Let us now introduce [13, 21]

$$x = \log(S) \tag{2.4.4.4}$$

$$W = \log(N(S)). \tag{2.4.4.5}$$

And let us assume that conditioned on a fixed value of $V < x$, $W|V$ has the pdf [13, 21]

$$f_{W|V}(w;\beta_0,\beta_1,\sigma,x,v) = \frac{1}{\sigma}\phi_{W|V}\left(\frac{w-[\beta_0+\beta_1\log(\exp(x)-\exp(v))]}{\sigma}\right). \tag{2.4.4.6}$$

The marginal pdf for $W$ can be obtained as [13, 21]

$$f_W(w;x,\theta) = \int_{-\infty}^{x}\frac{1}{\sigma\,\sigma_\gamma}\phi_{W|V}\left(\frac{w-\mu(x,v,\theta)}{\sigma}\right)\,\phi_V\left(\frac{v-\mu_\gamma}{\sigma_\gamma}\right)\,dv. \tag{2.4.4.7}$$

with $\theta \equiv (\beta_0,\beta_1,\sigma,\mu_\gamma,\sigma_\gamma)$ and $\mu(x,v,\theta) = \beta_0 + \beta_1\log(\exp(x) - \exp(v))$. The marginal cdf for $W$ can similarly be obtained as [13, 21]

$$F_W(w;x,\theta) = \int_{-\infty}^{x}\frac{1}{\sigma\,\sigma_\gamma}\Phi_{W|V}\left(\frac{w-\mu(x,v,\theta)}{\sigma}\right)\,\phi_V\left(\frac{v-\mu_\gamma}{\sigma_\gamma}\right)\,dv. \tag{2.4.4.8}$$

where $\Phi_{W|V}(\cdot)$ represents the cdf of $W|V$. We will refer to the stochastic model in Eq. (2.4.4.1) – Eq. (2.4.4.8) as the RELFLM. The pdf and the cdf for the RELFLM ($W$) do not have a closed-form representation, but can be easily evaluated numerically.

*1.4.3 Maximum Likelihood Estimation of Parameters of Random Endurance-Limit Models*

For a given set of test measurements, comprising of ($S_i$,$N_i$) combinations, or equivalently of

$$(x_i,w_i) = (\log(S_i), \log(N_i)) \tag{2.4.4.9}$$

combinations, where $i = 1, 2, \ldots, n$, the model parameters

$$\boldsymbol{\theta} = [\beta_0, \beta_1, \sigma, \mu_\gamma, \sigma_\gamma] \tag{2.4.4.10}$$

can be obtained by maximizing a likelihood function:

$$\widehat{\boldsymbol{\theta}} = \arg\max_{\boldsymbol{\theta} \in \boldsymbol{\Theta}} L(\boldsymbol{\theta}). \tag{2.4.4.11}$$

Here, the likelihood function is given by [13, 21]

$$L(\boldsymbol{\theta}) = \prod_{i=1}^{n} L_i(\boldsymbol{\theta}) = \prod_{i=1}^{n} [f_W(w_i; x_i, \boldsymbol{\theta})]^{\delta_i} [1 - F_W(w_i; x_i, \boldsymbol{\theta})]^{1-\delta_i} \tag{2.4.4.12}$$

where

$$\delta_i = \begin{cases} 1 & \text{if } w_i \text{ is a failure} \\ 0 & \text{if } w_i \text{ is a censored observation (a run} - \text{out).} \end{cases} \tag{2.4.4.13}$$

where $f_W(\cdot)$ is given by Eq. (2.4.4.7) and $F_W(\cdot)$ by Eq. (2.4.4.8), respectively. The integration in Eqs. (2.4.4.7) - (2.4.4.8) and the maximization in Eqs. (2.4.4.11) – (2.4.4.12) must be done numerically. The details can be found in [13, 28]. When the parameters have been estimated, one can calculate the fatigue life, for a given probability of failure, $p$, using Eq. (2.4.4.8).

The function, $L(\boldsymbol{\theta})$, can be interpreted as being approximately proportional to the probability of observing the data set in Eq. (2.4.4.9) for a given set of the model parameters, $\boldsymbol{\theta}$. With $\log(\cdot)$ being a monotonically increasing function, the ML estimates $\widehat{\boldsymbol{\theta}}$ from Eq. (2.4.4.11), which maximizes $L(\boldsymbol{\theta})$, and also maximizes [8]

$$\mathcal{L}(\theta) = \log[L(\theta)] = \sum_{i=1}^{n} \mathcal{L}_i(\theta), \tag{2.4.4.14}$$

where

$$\mathcal{L}_i(\theta) = \delta_i \log[f_W(w_i; x_i, \theta)] + (1 - \delta_i) \log[1 - F_W(w_i; x_i, \theta)] \tag{2.4.4.15}$$

represents the contribution of the $i$th observation [8]. Generally, it is numerically simpler to work with the log-version of the likelihood function, i.e., Eq. (2.4.4.14), as opposed to Eq. (2.4.4.12).

*1.4.4 Profile Likelihood Method for Obtaining Approximate Confidence Intervals for the Model Parameters*

Once the model parameters have been estimated, from the optimization problem in Eqs. (2.4.4.11) – (2.4.4.12), the confidence intervals for these parameters can be obtained by computing the profile ratios of these parameters together with the corresponding chi-square statistics [21]. These confidence intervals are obtained by inverting the likelihood ratio test [8].

Let us define $\theta = (\theta_1, \theta_2)$ as a partition of $\theta$, where $\theta_1$ represents a vector with $k$ quantities of interest [8]. We denote the ML estimate of $\theta$ as $\hat{\theta}$. The profile likelihood for $\theta_1$ is defined as [8]

$$R(\theta_1) = \max_{\theta_2} \left[ \frac{L(\theta_1, \theta_2)}{L(\hat{\theta})} \right]. \tag{2.4.4.16}$$

A large value for $R(\theta_1)$, i.e., a value close to 1, suggests that the data observed for that value of $\theta_1$ are quite probable, relative to the ML estimate [8]. A very small value for $R(\theta_1)$, i.e., a value very close to 0, on the other hand, suggests that the observed data are very unlikely, given that value of $\theta_1$.

The asymptotic distribution of $-2\log[R(\theta_1)]$ is a chi-squared distribution with $k$ degrees of freedom, when evaluated at the true value, $\theta_1$. Consequently, an approximate $100(1-\alpha)\%$ confidence region for $\theta_1$ is given by the set of $\theta_1$ that fulfills the relation [8]

$$-2\log[R(\theta_1)] \leq \chi^2_{(k;1-\alpha)}, \qquad (2.4.4.17)$$

or equivalently,

$$R(\theta_1) \geq \exp\left[-\frac{\chi^2_{(k;1-\alpha)}}{2}\right], \qquad (2.4.4.18)$$

where $\chi^2_{(k;1-\alpha)}$ denotes an (1 - α) quantile of a chi-squared distribution with $k$ degrees of freedom [8].

*1.4.5 Contrived Example (from [21]) - Supporting Observations*

The example in Figure S5 presents F-class and Category 71 design curves, which show a permissible stress taken from the code of BS 5400 for the design and construction of steel, concrete, and composite bridges as well as from the Eurocode 3 for the design of steel structures [21]. These are conventional bilinear S-N curves, for predicting fatigue life under a given load, and can be found in design codes and standards. The RELFLM model, on the other hand, gives rise to a nonlinear S-N curve on a log-log scale in the vicinity of the fatigue limit [21]. The RFELM model approaches the horizontal line asymptotically, instead of the abrupt knee point exhibited by the bilinear S-N curves [21]. The design curve has been drawn at the median value minus two standard deviations, in case of the BS 5400 standard, or alternatively in case of the Eurocode 3 standard at minus 1.5 standard deviations if the standard deviation has a 75% confidence level. Database 1 contains fatigue-life data primarily at relatively high stress levels, whereas Database 2 contains fatigue data mainly at relatively lower stress levels. For further specifics on the test set-up, collection and analysis of the experimental data, refer to [21].

The fatigue-endurance limit for the F-class curve in Figure S5 is 56 MPa, corresponding to the fatigue life of $10^7$ cycles. The F-class curve yields almost the same predictions as the Category 71 curve, except that the latter curve becomes horizontal at 5 x $10^6$ instead of $10^7$ cycles. This difference highlights the uncertainty in the stress region near the fatigue-limit regime (near the knee point of the bilinear S-N curves). Lassen et al. note that small alterations in the position of the knee point can have strong bearing on fatigue-life predictions, and as a result, on the fatigue design and on the final dimensions of the welded details [21]. Overall, the RELFLM model offers better fit to the experimental fatigue-life data at stress levels in vicinity of the knee point of the conventional bi-linear S-N models [21].

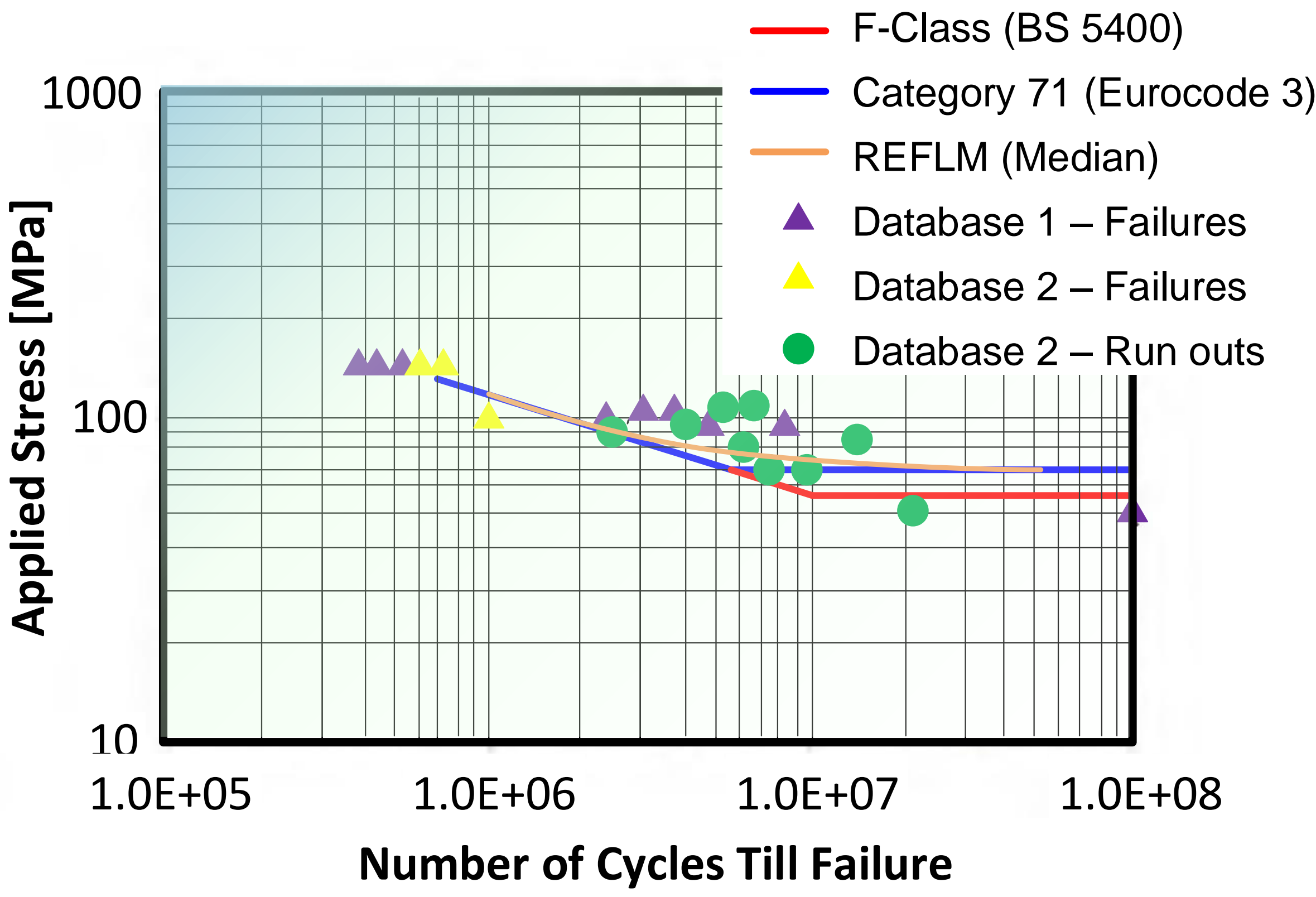

**Figure S5**: Contrived example offering a comparison of the RELFLM model with the two models yielding a bilinear S-N curve on the log-log scale (the F-class and Category 71 models), for the case of fatigue cracks in welded steel joints where cracks emanate from the weld toe [21].

Table S1 presents point estimates of the parameters comprising the random endurance fatigue-life model, along with the corresponding 90% confidence intervals. The confidence intervals are obtained from plots of the profile ratios for $\mu_\gamma$ and $\sigma_\gamma$, as shown in Figure S6. The point estimate for the fatigue limit, $\gamma$, for the RELFLM ($\mu_\gamma$) is near 60 MPa, according to Table S1. The BS 5400 fatigue limit of 56 MPa is well within the 90% confidence interval for $\mu_\gamma$, and is quite close to the point estimate for $\mu_\gamma$ of 60 MPa. The Eurocode fatigue limit of 70 MPa is quite outside the 90% confidence interval for $\mu_\gamma$ [21].

**Table S1**: Sample estimation of the parameters comprising a RELFLM (adapted from [21]).

| Parameter | Meaning | Point Estimate | 90% Confidence Interval | |
|---|---|---|---|---|
| | | | Minimum | Maximum |
| $\beta_0$ | Intercept of fatigue-life model | 22.48 | 22.407 | 22.555 |
| $\beta_1$ | Slope of fatigue life model | 2.10 | 2.084 | 2.118 |
| $\sigma$ | Width parameter for conditional pdf for $w\|v$ | 0.14 | 0.089 | 0.240 |
| $\mu_\gamma$ | Location parameter for pdf of $v$ | 4.10 (60.3 MPa) | 4.04 (57 MPa) | 4.15 (64 MPa) |
| $\sigma_\gamma$ | Width (scaling) parameter for pdf of $v$ | 0.16 | 0.120 | 0.216 |

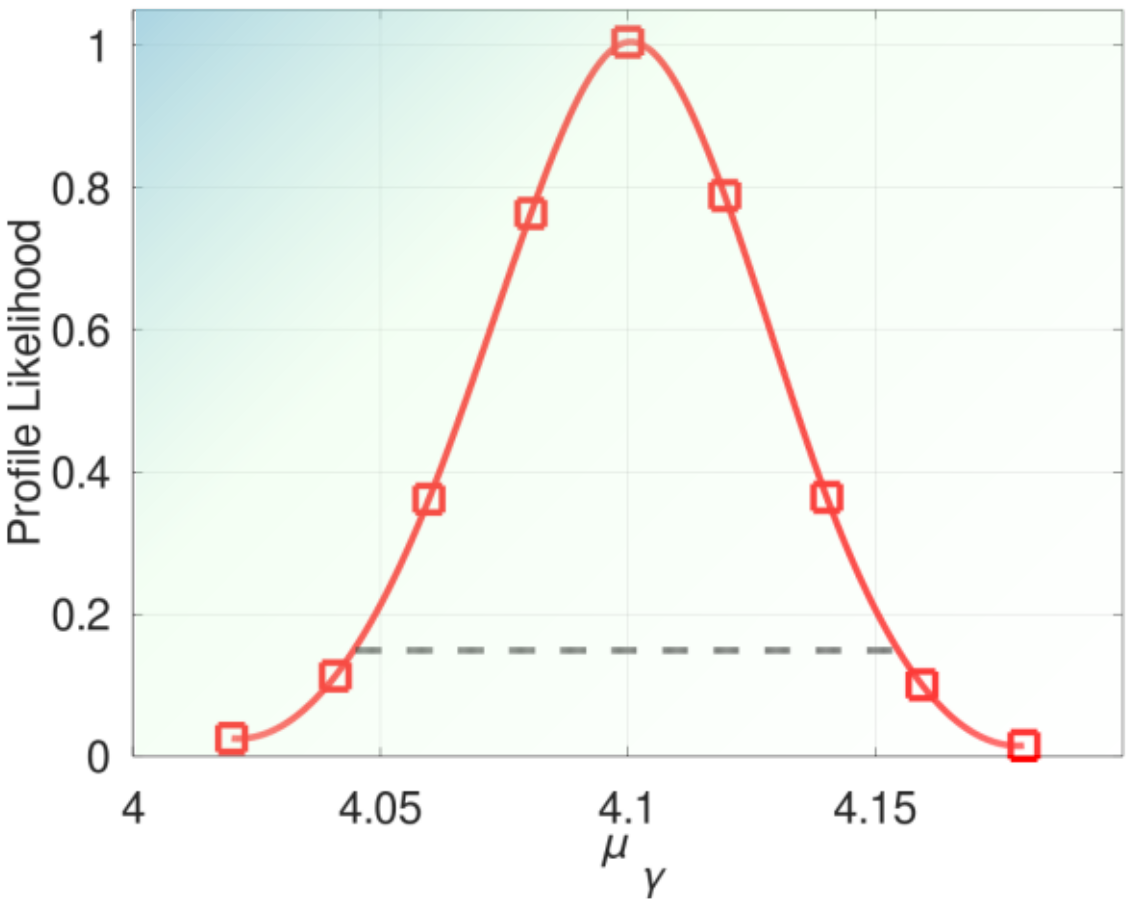

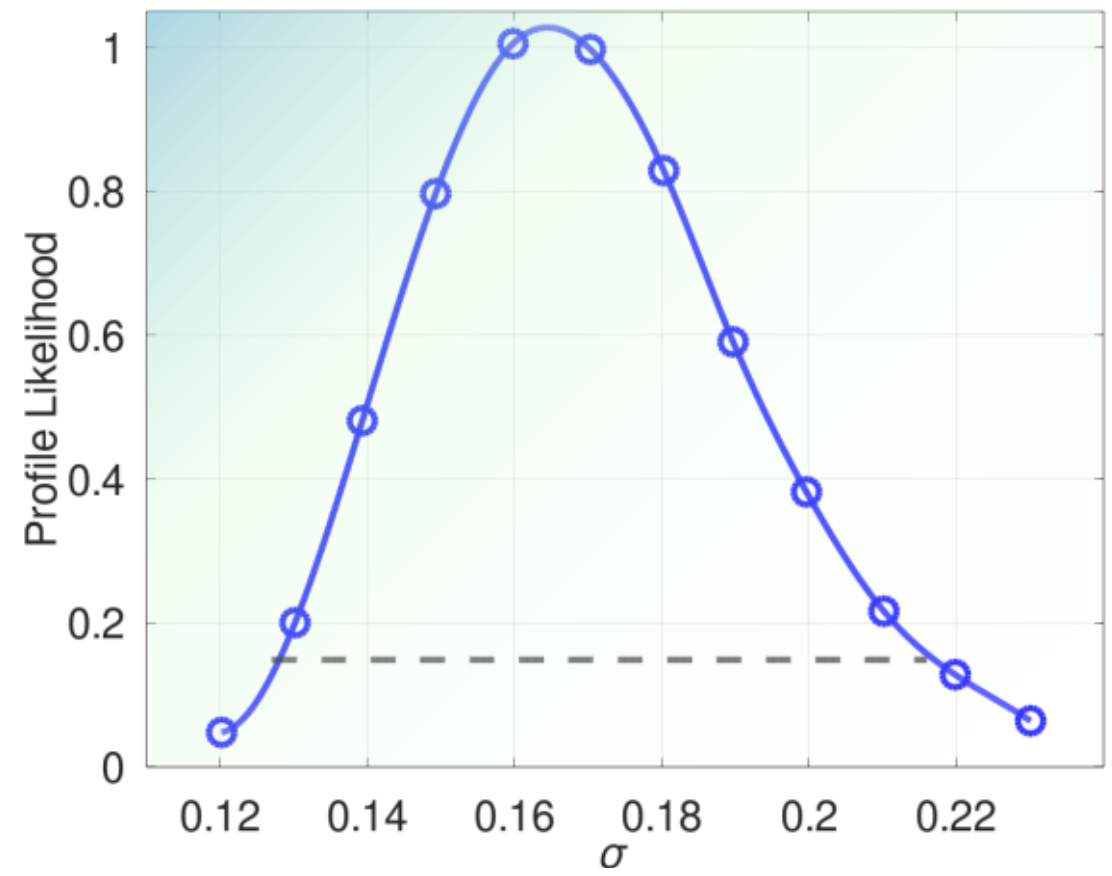

**Figure S6**: Profile likelihoods and 90% confidence intervals for the mean ($\mu_\gamma$) and standard deviation ($\sigma_\gamma$) of the fatigue-endurance limit ($\gamma$) of the random endurance-limit fatigue-life model (adapted from[21]).

## 2 More on Machine Learning Models

### 2.1 Mathematical Model for Low-Cycle Fatigue

In reference [30], the S-N curve for LCF is modeled as

$$N(\sigma, p_1, p_2, p_3, \cdots, p_N) = f_1(p_1, p_2, p_3, \cdots, p_N)\sigma^{-f_2(p_1, p_2, p_3, \cdots, p_N)}. \tag{2.5.1.3}$$

Here, $p_1$, $p_2$, $p_3$, …, $p_N$ can model the input parameters impacting the fatigue life of a HEA component. We assume that multiple effects cause failure and that these effects are close to be independent. The function, $f_1(\cdot)$, models a prior knowledge but the function, $f_2(\cdot)$, conditional probabilities [29]. In case of dependent events, one can apply a Bayesian model, with the same definition of $f_1(\cdot)$ and $f_2(\cdot)$ [30].

In the event of independent events, it makes sense to apply the direct linear regression to assess $f_1(\cdot)$ and $f_2(\cdot)$ [29]. But in the case of coupled failure modes, $f_1(\cdot)$ and $f_2(\cdot)$ may consist of complex Bayesian functions. We may not know these functions, and one may need to deduce them using regression analysis. These functions may be hard to derive. *So this is where ML comes in*. One can employ neural networks or support vector machines to effectively deduce these functions from the data. Even with 100 – 200 parameters impacting the fatigue life of AM metallic components, this is still a relatively small set by the standards of ML [30].

Our preferred approach to predicting the fatigue life, which again is based on the statistical fatigue life model of [12, 31], consists of the following steps [29]:

1. One infers the parameters, which impact the fatigue life ($p_1$, $p_2$, $p_3$, …, $p_N$), from the input data, using a model similar to Eq. (4.2.1) from the main manuscript, for example, using multi-variate linear regression;
2. One predicts variations for the individual parameters, using a model, such as

$$\begin{aligned} p_1 &= g_1[\text{UTS}, \text{process}, \text{defect (process)}, \text{grain (process)}, \text{microstructure (process)}, T. \ldots], \\ p_2 &= g_2[\text{UTS}, \text{process}, \text{defect (process)}, \text{grain (process)}, \text{microstructure (process)}, T. \ldots], \end{aligned} \tag{2.5.1.4}$$

$$p_3 = g_3[\text{UTS}, \text{process}, \text{defect (process)}, \text{grain (process)}, \text{microstructure (process)}, T. \ldots],$$

$$\ldots$$

$$p_N = g_N[\text{UTS}, \text{process}, \text{defect (process)}, \text{grain (process)}, \text{microstructure (process)}, T. \ldots];$$

3. One compares and contrasts the new parameter set with physics-based intuitions (see Figure S7); and then
4. One reconstructs the S-N curve based on the new set of parameters, using Eq. (2.5.1.3).

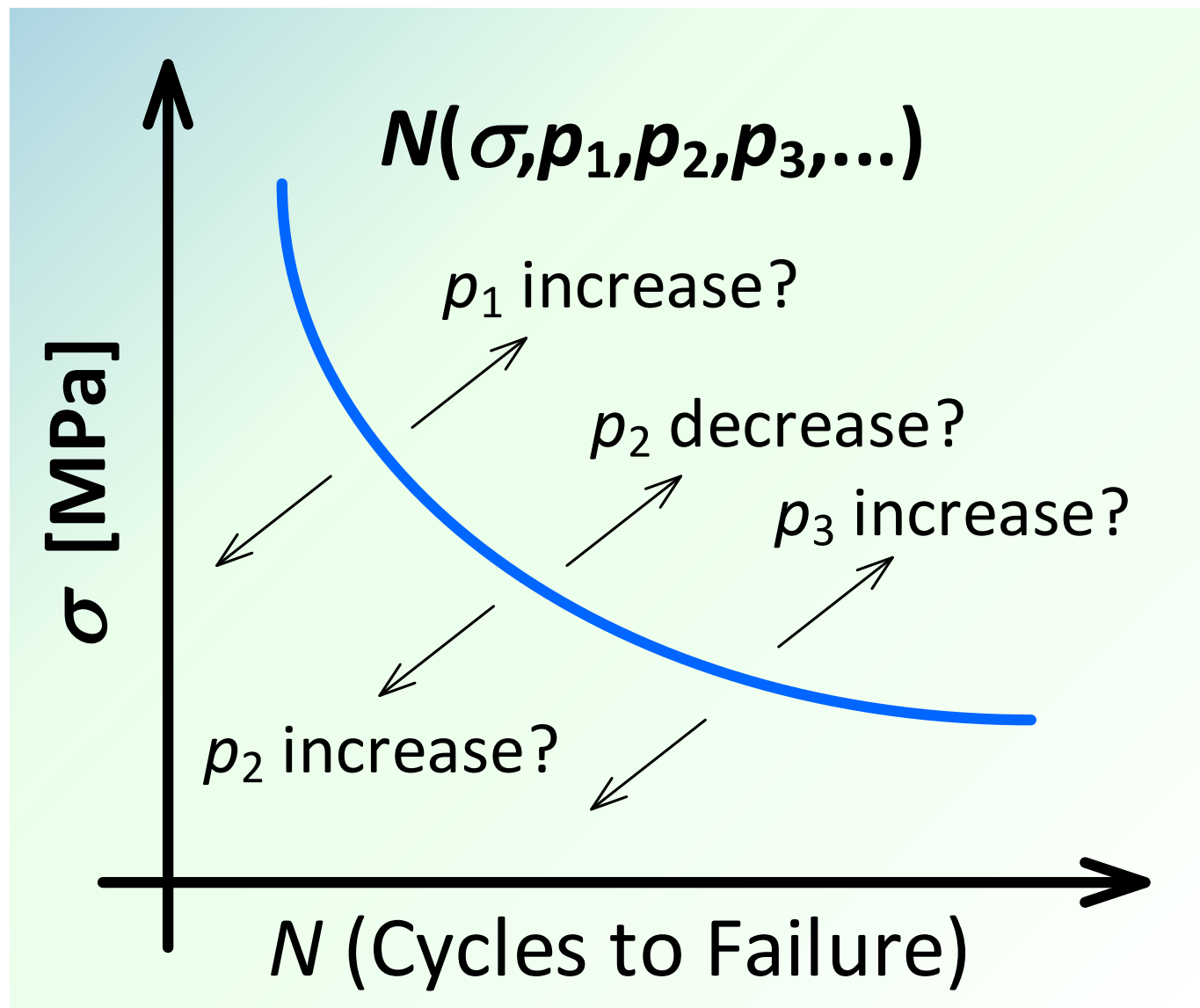

**Figure S7**: Prediction of a stress-life curve, using the model of Eqs. (2.5.1.3) and (2.5.1.4).

2.2 <u>Analysis of the Sources of Variations: Characterization of the Extent of These Variations</u>

Figure 4.34 of [29] presents a comparison of the HCF properties of HEAs to those of conventional alloys. The HEAs appear to generally result in the higher UTS and endurance limits, compared to conventional alloys. Nevertheless, there is much scatter in the data. But in spite of the scatter, the endurance limit seems primarily correlated with the UTS. Therefore, by identifying compositions with larger UTS, one can expect greater fatigue resistance. A key to accurate prediction involves having understanding, and properly accounting for (explaining) the sources of variations in the data. Note that Figure 4.34 of [29] assumes a very simple prediction model of the form

$$\text{Endurance_limit} = f(\text{ UTS }) \tag{2.5.1.5}$$

*The scatter in the data is caused by input sources, such as defect levels, process parameters, or grain sizes, which are not accounted for in the simple prediction model of Eq. (2.5.1.5).* The endurance limits in Figure 4.34 of [29] can be interpreted as multi-dimensional data points, which in addition to the UTS exhibit the dependence on the grain sizes, process parameters, and defect characteristics.

Table 4.9 of [29] *demonstrates that one can expect ~ 2x variations in the endurance limits, based on defect levels (defect size, density, and type) and raw material purity*, for a fixed UTS. Such a trend suggests that the explicit access to the information on the defect level may be needed in order to accurately predict the endurance limit. The samples in Table 4.9 of [29] have been homogenized at 1,000°C for 6 hour, water quenched, and then cold rolled. For Condition 1, shrinkage pores and macro-segregation may have remained in some portions. For Conditions 2 and 3, shrinkage pores and macro-segregation have been removed before cold rolling [29].

Table 4.10 of [29] *similarly shows that one can expect ~ 2x variations in the endurance limits, likely caused by variations in the grain size*, even for the same microstructure [Face-Centered-Cubic (FCC)] and similar process (hot-rolled and heat-treated). In combination, Tables 4.9 and 4.10 of [29] illustrate that the accurate prediction of the endurance limit is not possible, based on the UTS alone. For accurate predictions, one needs to know the defect levels (the defect size, density, and type), the grain size, and even parameters of the heat-treatment process, in addition to the UTS. These observations seem to be consistent with those of Hemphill et al. [12] as well as with those of Tang et al. [32].

### 3. More on Other Non-Stochastic Models

Material degradation, due to fatigue, in particular crack initiation and growth, is a multi-scale process, where atomic bonds rupture, defects and dislocations nucleate, which build up from the microscopic to the macroscopic scale, leading ultimately to a macroscopic fracture. Figure S8 highlights multi-scale modeling aspects associated with the fatigue-degradation process. Recent advances in fracture mechanics have generally been concerned with aspects of multi-physics, where one or more physical field enhances or degrades the fracture resistance of the material under a study[33]. Material scientists tend to be concerned with the microscopic mechanisms behind fatigue, and are likely to pay significant attention to the nucleation of micrometer-size flaws along slip bands and grain boundaries[34]. A practicing engineer, on the other hand, may be more interested in the limit of resolution of non-destructive crack-detection equipment (which is typically a fraction of a millimeter), with nucleation of fatigue cracks, and with the initial crack sizes used for design [34].

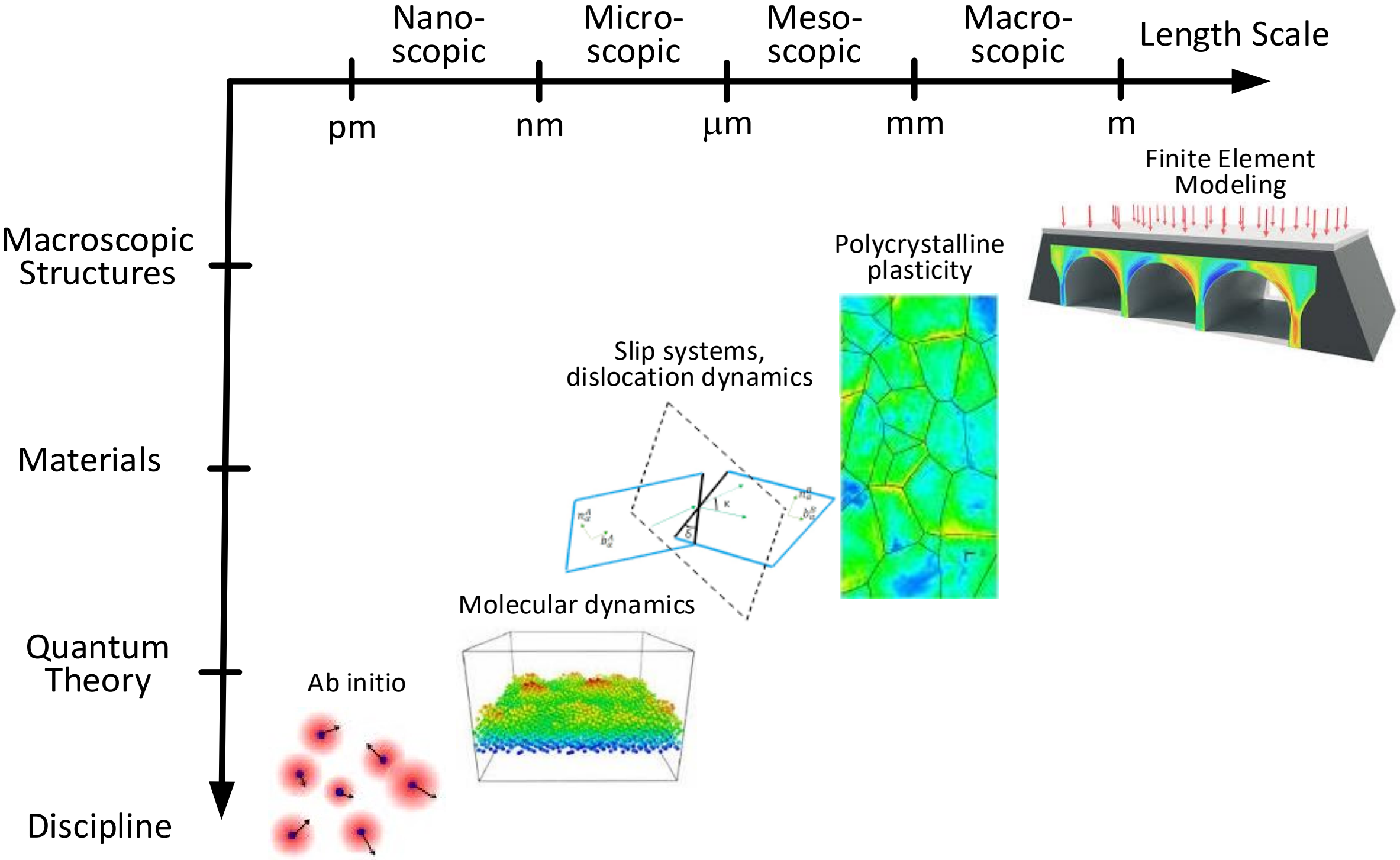

**Figure S8**: Multiscale aspects to the fatigue degradation process [33].

### 3.1 Macroscopic Models

#### *3.1.1. Stress-Life (S-N), Strain-Life (ε-N), Cyclic Stress-Strain, and Fatigue Crack Growth (da/dN-ΔK) Modeling*

The stress-life modeling is covered in Section 2.1 of the main manuscript and the strain-life modeling for HEAs similarly covered in Section 2.2. The fatigue-crack-growth modeling is similarly reviewed in Section 2.3 of the main manuscript.

#### *3.1.2. The Smith-Watson-Topper Relation*

Complementing the Basquin law for stress-life modeling and the Coffin-Manson law for strain-life modeling, mentioned in Section 1.2 above, the Smith-Watson-Topper relations associates fatigue quantities obtained at different stress ratios, $R_1$ and $R_2$, described as follows [35]:

$$\sqrt{\sigma_{max,1}\ \ \sigma_{a,1}} = \sqrt{\sigma_{max,2}\ \ \sigma_{a,2}} \qquad (2.6.1.1)$$

Here $\sigma_{max}$ denotes the maximum stress applied (the maximum of the stress amplitude), and the subscripts 1 and 2 reference data obtained at the respective stress ratios. The Smith-Watson-Topper relation can be employed to normalize stress measurements, obtained at different stress ratios, $R$, to a fixed stress ratio, $R_1$:

$$\sigma_{avg\ @\ R_1} = \sqrt{\frac{(1-R_1)}{(1-R)}}\ \sigma_{avg\ @\ R}. \qquad (2.6.1.2)$$

Here, $\sigma_{avg}$ represents the average stress amplitude:

$$\sigma_{avg} = \frac{(1-R)}{2}\sigma_{max}. \qquad (2.6.1.3)$$

#### *3.1.3. Classical Theory on Elastic Bending*

In applied mechanics, the theory of bending seeks to explain the macroscopic behavior of a sample of structural material subjected to an external load that is applied perpendicularly to a longitudinal axis of the sample. The elastic properties of single-crystal and polycrystal HEAs are important for the fundamental study of the mechanical behavior of HEAs and to the search for the next-general HEAs, such as for structural or biomedical applications.

Elastoplastic beam models and elastic beam theory can be used to estimate the stress in bending samples and generate a plot of the stress amplitude vs. cycles to failure [10, 36]. Specifically, for the case of the four-point bending fatigue experiment and a rectangular sample, the maximum stress, $\sigma_{max}$, on a tensile surface within the span of two outer pins can be calculated, using the classical beam theory [12, 37]

$$\sigma_{max} = \frac{3\ P\ (S_o - S_i)}{2\ B\ W^2}. \qquad (2.6.1.4)$$

Here, $P$ represents the applied load, $S_o$ is the outer span length, $S_i$ the inner span length, $B$ the sample thickness, and $W$ the sample height, shown in Figure S9.

Similarly, for the case of the three-point bending fatigue experiment and a rectangular sample, the maximum stress, $\sigma_{max}$, on a tensile surface can be computed as

$$\sigma_{max} = \frac{3\ P\ L}{2\ B\ W^2} \qquad (2.6.1.5)$$

where $L$ represents the length of the support span, as shown in Figure S10.

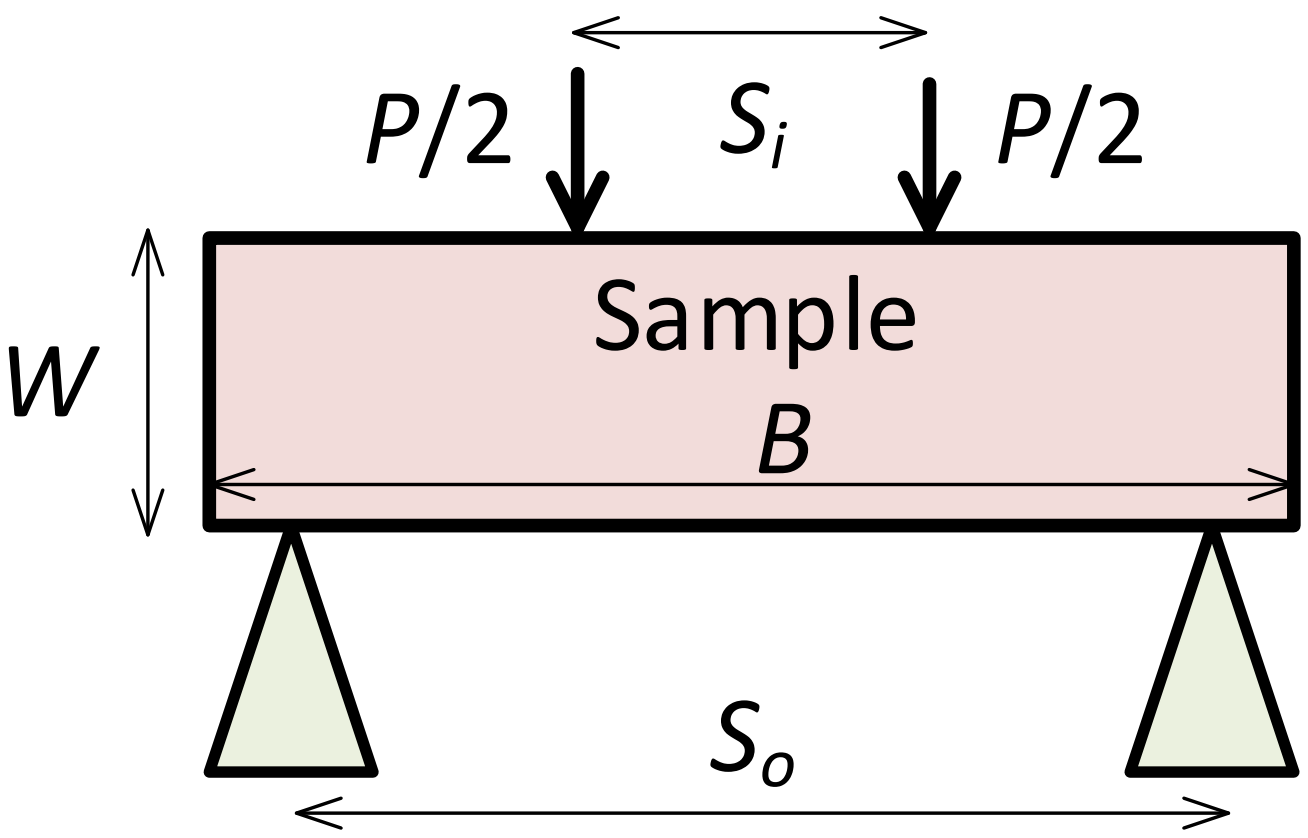

**Figure S9**: Definition of the quantities involved in the four-point bending fatigue test [10,36].

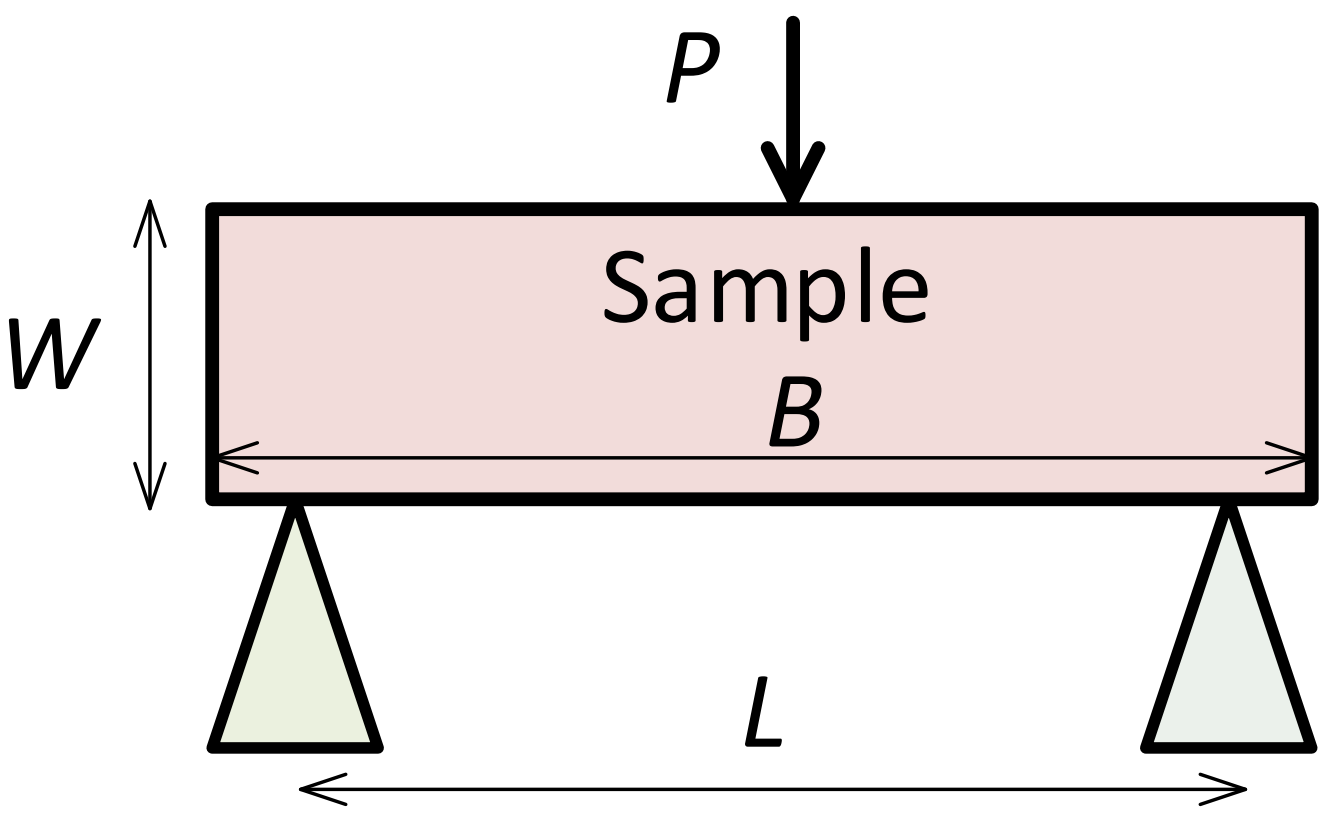

**Figure S10**: Definition of the quantities involved in the three-point bending fatigue test [10,36].

3.2 Mesoscopic Models

*3.2.1. On Multi-Scale Modeling in General*

Fracture crack growth, in general, is a multi-scale process, where the crack initiation (nucleation) is on or below microscale, but the crack propagation is on macroscale [34].

*3.2.2. On Multi-Scale CPFEM and Elastic-Visco-Plastic Self-Consistent (EVPSC) Modeling*

Multi-scale CPFEM modeling offers clear advantage in describing deformation behavior of materials based on microstructural evolutions, because it can account for different plastic deformation micro-mechanisms, such as dislocation slip, displacive transformation, and deformation twinning [38, 39]. CPFEM has been successfully used to describe the overall response of polycrystalline materials, when accounting for its assembly with different grain orientations [40, 41, 42, 39], apparently mostly under monotonic loading.

Figure S11 captures a representative multi-scale CPFEM modeling process from [43]. Here, CPFEM is used not to investigate fatigue properties, but to model the effect of temperature on tensile behavior of an interstitial HEA (iHEA) with the nominal composition of $Fe_{49.5}Mn_{30}Co_{10}Cr_{10}C_{0.5}$ (at.%). A thermodynamic model is presented in order to calculate the stacking fault energy of the iHEA at different temperatures.

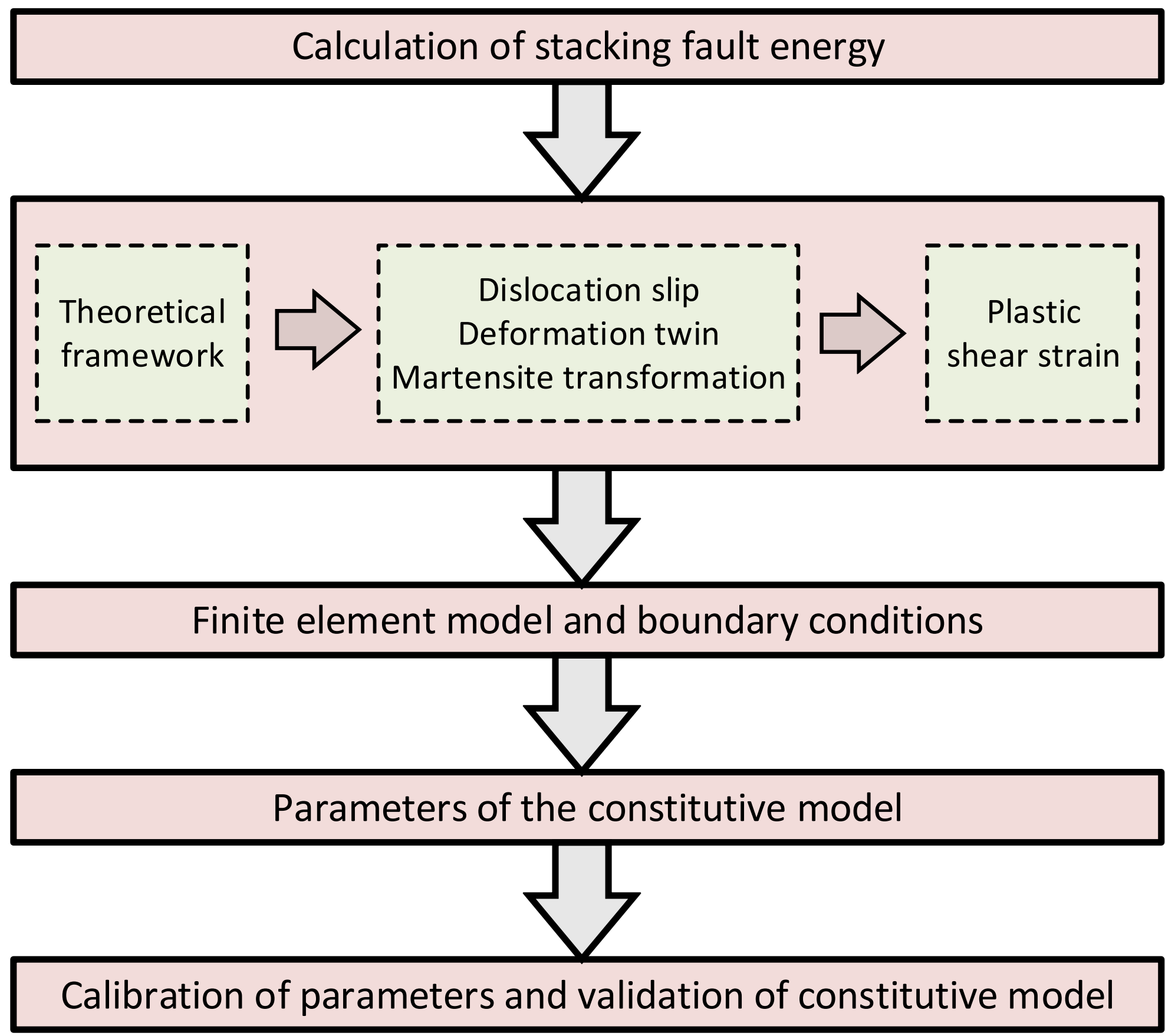

**Figure S11**: High-level overview of a representative CPFEM model [43].

A stacking fault refers to a planar defect (an error in the sequence of atomic layers) that can occur in crystalline materials. Stacking faults are characterized by a higher energy state, which is quantified by the formation enthalpy per unit area, and referred to as the stacking fault energy. In case of FCC crystals, the stacking fault energy can be calculated from the width of dislocation dissociation as [44]

$$SFE = \frac{G\, b_1 \cdot b_2}{2\,\pi\, d} = \frac{G\ \ b^2}{4\,\pi\, d}, \tag{2.6.2.1}$$

Here, $\mathbf{b}_1$ and $\mathbf{b}_2$ refer to Burgers vectors, $b$ to vector magnitude for the dissociated partial dislocations, $G$ to shear modulus, and $d$ to distance between partial dislocations. Some of the earliest descriptions of models of crystallographic systems date back to the work of Taylor in 1934 [45]. Then in 1972, Hill and Rice constructed a general time-independent constitutive model for crystallographic shearing. They presented a general finite deformation elastic–plastic framework for analyzing single crystals[46]. Then again in 1982, Peirce et al. numerically modeled deformations of ductile single crystals subjected to tensile loading[47]. Peirce et al. successfully formulated an elastic-plastic relation based on Schmid's law. The authors considered lattice rotations for the non-uniform and localized deformations, included self-hardening and latent hardening of the slip systems, and conducted comparison of the resolved shear stress vs. shear strain of the experimental points and their simulation profiles, both of which exhibited nonlinear behavior [48,47]. The models by Peirce et al. were successful in quantitatively describing the nonlinear behavior of crystal plasticity.

In 2010, Wang et al. introduced a large strain EVPSC model for polycrystalline materials [49]. At a single-crystal level, both rate-sensitive slip and twinning were included as plastic deformation mechanisms, whereas elastic anisotropy was accounted for in the elastic moduli. The transition for single-crystal to polycrystal plasticity was based on an entirely self-consistent approach [49]. Wang et al. demonstrated that differences in predicted stress-strain curves and texture evolutions, based on the EVPSC model and the viscoplastic self-consistent (VPSC), which was introduced by Lebensohn and Tome [50], were negligible at large strain levels and for monotonic loading. For deformations involving unloading and strain-path changes, the EVPSC predicts a smooth elasto-plastic transition, whereas the VPSC model yields a discontinuous response, due to lack of elastic deformation. Wang et al. also show that the EVPSC model can account for certain important experimental features, which cannot be simulated, using the VPSC model [49].

In 2010, Wang et al. further evaluated various self-consistent polycrystal plasticity models for hexagonal close packed (HCP) polycrystals by studying the deformation behavior of sheets of the magnesium alloy AZ31B under different uniaxial strain paths [51]. Both slip and twinning contribute to plastic deformations, in all of the polycrystal plasticity models employed. Material parameters for the various models are fitted to experimental results obtained under uniaxial tension or compression along the rolling direction. These various models are then used to predict uniaxial tension or compression along the transverse direction as well as uniaxial compression under the normal direction. Wang et al. carry out the assessment of predictive capability of polycrystal plasticity models, based on comparisons of the predicted and experimental stress responses and stress ratios [51]. The authors determine that amongst the models under examination, the self-consistent models with grain-interaction stiffness halfway between those of the limiting Secant (stiff) and Tangent (compliant) approximations yield the best results. An Affine self-consistent scheme results in the best overall performance, among the options considered. The authors demonstrate that the $R$ values under uniaxial tension or compression within the sheet plane exhibit strong dependence on the strain imposed. This trend implies that anisotropic yield functions developed using the measured $R$ values need to account for the strain dependence [51].

In 2013, Wang et al. proposed a physics-based twinning and de-twinning (TDT) model that possessed the ability to deal with both mechanisms during plastic deformation [52]. Deformation twinning and de-twinning are the plastic-deformation mechanisms in HCP crystals, together with slips, which strongly affect texture evolution and anisotropic response of the HCP crystals. As a result, several twinning models have been proposed and incorporated into existing models for polycrystalline plasticity [52]. Compared to twinning, de-twinning involves an inverse process, which is relevant to cycling, fatigue and complex loads, but which is rarely incorporated into polycrystalline plasticity models. The TDT model proposed by Wang et al in [52] is characterized by four deformation mechanisms: (1) twin nucleation, (2) twin growth, (3) twin shrinkage, and (4) re-twinning. Twin nucleation (1) and twin growth (2) are attributed to deformation twining, but twin shrinkage (3) and re-twinning (4) to de-twinning. The TDT model proposed is implemented within the confines of the EVPSC model. Wang et al. illustrate validity and capability of the TDT model by simulating cyclic loading of a AZ31B magnesium alloy plate and a AZ31 bar. Comparison with measurements suggests that the TDT model is capable of capturing the key features observed in experiments, which suggests that the mechanical response in the materials simulated is mainly associated with twinning and de-twinning[52].

In 2015, Khan et al. reviewed the history of combining crystal plasticity with FEM as CPFEM[53]. The authors demonstrated how CPFEM is capable of predicting finite plastic deformation of single crystalline metals over a wide range of strain rates.

In 2019, Ali et al. employed artificial neural networks (ANNs) in conjunction with CPFEM in modeling of AA6063-T6 aluminum alloys, in an effort to address increasing models, equations, and the number of parameters [54]. Although leading researchers, such as Peirce et al., have constructed comprehensive CPFEM models, a primary challenge relates to the balance between accuracy and computational efficiency, where the computational cost effectiveness decreases as the complexity of microstructure increases [55]. Ali et al. back-fitted the constitutive model of the systems under investigation, using experimental data, before introducing the ANNs. In the case of crystal plasticity, the microscopic material parameters in the formulas are considered in order to solve the constitutive model for connecting the macroscopic performances, such as the stress-strain curves, and the microscopic mechanisms, such as the textures. Ali et al. employed typical crystal plasticity simulations to successfully forecast experimental stress-strain and texture data. The results from the crystal plasticity simulations were used to train the ANN models that predicted the material behavior [54].

In 2020, Lu et al. describe the cyclic plasticity of a typical carbon-doped iHEA, with the nominal composition of $C_{0.5}Co_{10}Cr_{10}Fe_{49.5}Mn_{30}$, using a CPFEM framework [39]. Uniaxial stress-controlled cyclic tests with a non-zero mean stress were conducted, at different loading stress levels, to study the cyclic plasticity of the iHEA. Microscopic characterizations were, furthermore, employed to shed light on microstructural evolutions during the cyclic deformations. Observed relationships between the macroscopic ratcheting and microscopic behavior of the iHEA were utilized to develop a crystal plasticity constitutive model. CPFEM modeling was carried out under the same loading conditions as the cyclic experiments. Comparison of the modeling results with experimental data revealed that the CPFEM model developed could accurately predict the ratcheting strain of the iHEA under the study [39].

*2.2.3. On Continuum Solid Mechanics*

Continuum mechanics involves the study of the physics of continuous materials. Continuum solid mechanics includes the study of the physics of continuous materials with well-defined shape when in rest. Elasticity describes the ability of solid materials to return to their rest shape after the stresses applied have been removed. Plasticity describes behavior of materials that permanently deform after sufficient applied stress.

Continuum-mechanics models, in the context of fatigue behavior of alloys, generally describe multi-scale behavior, such as being associated with dislocation growth or crack nucleation or development, such as during Stage II (see Figure 2.3.5 of the main manuscript). The Paris law for Stage-II crack growth can be considered as a continuum-mechanics model. The classical theory of continuum solid mechanics assumes a continuous distribution of mass within a body and that all internal forces are contact forces that act across a zero distance [56]. In the classical theory, the total strain has been decomposed into an elastic strain, to describe elastic behavior, and a plastic strain, to measure plasticity [57].

Classical continuum solid-mechanics models, in the context of fatigue behavior of alloys, include:

1. The Griffith theory for spontaneous crack growth up to a critical crack size.
2. The *J*-contour integral for the elastic-plastic body of a solid material.
3. The Eshelby theory for estimating the elastic field produced by an inclusion in a solid.
4. The Rice-Thompson model (or Rice's model) for predicting how a pre-cracked body responds to the applied load.

5. The Hertz contact theory for relating the indentation force with the indentation depth.
6. The Taylor hardening model for relating the dislocation density, $\rho$, with the hardness, $H$.
7. The theory of continuum elasticity to determine the stress field for a crack in an infinite, anisotropic linear elastic medium, as has been done for magnesium [58].

In addition, according to another continuum-mechanics model (related to nanoindentation), pop-in phenomena, indicating the homogeneous dislocation nucleation, occur when the maximum shear stress, $t_{max}$, reaches the theoretical critical shear stress. The maximum shear stress during nanoindentation occurs below the indent tip, described as follows:

$$\tau_{max} \equiv 0.31\left(\frac{6E_r^2}{\pi^3 r^2}P_{pop-in}\right)^{1/3} \tag{2.6.2.2}$$

Here $P_{pop-in}$ denotes the pop-in load, but $E_r$, and $r$ represent the reduced modulus and the tip radius, respectively [59, 60].

In the Griffith theory, materials are assumed to have pre-existing cracks. Griffith considered a large plate with a central crack under a remote stress and calculated the change in the energy with a crack size [61]. The surface energy associated with a crack, $\Delta U_{surf}$, is given by

$$\Delta U_{surf} = 4\,a\,t\,\gamma \tag{2.6.2.3}$$

where $t$ represents the plate thickness, and $2a$ denotes the crack length. The surface energy results from the decrease in the stored elastic energy, $\Delta U_{elast}$, which is given by

$$\Delta U_{elast} = -\frac{\pi a^2 t \sigma^2}{E} \tag{2.6.2.4}$$

where $\sigma$ denotes the stress applied. Combining Eqs. (2.6.2.3) and (2.6.2.4) gives rise to the total energy

$$\Delta U_{total} = \Delta U_{surf} + \Delta U_{elast} = 4\,a\,t\,\gamma - \frac{\pi a^2 t \sigma^2}{E}. \tag{2.6.2.5}$$

As the crack length increases, the total energy first increases and then decreases. Therefore, under a fixed stress, there is a critical crack size above which crack growth lowers the energy. This critical crack size can be determined from the condition

$$\frac{d\,\Delta U_{total}}{da} = 0. \tag{2.6.2.6}$$

The condition in Eq. (2.6.2.6) yields

$$a_{crit} = \frac{2\,E\,\gamma}{\pi\,\sigma^2}. \tag{2.6.2.7}$$

Eq. (2.6.2.7) captures the Griffith criterion. According to the Griffith theory, a pre-existing crack will grow spontaneously until Eq. (2.6.2.7) is satisfied [62].

Griffith portrayed a thermodynamic equilibrium picture, back in 1921, where the mechanical stability of a crack was formulated as a balance between a cracking force, the energy release rate, $G$, and the surface energy, $\gamma_s$, of the two fracture surfaces [63]. The driving force on a brittle crack can be obtained from elastic theory

$$G = K^2/E' \tag{2.6.2.8}$$

where $E'$ represents an appropriate elastic modulus, and $K$ a stress intensity factor characterizing the strength of the stress singularity at the crack tip [63].

The *J*-contour integral for a linear-elastic body of solid material, introduced by Rice in 1968 is an expression for determining potential energy changes involved in crack growth in a nonlinear elastic material, under monotonic loading. Rice showed that the *J*-integral captured the strain energy release rate in a cracked body (a non-linear elastic material), or the energy available at the tip of a crack to form new crack surfaces, as the crack extended [64,65]. The *J*-integral has since been extensively employed to the study of crack propagation, in linear-elastic materials, in nonlinear elastic materials, and especially in ductile materials. The integral has been established as an essential tool for classic-continuum solid mechanics[64]. The *J*-integral of a plane homogenous body is defined as [64, 65]

$$J = \int_{\Gamma} \left( W\, dy - T_i \frac{\partial u_i}{\partial x}\, ds \right) \quad (2.6.2.9)$$

as shown in Figure S12. Here, *W(x,y)* represents the strain energy density along the path, Γ (*x*,*y*), refer to the coordinate directions, ***n*** is outward-directed unit vector normal to Γ, ***T*** denotes traction vector acting on Γ, and ***u*** is a displacement vector along Γ. It is assumed that the body contains a straight-through crack parallel to the *x*-axis, as shown in Figure S12 [64,65]. The *J*-integral can be interpreted both as a fracture energy parameter and as a stress intensity parameter because *J* uniquely characterizes crack-tip stresses and strains. The integral can be evaluated along any arbitrary path enclosing the crack tip. The integral can be considered analogous to the shear modulus, *G*, for a linear elastic material. The *J* integral has been widely deployed to compute the energy flow to the crack tip, to estimate crack opening and as part of failure criteria for ductile materials [66,67].

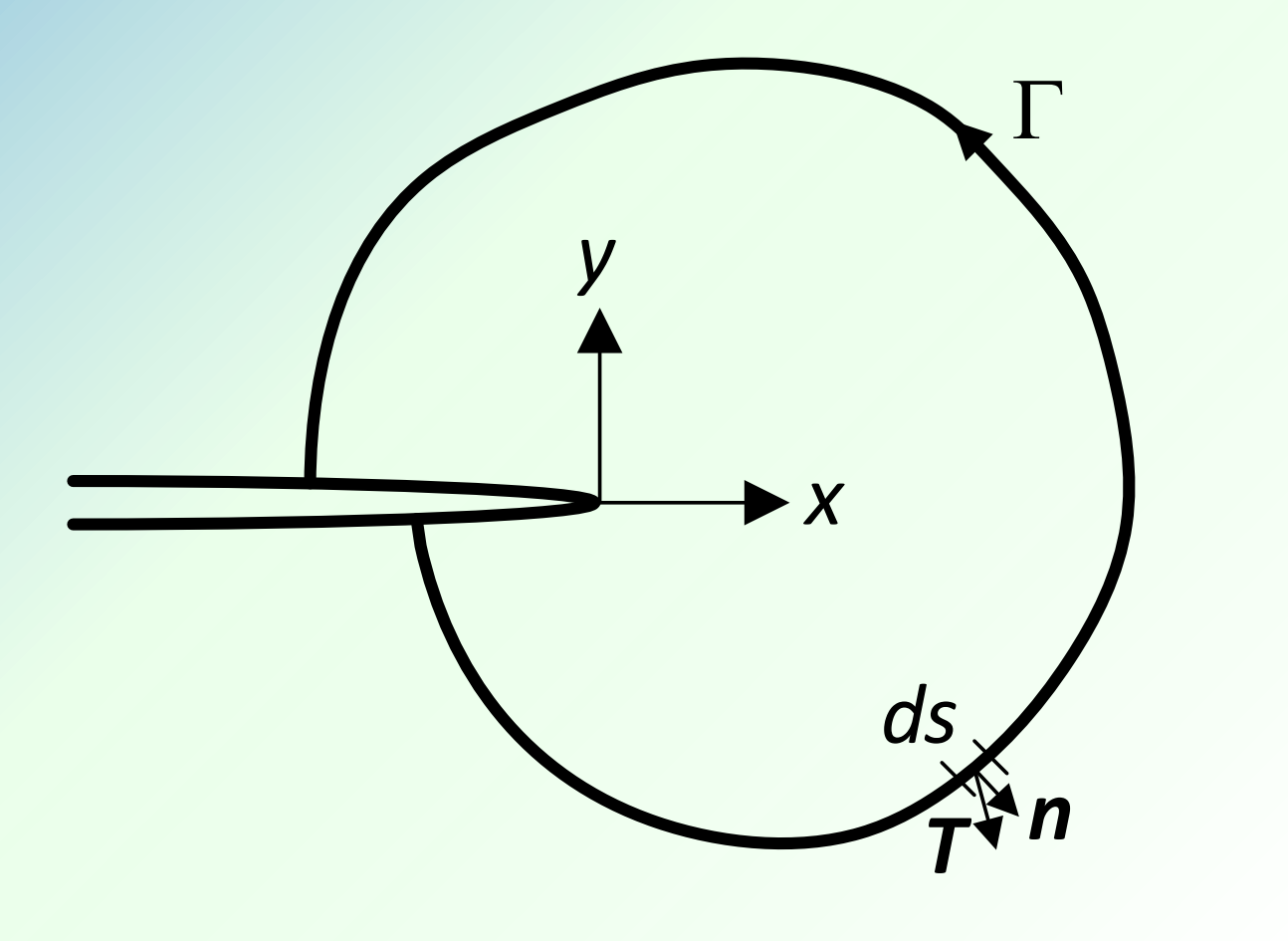

**Figure S12**: *J*-contour integral along the arbitrary path, Γ, enclosing a crack tip in a non-linear elastic material [66].

In [68], Dowling presented the analysis and discussion of existing estimates of the *J* integral for cracks in infinite bodies and extended to multiaxial loading. Dowling presented equations, which provide convenient estimates of the *J* integral for Ramberg-Osgood type elasto-plastic materials, containing cracks, and are subjected to multiaxial loading. The relationship between the *J* integral and the strain normal to a crack is noted to be weakly dependent on the state of the stress applied. But the relationship between the *J* integral and the stress normal to the crack is noted to be strongly dependent on the state of the stress applied [68].

The cyclic *J*-integral, the Δ*J*-integral, is a crack-tip parameter from elastic-plastic fracture mechanics, which can be used as a governing parameter for the description of fatigue-crack growth in metallic structures subjected to cyclic loading [69, 70,71]. The general expression for the cyclic *J* integral is given by [69, 70,71]:

$$\Delta J = \int_{\Gamma} \left( \Delta W\, dy - \Delta T_i \frac{\partial u_i}{\partial x}\, ds \right) \quad (2.6.2.10)$$

where changes in the strain energy density, $\Delta W$, is obtained from

$$\Delta W = \int_0^{\Delta\varepsilon_{ij}} \Delta\sigma_{ij}\, d\Delta\tilde{\varepsilon}_{ij}. \tag{2.6.2.11}$$

As usual, the Δ symbol, which precedes the stress tensor, $\sigma_{ij}$, the strain tensor, $\tilde{\varepsilon}_{ij}$, the components of the traction vector at the contour Γ, $T_i$, and the components of the displacement vector, $u_i$, designates changes of these quantities. The Δ*J* parameter has been successfully applied as an appropriate parameter for describing the cyclic crack-driving force in many studies of fatigue-crack propagation [72, 73, 74, 75, 71], provided that it has been defined in a proper way. Nevertheless, the Δ*J* parameter has been criticized both for fundamental as well as for practical reasons [71]. The most common objection against the Δ*J*-integral states that, since the *J* integral is based on the theory of nonlinear elasticity or (with limitation) deformation plasticity, it does not allow for unloading or non-proportional plastic deformation. Even so, it has been demonstrated that a properly defined Δ*J*-integral maintains path independence [71,73]. The Δ*J* integral also has some theoretical limitations, such as related to the stress-strain behavior of the material under the study. It has been demonstrated that path independence can be violated, if the material under the study is not completely cyclically stabilized [71,76]. Furthermore, in the presence of temperature gradients and temperature-dependent material behavior, strict compliance with the conditions of path independence may not be achieved. Some remedies have been proposed [71, 77, 78, 79, 80, 81]. A second category of limitations raised relates to crack closure [82, 71, 83, 84, 85, 86]. It may be difficult to establish a reference state as a necessary basis. There need to be no stresses at the crack flanks. Otherwise, there will be no path independence [71].

Figure S13 illustrates the use of the *J* integral as a fracture criterion [66].

Landes and Begley proposed the use of a *J*-integral-like parameter, a *C**-integral, for characterizing creep crack growth rates at elevated temperatures under steady-state creep conditions [66, 87, 88, 89]. The C*-integral is a path independent integral, similar to the *J* integral. Analogous to the *J*-integral, C* may also be interpreted as an energy release rate:

$$C^* = \int_\Gamma \left( W^*\, dy - \sigma_{ij} n_j \frac{\partial \dot{u}_j}{\partial x}\, ds \right) \tag{2.6.2.12}$$

where

$$W^* = \int_0^{\varepsilon} \sigma_{ij}\, d\dot{\varepsilon}_{ij} \tag{2.6.2.13}$$

Here, $\sigma_{ij}$ represents stresses, $\dot{\varepsilon}_{ij}$ strain rates, and $\dot{u}_j$ displacement rates.

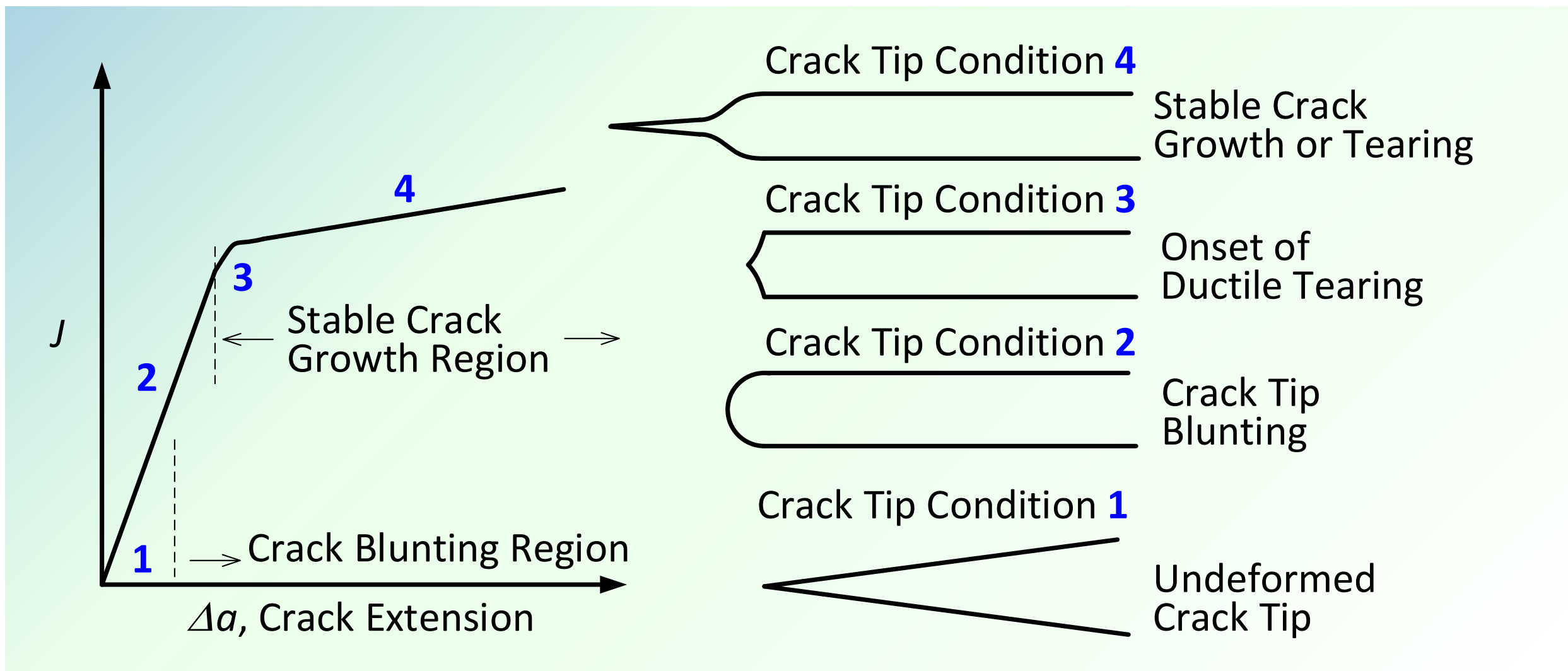

**Figure S13**: *J* integral as a fracture criterion [66].

Saxena has applied the *C** integral to experimentally characterize creep-crack growth of alloys subject to elevated temperatures [90]. Saxena has also developed a *C*(*t*)-parameter for characterizing the non-linear creep crack growth behavior over a wide range of creep and creep-fatigue conditions (small scale to steady-state creep) [90]:

$$C(t) = -\frac{1}{B}\frac{\partial U_t^*}{\partial a} \tag{2.6.2.14}$$

Here, *B* represents specimen thickness, but *a* crack length. $U_t^*$ is the instantaneous value of the stress-power obtained from considering the area under the load-deflection rate relationship. The quantity $\partial U_t^*$, with the subscript denoting that this value is at a fixed time *t*, represents the difference in energy rates (or power) supplied to two cracked bodies, with identical creep deformation histories, as they are loaded to different load or deflection-rate levels.

The $C_t$-parameter generalizes the stress power dissipation rate interpretation of *C** into a transient creep regime [91]. The $C_t$-parameter represents the instantaneous rate during a transient period of creep crack growth, whereas C* represents a steady state rate [91]. Under steady-state creep conditions, *C*(*t*) has been shown to reduce to the *C**-integral [90].

Saxena and his coworkers have also proposed a creep-fatigue crack growth (CFCG) model for the condition in which the time-dependent crack growth is mainly caused by creep deformation and cavitation damage [92, 93, 94, 95, 91]. The most widely used method for characterizing CFCG behavior involves partitioning of the total crack growth rate into a cycle-dependent part and a time-dependent part (see e.g., [96]). Saxena and Gieseka applied the concept of $C_t$ for characterizing crack growth rate during hold time [92,95]. Due to experimental limitations, it can be difficult to obtain the instantaneous values for the crack-growth rates, *da*/*dt* and $C_t$, during the hold period. Therefore, the measured average values, during the hold time, have been used:

$$\left(\frac{da}{dt}\right)_{avg} = \frac{1}{t_h}\left(\frac{da}{dN}\right)_{hold} \tag{2.6.2.15}$$

$$(C_t)_{avg} = \frac{1}{t_h}\int_0^{t_h} C_t \, dt \tag{2.6.2.16}$$

In [91], Yoon, Saxena, and Liaw further extend the model for predicting both the creep crack growth behavior and the CFCG behavior under trapezoidal loading. Here, $\Delta K$ and $(C_t)_{avg}$ are employed as correlating parameters for the cycle-dependent crack growth rate and for the time-dependent crack growth rate, respectively. The model is an extension of the one proposed by Saxena et al. for elastic-secondary creeping materials [92], now extended to the case when the effect of instantaneous crack tip plasticity of the material cannot be ignored. The model is considered appropriate for assessing the residual life and/or safe inspection interval of high-temperature components experiencing creep or creep-fatigue load [91].

The Eshelby inclusion problems refers to a set of problems involving ellipsoidal-elastic inclusions in an infinite elastic body. Analytical solutions to such problems were first derived by Eshelby in 1957, within the framework of continuum elasticity [97, 98]. The Eshelby inclusion theory elegantly describes the elastic field produced by an inclusion in a solid [99]. As noted in the main manuscript, the Eshelby theory predicts that the dependence of the displacement field, $u(r)$, on the distance away from an inclusion, $r$, is given by [99]

$$u(r) \propto \frac{1}{8\pi}(1-\nu)r^2, \qquad (2.6.2.17)$$

where $\nu$ represents the Poisson's ratio.

Kelly et al.[100], and later Rice and Thomson [101], developed models, based on continuum mechanics, for predicting how a pre-cracked body responded to an applied load. The Rice model addresses dislocation emission from a sharp crack tip within a so-called Peierls framework [102, 103]. The Peierls stress captures the force that is needed to move a dislocation through a lattice. The Peierls stress depends on the width of the dislocation, $W$, which represents a measure of the distance over which the lattice is distorted because of the presence of the dislocation, as well as on the distance between similar planes, $a$. The Peierls stress has been shown to depend on $W$ and $b$ as follows:

$$\tau_{p-n} \propto G\ e^{-2\pi W/b}, \qquad (2.6.2.18)$$

where $W = a / (1 - \nu)$, and $b$ represents the distance between equilibrium positions in the lattice (a period). The Rice model assumes that the displacement field across a slip plane emanating from a crack tip follows a periodic relation between the shear stress and atomic displacement [104]. Rice greatly advanced the fundamental understanding of brittle versus ductile fracture behavior by incorporating the Peierls framework into the description of the nucleation of dislocations [102]. The associated theoretical analysis is well established [100, 101, 102, 103, 105, 106, 107, 108, 109, 110]. The Rice theory addresses the nucleation of two-dimensional (2D) dislocation geometries [102, 103, 104]. According to the Rice theory [111], void (crack) growth can be described as follows:

$$\frac{dR}{d\varepsilon_p} = 0.28\ \exp\left[\frac{1.5\,\sigma_m}{\sigma_{eff}}\right], \qquad (2.6.2.19)$$

Here, $R$ represents the average radius of a void, $\varepsilon_p$ denotes the plastic strain, but $\sigma_m$ and $\sigma_{eff}$ are the mean stress and effective stress, respectively. Over the years, the Rice-Thomson model, and the Rice framework, have been extended, such as to account for the elastic anisotropy, the effect of crack blunting, 3D dislocation nuclei, successive nucleation events, and surface ledges formed at the crack tip [103, 106, 107, 108, 109, 110, 104].

According to the classic Hertz contact theory, the indentation force exhibits the following dependance on the indentation depth [112, 113]:

$$P = \frac{4}{3}\ E^*\ R^{1/2}\ h^{3/2}, \qquad (2.6.2.20)$$

Here, $P$ represents the indentation force, $R$ the radius of the spherical indenter, and $h$ its height. $E^*$ represents the reduced elastic modulus.

The Taylor work-hardening model encapsulates conventional continuum mechanics work-hardening theory (a continuum mechanics treatment of the stress-strain response). The Taylor hardening model specifies a relationship between the dislocation density, $\rho$, and hardness, $H$ [112, 114]:

$$H = 3\sqrt{3}\,\alpha\,\mu\,b\,\sqrt{\rho}, \qquad (2.6.2.21)$$

Here, $\alpha$, $\mu$, and $b$ represent an empirical constant, the shear modulus, and the Burgers vector, respectively. The value of the empirical constant is in the range of 0.3 – 0.5, in general. It is influenced by factors, such as the strain rate, dislocation density, alloy composition, and temperature [112].

Beltz et al. presented continuum analysis of isotropic elastic media, which suggested that the energy release rate for crack advance was somewhat larger than the Griffith value of 2 $\gamma_s$ at above a few Burgers vectors crack-tip radius, consistent with findings from atomistic simulations [104, 115].

Finally, there have been studies published, utilizing simulation and modeling, based on continuum mechanics, to decipher fracture surfaces and offer quantitative assessment of damage evolution characteristics [116,117]. Reference [116] addresses the role of martensitic transformation on damage and crack resistance in TRIP-assisted multiphase steels, whereas Ref. [117] seeks to model the damage and failure in multiphase high-strength DP and TRIP steels. However, according to [118], the complexity of phase constituents and microstructural characteristics with respect to various deformation levels hinders the development of reliable constitutive laws.

*3.2.4. On Peridynamic Modeling*

Peridynamics involves a nonlocal formulation of classical solid mechanics capable of unguided modeling of crack initiation, propagation, and fracture [66]. The peridynamic theory was introduced for the purpose of offering natural handling crack initiation, extension, and final failure of a body, without the need of supplementary methods [119, 120]. The theory was introduced to handle problems in classical solid mechanics with discontinuous fields. In contrast with continuum solid mechanics, which is based on partial differential equations, peridynamics is based upon integral equations [66], [64]. Thereby, one can avoid spatial derivatives, which are not defined at discontinuities, such as at crack surfaces. Peridynamics has been successfully applied to damage analysis of viscoplastic materials, to dynamic fracture and crack branching in glass, to damage in composite materials from the impact or shock loading and to nano-scale structures [67].

Through state-based peridynamics modeling, and utilizing an energy balance approach, Silling et al. derived a nonlocal $J$ integral with the following structure [64,56]:

$$J = \int_P \int_{B/P} (\boldsymbol{u}_x^T \boldsymbol{t}' - (\boldsymbol{u}_x')^T \boldsymbol{t})\, dV' dV + \int_{\partial P} \psi \mathbf{n}\, dA \qquad (2.6.2.23)$$

where

$$\mathbf{u}_x = \text{grad}\, \mathbf{u}(\mathbf{x} - \mathbf{V}t), \qquad (2.6.2.24)$$

$$\boldsymbol{u}_x' = \text{grad}\, \mathbf{u}(\mathbf{x}' - \mathbf{V}t), \qquad (2.6.2.25)$$

$\psi$ represents the free energy, $P$ denotes a closed, bounded subregion with a constant shape that translates through a reference configuration, $B$, with a velocity, $\mathbf{V}$, where $\partial P$ represents the boundary of $P$, and $\mathbf{n}$ is an outward-directed unit normal to $\partial P$. The subregion, $P$, contains points where there is energy dissipation that moves to the right with $\mathbf{V}$. There is a flux of material through the boundary, $\partial P$. We assume a steady-state motion of the form,

$$\mathbf{y}(\mathbf{x}, t) = \mathbf{x} + \mathbf{u}(\mathbf{x} - \mathbf{V}t), \qquad (2.6.2.26)$$

where **u** is some differentiable function [56]. Equation (2.6.2.9) relates the free energy lost in some dissipative process to quantities along the surface of a subregion, $P$, which contains the material where the dissipation is occurring.

The main motivation for studying the $J$ integral pertains to its ability to enable the computation of the energy consumed by a growing crack for any bond failure criterion, and for any dissipative mechanisms and nonlinearities that may be occurring, provided that these trends occur sufficiently close to the crack tip.

Figure S14 - Figure **S15** capture peridynamic modeling results from [64] for a central crack in a remote loading of $\sigma_0 = 1$ MPa and a half crack length of $a = 0.05$ m in an infinite plate. For consistency with [67], it is assumed that the plate material has Young's modulus of 72 GPa and Poisson's ratio of 1/3. Given the remote loading of 1 MPa in an exact analytical solution of the central-crack problem, one can calculate boundary stresses for the specimen shown in Figure S14 [64]. Figure S15 (left) visualizes the strain-energy densities for exact analytical displacements. Figure S15 shows the corresponding peridynamic formulation. The color map highlights the close agreement between the strain energy density for the exact analytical displacements and from the peridynamic modeling.

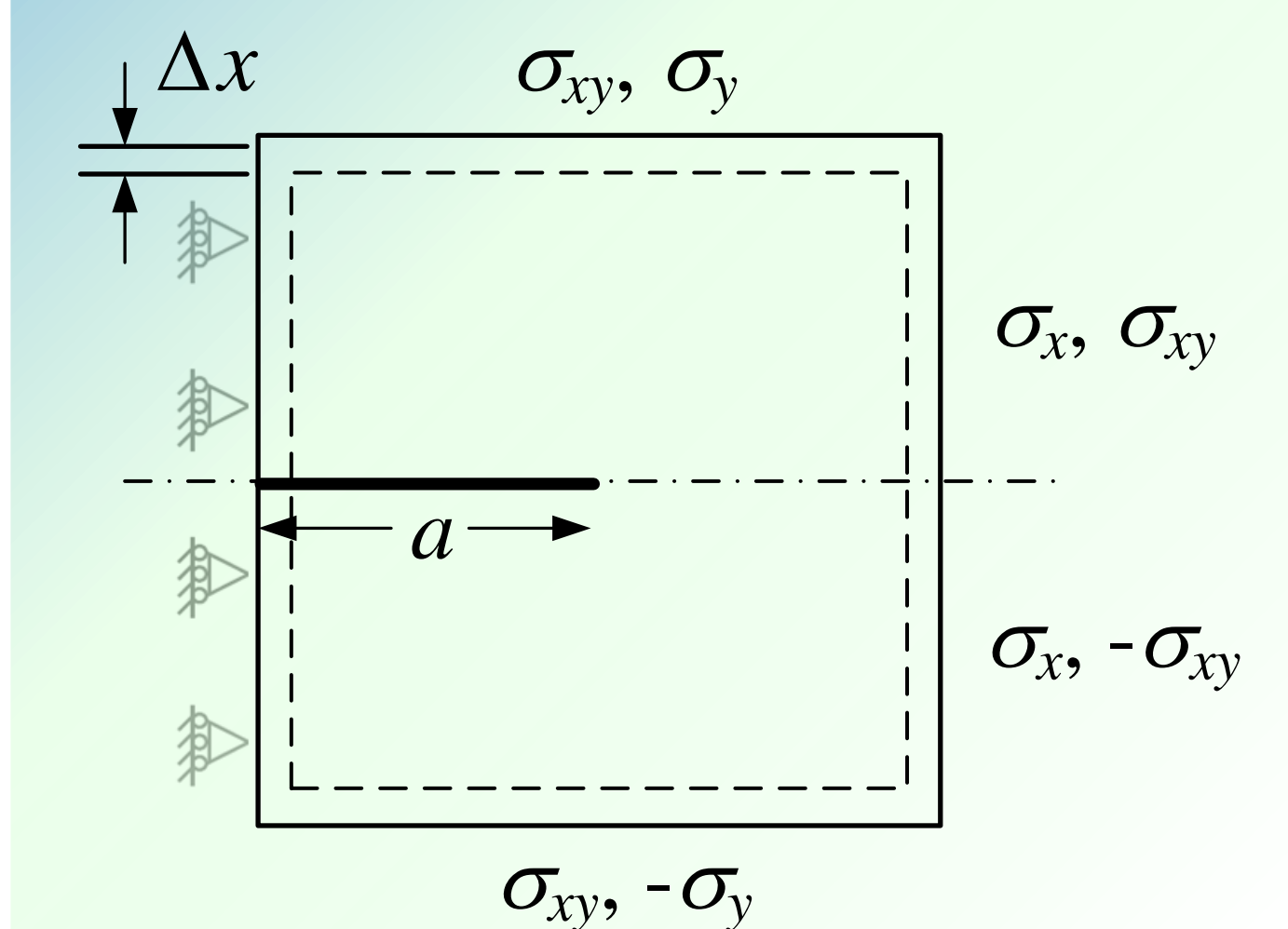

**Figure S14**: Analytical stresses and symmetry boundary conditions [64].

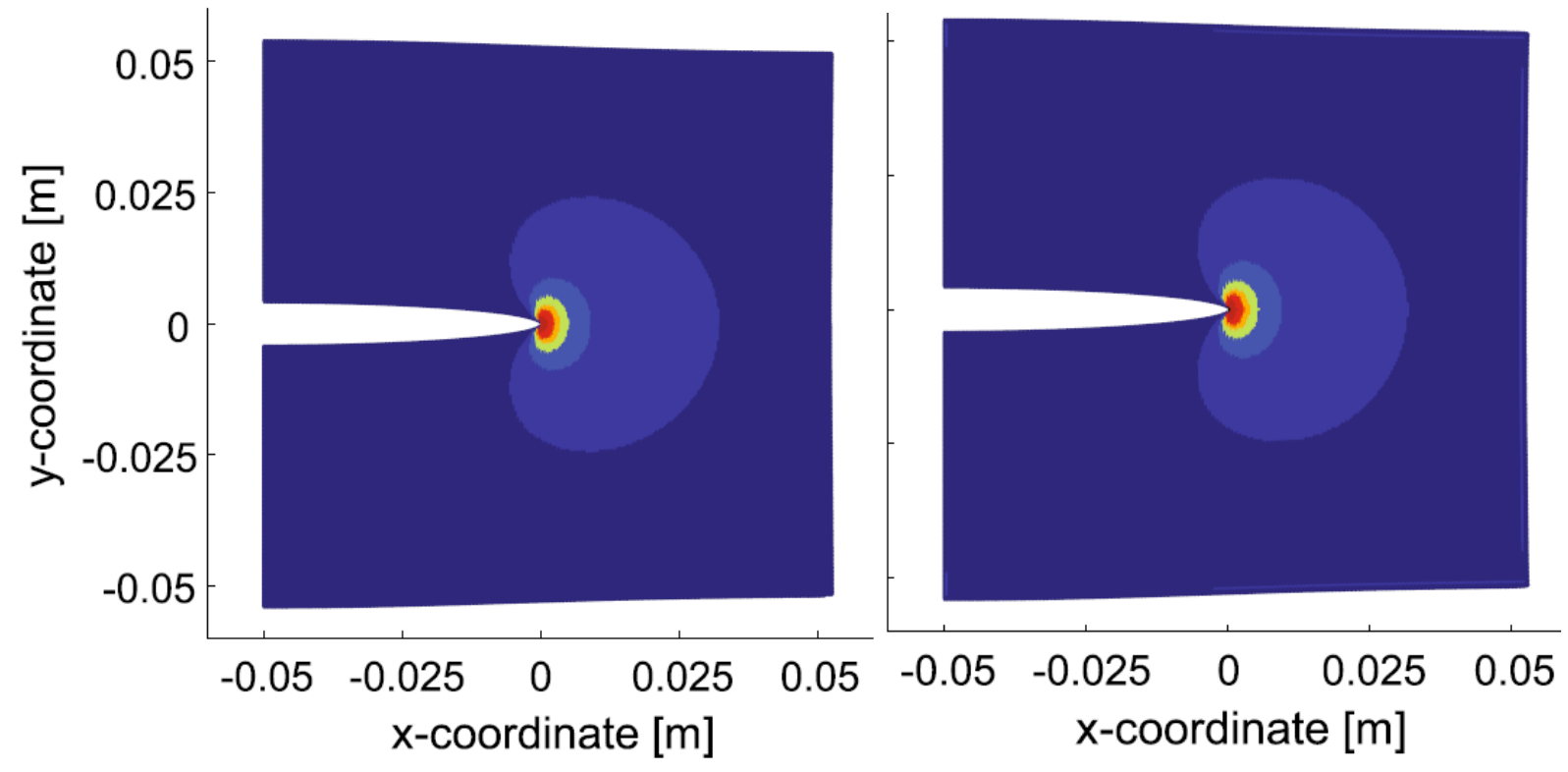

**Figure S15:** Strain-energy density for an infinite plate with a straight crack, under a uniform, uniaxial remote stress perpendicular to the crack [64]. Left: Exact analytical displacements with $250^2$ material points. Right: Peridynamic model with $250^2$ material points.

*3.2.5 On Phase-Field Modeling and Calculations of Phase Diagrams (CALPHAD)*

Phase-field simulations enable the estimation of elemental segregation (diffusion) and composition profiles, across individual phases. The segregation profiles of constituent elements across phases can be estimated, using phase-field modeling. Phase-field modeling seeks to predict such outcomes from the solidification process, esp. for polycrystalline materials with more than one grain. Phase-field modeling attempts to account for interfaces at grain boundaries, while still keeping the model complexity reasonable (smaller than for CPFEM or ab initio simulations). Figure S16 provides a high-level overview of the primary inputs and outputs to phase-field modeling. Table S2 presents a representative example with additional details.

CALPHAD modeling enables thermodynamic and phase-diagram calculations for multi-component alloy systems and provides input to microstructures (phase-field) modeling, as illustrated in Figure S16 [121, 122]. The CALPHAD calculations seek to minimize the Gibbs free energy, which is a function of temperature, pressure, and composition. The Gibbs free energy provides an estimate of the relative stability of phases for a given alloy composition [121]. Phase transformations can generally be determined from the minimization of the total free energy, which comprises of chemical, interfacial, elastic, plastic, and magnetic components [121].

Phase-field simulations can be used to model and predict microstructural evolution, such as morphological evolution of second-phase precipitates [123,124]. Phase-field (microstructure) simulations can be carried out for various processing conditions, such as laser processing, arc melting, or welding[123, 125, 126].

Prediction of segregation profiles and morphology variations with respect to processing conditions has been reported in the literature [123, 127]. For information on the mechanisms underlying solidification and solid-state phase transformations in alloys, and their association with phase-field modeling, refer to [128, 129, 130, 131, 132, 133, 134, 135]. Realistic predictions of microstructure evolution, using phase-field modeling, has been obtained for many commercial alloys, such as Al-based alloys, steels, and superalloys [121, 125, 127, 136, 137]. Phase-field methods have been widely applied to the Ni-base superalloys to simulate the morphological evolution of γ-precipitates, including the particle coarsening during heat treatments and the particle rafting under applied stress fields [123, 138, 139, 140, 141, 142, 143, 144]. Phase-field modeling is expected to provide an important theoretical basis for predicting microstructural evolution and for assisting with composition design of new HEAs. The prediction of microstructure evolution using phase fields is expected to help with reducing the number of experimental trials needed for alloy design [123].

Prediction of microstructure evolution using phase fields can be accomplished using a TQ interface with Thermo-Calc, the COMSOL Multiphysics software, or the MICRostructure Evolution Simulation Software (MICRESS®).

| | **Primary Inputs** | **Primary Outputs** | **Association** |
|---|---|---|---|
| **CALPHAD** | Temperature, pressure, composition | Plots showing phase fraction vs. temperature | Solidification pathway: Scheil's solidification simulation |
| **Phase Field Modeling** | Diffusion constants & kinetic coefficients of composition elements | Elemental segregation & composition profiles across phases | Modeling of dendritic & interdendritic regions in microstructures |

**Figure S16**: CALPHAD compared and contrasted with phase-field modeling (adapted from [121]). For additional information on linkage between CALPHAD and phase-field modeling, refer to [121, 124].

**Table S2**: Representative inputs and outputs from phase-field simulations of a eutectic HEA (adapted from [254]).

| Input | | Output |
| --- | --- | --- |
| **Parameter** | **Value and Unit** | |
| Diffusion constant of Co in Melt | $3.80 \times 10^{-5}$ cm$^2$/sec | Elemental segregation / Composition profiles across phases |
| Diffusion constant of Co in Dendrite | $1.10 \times 10^{-10}$ cm$^2$/sec | |
| Diffusion constant of Cr in the Melt | $1.98 \times 10^{-5}$ cm$^2$/sec | |
| Diffusion constant of Cr in Dendrite | $2.69 \times 10^{-10}$ cm$^2$/sec | |
| Diffusion constant of Ni in the Melt | $2.82 \times 10^{-5}$ cm$^2$/sec | |
| Diffusion constant of Ni in Dendrite | $4.25 \times 10^{-10}$ cm$^2$/sec | |
| Diffusion constant of Ta in the Melt | $7.27 \times 10^{-6}$ cm$^2$/sec | |
| Diffusion constant of Ta in Dendrite | $7.72 \times 10^{-10}$ cm$^2$/sec | |
| Diffusion constant of Mn in the Melt | $5.29 \times 10^{-6}$ cm$^2$/sec | |
| Diffusion constant of Mn in Dendrite | $3.23 \times 10^{-10}$ cm$^2$/sec | |
| Diffusion constant of V in the Melt | $9.66 \times 10^{-6}$ cm$^2$/sec | |
| Diffusion constant of V in Dendrite | $2.60 \times 10^{-10}$ cm$^2$/sec | |
| Kinetic coefficient ($\mu$) between phases liquid and primary dendritic phase | $30.00 \times 10^{-4}$ cm$^4$/(J sec) | |
| Surface energy between phases liquid and primary dendritic phase | $1.00 \times 10^{-3}$ J/cm$^2$ | |
| Domain size, grid size | $300 \times 300$ μm, 0.1 μm | |

The Cahn-Hilliard diffusion model was developed from a spinodal phase decomposition in binary alloy systems to describe the field of conserved concentration [144, 145, 146]. Spinodal decomposition is a mechanism by which a single thermodynamic phase spontaneously separates into two phases, i.e., without nucleation. Such decomposition occurs when there is no thermodynamic barrier to phase separation. The classical Cahn-Hilliard nonlinear diffusion model describes the process of phase separation, by which the two components of a binary fluid spontaneously separate and form domains pure in each component. If $c$ represents the concentration of the fluid, with $c = \pm 1$ indicating domains, then the Cahn-Hilliard equation can be written as

$$\frac{\partial c}{\partial t} = D\, \nabla^2 (c^3 - c - \gamma \nabla^2 c) \tag{2.6.2.27}$$

where $D$ is a diffusion coefficient (with units of length$^2$/time) and $\sqrt{\gamma}$ specifies the length of the transition region between the domains. $\gamma$ could be used for another definition befote (e.g., 2.6.2.5). Here, $\partial/\partial t$ represents partial derivative with respect to time, and $\nabla^2$ denotes the Laplace operator in $n$ dimensions. Moreover, the quantity, $\mu = c^3 - c - \gamma\nabla^2 c$, is recognized as a chemical potential. In [123], Li et al. applied the phase-field method to analyze coherent BCC/B2 microstructures that exist in Al-Ni-Co-Fe-Cr HEAs. These coherent BCC/B2 microstructures, spherical or cuboidal nanoprecipitates, or lend themselves to spinodal decomposition. Utilizing the Chan-Hilliard nonlinear diffusion equation, the authors constructed a 2D phase-field model using the COMSOL Multiphysics software to shed light on the coherent microstructural evolution of the BCC/B2 HEAs. Since the disordering-ordering transition of BCC and B2 phases is attributed to spinodal decomposition [147], the authors evoke the Chan-Hilliard equation to describe microstructural evolution in the elastically inhomogeneous HEA system [123]. The total free energy of the system includes the chemical free energy, $F^{ch}$, and elastic energy, $F^{el}$, where the chemical free energy consists of a bulk free energy and gradient interfacial energy [148]. The chemical free energy, $F^{ch}$, is expressed as

$$F^{ch} = N_V \int_V \left[ f(c) + \kappa(\nabla c)^2 \right] dV \qquad (2.6.2.28)$$

where $N_V$ represents the number of atoms per unit volume, $f(c)$ denotes the bulk free energy density, and $\kappa$ is a gradient energy coefficient that depends both on the composition, $c$, and temperature. The elastic contribution to the total free energy, $F^{el}$, is expressed, using the classical Khachaturyan's micro-elasticity theory [149, 150]. It was discovered that the spherical / cuboidal nanoprecipitations and the weave-like spinodal decomposition modeled were consistent with experimental results [123].

In [121], Shah et al. present a framework, which includes thermodynamic predictions, using CALPHAD, microstructure simulation using phase fields, and experimental validation, and apply the framework to the design of eutectic HEAs with seven components. Segregation of elements, in the interdendritic region, for the eutectic HEAs, during the solidification of the primary phase, is quantified and plotted [121].

*3.2.6 On Entropy-Based Models*

Unification of Newtonian mechanics with thermodynamics, using entropy as a link, eliminates the need for continuum mechanics models, where the second law of thermodynamics is usually imposed as an external constraint, but is not fulfilled at the material level, because the derivative of displacement relative to entropy is assumed to be zero [151]. The theory of elasticity, for example, assumes that there is no entropy generation at the material level. As a consequence, the deformation processes are reversible, which violates the second law of thermodynamics [151].

In[152], Jamal et al. applied the theory of unified mechanics to the prediction of fatigue life of $Ti_6Al_4V$ alloys. The authors show that using a theory of unified mechanics, the fatigue life can be predicted using physics, as opposed to empirical curve fitting methods [151].

In [153], Sosnovskiy et al. state main principles for the discipline of mechano-thermodynamics, a physical discipline that unites Newtonian mechanics and thermodynamics. The authors present analysis of more than 600 experimental results on polymers and metals, which are used for determining a unified mechano-thermodynamics function of limiting states, also known as fatigue-fracture entropy (FFE) states [151].

In [154], Young et al. propose using the cumulative distribution function derived from the maximum entropy formalism, and utilizing the thermodynamics entropy as a measure of damage, to fit LCF data of metals. The authors measure the thermodynamic entropy from hysteresis loops of cyclic tension-compression fatigue tests on aluminum 2024-T351[151]. In [155], Yun et al. demonstrate that fatigue-fracture entropy (FFE) is a material property independent of geometry and loading. The authors present entropy-damage indicators for metallic material fatigue processes obtained from three associated energy dissipation sources [151]. In [156], Osara et al. present a new fatigue-life predictor, based on ab initio irreversible thermodynamics, as a part of a degradation-entropy generation methodology for the characterization of systems and processes and for failure analysis [151]. In [157], Wang et al. propose an entropy-based failure prediction model for creep and fatigue, based on Boltzmann probabilistic theory and continuum damage mechanics. The authors present a new method to determine the entropy increment rate for creep and fatigue processes, develop a relationship between the entropy increase rate during creep processes and normalized creep failure time, and compare with experimental results [151].

In [158], Idris et al. present a study involving assessment of fatigue-crack growth rate for a dual-phase steel under spectrum loading, based on entropy generation. The authors simultaneously measure temperature evolution and crack length during fatigue-crack-growth tests until failure, to ensure the validity of the assessment, and develop a model that can determine the characteristics of fatigue crack growth rates, particularly under spectrum loading [151]. In [159], Sun et al. propose a study on the use of copula entropy for quantifying dependence among multiple degradation processes. The authors studied multivariate degradation modeling, to capture and measure dependence among multiple features, and adopted copula entropy, which was a combination of a copula function and information entropy, to measure dependence among different degradation processes [151].

Last but not least, in [160], Liang et al. report on effective surface nano-crystallization of an $Ni_2FeCoMo_{0.5}V_{0.5}$ HEA by rotationally accelerated shot peening. Through application of transmission electron microscopy, the authors noticed that deformation twinning and dislocation activities were responsible for effective grain refinement in the $Ni_2FeCoMo_{0.5}V_{0.5}$ HEA.

*3.2.7 Mesoscopic Modeling of Hydrogen-Induced Embrittlement*

In regards to continuum modeling in context with hydrogen-induced effects, there are two primary models to consider [161]: The hydrogen-enhanced localized plasticity (HELP) model, originally introduced in [162], and the hydrogen-enhanced decohesion model (HEDE) model, introduced in [163]. Origination and motion of dislocations, combined with the HELP effect, in the vicinity of a crack tip, can lead to local plasticity, because of the high concentration of dislocations. The difference between the HELP and the HEDE models relates to the fact that the HEDE model takes into account that the energy of appearance of the free surfaces of fracture reduces with the increase of the local hydrogen concentration. For solving many practical problems, the HELP approach tends to require great computational cost, as noted in [164]. In [161], Belvaev et al. also present a bi-continuum model for describing material fracture without any preliminary assumptions about the existence of microcracks or about a certain concentration of dislocation and their orientation. Their approach also differs from previous modeling of hydrogen embrittlement in that a parameter of crack resistance is introduced [161, 165].

The nucleation of micro-voids (the crack nucleation) can be considered as a continuum process, as noted above. In [166], Xie et al. address the hydrogen-induced slow-down of spallation (embrittlement) in CrMnFeCoNi HEAs under impact loading. A multi-scale statistical micro-damage mechanics model is introduced, based on the microstructural characterization and first-principles calculations. Hydrogen-retarded nucleation of micro-voids is attributed to hydrogen-vacancy complexes with great migration energy, but the formation of nano-twins with high resistance is noted to reduce their growth rate [166]. The authors note that under a hydrogen-rich environment, hydrogen affecting spallation behavior consists of interactions between hydrogen and several micro- and meso-scale time-dependent processes, such as the nucleation and evolution of microstructural defects like vacancies, twins, and micro-voids. This trend gives rise to a typical multi-scale coupling and collective evolution dynamics problem. The statistical micro-damage model is developed in an effort to gain the basic understanding of the underlying mechanism. An approximation is introduced at the macroscopic level, in order to provide an appropriate representation for multi-scale coupling of micro-damage and continuum damage [166]. The statistical micro-damage model draws upon a statistical micro-damage phase-space model introduced by Bai et al. [167, 168, 169]. By introducing the hydrogen effect on the nucleation and growth of micro-voids, the evolution of the micro-void density, $n(a,t)$, can be represented as

$$\frac{\partial n}{\partial t}+\frac{\partial(n\dot{a})}{\partial a}=\dot{n}_c, \qquad (2.6.2.29)$$

Here, $a$ represents the radius of a micro-void, but $\dot{n}_c$ and $\dot{a}$ denote the nucleation and growth rates related to hydrogen [166].

## 3.3 Microscopic Models

### *3.3.1 Modeling of Dislocation Dynamics*

With modeling of dislocation dynamics being multi-scale in nature, a comprehensive understanding of dislocation dynamics will need to involve some degree of atomistic modeling, and will need to reconcile continuum models with results of atomistic modeling [170, 171, 172]. In this subsection, we first describe models addressing crack initiation (nucleation of dislocations) and then present models addressing movement or propagation of dislocations.

For a nice overview of the elements of dislocation theory, such the definitions of full (perfect) dislocations, of partial dislocations, Shockley partial dislocations, edge dislocations, screw dislocations, or of circular dislocations, refer to Chapter 2 of [173]. According to analysis by Frenkel [174] from 1926, shear stress can be expressed in the form of sinusoidal variation in energy throughout a crystal lattice:

$$\tau=\tau_{max}\sin\left(\frac{2\,\pi\,x}{b_{eq}}\right) \qquad (2.6.3.1)$$

Here, $\tau$ refers to the applied shear stress, $\tau_{max}$ to the maximum theoretical strength of the crystal, $x$ to the distance atoms are moved, and $b_{eq}$ to the distance between equilibrium positions. In 1934, Taylor, Orowan and Polanyi independently proposed the existence of a lattice defect that would allow a lattice sub-cube to slip at much lower stress levels than prescribed by $\tau_{max}$ in Eq. (2.6.3.1). By introducing an extra half plane of atoms into the lattice, they demonstrated that atom bond breakage on the slip plane could be restricted to the immediate vicinity of the bottom edge of the half plane (referred to as a dislocation line) [173].

For an essential background on the dislocation theory, in context with HEAs, refer to nice exposition by Cantor [175]. Cantor notes that dislocations tend to pile up at grain boundaries and other obstacles, initiating secondary slip, creating dislocation intersections and increasing the number of obstacles to slip, leading to the formation of extensive dislocation tangles and forest dislocations [176, 177, 178, 179, 180,181]. According to Cantor, dislocation multiplication and interactions lead to classical parabolic Taylor work hardening, with a square root dependence of the flow stress on dislocation density, and a roughly inverse parabolic dependence on plastic strain [178, 179, 180,181]. The core structure tends to determine what type of dislocations are observed. In an FCC material, partial dislocations tend to be energetically favorable, compared to full dislocations. According to the classic dislocation theory, the free energy, $\Delta G$, required to nucleate a circular dislocation with a radius, $r_o$ amounts to [182]

$$\Delta G = 2\pi r_o W_{dis} + \pi r_o^2 \gamma' = \pi r_o^2 b \tau_{max} \qquad (2.6.3.2)$$

Here, $W_{dis}$ represents the dislocation-line energy, $\gamma'$ represents the stacking fault energy, and $b$ the Burgers vector (which comes out as 0.255 nm in case of the CoCrFeMnNi HEA [60]).

As noted above, Rice greatly advanced the understanding of brittle versus ductile fracture behavior by incorporating the Peierls framework into the description of the nucleation of dislocations [102]. The theoretical analysis associated with the *Rice model* is well established [100, 101, 102, 103, 105, 106, 107, 108, 109,110]. Rice derived that the instability for a complete lattice dislocation occurs when

$$G = \gamma_{usf}, \qquad (2.6.3.3)$$

that is, the energy release rate equals the unstable stacking fault (USF) energy, $\gamma_{usf}$ [104].
Patriarca et al. addressed slip nucleation in single-crystal FeNiCoCrMn HEA and presented an atomistic-modified Peierls-Nabarro modeling formalism for a dislocation core [183]. It is of fundamental importance to precisely determine the minimum stress required to move a dislocation, defined as the critically resolved shear stress (CRSS), but also to accurately predict this CRSS through the theory and simulation and without empirical constants. The CRSS is an important parameter in the crystal plasticity, dislocation dynamics, or other continuum mechanics treatments of the stress-strain response, and of fundamental importance to multi-scale modeling [183]. According to the Peierls-Nabarro formalism, the CRSS for the slip, $\tau$, is given by [183]

$$\tau = \frac{1}{b} \max\left\{\frac{d\,E_{misfit}}{du}\right\}, \qquad (2.6.3.4)$$

Here, $b$ represents the magnitude of the Burgers vector, but $E_{misfit}$ a misfit energy term, which depends on the equilibrium lattice constant, $a_0$, the unstable stacking fault energy, $\gamma_{us}$, and the intrinsic stacking fault energy, $\gamma_{isf}$ [183]. For complete description of the Peierls-Nabarro formalism, and the energy terms involved in Eq. (2.6.3.4), refer to [184, 185, 186, 187,188].
Patricia et al. pointed out that since most of the previous experiments had been conducted on polycrystalline samples, it was difficult to determine the CRSS, again a fundamental quantity, because the stress state at the grain level was not known. However, a way to precisely pinpoint the CRSS was provided (an intrinsic measure of the CRSS for clip obtained), by choosing the $[5\bar{9}1]$ crystal orientation such that the {111}<110> slip system with the highest Schmid factor nucleated. Further, Patricia et al. undertook atomistic calculations to obtain the lattice constant, $a_0$, the unstable stacking fault energy, $\gamma_{us}$, and the intrinsic stacking fault energy, $\gamma_{isf}$, associated with the generalized stacking fault energy (GSFE) curve. Using the calculated lattice parameter, $a_0$, the stacking fault energy parameters of $\gamma_{us}$ and $\gamma_{isf}$, the shear moduli, $G$, Patriarca et al. established CRSS for slip to be 178 MPa, which was in close agreement with experiments [183]. Patriarca et al. note the close agreement between the experiment and theory and confirm the efficacy of their modeling as well as its potential to be applied to HEAs of interest other than the single-crystal FeNiCoCrMn HEA [183].
The critical shear stress for the dislocation motion in HEAs has been shown to be significantly higher than that in conventional metals due to strong solute strengthening [189]. Further, in terms of the nucleation behavior, dislocations in HEAs have been shown to lend themselves to multi-scale bow-out configurations, which look different from the simple bow-out configurations in conventional metals, such as Cu [190,191].
As Xiao et al. note, in context with dislocation nucleation in the HEA[190], although a solid-solution model can well fit the experimentally measured yield strength in HEAs [192, 193, 194], evidence from dislocation nucleation is still crucial to validate the modeling work and reveal the physical origin of the high strength in HEAs, where nucleation is the limiting process. Dislocation nucleation can in general be characterized both by thermal (e.g., athermal stress) and thermal activation parameters (e.g., activation energy and volume) [190, 195, 196]. In the case of nanoindentation, homogeneous dislocation nucleation is believed to occur, and is believed to provide an upper bound to the ideal strength of HEAs [190, 197]. For HEAs with FCC structures, there are generally four nucleation scenarios to consider: (1) the nucleation of a single leading partial dislocation, (2) the nucleation of a full dislocation with complementary leading and trailing partial dislocations, (3) deformation twinning, and (4) FCC – HCP martensite transformation [190, 198]. In the case of these four scenarios, the nucleation of the first leading partial dislocation is usually the most difficult and rate-limiting stage [190, 199]. Xiao et al. studied the homogeneous and heterogeneous nucleation of dislocations, in particular, of the 1/6 <112> partial dislocation, in the representative FCC HEA of CoNiCrFeMn by combining molecular dynamics simulations and continuum mechanics modeling. Xiao et al. apply the dislocation theory based on continuum mechanics to explore the strain rate, temperature, and diameter-dependent yield strength of the FCC CoNiCrFeMn HEA nanowires by directly fitting the activation energy and athermal stress for the surface dislocation nucleation from the atomistic simulations. The authors conclude that as compared to conventional metals, such as Cu, HEAs have a relatively low activation energy but require a high athermal stress for dislocation nucleation [190].

The *Frenkel model* addresses the prediction of homogeneous nucleation of dislocations [174, 60]. According to the classic dislocation theory, the free energy, $\Delta G$, required to nucleate a circular dislocation with a radius of $r_o$ amounts to [182]

$$\Delta G = 2\pi r_o \, W_{dis} + \pi r_o^2 \gamma' = \pi r_o^2 b \tau_{max} \qquad (2.6.3.5)$$

Here, $W_{dis}$ represents the dislocation-line energy, $\gamma'$ denotes the stacking fault energy, and $b$ the Burgers vector (which comes out as 0.255 nm in the case of the CoCrFeMnNi HEA [60])

According to the *Defactant model*, the dissolved hydrogen can be assumed as defactants that reduce the formation energy of defects [60, 200, 201, 202].

An accurate description of dislocation motion, and their interactions with internal boundaries, is more complicated than prescribed by classical continuum mechanics [203]. One reason may pertain to the role of the lattice structure at incoherent interfaces [203,204]. To obtain further insight into interactions of dislocations with internal boundaries, various approaches have been pursued, based on both experiments and simulations. Broadly speaking, dislocations and internal boundaries provide local short-circuit diffusion channels to sweep solutes into or out of volume [205, 206, 203].

In terms of the effect of hydrogen-induced surface steps on the nanomechanical behavior of a CoCrFeMnNi high-entropy alloy, it has been shown that for a linear dislocation, dissolved hydrogen atoms can segregate around dislocations and effectively increase the dislocation core radius and therefore reduce the dislocation-line energy [60, 207,208].

The *Paris law* [Eq. (2.3.5) from the main manuscript)] is an example of a multi-scale model addressing the movement or propagation of dislocations (cracks), In addition, Reference [112] addresses the nucleation of Shockley partial dislocations, Hurth and Lomer-Cottrell dislocations lock, and movements of stacking faults. In Reference [112], the occurrence and development of plasticity were analyzed by quantifying the dislocation density and determining the mechanism of dislocation evolution. Moreover, dislocation emission has shown to be critical in the deformation and fracture process of metallic nanowires [57].

Dislocation motion in HEAs has been further studied, through the experimental or theoretical work, in [190, 209, 210].

*3.3.2 Micromechanics Models for Elastic Constants of Single-Phase Polycrystals*

According to the *Voigt model*[211], the strain in each crystal comprising the polycrystal is assumed to be uniform and equal to the macrostrain of the polycrystal[212]. While perhaps only peripherally related to fatigue, the diffraction values for the Young's modulus and the Poisson's ratio, $E_x$ and $v_x$, are assumed equal to the mechanical values, $E$ and $v$ [212]. The diffraction compliances, $S_1$ and $S_2$, are assumed to be related to the single-crystal elastic constants, $c_{ii}$, as follows[212]:

$$\frac{S_2}{2} = \frac{1+v_x}{E_x} = \frac{15}{7\,c_{11}+2c_{33}-5c_{12}-4c_{13}+12c_{44}}, \tag{2.6.3.6}$$

$$S_1 = -\frac{v_x}{E_x} = \frac{3}{(2\,c_{11}+c_{33}+2c_{12}+4c_{13})} \times \frac{(4c_{44}-c_{11}-c_{33}-5c_{12}-8c_{13})}{(7c_{11}+2c_{33}-5c_{12}-4c_{13}+12c_{44})}, \tag{2.6.3.7}$$

In the *Reuss model*[213], the stress in each crystal comprising the polycrystal is assumed to be uniform and equal to the macro-stress of the polycrystal[212]. The Reuss model expresses the diffraction compliances of $S_1$ and $S_2$, in terms of the compliances of the individual crystals, $s_{ij}$, as follows[214]:

$$\begin{aligned} \frac{S_2}{2} = \frac{1+v_x}{E_x} = \; & \frac{1}{2}(2\,s_{11}-s_{12}-s_{13}), \\ & -\frac{1}{2}(5s_{11}-s_{12}-5s_{13}+s_{33}-3s_{44})\ \cos^2\phi \\ & +\frac{3}{2}(s_{11}-2s_{13}+s_{33}-s_{44})\ \cos^4\phi \end{aligned} \tag{2.6.3.8}$$

$$\begin{aligned} S_1 = -\frac{v_x}{E_x} = \; & \frac{1}{2}(s_{12}+s_{13}), \\ & +\frac{1}{2}(s_{11}-s_{12}-s_{13}+s_{33}-s_{44})\ \cos^2\phi \\ & -\frac{1}{2}(s_{11}-2s_{13}+s_{33}-s_{44})\ \cos^4\phi \end{aligned} \tag{2.6.3.9}$$

where $\phi$ represents the angle between the normal of the diffraction plane (*hkl*) and the *c*-axis[212]. The mechanical elastic constants can be calculated, by taking the average, which comes out as 1/3 for $\cos^2\phi$ but as 1/5 for $\cos^4\phi$ [212].

The *Voigt-Reuss average model* is obtained as the average of the Voigt and the Reuss models. Hill has demonstrated that the average of the Voigt and Reuss models represents bounds on the elastic modulus of polycrystalline materials [215].

For the information about the *Kröner's model* [212, 216], refer to Section 2.6.3.3 below.

In Ref. [217], Diao et al. review and summarize the elastic properties of HEAs, which is essential for purpose of providing a fundamental study of the mechanical and fatigue behavior of HEAs and for identifying next-generation biomaterials based on HEAs[218, 219, 220].

*3.3.3 Micromechanics Models for Elastic Constants of Multi-Phase Polycrystals*

Hill has demonstrated that the *Voigt-Reuss average model* provides bounds for the elastic moduli of polycrystalline materials, as noted above[215, 212].

Again, while perhaps only peripherally related to fatigue, the strain tensor for a crystal in a polycrystal, subjected to applied stress, $\sigma_{\widehat{kl}}$, can be specified, in accordance with *Kröner's model*[216], as

$$\varepsilon_{ij} = \sum_{k=1}^{3}\sum_{l=1}^{3}(s_{ijkl}+t_{ijkl})\sigma_{\widehat{kl}}, \tag{2.6.3.10}$$

were $s_{ijkl}$ represents compliances for the single crystals. But the terms, $t_{ijkl}$, denote an additional term resulting from constraints imposed by neighboring grains. The terms, $t_{ijkl}$, are obtained as a function of single-crystal compliances through the application of the Eshelby inclusion theory [212, 221].

One can also compute the mechanical elastic constants of polycrystals from the single-crystal elastic constants by employing an equation derived by Kneer [212, 222].

*3.3.4 Molecular Dynamics*

Figure S17 provides high-level overview of a molecular dynamics (MD) simulator [223]. The Large-scale Atomic/Molecular Massively Parallel Simulator (LAMMPS), developed by Plimpton et al., is a classical MD simulator, with the high-level structure illustrated in Figure S17, aimed at materials modeling [223,224]. LAMMPS more specifically represents a collection of programs usually used for molecular dynamic simulations. The LAMMPS software supports potentials for solid-state materials, such as metals or semiconductors, and soft matters, such as biomolecules or polymers and coarse-grained or mesoscopic systems. The LAMMPS software can be used to model an ensample of atoms or, more generically, as a parallel particle simulator at the atomic, meso or continuum scale. The programs comprising the LAMMPS software can be used to calculate diffusion and rearrangement of materials. These tend to be microscopic, short-time-frame information [223, 224].

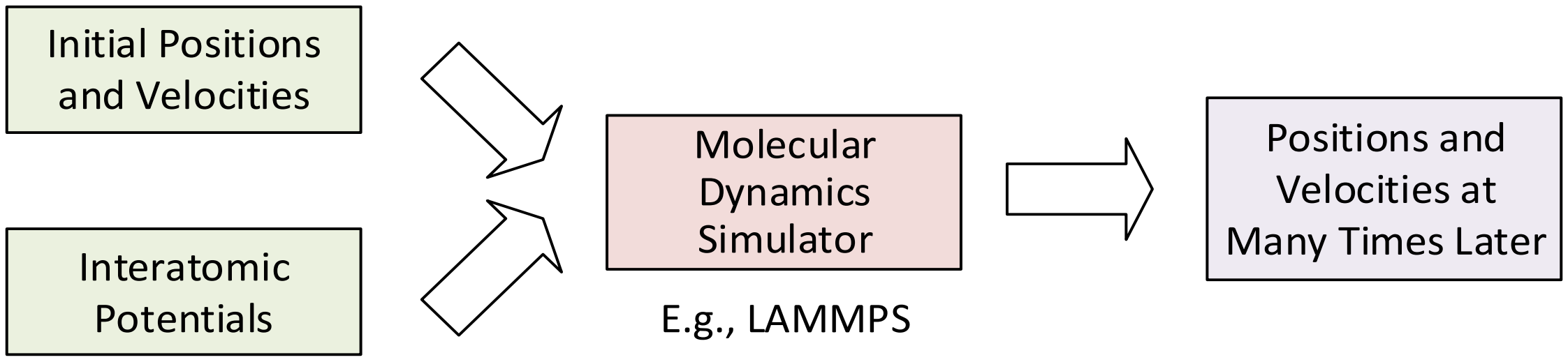

**Figure S17**: High-level overview of a molecular dynamics simulator (adapted from [225]).

As many classical MD simulators, LAMMPS expects the initial positions of the particles to be provided as input. LAMMPS can automatically generate initial positions for simple crystal structures, or accept data files provided by user for materials with complex atomic structures. When presented with initial positions and velocities of the particles, together with corresponding interatomic potentials, LAMMPS can simulate a wide variety of materials. When simulating a specific material, LAMMPS can utilize statistical mechanics to convert atomic trajectories into macroscopic properties such as temperature, volume, pressure and density. LAMMPS provides the user with the ability to specify which properties are desired as outputs. Statistical mechanics equations, built into the LAMMPS software, allow LAMMPS to output of a variety of macroscopic properties. LAMMPS offers tools for calculating elastic constants of materials both at 0 K as well as at finite temperatures. The elastic constants can be calculated by deforming a simulation box in six directions and measuring the resulting changes to a stress tensor. LAMMPS is capable of determining the elastic constants of simulated materials at a variety of temperatures. Properties, such as the bulk modulus, shear modulus, Young's modulus and the Poisson ratio, can be obtained from the elastic constants. Accuracy of the simulated properties depends mostly on the quality of the atomic potential that is provided as input. Determining the elastic constants through LAMMPS can provide access to a variety of mechanical or fatigue properties. LAMMPS can simulate most materials for which the crystal structure and interatomic potential file can be obtained. From the simulation, important thermodynamic, structural and mechanical properties of the materials can be estimated [225].

ReaxFF is a bond-ordered reactive force field that captures a practical approach to molecular dynamics simulations of large-scale, such as 1,000s of atoms, reactive chemical systems [226]. Classical MD makes use of classical force fields (potentials) that capture interactions between pairs of atoms or molecules. These force fields are generally non-reactive, which means that the bonds can vibrate and stretch but without breaking. But reactive force field potentials support making and breaking of bonds, and therefore can support reactions. Towards such end, traditional force fields are unable to model chemical reactions because of the requirement of breaking and forming bonds (a force field's functional form depends on having all bonds defined explicitly). ReaxFF, in contrast, utilizes a general relationship between bond distance and bond order, on one hand, and between bond order and bond energy, on the other hand, that leads to proper dissociation of bonds to separated atoms. Other valence terms present in the force field, such as angle and torsion, are specified in terms of the same bond orders. So all these terms can gradually go to zero during bond breakage. To be specific, ReaxFF eschews explicit bonds in favor of bond orders, which allows for continuous bond formation/breaking. ReaxFF supports a class of potentials, a class that includes Coulomb and van der Waals potentials, for purpose of describing nonbond interactions between all atoms, with no exclusions [226]. The modified embedded-atom-method (MEAM), the angular dependent potential (ADP), the charge optimized many body (COMB) and the embedded-atom method (EAM) entail other classes of potentials used for molecular dynamics simulations [227]. Even broader class of potentials for molecular dynamics simulations can be obtained through the NIST's Interatomic Potentials Repository [225,227]. Farkas et al. developed a set of EAM interatomic potentials to represent highly idealized FCC mixtures of Fe-Ni-Cr-Co-Al at near-equiatomic compositions [228]. Potential functions for the transition metals and their crossed interactions were taken from the authors' previous work on Fe-Ni-Cr-Co-Cu [229], whereas cross-pair interactions involving Al were developed, using a mix of the component pair functions fitted to know intermetallic properties. The authors noted that significant short-range ordering appeared in binary equiatomic random FCC mixtures containing Al. The potentials developed were employed for predicting the relative stability of FCC quinary mixtures, as well as ordered $L1_2$ and B2 phases, as a function of the Al content. The authors indicated that their predictions were in qualitative agreement with experiments.

When examining elastic fields at the atomic level, the validity of the assumptions of continuum mechanics needs to be verified, especially if the atoms are modeled as discrete particles. In regards to the extension of the Eshelby theory, summarized in Eq. (2.6.2.17), to atomic-level stresses in liquid without external stresses, Wu et al. conducted pioneering work through the application of MD simulations [230]. Wu et al. demonstrated that atomic-level stresses in liquid comply with Eshelby theory, but have an additional component, one subjected to an exponential decay with distance [230].

In regards to applications of molecular dynamics to analyze the properties of alloys, in particular of HEAs, Tong et al. probed the local lattice distortion in MEAs and HEAs, and experimentally inspected the continuum assumption for the local strain field induced by a relatively large-size mismatch in Mn- and Pd-HEAs free from external perturbations [99]. Analogous to Eq. (2.6.2.17), Tong et al. demonstrate that the strain field caused by the atomic-size mismatch follows the law of $1/r^2$ with $r$ here representing radius of the strain field [99].

In regards to applications of MD to HEAs, in context with ML, the work of Schmidt et al. suggests that ML performs better than traditional MD for phase predictions [231]. Reference [232] also presents a model that through the use of the high-throughput MD computation of the enthalpies of formation of binary compounds, predicts specific combinations of elements most likely to form single-phase HEAs.

Using MD simulation, Qi et al. explored the influence of crystallographic orientation on the mechanical properties and microstructure evolution of the single-crystal CoCrFeMnNi FCC HEA and quantified by nanoindentation [112]. The authors simulated nanoindentation using MD to study the effects of crystallographic orientation on the mechanical properties of the single-crystal FCC HEA structure. Qi et al. employ a MEAM potential proposed by Choi et al. [233]. In the MEAM potential method, the interaction force between the nano-indenter and the model is purely repulsive[234]:

$$F(r') = \begin{cases} -K\,(R - r')^2 & r' < R \\ 0 & r' \geq R \end{cases}, \tag{2.6.3.11}$$

Here, $K = 10$ eV / A$^{o3}$ represents the stiffness of the nano-indenter, $r'$ is the distance between the atom in the model and the center of the indenter, and $R$ denotes the radius of the indenter (for additional specifics, refer to Table 1 of [112]). The authors find that there exist clearly differences in mechanical behaviors for the various crystallographic orientations. In case of the elastic stage of the P-h curve, the results of MD simulation are fairly consistent with the Hertz contact fitting theory, per Eq. (2.6.2.20), which speaks to the accuracy of the results [112].

In regards to the application of molecular dynamics to study the multi-scale degradation process that ultimately leads to fatigue failure in HEAs, Xiao et al. similarly studied the homogeneous and heterogeneous nucleation of 1/6 <112> partial dislocations in a representative CoCrFeMnNi FCC HEA, by combining MD simulations and continuum mechanics modeling [190]. Models with different dislocation radius, under various applied shear stresses, are optimized, using LAMMPS with a second nearest-neighbor MEAM potential for the CoCrFeMnNi [233] and an EAM potential for Cu [235]. A dislocation loop with initial radius of r = 4, 8, 12, 16, 20, 24, 28, 32, 36, or 40 Å is provided to each model [190]. A shear stress is then applied to the dislocation loop by adjusting the shear strain in each of the models [195,196]. The athermal stress for the nucleation of the dislocations is then determined as the shear stress corresponding to the activation energy equaling zero [190, 196].

## 3.4 Nanoscopic Models

### *3.4.1 Atomistic Aspects of Fracture*

The first atomistic studies of fracture, by Chang in 1970 and Sinclair et al. in 1972, demonstrated that crack propagation is only possible on certain crystallographic planes within a crystal and that crack tip bonds break after being stretched nearly to their elastic limit [236]. Atomistic analysis has also revealed that discreteness of the lattice can manifest itself in a so-called lattice trapping effect[237]. The magnitude of the trapping can strongly depend on the force law, that characterizes the atomic interactions [238], and can explain thermally activated subcritical crack growth [239, 240].

Since the brittle-to-ductile transition (BDT) in semi-brittle materials is largely affected by the relation between loading rate of a crack and the rate of plastic deformation near the crack tip, the BDT tends to be influenced by dislocation mobility [241, 242], which for semi-brittle materials tends to be too sluggish to be reasonably dealt with by atomistic simulations [243]. With that said, many processes, like dislocation nucleation and dislocation multiplication, are involved in the BDT, and some of these processes can be investigated through atomistic modeling [63]. For instance, atomistic aspects of dislocation crack tip interaction can be studied for inherently ductile materials, like FCC materials, where dislocation motion can be captured in time scale accessible to molecular dynamics [63].

In contrast, with the exception of [244], not much has been reported, regarding the atomistic mechanisms of dislocation nucleation in HEAs, despite the importance of atomistic mechanisms in determining the mechanical behavior in HEA materials[183]. In [190], Cui et al., moreover, employed large-scale atomistic simulations, in order to gain understanding of dislocation-mediated deformation mechanism in cold-welded nanowires. But in [57], Patriarca et al. studied slip nucleation in single-crystal FeNiCoCrMn HEAs, and presented an advanced atomistic modified Peierls–Nabarro modeling formalism, as previously mentioned above. Patriarca et al. undertook atomistic calculations to obtain a lattice constant, $a_0$, and important energy terms, such as the unstable stacking fault energy, $\gamma_{us}$, and the intrinsic stacking fault energy, $\gamma_{isf}$, associated with a generalized stacking fault energy (GSFE) curve. The GSFE curve captures free energy differences between a crystal fault and the bulk lattice with various degrees of shear displacements [183,183]. The unstable stacking fault energy, $\gamma_{us}$, represents the fault energy per unit area required to nucleate a slip. The intrinsic stacking fault energy, $\gamma_{isf}$, is equivalent to the differential of the HCP and FCC free energy per unit area [245]. By utilizing a well-developed modified Peierls–Nabarro (PN) formalism for cubic metals [183, 184, 185, 186, 246, 187], Patricia et al. calculated the critically resolved shear stress theoretically, producing close agreement with the experimentally observed value [188].

*3.4.2 Monte Carlo Modeling*

Monte Carlo methods comprise a broad class of computational algorithms that rely on repeated random sampling to obtain numerical results. The core idea entails using the randomness to solve problems that may be deterministic in principle. Monte Carlo simulation methods tend to find the greatest use, when applied to problems from mathematics or the physical sciences that are very difficult – or even impossible – to solve using other methods. Monte Carlo simulation methods do vary, but tend to include the following key steps:

1. Define a domain for the inputs allowed.
2. Generate inputs randomly from a probability distribution defined over the domain.
3. Perform deterministic computations on the inputs.
4. Aggregate the results.

In[183], Feng et al. employed Monte Carlo (MC) modeling to determine atomic arrangements in B2 precipitates and the FCC matrix of $Al_{0.5}$CoCrFeNi HEA. Feng et al. start with simulation cells representing systems of interest according to experimental results. The authors then determine the most energetically-favorable atomic arrangement by applying a Metropolis MC60 algorithm, during which randomly-selected pairs of atoms are swapped. Such swaps are accepted with the probability of [247]

$$p = \min\left\{1, \exp\left(\frac{-\Delta E}{k_B T}\right)\right\}, \qquad (2.6.4.1)$$

Here, $\Delta E$ represents change in the energy associated with the MC move, $k_B$ is the Boltzmann constant, and $T$ denotes temperature. The energy and configuration, at a given step is recorded, and the process repeated until the convergence is achieved. In the case of MC modeling of the FCC matrix, the MC simulation cell consists of 6 {111} planes with a total of 54 atoms, corresponding to a composition of 4Al-13Co-14Cr-13Fe-10Ni (in terms of the number of atoms). A BCC cell was also simulated, consisting of 54 atoms in the form of 6 {110} layers, and corresponding to a composition of 14Al-9Co-5Cr-7Fe-19Ni, for purpose of calculating atomic bonding information for the B2 structure. As known, B2 is an ordered structure, based on a disordered BCC structure. In the MC simulation, Feng et al. needed to initially define a random BCC structure, and then employ potential interactions among the constituent atom elements. When the energy has been minimized, one has obtained the B2 structure. In other words, through the understanding of changes in the bonding environment of the initial BCC structure, Feng et al. estimated the configuration of the B2 structure observed. A plane-wave-based DFT code, the Vienna Ab initio Simulation Package (VASP), was used to calculate all energies [247, 247, 248], utilizing pseudopotentials based on a projector augmented wave method [249], exchange correlations described by Perdew, Burke, and Ernzerhof [250], and 3 × 3 × 3 Monkhorst-Pack mesh for the integration of the Brillouin zone [251]. During a relaxation stage, all degrees of freedom were allowed to relax, until forces on all atoms were smaller than 0.01 eV/Å. Furthermore, spin polarization was enabled in the MC calculations, due to magnetic nature of the constituent elements [252].

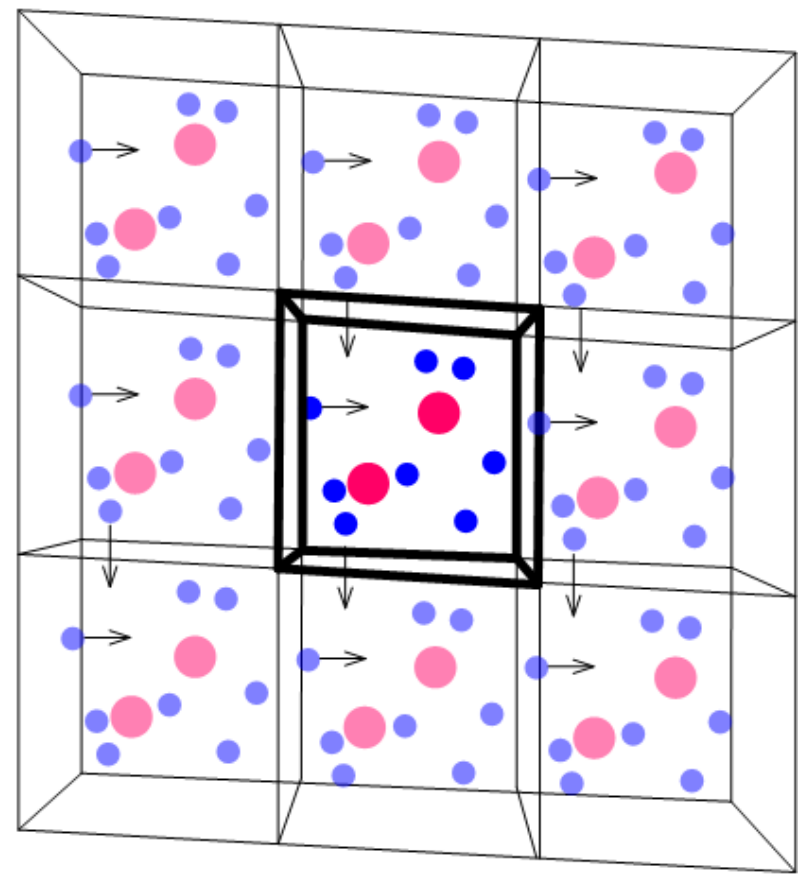

**Figure S18**: Schematic illustration of a supercell for DFT calculations with periodic-boundary conditions [225, 253,254].

*3.4.3 Density Functional Theory (An Ab-Initio Method)*

Investigations of alloy properties from the perspective of bottom-up analysis, starting with the first principles that apply to each fundamental element in the system model, i.e., ab initio (or from the beginning), involve computational modeling at atomic scale. The computational modeling relies on quantum mechanics, but not on empirical results. Such investigations can start at the level of bond energies for each electron in each atom for the system under study. Density functional theory (DFT) seeks to find the ground state for a collection of atoms by solving the Schrödinger for the electrons and nuclei in the collection. First-principles studies leverage calculations of quantum mechanics, usually with DFT approximating the many-body Schrödinger equation, to simulate the electronic properties and stability of alloy candidates, from which the most promising candidates can be identified and later confirmed experimentally [247].

In DFT modeling, energy terms of atoms (electrons) comprising a supercell, as shown in Figure Figure S18, are accounted for. Periodic-boundary conditions are usually assumed, but non-periodic boundary conditions can also be treated. One can implement a unit cell or a supercell. Either way, the cell implemented is periodically repeated in the calculations. For explanations of fundamental concepts in DFT, such as Brillouin zones or pseudopotentials, refer to [225, 253,254]. Chemical bonding and other characteristics of materials are mainly characterized by valence (outer shell) electrons. Core electrons are, in this sense, less important. Pseudopotentials replace the electron density from a chosen set of core electrons with a smoothed density. This approximation, referred to as the "frozen core approximation" greatly simplifies the overall computation process [255, 253,254]. Current DFT codes typically provide a library of pseudo-potentials for different elements.

VASP is a software package for atomic scale materials modelling, e.g., electronic structure calculations and quantum-mechanical molecular dynamics, from first principles. The VASP software is capable of preforming DFT calculations. The VASP software computes an approximate solution to the many-body Schrödinger equation, either within DFT, by solving a Kohn-Sham equations, or within a Hartree-Fock (HF) approximation, by solving Roothaan equations [29, 255, 256]. Hybrid functionals that mix the Hartree-Fock approach with the density functional theory are also implemented in VASP. In addition, Green's functions methods and many-body perturbation theory (2nd-order Møller-Plesset) are available as well [225].
NWChem is another package for computational chemistry incorporating first-principle (quantum chemical) calculations. NWChem offers capabilities related to molecular mechanics, molecular dynamics, Hartree-Fock (self-consistent field method), density functional theory, time-dependent density functional theory as well as post-Hartree-Fock methods [257].
The multi-scale degradation process, which ultimately leads to fatigue failure at the macroscale, starts with the breakage of atomic bonds at the nanoscale, as noted above. DFT can be used to calculate bond lengths and angles as well as activation barrier for bond breakage and for chemical kinetics, with relatively high accuracy. But it is worth noting that the DFT calculations can become quite computationally expensive, often taking hours to days for a single molecular structure (and for a single alloy system). More accurate results are often associated with lengthier calculations performed with higher levels of granularity. Software packages involving first-principle calculations, DFT and quantum chemistry include Gaussian, Spartan, Jaguar and Molpro [225].
In regards to other applications of ab initio DFT to HEAs, Xie et al. address hydrogen-induced slow-down of spallation in HEA under shock loading [225]. Xie et al. reported an unexpected phenomenon of hydrogen-retarded spallation in CrMnFeCoNi HEA under plate impact loading. In an effort to advance understanding of hydrogen embrittlement in chemically complex HEAs, The authors developed a trans-scale statistical damage mechanics model, based on microstructural characterization and first-principle calculations. Xie et al. attributed hydrogen-retarded nucleation of micro-voids to hydrogen-vacancy complexes with high migration energy, while formation of nano-twins with high resistance reduced the growth rate of the micro voids [166].

### 3.5 Sample Application of Macroscale, Mesoscale and Microscale Modeling: Prognostics of High-Temperature Component Reliability (Thermomechanical Fatigue or Fracture)

Since alloy material, used for high-temperature components, such as natural gas-fired turbines, air craft turbines and steam turbines, are being to pushed to the limits of their capabilities, accurate mathematical models are needed to predict the fatigue life of high-temperature components, to prevent unscheduled outages[166]. Thermomechanical fatigue (TMF) and fracture results from a combination of thermal and mechanical loads that contribute to a fracture. Figure S19 offers an illustration of an integrated, multi-scale methodology for high-temperature component reliability, one that combines the use of constitutive equations, crack formation, and crack growth models[258,258]. Figure S20 similarly presents a high-level approach to implementing an integrated, multi-scale methodology for prognostics of high-temperature component reliability.

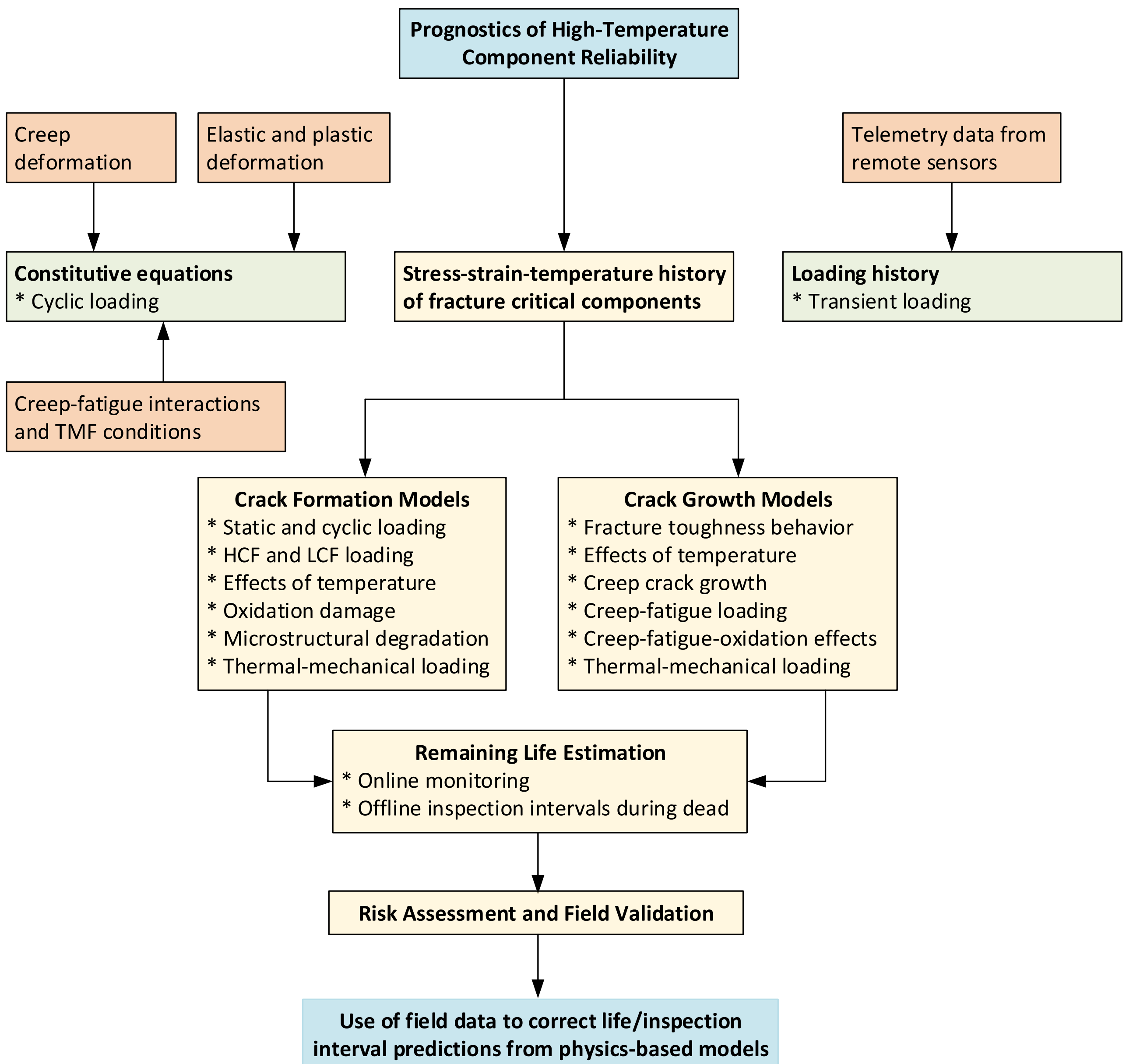

**Figure S19**: An integrated, multi-scale methodology for prognostics of high-temperature component reliability (adapted from [259]).

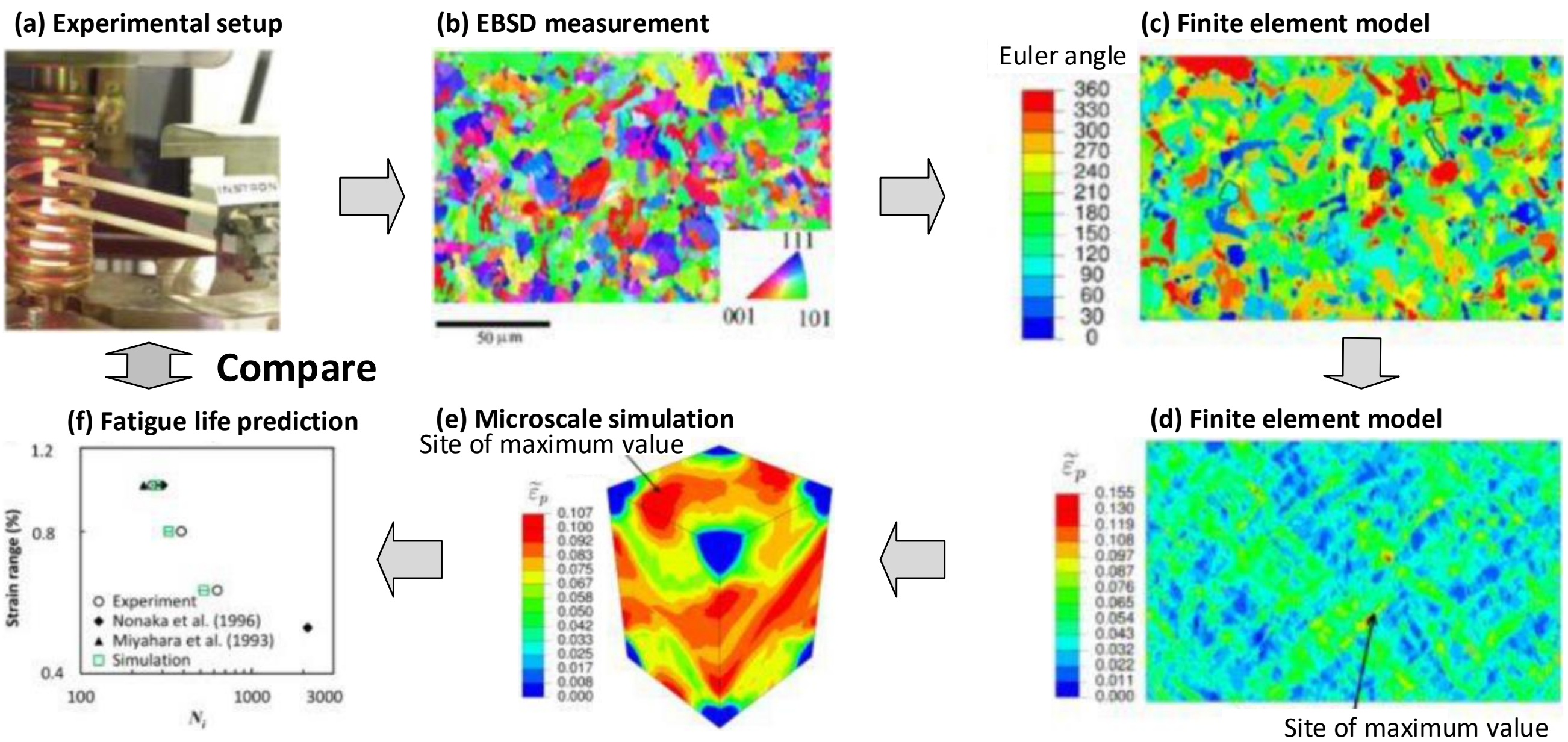

**Figure S20**: A high-level approach to implementing an integrated, multi-scale methodology for prognostics of high-temperature component reliability (adapted from [258]).

## Supplementary References